\documentclass[a4paper,11pt]{article}
\pdfoutput=1
\usepackage{jheppub}
\usepackage{xurl}
\usepackage{microtype}
\usepackage{enumitem}
\usepackage{cleveref}
\usepackage{lineno}
\usepackage{amsmath}
\usepackage[x11names]{xcolor}
\usepackage{graphicx}
\usepackage[export]{adjustbox}
  \makeatletter
\renewcommand{\@fpheader}{}
\makeatother
\graphicspath{ {./images/} }
\usepackage{xcolor}
\usepackage{hyperref}
\hypersetup{
colorlinks=true, linkcolor=blue, filecolor=magenta, urlcolor=teal}
\colorlet{darkred}{red!70!black}
\colorlet{darkgreen}{green!70!black}

\usepackage{bbold}
\usepackage{tikz}
\usetikzlibrary{decorations.markings}
\usetikzlibrary{decorations.pathmorphing}
\usetikzlibrary{decorations.markings}
\usetikzlibrary{decorations.pathmorphing}
\usetikzlibrary{patterns.meta}
\usepackage[utf8]{inputenc}
    \usepackage{tikz}
    \usetikzlibrary{patterns}
\renewcommand{\geq}{\geqslant}
\renewcommand{\leq}{\leqslant}

\preprint{TIFR/TH/25-19}
\title{\boldmath Semi-universality of CFT$_d$ entropy at large spin}

\author[a,1]{Harsh Anand}
\emailAdd{harsh.anand@tifr.res.in}
\author[b,2]{Nathan Benjamin,}
\emailAdd{nathanbe@usc.edu}
\author[a,3]{Vipul Kumar,}
\emailAdd{vipul.kumar@tifr.res.in}
\author[a,4]{Shiraz Minwalla,}
\emailAdd{minwalla.theory@tifr.res.in}
\author[a,5]{Jyotirmoy Mukherjee,}
\emailAdd{jyotirmoy.mukherjee\_119@tifr.res.in}
\author[c,6]{Sridip Pal,}
\emailAdd{sridip@ihes.fr}
\author[a,7]{Asikur Rahaman}
\emailAdd{asikur.rahaman@tifr.res.in}
\affiliation[a]{Department of Theoretical Physics,
Tata Institute of Fundamental Research, Homi Bhabha Rd, Mumbai 400005, India}
\affiliation[b]{Department of Physics and Astronomy, University of Southern California, Los Angeles, CA 90089, USA}
\affiliation[c]{Institut des Hautes Études Scientifiques (IHES), 91440 Bures-sur-Yvette, France}

\abstract{The thermal partition function,  $Z$, of a $CFT_d$ on $S^{d-1}$ is parameterized by the inverse temperature $\beta$ along with $\lfloor d/2\rfloor$ angular velocities $\omega_i$. In this paper, we investigate the behaviour of this partition function when $n$ of the $\omega_i$ are scaled to unity (the largest allowed value) at fixed values of the other $(\lfloor d/2\rfloor-n)$ angular velocities. We argue that $\ln Z$ develops a simple pole in $(1-\omega_i)$ for each $\omega_i$ that is scaled to unity. The residue of this product of poles is a theory-dependent (so non-universal) function of $\beta$ and the fixed angular velocities. The inverse Laplace transformation of this partition function constrains the functional form of the field theory entropy as a function of charges in a limit in which angular momenta and the twist are scaled as follows. While $n$ special angular momenta $J_1\ldots J_n$ are scaled to infinity, the twist and the other angular momenta -- collectively denoted $x_i$ -- are also taken to infinity but at the slower rate that ensures that the scaled charges $x_i/(J_1 J_2 \ldots J_n)^{\frac{1}{n+1}}$ are held fixed. In this limit, we demonstrate that the scaled entropy $S/(J_1 J_2 \ldots J_n)^{\frac{1}{n+1}}$ depends only on the  $\lfloor d/2\rfloor-n+1$ scaled charges defined above (the precise form of this dependence is non-universal). We verify our predictions (and compute all non-universal functions) in the case of free scalar theories (which show surprisingly rich behaviour) as well as large $N$, strongly coupled ${\cal N}=4$ Yang-Mills theory. The last theory is analyzed in the bulk via the AdS/CFT correspondence. In the scaling limit described above, its phase diagram displays sharp phase transitions between black hole, grey galaxy, and thermal gas phases.}

\begin{document}
\maketitle
\date{}
\flushbottom

\newpage
\let\svthefootnote\thefootnote
\newcommand\blfootnotetext[1]{
  \let\thefootnote\relax\footnote{#1}
  \addtocounter{footnote}{-1}
  \let\thefootnote\svthefootnote
}

\let\svfootnotetext\footnotetext
\renewcommand\footnotetext[2][]{
  \if\relax#1\relax
\ifnum\value{footnote}=0\blfootnotetext{#2}\else\svfootnotetext{#2}\fi
  \else
\ifnum\value{footnote}=0\blfootnotetext{#2}\else\svfootnotetext{#2}\fi
    \else\svfootnotetext[#1]{#2\fi
  \fi
}
}

\section{Introduction}

The $S^{d-1} \times S^1$ partition function ($Z$) of a $d$ spacetime dimensional conformal field theory completely characterizes its operator spectrum, and so is an object of considerable interest. While the partition functions of distinct CFTs differ in their details, they  share common structural features. 

A familiar universal structural feature of CFT partition functions concerns the behaviour of $\ln Z$ in the high-temperature limit. When all angular velocities $\omega_i$ (see around \eqref{thermpf} for a definition) are set to zero, it follows immediately from dimensional analysis and extensivity that $\ln Z  \propto T^{d-1}$  at high temperatures. This result has a remarkably simple generalization to the limit in which the temperature is scaled to be large at fixed but nonzero values of  $\omega_i$. It was demonstrated almost 20 years ago \cite{Bhattacharyya:2007vs} (see also \cite{Banerjee:2012iz, Jensen:2012jh, Shaghoulian:2015lcn, Benjamin:2023qsc}) that $\ln Z$ is once again constrained to take a fixed universal form \cite{Bhattacharyya:2007vs} up to an undetermined proportionality constant (see \S \ref{semiuniint} for details). The ($\beta$ and $\omega_i$ independent) proportionality constant that multiplies  this universal structure is the only dynamical information in the partition function at leading order in the high-temperature limit. The Legendre transformation of this universal partition function completely determines the form of the entropy as a function of charges \cite{Shaghoulian:2015lcn, Benjamin:2023qsc} at large energies. More precisely, this universal entropy formula holds when both energies and angular momenta are large, but the energy is scaled to infinity faster than a certain (dimension dependent, see below) power of the angular momenta.

 The universal behaviour of the previous paragraph emerges as we scale $\beta$ to its lower bound -- namely zero -- from above. Recent studies of so called grey galaxy solutions \cite{Kim:2023sig, Bajaj:2024utv, Choi:2024xnv, Choi:2025lck} have highlighted the fact that the angular velocities $\omega_i$ are also  bounded quantities; the partition function of any CFT diverges as $|\omega_i| \rightarrow 1$ from below. It is thus natural to ask whether CFT partition functions also have universal properties in the limit that one (or a few) $|\omega_i|$ are scaled to unity, at fixed values of $\beta$ and the other $\omega_j$. In this paper we argue that this is indeed the case. A central claim of this paper is that $\ln Z$, for every CFT, develops a simple pole in $(1-\omega_i^2)$ \footnote{Consequently, $\ln Z$ has a pole both in $1-\omega_i$ and $1+\omega_i$.  While we focus on the pole at $\omega_i=1$ in much of this paper, our final results (with obvious small modifications) also apply to the pole at $\omega_i=-1$.} (separately for each value of $i$) in the limit described above. The residue of each of these poles is a function of $\beta$ and the other (fixed) $\omega_j$. Unlike the situation in the $\beta \to 0$ limit, however, the functional form of this residue is non-universal and so dynamically determined. In this sense the $|\omega_i| \to 1$ limit of CFTs is only semi-universal; while the structure of the singularities of $\ln Z$ (in the limit $|\omega_i| \to 1$) is universal, the residue of these singularities is non-universal.

The Legendre transformation of our semi-universal partition function yields striking  constraints on the form of the CFT entropy as a function of the twist and angular momenta, in the large angular momentum limit that we now describe. Recall that in the previous paragraph we scaled $n$ of the $|\omega_i|$ to unity. The corresponding microcanonical limit is obtained by scaling the corresponding  $n \leq \lfloor d/2\rfloor$ angular momenta
($J_1, J_2 \ldots J_{n}$) to infinity. The various $J_i$, $i=1\ldots n$, are taken to infinity at comparable rates. On the other hand, the twist and other angular momenta are also scaled to infinity,  but at the slower rate determined by the requirement  that the scaled charges $x_i/(J_1 J_2 \ldots J_n)^{\frac{1}{n+1}}$ are held fixed in the limit (here $x_i$ is either the twist, or one of the remaining angular momenta). 

The microcanonical version of semi-universality is the following constraint on the form of the entropy as a function of charges, in the limit described in the previous paragraph: the scaled entropy $S/(J_1 J_2 \ldots J_n)^{\frac{1}{n+1}}$ is a function only of the $(\lfloor d/2\rfloor +1 -n)$ scaled charges $x_i/(J_1 J_2 \ldots J_n)^{\frac{1}{n+1}}$.

The constraint on the entropy, described above, can be worded in the following manner. The  entropy takes an effectively `extensive' form, with $(J_1 J_2 \ldots J_n)^{\frac{1}{n+1}}$, playing the role of the volume, the scaled entropy playing the role of the entropy density, and scaled charges playing the role of charge density. With these definitions, the constraint on the entropy function is the fact that the entropy density, as a function of scaled charges, is independent of volume. Note that entropy (naively a function of $\lfloor d/2\rfloor+1$ variables) is completely specified by a (non-universal) function of only $\lfloor d/2\rfloor +1-n$ 
`densities'.

The semi-universality of the asymptotic forms of the partition function and entropy functions described above are the central claims of this paper. As we do not have a fully watertight proof, it is safest to regard our central claims as conjectural.  In this paper, we do, however, offer one (rather substantial) argument in favour of -- and two somewhat nontrivial checks of -- our central claims. In the rest of this introduction we present a very brief outline of our `derivation' and two consistency checks. 

Our argument, presented in \S \ref{su} and \S \ref{gad}, proceeds as follows. We use the equilibrium partition function (or thermal effective action) studied in \cite{Banerjee:2012iz, Jensen:2012jh, Benjamin:2023qsc}, to
expand $\ln Z$ in a power series expansion in small $\beta$. The coefficient of each $\beta^n$ (in this expansion) is an independent function of the angular velocities $\omega_i$. In \S  \ref{su} and \S \ref{gad} we study these coefficients at every value of $n$, and prove that each of these coefficients develops a simple pole in $(1-|\omega_i|)$ (separately for each $\omega_i$ whose modulus is scaled to unity). Since this result holds at every order in $n$, it {\it almost} establishes our central claim, namely that $\ln Z(\beta, \omega_i)$ itself develops the same poles (and no higher singularities), with a residue that is complicated and non-universal.
However our results fall short of a complete proof of our central claim for the following two related reasons
\begin{itemize}
\item Our order-by-order argument does not rule out the possibility that terms that are non-perturbative in the small $\beta$ expansion, e.g. terms that scale as $e^{-a/\beta}$ violate the scaling proposed in this paper (note, however, the analysis of  \S \ref{intinst} which suggests that such effects are unlikely in interacting theories).
\item Our argument does not account for potential `order of limit' complications. We work order-by-order in $\beta$, take $\omega_i \to 1$  at each coefficient, and then resum the result. We cannot rule out that the result of this series of operations might differ from the result obtained simply by taking $\omega_i \to 1$ at finite $\beta$.
\end{itemize}
Despite these caveats, the all orders argument described above clearly lends substantial weight to the central conjecture of this paper.

Our order-by-order argument for the semi-universality of the partition function is actually local in nature; we first prove that the local value of the `thermal effective action' takes a semi-universal form in the limit under study, and then integrate this to obtain a semi-universal form of $\ln Z$. 
The local nature of this argument tempts us to speculate that the one point function of the stress tensor (and not just the value of the partition function $\ln Z$) is semi-universal in the limit \eqref{limnew}. See 
\ref{cust} for our precise conjecture. 

In sections \ref{free} and \ref{lgym}, we then proceed to present two independent checks of our central claims. In \S \ref{free} we recall the well known result for the $S^{d-1} \times S^1$ thermal partition function of free scalars in arbitrary dimensions, and verify that our central claims do indeed hold in these examples. We further  encounter some surprises along the way (see around \eqref{converem} for a discussion). 

As a second check, in \S \ref{lgym} we use the AdS/CFT correspondence to verify that our central result also applies to the $S^{3} \times S^1$ partition function of large $N$ ${\cal N}=4$ Yang-Mills theory at strong coupling. Once again we verify that this partition function takes 
a form consistent with our central conjecture. In this case, however, the non-universal function of temperature (or scaled twist) turns out to be rather interesting; it displays sharp phase transitions that take us from a black hole to a grey galaxy to a thermal gas phase; 
our conjecture holds in each of these phases. The analysis of \S \ref{lgym} is also independently interesting from the viewpoint of holography; it highlights a new interesting scaling limit of ${\cal N}=4$ Yang-Mills, one in which both black hole and grey galaxy solutions simplify and develop novel features.

Our analysis is related to, but qualitatively different from, the lightcone bootstrap \cite{Komargodski:2012ek,Fitzpatrick:2012yx,Pal:2022vqc,vanRees:2024xkb}. First of all, we are studying the partition function $Z(\beta, \omega_i)$, rather than a four-point correlation function. Second and more importantly, we do not need to take $\beta\rightarrow\infty$ after taking $\omega_i \rightarrow 1$ (or equivalently, we are able to take twist to infinity, not only to a finite result). As such, we describe additional regimes beyond the ``multi-particle" spectrum at large spin, finite twist\footnote{Note that multiparticling a basic set of seed fields already takes the semi-universal form described in this paper (see Appendix \ref{lcform}).}. For example, our semi-universal form accurately applies to black hole physics or grey galaxy physics in $\mathcal{N}=4$ Yang-Mills Theory -- phases that cannot, in any sense, be thought of as a collection of highly spinning, non-interacting free bulk particles. We discuss the relation of our results with lightcone bootstrap more in \S \ref{sec:LCboot} and Appendix \ref{lcform}.

While the focus of this paper has been on the study of conformal field theories, our results effectively constrain the spectrum in the deep UV of the theories. So therefore (we conjecture) our results also apply at leading order to massive theories (see \S \ref{mt} for a discussion).

The general considerations of this paper apply to every dimension $d \geq 3$. While there is a sense in which the analysis of this paper also applies to $d=2$, this dimension is distinct in several respects, and is best regarded as a special case (see \S \ref{d2int} for a discussion).

In \S \ref{intro} below, we now turn to a more detailed  presentation of the background to our work and a more complete summary of our principal results.

\section{Outline and summary of principal results}\label{intro}

In this section we review key background material, and then present a summary of the key arguments and principal results of this paper. The reader can view this section as a short(er) `letter' version of the full paper.

\subsection{Thermal partition functions on general manifolds} \label{tpgm}

Consider a QFT on a $d$ dimensional Lorentzian manifold $M$ that has a timelike Killing vector. As the QFT operator $H$ that generates translations in the Killing direction is conserved, one can formulate and study the QFT partition function $Z={\rm Tr}\ e^{-\beta H}$  on any such manifold. $Z$ may be computed by evaluating the field theory path integral on the Euclidean continuation of $M$, with Euclidean time compactified. It was pointed out in  \cite{Banerjee:2012iz,Jensen:2012jh} that such a path integral is generically massive from a $(d-1)$-dimensional viewpoint, and so admits a local expansion in derivatives. More concretely, any such partition function is given by an expression of the form 
\begin{equation}\label{ddimpf}
\ln Z= \int d^{d-1}x \sqrt{g_{d-1}}~ {\cal L}
\end{equation} 
with 
\begin{equation}\label{psexpl}
{\cal L}= \sum_{n=0}^\infty {\hat {\cal L}}_n\,, 
\end{equation}	
where ${\hat {\cal L}}_n$ is a local geometrical invariant of $n^{th}$ order in derivatives.

In this paper we focus, almost entirely, on the study of the partition function of CFTs. Since CFTs have no intrinsic dimensionful parameters, it follows immediately from dimensional considerations  that 
${\hat {\cal L}}_n=\beta^{n-d+1} {\tilde {\cal L}}_n$ (where $\beta=1/T$ is the inverse temperature, or the circumference of compactification of the Euclidean time circle, and ${\cal \tilde L}_n$ is now a $\beta$ independent local geometric expression of $n^{th}$ order in derivatives). Eqn.\!~\eqref{psexpl} can thus be rewritten as 
\begin{equation}\label{psexplbeta}
{\cal L}= \sum_{n=0}^\infty \beta^{n-d+1}{\tilde {\cal L}}_n\,. 
\end{equation}	
On inserting \eqref{psexplbeta} into \eqref{ddimpf},  we obtain a power series expansion of $\ln Z$ in the inverse temperature $\beta$. We generically expect the expansion \eqref{psexpl} to be asymptotic rather than convergent. Relatedly, the full expansion of the partition function also has `non-perturbative' corrections, i.e. corrections that go like $e^{-a/\beta}$ where $a$ is a positive number. Such corrections were discussed in \cite{Benjamin:2023qsc,Benjamin:2024kdg}, and will also appear later in this paper.

It was demonstrated in \cite{Banerjee:2012iz,Jensen:2012jh} 
that the set of local geometric coordinate and Weyl invariants of the manifold $M$ form a finite dimensional vector space at any given order in the derivative expansion. Let us denote the dimensionality of this space at $n^{th}$ order in derivatives by $N(n)$. It turns out that $N(0)=1$ (i.e. there is a unique coordinate and Weyl invariant structure at $0^{th}$ order in derivatives). It follows that  ${\tilde{\cal L}}$ (and $\ln Z$) take a universal form at leading order in the small $\beta$ limit (up to an  overall proportionality constant which is non-universal). It also turns out that $N(1)=0$, so the universal leading order form described above receives no corrections at first subleading order in $\beta$ (this assertion ignores the contribution of anomalies; see 
\cite{Banerjee:2012iz,Jensen:2012jh} for a more complete discussion). On the other hand, it turns out that $N(2)=2$,  (i.e.\!~there are two independent two-derivative local Weyl invariants built out of the geometrical data of $M$), so complete knowledge of ${\tilde{\cal L}}$ (and 
$\ln Z$) to second subleading order in $\beta$ requires the specification of two additional non-universal dimensionless real numbers. This discussion continues to higher values of $n$.  ${\tilde{\cal L}}_n$ is a particular vector in a universal $N(n)$ dimensional vector space of $n$ derivative  coordinate and Weyl geometric invariants, and complete information about ${\tilde{\cal L}}$ (and $\ln Z$) at $n^{th}$ subleading order requires the specification of $N(n)$ additional nonuniversal real numbers. See \S 1.1 of \cite{Banerjee:2012iz} for a more detailed (four page long) executive summary of this derivative expansion which dwells on features we have glossed over in our lightning review: in particular the fact that the variation of the partition function of \cite{Banerjee:2012iz, Jensen:2012jh} w.r.t. the background metric yields a complete prediction for the equilibrium stress tensor of the field theory under consideration,  the generalization of this structure to include chemical potentials in theories with additional conserved charges, and the role of anomalies.

The derivative expansion of $\ln Z$, described above, can also be obtained directly in  Lorentzian space. At high enough temperatures, the  Lorentzian dynamics of any CFT is well-approximated by the equations of perfect fluid hydrodynamics. These equations admit an equilibrium solution in which the fluid velocity is proportional to the timelike Killing vector of the manifold. The energy and entropy, and hence $\ln Z$, of this solution can be read off from the fluid stress tensor and entropy current.  Indeed, the authors of  \cite{Banerjee:2012iz,Jensen:2012jh} were motivated to develop the Euclidean formalism described earlier in this subsection, partly in order to determine a class of universal constraints on the non-dissipative transport coefficients in hydrodynamics (an exercise that was then taken forward in several subsequent papers, see e.g. \cite{Bhattacharya:2011tra,Bhattacharyya:2012nq,Bhattacharyya:2012xi,Jensen:2012kj,Bhattacharya:2012zx,Jensen:2013kka,Bhattacharyya:2013lha,Haehl:2013hoa,Bhattacharyya:2014bha,Haehl:2014zda,Haehl:2015pja,Crossley:2015tka,Haehl:2015foa,Crossley:2015evo,Glorioso:2017fpd}). 
More recently, the  paper \cite{Benjamin:2023qsc} has reviewed the formalism of \cite{Banerjee:2012iz,Jensen:2012jh}, named this approach ``thermal effective theory" and emphasized novel applications of this formalism, a development that has been taken forward in,  e.g., the recent papers \cite{Kang:2022orq,Luo:2022tqy,Benjamin:2024kdg,Allameh:2024qqp,Diatlyk:2024qpr,Kravchuk:2024qoh,Kusuki:2025pgx,Banihashemi:2024yye,Banihashemi:2025qqi}. As explained above, `thermal effective field theory', can be thought of as the specialization of  hydrodynamics to equilibrium configurations.

\subsection{Twisted thermal partition functions on $S^{d-1}\times S^1$ at high temperature}
\label{ttpf}

While the methods of \cite{Banerjee:2012iz, Jensen:2012jh} can be used to study CFT partition functions on any manifold with a timelike Killing vector, the particular case of (twisted) $S^{d-1}\times S^1$ is of particular interest, as the spectrum of states of any $CFT_{d}$ on $S^{d-1}$ is isomorphic to its spectrum of local operators.
This paper is dedicated to the study of partition functions on this special manifold. 

While we study every value of $d\geq 2$ in this paper, in the interests of clarity we focus -- in this subsection and the next -- on the special case $d=4$, postponing a discussion of generalization to \S\ref{gadi} below.

Consider a Lorentzian CFT on a spatial $S^3$. Let $t$ denote Lorentzian time, and let $\theta$, $\phi_1$ and $\phi_2$ be angles on the $S^3$. We picture the $S^3$ as embedded in $C^2$, $\phi_1$ and $\phi_2$ are the polar angles in the two orthogonal planes of $C^2$, so that the metric on $S^3$ is given by 
$ds^2= \sin^2 \theta d\phi_1^2+ \cos^2\theta d \phi_2^2 + d\theta^2$.

Let $J_1$ and $J_2$ denote the generators of angular translations in $\phi_1$ and $\phi_2$ respectively. We wish to study the partition function 
\begin{align}\label{thermpf}
Z&:= {\rm Tr}\ e^{- \beta( E-\omega_1 J_1 -\omega_2 J_2)}\\
&={\rm Tr}\ e^{- \beta \tau- \nu_1 J_1 - \nu_2 J_2} \label{thermpfnew} \\
&{\rm where} ~~~\nu_1:= \beta(1-\omega_1), ~~~~\nu_2 :=  \beta(1-\omega_2), ~~~\tau:= E-J_1-J_2 \,.\label{redchempot}
\end{align} 
Throughout this paper we will refer to the quantity $\tau$  as the twist. We call $\omega_1$ and $\omega_2$ `angular velocities', and sometimes refer to the quantities  $\nu_1$ and $\nu_2$ as reduced angular velocities.

We pause to emphasize that,  unlike the quantities $E -|J_1|$ and $E-|J_2|$, $\tau$ defined above is not guaranteed to be a positive quantity by the unitarity of representation theory alone. Primary operators in a $d=4$ CFT are labelled by the highest weights $(E, j_L, j_R) = (E, \frac{j_1+j_2}{2}, \frac{j_1-j_2}{2})$, where without loss of generality, we choose $j_1\geq j_2\geq 0$. The unitarity bound in $d=4$ asserts \cite{Mack:1975je,Minwalla:1997ka}
\begin{equation}\label{unibound} E\geq \begin{cases}
  {\rm max}(j_1, j_2) +2& \text{if} ~~j_1 \neq j_2 \\
  j+1 & \text{if} ~~j_1=j_2=j >0 \\
  0      & \text{if}~~ j_1=j_2=0,
\end{cases}
\end{equation}
a bound that (by itself) permits $E<J_1+
J_2$ (note that $J_i$ runs from $-j_i$ to $j_i$) and so $\tau<0$. Despite the fact that unitarity of conformal representation theory alone does not demand this, there is  evidence that CFTs 
actually obey stricter inequalities. In particular ref.~\!\cite{Cordova:2017dhq} has used ANEC based conformal collider bounds (see \cite{Hofman:2008ar}), to conjecture that all CFTs are subject to the stricter bounds 
\begin{equation}\label{confcol}
E\geq j_1+j_2, ~~~~{\rm if} ~{\rm min}(j_1, j_2) >2
\end{equation}
Taken together, \eqref{unibound} and \eqref{confcol} guarantee that all states have $\tau\geq 0$, except those in representations $(E, j_1=\frac{3}{2}, j_2=\frac{3}{2})$ with $\frac{5}{2}\leq E < 3$, and those in representations $(E, j_1=2, j_2=2)$ with 
$3\leq E <4$. Thus even though $\tau$ has not (yet) been shown to be a strictly non-negative quantity, it has been shown to be bounded from below (by $-1$). Consequently, the partition function \eqref{thermpf} converges when $\beta$, $\nu_1$ and $\nu_2$ are all positive (more generally, i.e. when $\beta$ is positive and $|\omega_i|<1$ for $i=1,2$).

As explained in the previous subsection, $\ln Z$ in \eqref{thermpf} may be analyzed in two ways; it was demonstrated, in 
\cite{Banerjee:2012iz, Jensen:2012jh} that these two methods are completely equivalent. One could either 
analyze equilibrium solutions of hydrodynamics, with fluid velocity proportional to 
$\partial_t +\omega_1 \partial_{\phi_1} +\omega_2 \partial_{\phi_2}$ or one could  continue to Euclidean space and study 
$\ln Z$ on the appropriately twisted $S^3 \times S^1$, using the methods of 
\cite{Banerjee:2012iz, Jensen:2012jh}.

In this subsection we focus on the high-temperature limit 
\begin{equation}\label{limnorm}
\beta \to 0, ~~~\omega_i~~~{\rm fixed}\,.
\end{equation}
For reasons explained in the previous subsection, $\ln Z$ takes a universal form in this limit (up to one overall dynamically determined numerical constant which we call $c'$ below). Almost 20 years ago, the authors of \cite{Bhattacharyya:2007vs} \footnote{The paper \cite{Bhattacharyya:2007vs} (together with \cite{Aharony:2005bm, Lahiri:2007ae}) can be viewed as a precursor to the \textit{Fluid-Gravity} correspondence.  In particular the ref.~\!\cite{Bhattacharyya:2007vs} used the universal hydrodynamical prediction, derived in that paper, to provide an explanation of the properties of black holes in $AdS$.} used the fluid method to demonstrate that, to leading order in the limit \eqref{limnorm},  
\begin{equation} \label{uniform}
\ln Z(\beta, \omega_1,  \omega_2)= c' \frac{ T^3}{(1-\omega_1^2)(1-\omega_2^2)}\left[1+O(T^{-2})\right], ~~~~~~T:=\frac{1}{\beta}.
\end{equation}

Eqn.~\!\eqref{uniform} captures the trace over the spectrum of states of the CFT on the spatial slice $S^3$  (or over the space of local operators of the theory) in the canonical ensemble. The inverse Laplace transform of eq.~\!\eqref{uniform} yields a (universal) microcanonical formula for the entropy as a function of energy $E$ and angular momenta $J$. It was explained in \cite{Benjamin:2023qsc} (see also \cite{Shaghoulian:2015lcn}), that this Legendre transform accurately captures the entropy of the field theory in the limit 
\begin{equation}\label{limhold}
E \to \infty, ~~~\tau \to \infty,~~~~\tau \gg 
\left(c' J_1J_2\right)^{1/3}.
\end{equation}

Consider setting $\omega_1=\omega_2=\omega$, and slowly increasing $|\omega|$ starting from zero. \eqref{uniform} (plus the implicit claim that all subleading corrections to this formula are smooth -- i.e. non-divergent -- functions of the $\omega_i$ for $|\omega_i|<1$) tells us that the partition function $Z=\mathrm{Tr}\ e^{-\beta( H-\omega(J_1+J_2))}$ remains smooth (nonsingular) until we hit $|\omega|=1$. It follows that the spectrum of $E-\omega(J_1+J_2)$ is bounded from below for all $\omega <1$. This result is already much stronger than the unitarity bound\footnote{If, for instance, our theory were to host (unitarity allowed) states with $(E=j+1, j_1=j, j_2=j)$ at arbitrarily large values of $j$ then  $E-\omega(j_1+j_2)$ would be unbounded from below for any $\omega>1/2$.}, but perfectly in tune with -- and in fact implied by -- the results of \cite{Cordova:2017dhq}. There is some sense, in other words, in which fluid dynamics already knows about (and almost implies) the  `stronger than unitarity' ANEC bounds of \cite{Cordova:2017dhq}. In fact, the ANEC bound proved in \cite{Cordova:2017dhq} is tightly related to an elementary hydrodynamical fact, namely that a relativistic fluid on a sphere is allowed to whirl around at speeds that extend all the way to the speed of light, but no faster. It would be interesting to investigate if the same bound can also be obtained starting from other elementary
physical principles, like  causality (see e.g. \cite{Hartman:2015lfa}), the chaos bound \cite{Maldacena:2015waa} or (relatedly) the CRG conjecture \cite{Chowdhury:2019kaq, Chandorkar:2021viw}.

We have explained above that the fluid results \eqref{uniform} cannot correctly capture the high-temperature limit of $\ln Z$ unless $E-\omega(J_1+J_2)$ is bounded from below for all $\omega <1$, because, 
in this situation, $\ln Z$ diverges at a value of $\omega_1=\omega_2=\omega<1$. While this point is true, it can be masked (and so hard to see) in the high-temperature (small $\beta$) 
expansion described in this section, but is less masked -- so easier to 
see -- in the small $\nu_i$ expansion that we turn to in the next subsection, and most of the rest of this paper. We discuss this point further in the discussion section  \S \ref{disc} (see the fourth and fifth paragraphs of that section) as well as in Appendix \ref{exotic}, in the context of the example of exotic free theories studied in 
\cite{Loganayagam:2012zg}.

\subsection{Twisted thermal partition functions on $S^3\times S^1$ as $\omega_i\to 1$}\label{semiuniint}

Still focussing on $d=4$, in this subsection we present principal assertions of this paper in the context of a symmetric large $J$ limit. In this subsection we simply state our main claims, postponing a discussion of our arguments for these claims to the next subsubsection.

As in the previous subsection, we are interested in the thermal partition function, \eqref{thermpf}, of a four-dimensional CFT. However, we now investigate this partition function in the limit 
\begin{equation}\label{limnew}
\nu_i \to 0, ~~~\left(\beta,\  \nu_1/\nu_2\right)~~{\rm fixed}.
\end{equation}
In this paper, we argue that the partition function of every CFT, in the limit \eqref{limnew}, takes the form 
\begin{equation} \label{sn}
\ln Z=\frac{\beta^2 h(\beta)}{\nu_1\nu_2}\left(1+O(
\nu_1)+O(\nu_2)\right)=\frac{4h(\beta)}{(1-\omega_1^2)(1-\omega_2^2)}\left(1+O(
1-\omega_1)+O(1-\omega_2)\right),
\end{equation}
where $h(\beta)$ is a non-universal (theory-dependent) function of temperature.  

We claim that  eq.~\!\eqref{sn} applies at all values of the temperature. Further note that the function $h(\beta)$ admits a small $\beta$ expansion of the form 
\begin{equation}\label{texp_orig}
h(\beta)= \frac{1}{\beta^3}\left( c + c_1 \beta^2 +\ldots \right)\,,
\end{equation}
where the coefficient $c$ in this expansion is determined by the leading term in the derivative expansion of \cite{Banerjee:2012iz,Jensen:2012jh}; the coefficient $c_1$
is determined by the first subleading term, and
so on.  Hence, in the high-temperature limit 
(where eqs.~\!\eqref{sn} and \eqref{uniform} both apply) the function $h(\beta)$ reduces to $h(\beta)\sim \frac{c'}{4\beta^3}$ in agreement with eq.~\!\eqref{uniform}.

Note that \eqref{sn} only determines $\ln Z$ in terms of an unknown (theory-dependent) function of $\beta$. It follows that \eqref{sn} does not assert that the partition function $\ln Z$ takes a completely universal form in the limit \eqref{limnew}.
Nonetheless, \eqref{sn} contains a great deal of information, as it determines $\ln Z$ (a function of $3$ variables) in terms of an unknown function of only a single variable. Keeping both these points in mind, we refer to eq.~\!\eqref{sn} as a \textit{`semi-universal form'} of the $S^3 \times S^1$ partition function in the small $\nu_i$ limit \eqref{limnew}. In contrast,  eq.~\!\eqref{uniform} is a universal prediction for the form of the partition function in the high-temperature limit, as the functional dependence of \eqref{uniform} on $\beta$ and 
$\nu_i$ is completely determined up to a single theory-dependent dimensionless constant.

The physical implications of the assertion \eqref{sn} are clearest in the microcanonical ensemble.
 In \S \ref{subsec:semiuniversal}, we demonstrate that the  inverse Laplace transformation (effectively Legendre 
transformation) of \eqref{sn} yields an entropy $S(E, J_1, J_2)$ as a function of charges. In particular, $S(E, J_1, J_2)$ takes the following `extensive' form 
\begin{equation}\label{finansforentint}
S(\tau, J_1, J_2)= (J_1 J_2)^{\frac{1}{3}}S^{\rm int}\left(\frac{\tau}{(J_1 J_2)^{\frac{1}{3}}}\right) \left(1+o(1)\right)\,.
\end{equation}
 in the limit 
\begin{equation}\label{lj}
J_i \to \infty, ~~~\tau \rightarrow \infty, ~~~{\rm with}~~~\left(\frac{J_1}{J_2}~,~~\frac{\tau}{(J_1 J_2)^{\frac{1}{3}}}\right)~~~{\rm fixed.}
\end{equation}

Eqn.~\!\eqref{finansforentint}, like \eqref{sn}, applies to every CFT$_4$ in the limit \eqref{lj} and may be viewed as the key assertion of this paper (as applied to 4d CFTs in the limit \eqref{lj}, see the next subsection for generalizations) from the microcanonical point of view.  $S^{\rm int}$ in \eqref{finansforentint} is a sort of Legendre transform of the function $h(\beta)$ 
(see eq.~\!\eqref{intent} for the precise relationship). Like $h(\beta)$, $S^{\rm int}$ is a theory-dependent function of its argument. As in our discussion of the partition function, \eqref{finansforentint} is a significant constraint on the form of the entropy as a function of the three variables  $\tau$, $J_1$, $J_2$, as it is determined in terms of a single non-universal function of a single variable. Once again we see semi-universality.

\subsection{Resummation of the small $\beta$ expansion} \label{sbe}

In this subsection we summarize our principal argument for our  key assertion \eqref{sn}. Our argument uses the small $\beta$ expansion \eqref{psexplbeta} of $\ln Z$. 

Working in the special case of a twisted $S^{3} \times S^1$ compactification, we expand the partition function in the form \eqref{ddimpf}. On plugging in the explicit formula for 
$g_3$ into \eqref{ddimpf}\footnote{The metric $g_{3}$ that appears in \eqref{ddimpf} is the base metric $g_{ij}$ that appears in 
\eqref{eq:lorstationary_metric}. In the special case of the twisted $S^3\times S^1$ compactification, this metric is listed in  \eqref{eq:fields}, for which $\sqrt{g_3}= \gamma(\theta) \sqrt{\Omega_3}$, see \eqref{eq:volumeform}.}, we find that this equation can be rewritten as 
\begin{equation}\label{newexp}
\ln Z = \int d\Omega_3\  \gamma(\theta) 
\sum_{n=0}^\infty \beta^{n-3} {\tilde{\cal L}}_n
\end{equation} 
where $d\Omega_3$ is the usual round measure on the three-sphere and $\gamma(\theta)$ is the (position-dependent) special relativistic time dilation factor for fluid motion on $S^3$ (see \eqref{gammadef}). 
We can simplify \eqref{newexp} by defining 
\begin{equation}\label{lnnn}
{\cal L}_n
= \gamma(\theta) {\tilde {\cal L}}_n
\end{equation}
so that the formula for $\ln Z$ now takes the form 
\begin{equation}\label{newexpn}
\ln Z = \int d\Omega_3 
\sum_{n=0}^\infty \beta^{n-3} {\cal L}_n
\end{equation} 
In \S \ref{su} below we carefully study the behaviour of ${\cal L}_n$, for each $n$ (see \eqref{psexplbeta}) in the limit \eqref{limnew}. We demonstrate that 
\begin{equation}\label{lnlim}
{\cal L}_n = a_n \gamma^4(\theta) + ... 
\end{equation}
where a formula for $\gamma(\theta)$ is presented in \eqref{gammadef}), $a_n$ are non-universal numbers of order unity, and the $\ldots$ represent contributions to ${\cal L}_n$, that are subleading (in the limit \eqref{limnew}) as compared to the term retained. 

Eqn \eqref{lnlim} should be understood as follows. Abstractly, the  quantities ${\tilde {\mathcal L}}_n$ that appear in \eqref{psexplbeta} are 
functionals of data that appear in the general metric \eqref{eq:stationary_metric}. When evaluated on the twisted $S^3 \times S^1$
metric of interest to this subsection, however, each of the ${\tilde {\mathcal L}}_n$ simply reduces to a function on the base $S^3$. \eqref{lnlim} asserts that 
these functions (times $\gamma(\theta)$, see \eqref{lnnn})
are all proportional to $\gamma^4(\theta)$ (at leading order in the limit \eqref{limnew}).  

Since we have shown that each of the ${\cal L}_n$ -- for each of the infinite possible values of $n$ -- takes the same functional form (in the coordinates on the $S^3$, but also in the variables $\omega_1$ and $\omega_2) $, we have almost proved that ${\cal L}$ (the sum of the ${\cal L}_n$) also takes the same form, i.e. that 
\begin{equation}\label{lnu}
{\cal L} = \left(\frac{4}{2 \pi^2}\right) \gamma^4(\theta) h(\beta) + \ldots, ~~~~
h(\beta)= \left( \frac{\pi^2}{2} \right) \sum_{n=0}^\infty 
\beta^{n-3} a_n \,.
\end{equation}
The normalization factor $\left(\frac{4}{2 \pi^2}\right)$ has been introduced for later convenience.
Integrating \eqref{lnu} over the $S^3$ yields our central result \eqref{sn}. 

While the arguments presented above come near a `physicist's proof' of \eqref{sn}, they are incomplete from a mathematical standpoint. The manipulations that take us from \eqref{lnlim} to \eqref{lnu} involve a potentially tricky exchange of the limit \eqref{limnew} and the sum over $n$ \footnote{At a formal level, this potentially tricky interchange is performed writing the second equality below
\begin{equation}\label{psexplnn}
\begin{split}
{\rm lim}_{\nu_i \to 0} {\cal L}
&={\rm lim}_{\nu_i \to 0} \left( \sum_{n=0}^\infty 
\beta^{n-3} {\cal L}_n \right) = \sum_{n=0}^\infty 
\beta^{n-3} \left( {\rm lim}_{\nu_i \to 0}{\cal L}_n \right) =\gamma^4(\theta) \sum_{n=0}^\infty 
\beta^{n-3} a_n \beta^{n-3}= \gamma^4(\theta) h(\beta)
\end{split}
\end{equation}}.
Moreover, the rigorously minded readers may find themselves questioning the status of the infinite sum on the RHS of \eqref{psexplbeta}, the starting point of our analysis\footnote{They might wonder whether the summation on the RHS of \eqref{psexplbeta} is convergent or asymptotic, and (if the latter) whether it receives non-perturbative corrections (we will have more to say on this point later in this paper). They may also find themselves asking similar questions about the summation over $n$ in the equation \eqref{lnu} that defines $h(\beta)$.}.
For these reasons it is safest to view  \eqref{lnu} as a conjecture rather than a clearly established result, albeit a conjecture for which we believe we have supplied a rather substantial argument, at least by physicists' standards.

\subsection{Generalization to arbitrary $d$ and asymmetric scaling limits}\label{gadi}

The generalization of eqs.~\!\eqref{uniform} and \eqref{finansforentint} to higher dimensions (and to more asymmetric scaling limits) is rather straightforward. In \S \ref{gad} below, we work in $d=2n$ or $2n+1$ dimensions and study the limit in which $p$ out of its $n$ rotational angular velocities are scaled to unity while the other angular velocities, and the temperature, are held fixed; i.e.
\begin{equation}\label{limnewdfour1}
\begin{aligned}
\nu_i \to 0,& ~~~(i=1\ldots p)\\
&{\rm while}~\left(\beta,~\nu_j ~{\rm for}~(j=p+1 \ldots n)~,~~\left(\frac{\nu_i}{\nu_k}\right)~{\rm for} ~(i,k=1\ldots p)\right)~~{\rm fixed}.
\end{aligned}
\end{equation}
In this limit, we argue in \S \ref{gad} that  $\ln Z$ (for any CFT$_d$) takes the form 
\begin{equation}\label{zint}
    \ln Z=\frac{\beta^p h(\beta,\omega_{p+1},\ldots,\omega_{n})}{\nu_1\ldots\nu_p}\,,
\end{equation}
where $h(\beta,\omega_{p+1},\ldots,\omega_{n})$ is a non-universal, theory-dependent function of its $n-p+1$ arguments. As \eqref{zint} determines $\ln Z$ (a function of $n+1$ variables) in terms of an arbitrary function $h$ of 
$n-p+1$ variables, this equation, once again, has a semi-universal character.

Upon taking the inverse Laplace transform of eq.~\!\eqref{zint}, we find that in the limit 
\begin{equation}\label{limint}
\begin{aligned}
&\mathrm{with}\quad i,j=1,\ldots p\,,\quad \quad k=p+1,\ldots n\,,\\
    &\tau \rightarrow \infty, ~~J_i \rightarrow \infty, ~~ \left[\left(\frac{J_i}{J_j}\right)~,~\frac{\tau}{(J_1\ldots J_n)^{\frac{1}{n+1}}}~,~\frac{J_{k}}{(J_1\ldots J_n)^{\frac{1}{n+1}}}\right]~~\mathrm{fixed}\,,
    \end{aligned}
\end{equation}
the entropy of every $CFT_d$ takes the scaling form
\begin{equation}\label{genformint}
    S\left(\tau, J_1\ldots J_{[\frac{d}{2}]}\right)=(J_1\ldots J_p)^{\frac{1}{p+1}}S^{\mathrm{int}}\left( \frac{\tau}{(J_1\ldots J_p)^{\frac{1}{p+1}}}, \frac{J_{p+1}}{(J_1\ldots J_p)^{\frac{1}{p+1}}},\ldots , \frac{J_{[\frac{d}{2}]}}{(J_1\ldots J_p)^{\frac{1}{p+1}}}\right) 
\end{equation}
where the $n-p+1$ variable function $S^{\rm int}$ is, once again, a sort of Legendre transform of the function $h$ (see \eqref{lgen}).

\subsection{New elements at arbitrary $d$}

 Although the final result for $\ln Z$, \eqref{zint}, presented in the previous subsection is a straightforward generalization of its $d=4$ counterpart \eqref{sn}, the argument that leads to this result has some new elements that we now pause to review. 

As we have explained in \S \ref{sbe}, the key element in our argument for the partition function \eqref{sn} was the result that ${\cal L}_n \propto \gamma^4(\theta)$ at leading order in the limit 
\eqref{limnew}. The naive generalization is the claim that ${\cal L}_n \propto \gamma^d$ in the limit \eqref{limnewdfour1}. While this guess does indeed turn out to be correct when $d=2n$ and $p=n$, it turns out to be incorrect in every other case (i.e. whenever $d$ is odd, or in even dimensions but with $p<n)$. 

The qualitative difference between generic cases and the special case  $d=2n$ and $p=n$ can be seen by studying the function $\gamma^d(\theta)$. When $d=2n$ and $p=n$, it is easy to verify that the function $\gamma^d$ on $S^{d-1}$ is smooth in the limit \eqref{limnewdfour1} (this was the case for $d=4$ in the limit \eqref{sn}). However, in all other cases, the function $\gamma^d(\theta)$ is not smooth in the limit \eqref{limnewdfour1}, but develops a highly peaked structure  (with angular localization scales of order $\sqrt{\nu_i}, ~i=1\ldots p$) around a particular submanifold, namely a $S^{2p-1}$ in $S^{d-1}$. In this situation, the large values of derivatives in the directions normal to the $S^{2p-1}$ provide a new source of singular scaling in the limit \eqref{limnewdfour1}. As a consequence,  expressions like  $\gamma' \gamma^{d-2}$ (which were subleading compared to $\gamma^{d}$ when $d=2n$ and $p=n$) become comparable, and so contribute to ${\cal L}_n$ at leading order in the limit \eqref{limnewdfour1} in the general case.
An infinite number of such leading order terms exist, and the precise functional form of ${\cal L}_n$ in the directions normal to the $S^{2p-1}$ is therefore non-universal \footnote{A similar phenomenon was observed in the study of grey galaxies. The boundary stress tensor of the gas component of an $AdS_4$ grey galaxy was computed in \cite{Kim:2023sig}. It was pointed out there that the stress tensor is highly localized around an equator; the explicit form of this localization function was presented in equation 5.39 of that paper. The generalization of this localization phenomenon to grey galaxies in higher dimension was 
discussed in \cite{Bajaj:2024utv}.}.
On the other hand, derivatives of $\gamma$ in directions that are tangent to the $S^{2p-1}$ continue to be suppressed, so all leading order terms in ${\cal L}_n$ have a universal dependence on the angular coordinates of $S^{2p-1}$. \footnote{Once again, it was previously noted that the stress tensor distribution in these tangent directions should be universal (and governed by hydrodynamics) in the case of grey galaxies; see e.g. the third paragraph in the discussion section of \cite{Bajaj:2024utv}.}

Though the angular profile of  ${\cal L}_n$ is non-universal in the directions normal to $S^{2p-1}$, all details of this non-universal profile are concentrated around an angular width of angle $\sqrt{\nu_i}$ in the $S^{2p-1}$, a width that goes to zero in the limit \eqref{limnewdfour1}. Focussing on angular scales larger than $\sqrt{\nu_i}$, therefore, we can simply replace this distribution by a delta function in the normal directions. The coefficient of this delta function turns out to be universally proportional to $\gamma^{2p}$ (where $\gamma$ is now constructed only out of the velocity tangent to the $S^{2p-1}$, and is evaluated only as function of the coordinates on $S^{2p-1}$), times a function of $\beta$ and angular velocities that are not scaled to unity. Note that 
$\gamma^{2p}$ is a smoothly varying function on $S^{2p-1}$. 

In this coarse-grained sense, therefore, the Lagrangian ${\cal L}$ once again takes a universal form;  ${\cal L}$ equals $\gamma^{2p}$ times a constant independent of $\nu_i$($i=1 \ldots p$). Integrating this expression over $S^{2p-1}$ yields the expression \eqref{zint}.

In all dimensions, we can also analyze scenarios where $p<n$ in the same way as above. We do this analysis explicitly in Sec. \ref{adgen}.

\subsection{Special features in $d=2$}\label{d2int}

\subsubsection{The high-temperature limit}

While the analysis presented above applies to every $d\geq 2$, the case $d=2$ is special. This is already true (in a limited way) in the high-temperature limit reviewed in \S \ref{ttpf}. The leading order generalization of \eqref{uniform} to $d=2$ is 
\begin{equation}\label{hgy}
\ln Z(\beta, \omega)= c'\frac{T}{1-\omega^2}
\end{equation}
and applies in the limit $T \to \infty$ at fixed $\omega.$ Now the result \eqref{hgy} is familiar from another point of view; \eqref{hgy} is simply the famous Cardy formula, which also tells us that the 
quantity $c'$ is related to the central charge of the 2d CFT via $c'= \frac{\pi^2 c}{3}$. The central charge $c$ parametrizes a Weyl anomaly, and so is often effectively computable in a simple manner. 

We see from this discussion that the formula \eqref{uniform} (seen in \cite{Bhattacharyya:2007vs}) should be viewed as a higher dimensional generalization of the Cardy formula (this point was emphasized in \cite{Benjamin:2023qsc, Shaghoulian:2015lcn}). However there is one difference; when $d>2$ there is no known way of relating the constant $c'$ to an anomaly, so the computation of $c'$ is a hard dynamical problem in $d>2$ (as illustrated by the famous factor of $\frac{3}{4}$ between the weak and strong coupling partition function in ${\cal N}=4$ Yang-Mills theory \cite{Gubser:1996de, Gubser:1998nz}).

\subsubsection{The small $\nu$ limit}

The differences between $d=2$ and all higher values of $d$ are even more pronounced in the limit 
\begin{equation}\label{lim2d}
\nu \to 0, ~~~\beta ~~~{\rm fixed}\,,
\end{equation}
studied in this paper. Setting $d=2$ in ~\!\eqref{zint} yields 
\begin{equation}\label{dtwo}
\ln Z= \frac{\beta h(\beta)}{\nu}\,.
\end{equation}
In $d=2$ however, one can do more; one can actually use the methods of 
\cite{Banerjee:2012iz, Jensen:2012jh} to completely determine the functional form of $h(\beta)$, in a manner we now describe.

In every dimension, the local density ${\cal L}$ in \eqref{ddimpf} is a coordinate, gauge and Weyl invariant local functional of the relevant local quantities, namely the local size of the thermal circle (which we denote by $\beta e^{\sigma}$, see \eqref{eq:stationary_metric}) and curvatures of the $d-1$ base manifold and the $d-1$ dimensional Kaluza-Klein gauge field (see \eqref{eq:stationary_metric}). In $d=2$, however,  $d-1=1$, and all curvatures vanish identically. Consequently, ${\cal L}$ can depend only on $\sigma$ and its derivatives, and it is easy to see that the only Weyl invariant local functional of this form is, in fact, $ {\cal L} \propto  T e^{-\sigma} $. \footnote{The argument goes as follows. One can use Weyl transformations to `gauge $\sigma$ away', i.e. to simply set $\sigma=0$. In this Weyl gauge the only local coordinate and gauge invariant is a constant, leading to a local contribution to $\ln Z \propto \int \sqrt{g_1}$. Moving back to a general Weyl frame turns this into $\ln Z \propto \int e^{- \sigma} \sqrt{g_1}$.}
An ${\cal L}$ of this form gives rise to  $h(\beta)$ proportional to $\beta^{-1}$. The proportionality constant can be read off by matching with \eqref{hgy} in the high-temperature limit, and one obtains 
\begin{equation}\label{eq:univ2d}
h(\beta)=\frac{2\pi}{\beta}\,\frac{2\pi c}{24}\, ,
\end{equation}
so that, at leading order in the small $\nu$ expansion 
\begin{equation}\label{lnztd}
\ln Z= \frac{\pi^2 c}{6 \nu}\,.
\end{equation}
The result \eqref{lnztd} was, in fact,  first obtained by different methods (using modular invariance) in  \cite{Kusuki:2018wpa,Benjamin:2019stq,Ghosh:2019rcj,Pal:2023cgk,Pal:2025yvz}. 

\subsubsection{The microcanonical ensemble}\label{subsec:MC2d}

Note that $\ln Z$ in \eqref{lnztd} is independent of $\beta$. As a consequence, the Legendre transform of \eqref{lnztd} yields
$$\tau= \partial_{\beta} \left( \ln Z \right) = 0 $$ and 
\begin{equation}\label{enttd}
S=2\pi \sqrt{ \frac{c J}{6} }
\end{equation}
\eqref{enttd} was first derived in Appendix B of \cite{Kusuki:2018wpa} using modular invariance and the modular kernel, and explored further in \cite{Benjamin:2019stq}, and later established as a theorem in \cite{Pal:2025yvz} for unitary CFTs with $c>1$.\footnote{While these papers studied the entropy associated with Virasoro primaries for $c>1$ CFTs with only Virasoro symmetry, one can argue that the leading result holds for density of all states i.e. entropy associated with all states irrespective of whether we have only Virasoro symmetry or presence of a bigger chiral algebra, for example, see \S\ref{sec:MI}.}

The fact that $\tau$ vanishes at leading order is a surprise from the viewpoint of our general $d$ formalism.  Extrapolation from general dimensions would have led us to expect $\tau = g(\beta) \sqrt{J}$ for an appropriate function $g(\beta)$. However \eqref{lnztd} tells us that the function $g(\beta)$ actually vanishes identically.

Of course $\tau$ only vanishes at leading order in $\sqrt{J}$; $\tau$ is non-vanishing at first subleading order -- ${\cal O}(1)$ in $J$. In \S \ref{mme} we explain that at this order, and under some mild assumptions, the dependence of the twist on $\beta$, together with the first correction to \eqref{enttd}, is universal. (This correction occurs at ${\cal O}(J^0)$ and is a function of $\tau$.)

\subsection{Semi-universality in free massless scalar field theory}\label{fmsint}

The prediction of semi-universality for all CFTs (namely the prediction 
\eqref{zint} for the thermal partition function at small $\nu_i$, and \eqref{genformint} for the entropy in the large $J$ limit \eqref{lj}), is the main result of this paper. In this subsection and the next we present two checks of this semi-universal prediction, both of which also teach us qualitative lessons. 

In this subsection, we consider the theory of a single free scalar field in arbitrary dimensions. The thermal partition function of the free theory on $S^{d-1}$ is, of course, very well known. In  \S \ref{free}, we specialize this exact result to the limit \eqref{limnew}, and demonstrate that it does, indeed, simplify to the form predicted in \eqref{zint} even at finite values of $\beta$. This fact provides some reassurance that the subtle order of limit and convergence issues flagged under \eqref{lnu} are benign -- at least in this particular example. Upon scaling all $\nu_i$ to zero,  it is easy to read off the function $h(\beta)$ (see \eqref{evenh} and \eqref{oddh}). While our exact formulae for $h(\beta)$ apply even at finite values of $\beta$, they can also be expanded around small $\beta$. In \S \ref{free} we present the relevant expansions in $d=3, 4, 5, 6$ and a general one valid for any even dimension in appendix \ref{app:freeScalarComp}. It turns out that this expansion takes a qualitatively different form at odd and even values of $d$. 

At odd values of $d$ (e.g. $d=3$ and $d=5$), $h(\beta)$
admits an expansion of the form \eqref{lnu}, with 
coefficients $a_n$ that are (generically) nonzero at every value of $n$, and whose modulus grows with 
$n$ in a manner proportional to $n!$. It follows immediately that the small $\beta$ expansion, in these dimensions, is asymptotic rather than convergent, and so is only really well defined after a Borel resummation. The Borel transform of $h(\beta)$ also displays instanton-like singularities. In short, this small $\beta$ expansion displays all the usual properties of a generic perturbative expansion.

The small $\beta$ expansion of $h(\beta)$ is much more surprising when $d$ is even.
As an illustration, consider the case $d=4$. In this case it turns out that the formula for $h(\beta)$ can be (exactly) recast as 
\begin{align}\label{converem}
    h^{d=4}(\beta)= \frac{\pi^4}{180\beta^3} - \frac{\pi^2}{72\beta} + \frac{3\zeta(3)}{8\pi^2} - \frac{7\beta}{1440}+ \sum_{n=1}^\infty \sum_{d|n} \frac{(-1)^{d+1} }{\pi^2 d^3} e^{-\frac{2\pi^2 n}{\beta}}.
\end{align}
The three things that stand out about \eqref{converem} are 
\begin{itemize}
\item The `perturbative' part of \eqref{converem} (i.e. the part that is a power series in $\beta$) is finite (it truncates at order $\beta$). 
\item The exact answer, nevertheless, has an infinite set of corrections that are  `non-perturbative' in the small $\beta$ limit. 
\item There is a constant piece ($\beta^0$) that does not fit in the framework of thermal effective field theory, which only generates powers of the $\beta^{-3+2n}$ for integer $n$. (This last point is explained by the fact that the free boson dimensional reduces to another CFT as opposed to a massive theory, namely to a free, massless theory in one fewer dimension).
\end{itemize}
Instanton-like non-perturbative corrections to power series expansions are often intimately related to lack of convergence of the corresponding series (signaling that the series in question is asymptotic rather than convergent, and sometimes determining the radius 
of convergence of the Borel resum of the perturbation series. In fact, as we have reviewed immediately above, this was the situation for the small $\beta$ expansion of $h(\beta)$ in odd dimensions). This is clearly not the case for \eqref{converem}, as
the perturbation series is finite. The `non-perturbative' corrections 
in \eqref{converem} are completely disconnected from the perturbative expansion; they explore genuinely new physics, and are therefore 
non-constrained by the arguments presented in \S \ref{sbe}. It is thus particularly gratifying that these non-perturbative terms nonetheless continue to respect the scaling \eqref{zint}, as predicted by our general conjectures. 

In \S \ref{free} we provide a physical interpretation of the non-perturbative corrections to $h(\beta)$, identifying them with instantons in the worldline representation of the Euclidean path integral (along the lines of the analysis presented in \cite{Benjamin:2023qsc,Benjamin:2024kdg}). The relevant instantons turn out to be very finely tuned (they exist in the free theory, but -- we conjecture -- are generically non-perturbatively suppressed in $\nu^{-1}$ once interactions are turned on). This observation is, presumably, tightly connected with the fact that the  truncation of the perturbative series in \eqref{converem} is almost certainly an artifact of the free nature of the theory. As we explain in \S \ref{free} we expect the small $\beta$ expansion to always continue to infinite order in interacting theories, and that any non-perturbative corrections to $h(\beta)$ in such theories are always intimately connected to the large order power series expansion of $h(\beta)$. We leave a careful investigation of these expectations to future work. 

\subsection{Semi-universality in large $N$ ${\cal N}=4$ Yang-Mills at strong coupling}

We also obtain a second -- and much more nontrivial -- check of our predictions (with one caveat; see below) by studying ${\cal N}=4$ Yang-Mills theory at large
$N$ and strong coupling. Of course, the strongly coupled nature of this theory makes it difficult to analyze directly as a field theory. However, the AdS/CFT 
correspondence allows us to use the bulk dual gravitational description 
to compute the thermodynamics of this theory. We find that the entropy function of ${\cal N}=4$ Yang-Mills theory does, indeed, agree with the form predicted in \eqref{finansforentint}. However, the function $S^{\rm int}$, in this case (in the large $N$ limit), is not an analytic function of its variable (as was the case for the free theory). Instead, the limit \eqref{lj} explores three different microcanonical phases as a function of the variable $\tau/(N^2J_1J_2)^{1/3}$, and the function $S^{\rm int}$ exhibits sharp phase transitions as a function of this variable. Moreover,  the temperature turns out to be a monotonic function of the scaled twist in the scaling limit under study. Consequently the canonical phase diagram of the theory also exhibits three phases, and two sharp phase transitions upon varying the temperature in the limit \eqref{limnew}. We now describe the resultant phase diagrams in more detail. 

\subsubsection{The black hole phase}

When 
\begin{equation}\label{oninint}
    T \geq \frac{1}{2\pi}, ~~~~{\rm equivalently}~~~~\frac{\tau}{(J_1 J_2)^{\frac{1}{3}}} > N^{\frac{2}{3}}\left(\frac{1}{2}\right)^{\frac{1}{3}}\,,
\end{equation}
 the dominant bulk phase (in the limit \eqref{lj}, equivalently \eqref{limnew}) is a large $AdS_5$ black hole. In this phase the entropy is given by the  thermodynamical formulae for 5d Kerr-AdS black holes. Satisfyingly, we find that 
these formulae do indeed reduce to the semi-universal form \eqref{finansforentint} and \eqref{sn}. $h(\beta)$ turns out to be given by the strikingly simple expression
\begin{equation}\label{hbhint}
    h_{BH}(\beta)  = \frac{N^2}{64}\left(\frac{2\pi}{\beta^3} \left(2\beta^2+\pi^2\right)^{\frac32} + \frac{2\pi^4}{\beta^3}-\frac{10\pi^2}{\beta}-\beta\right)\,,
\end{equation}
while the expression for $S^{\rm int}$ is listed in  \eqref{eq:sintBH}. We have obtained our final thermodynamical formulae \eqref{hbhint} and \eqref{sacalenthb} using the formulae of classical black hole thermodynamics. The reader may wonder whether classical thermodynamics, which certainly holds for any value of $J_i$ of order $N^2$, continues to be reliable in the strict large $J_i$ limit \eqref{lj} (and so, in particular, when $J_i$ scales like $N^a$ for $a>2$). We believe this is indeed the case; quantum effects (e.g. those associated with Hawking radiation) all seem small in the limit under study, as the black hole temperature stays fixed in the limit \eqref{lj}. However we leave a more careful verification of this expectation to future work.

\subsubsection{The gas phase}

In contrast, when the scaled twist of the theory is low enough so that, 
\begin{equation}\label{casenowint}
 T \leq \frac{1}{2\pi}, ~~~~{\rm equivalently}~~~~\frac{\tau}{\left(J_1J_2\right)^{\frac{1}{3}}} < A
 \end{equation}
(where the order unity number $A$ is listed in \eqref{adef}), 
the bulk theory lies in a 10-dimensional supergravity gas phase.
In this phase, $h(\beta)$ is given by $h_{YM}(\beta)$, defined in Eq. 4.13 of \cite{Bajaj:2024utv}. The function 
$S^{\rm int}$ is then obtained by performing a Legendre transformation, in particular, by inserting this result for $h(\beta)$ into \eqref{intent}. 

The formulae presented above hold for a free 10-dimensional gas. Around \eqref{gaseffects} and \eqref{jojt} below, we estimate under what circumstances interactions can be ignored so that the free approximation is valid; we find that this is the case when $J_i \ll N^6$. When this condition is violated, we expect the entropy formula to deviate from the free results presented above. In this paper, however, we leave the determination of the corrected entropy formula
of the gas phase (at values of $J_i$ of order or larger than $N^6$) to future work. Similar remarks apply to the gas component of the grey galaxy phase of the next subsubsection.

\subsubsection{The grey galaxy phase}

When the scaled twist lies in the range
\begin{equation}\label{ggint}
A<\frac{\tau}{\left(J_1J_2\right)^{\frac{1}{3}}}<N^{\frac{2}{3}}\left(\frac{1}{2}\right)^{\frac{1}{3}},
\end{equation}
our system lives in a rank four grey galaxy  \cite{Kim:2023sig, Bajaj:2024utv} phase.
As explained in \cite{Kim:2023sig, Bajaj:2024utv}, a grey galaxy can, roughly, be thought of as a non-interacting mix of a 5d Kerr-AdS$_5$ black hole and a gas. At least one of the angular velocities, $\omega_i$, of the black hole component of a grey galaxy is parametrically close to unity (this allows the gas surrounding the black hole to carry a macroscopic amount of energy and angular momentum). A grey galaxy is said to be of rank four when both $\omega_i$ are parametrically close to unity, but of rank two when only one of the two $\omega_i$ has this property. The function $S^{\rm int}$ for the rank four grey galaxy phase, in the limit relevant to this paper, turns out to take the simple form presented in \eqref{entropygg}.

As the scaled twist of the grey galaxy approaches the upper bound of \eqref{ggint}, its gas component vanishes, and the grey galaxy reduces to a pure black hole. Consequently, the phase transition between the black hole and grey galaxy phase is of the continuous (second order) kind. Interestingly enough, this phase transition temperature also coincides with the Hawking-Page transition temperature in the small $\nu_i$ (large $J$) limit under study (at finite $J$, in contrast, the Hawking-Page temperature is always higher than the black hole/grey galaxy superradiant phase transition temperature). 

On the other hand, as the scaled twist of the grey galaxy approaches the lower limit of \eqref{ggint}, the black hole shrinks to zero size, and the grey galaxy reduces to a pure gas. The phase transition between the grey galaxy and the gas phase is also continuous.

\subsubsection{Analogy of the phase diagram with boiling water}\label{abw}

In the scaling limit under consideration, the grey galaxy has a fixed temperature equal to $(2\pi)^{-1}$.
In fact, the phase diagram described in this subsubsection displays a close similarity with that of water boiling \footnote{We thank C. Patel for a useful discussion on this point.}. Within this analogy, a mass of water with a fixed number of water molecules $n$ is analogous to ${\cal N}=4$ Yang-Mills at fixed angular momentum. 
The internal energy $U$ of the water is analogous to the scaled twist of the ${\cal N}=4$ system. The extensivity of the water -- the fact that $S_{water}= n g(U/n)$ -- mirrors the extensivity of \eqref{finansforentint}. The black hole phase is analogous to the steam phase. Just as we reach the condensation point (100 degrees Celsius) of steam upon lowering the internal energy, we hit the superradiant instability of the black hole ($\frac{1}{2\pi}$) upon lowering the scaled twist. Just as further lowering  the internal energy leads to a half-steam, half-water phase at the fixed temperature 100 degrees Celsius, further lowering of the ${\cal N}=4$ scaled twist leads to a half-black-hole, half-gas grey galaxy phase at a fixed temperature $\frac{1}{2\pi}$. Upon further lowering the internal energy the proportion of the steam in the water (fraction of the black hole to total angular momentum in the grey galaxy) decreases, and eventually
we reach a pure water (pure $10$D SUGRA gas) phase. Further lowering the internal energy keeps us in the water (SUGRA gas) phase, but now at lower temperatures. 

\subsubsection{Generalizations}

We have described the phase diagram above, in terms of $N$, the rank of the SYM gauge group. $N$ is simply related to the five-dimensional Newton constant $G_5$, and all our results above can be recast in terms of $G_5$ by making the replacement $N\rightarrow \sqrt{\frac{\pi}{2 G_5}}$. Once this has been done, all the formulae presented above apply to all large $N$ strongly coupled 4d CFTs that have a two-derivative gravity dual description, with small modifications
\footnote{Such theories include those whose bulk dual description is IIB string theory on  $AdS_5 \times Y^{p,q}$ (rather than $AdS_5\times S^5$). The main modification (compared to the detailed analysis for ${\cal N}=4$ Yang-Mills theory presented in this paper) lies in the detailed thermodynamical formulae of the gas (in the gas phase or the gas component of the grey galaxy), which depends on the details of the low energy spectrum of single-trace operators (gravitons) of the theory. However the leading large $N$ phase diagram of these theories is identical to that of ${\cal N}=4$ Yang-Mills, as the details of thermodynamics play no role in obtaining this phase diagram. }. Thus the phase diagram outlined in this subsection has a certain degree of universality. 

Moreover, while the details of our phase diagram depend on the fact that $\lambda$ is large, we expect the general structure of this phase diagram to persist to finite values of $\lambda$ (and, perhaps, 
down to any nonzero value of $\lambda$, no matter how small). In this sense, the structure of the phase diagram described above may extend to beyond the strong coupling limit (in contrast, retreating away from large $N$ presumably smoothes out all sharp phase transitions).

\subsection{Relationship to the lightcone bootstrap}\label{sec:LCboot}

A famous result of the so called `lightcone bootstrap' \cite{Komargodski:2012ek,Fitzpatrick:2012yx} asserts that there is a sense in which every $(d>2)$-dimensional CFT has a free sector at large angular momentum\footnote{The lightcone bootstrap states: The OPE of two operators $\phi_1$ and $\phi_2$ with twist $\tau_1$ and $\tau_2$ has primary operators with twist arbitrarily close to $\tau_1+\tau_2+n$ for large enough spin. A natural question is whether this is true \textbf{every} spin after some threshold $J_*$ or whether we need to average over $J$. Here averaging over $J$ means that we collect all the primary operators, appearing in the OPE, with $0\leq J\leq \Lambda$, and show that there exists, in this set, primary operators with twist arbitrarily close to $\tau_1+\tau_2+n$ for large enough $\Lambda$. For $n=0$, the averaged version is rigorously established in \cite{Pal:2022vqc}, and for $n\in \mathbb{N}$, it is rigorously proven recently in a remarkable paper \cite{vanRees:2024xkb}. However, we expect a stronger result to be true: for every spin $J$, we collect the primary operators, appearing in the OPE with spin $J$ and bin them in a set labeled by spin $J$. For sufficiently large spin $J_*$, \textbf{every} set labelled by $J>J_*$ has primary operators with twist $\tau_1+\tau_2+n$. The lightcone bootstrap for correlators in $2$d is qualitatively different, see the appendix \ref{sec:lightconeboot2d} for a brief account.}. More precisely, every CFT is believed to host a basic set $\{O_i \}$ of `seed primaries' defined so that the OPEs of (arbitrary numbers of) these seed primaries (and their descendants) generate the full operator spectrum of the theory at large enough $J$. The assertion is that products of high angular momentum descendants of any collection of the $\{ O_i \}$ are effectively free, in the sense that the scaling dimension of this product operator is simply the sum of the scaling dimensions of its components. From the viewpoint of the AdS/CFT correspondence, the seed operators correspond to particles in the bulk, and their non-interacting nature is a consequence of the fact that particles in high angular momentum states are typically very far from each other.

Since this paper is dedicated to the study of high angular momentum configurations, and in light of the results above, it is natural to study the partition function of a free Fock space of an arbitrary collection of seed primaries and their descendants.
Such a partition function is easy to construct (see Appendix \ref{lcform}). In the limit \eqref{limnew}, this partition function does, in fact, take the universal form \eqref{sn}, with an $h(\beta)$ whose form is determined by the spectrum of seed primaries. 

While it is satisfying that the `free Fock space' partition function 
takes the universal form \eqref{sn}, this observation, unfortunately, does not constitute an explanation for \eqref{sn} or its higher dimensional cousins. The problem is that while interactions between large angular momentum descendants of a fixed number of seed primaries are small, the same is not true of a gas of such descendants of primaries, whose number itself scales with a positive power of the total angular momentum. Indeed, we have already seen this in the example of ${\cal N}=4$ Yang-Mills theory above. The free gas phase of this theory is a particular example of a Fock space of descendants of seed primaries (the seed primaries, in this case, are all single-trace chiral primaries of ${\cal N}=4$ Yang-Mills). As we have explained above, even in this phase, interactions are only negligible (at temperatures of order unity) at angular momenta that obey $J\ll N^6$ (note this is a severe restriction at small values of $N$). \footnote{It was already argued in section 4.3 of \cite{Cuomo:2022kio} that the free field picture of the $d=3$ lightcone bootstrap will break down in the large $J$ limit if the number of seed primaries (effectively the quantity  $Q$ in  \cite{Cuomo:2022kio}) is taken to be of  order $\sqrt{J}$ as $J$ is taken to infinity. $Q\sim \sqrt{J}$ is precisely the scaling we see from a free thermal bulk gas in $AdS_4$, in the limit $\nu \to 0$ (see under \eqref{entexp} for a similar count for the 10d sugra gas dual to $AdS_5 \times S^5$).}

As we raise the temperature, the situation gets worse. There is simply no useful sense in which the black hole component of a black hole or grey galaxy phase can be regarded as a weakly interacting mix of seed primaries. We expect that the free Fock space picture spelt out in this subsection 
applies quantitatively in the limit \eqref{lj} only when the temperature is scaled to zero in a manner that is coordinated with the 
limit $J \to \infty$.

That the ``free Fock space" i.e.\ the generalized free field (GFF) approximation can never hold at arbitrarily high twists (equivalently arbitrarily high temperatures) can also be seen as follows. Working in $d=4$ dimensions, within the GFF approximation,  $h(\beta)$ scales like $\beta^{-4}$ in the $\beta\to 0$ limit if the number of basic seed fields is finite (see \S \ref{lcform}). The high-temperature behaviour predicted by the GFF approximation can be even more singular if we have an infinite number of basic seed fields. However, we know that $h(\beta)$, appearing in eq.~\!\eqref{sn} actually scales like $\beta^{-3}$ in the $\beta\to 0$ limit, establishing that the GFF approximation can certainly never be valid at high temperatures. 

In summary, the `free Fock space' picture that the lightcone bootstrap might originally suggest does not quantitatively apply at any finite temperature in the limit \eqref{lj}, and does not constitute an explanation of the scaling behaviours predicted in this paper. The results of this paper go far beyond the `finite particle number' regime in which the free picture of the lightcone bootstrap applies.

\section{Semi-universality at large angular momentum on $S^3 \times S^1$}\label{su}

In this section we focus entirely on the special case of CFT$_4$ (and, briefly, their massive deformations). We also focus on a scaling limit in which both angular momenta of the theory are taken large at a fixed ratio. We discuss the generalization to other dimensions and more asymmetric distributions of angular momenta in the next section (\S\ref{gad}). As our analysis is performed using the framework developed in \cite{Banerjee:2012iz, Jensen:2012jh}, we begin this section with a brief review of that formalism.

\subsection{Partition function in the derivative expansion}

In this subsection we pause to present a brief review of the general framework for the derivative expansion of the thermal partition function presented in \cite{Banerjee:2012iz, Jensen:2012jh}. 

Consider a CFT on the Lorentzian spacetime 
\begin{equation}\label{eq:lorstationary_metric}
    ds^2 = -e^{2\sigma(x)}\left(d t + a_i(x)\, dx^i\right)^2 + g_{ij}(x)\, dx^i dx^j,
\end{equation}
Note that $\partial_t$ is a Killing vector of this spacetime. 
Let $H$ denote the charge that generates the `time translation' $\partial_t$, and consider the partition function, $Z={\rm Tr}\ e^{- \beta H}$. $Z$ can be evaluated by moving to Euclidean space (i.e.\ by setting $ t= -i \tau$), and compactifying $\tau$.
\begin{equation} \label{ident}
\tau \sim \tau + \beta, 
\end{equation} 
Consequently, $Z$ is given by the Euclidean field theory path integral on the 
`Kaluza-Klein' type spacetime
\begin{equation}\label{eq:stationary_metric}
    ds^2 = e^{2\sigma(x)}\left(d\tau + i a_i(x)\, dx^i\right)^2 + g_{ij}(x)\, dx^i dx^j,
\end{equation}
In \eqref{eq:stationary_metric}, the Euclidean time circle is  fibred over a 3d spatial manifold with coordinates $x^i$ and metric $g_{ij}$. Functions $\sigma(x)$, $a_i(x)$, and $g_{ij}(x)$ are arbitrary smooth functions of the spatial coordinates ($i a_i$ is the Kaluza-Klein gauge field). 

As explained in \cite{Banerjee:2012iz, Jensen:2012jh}, the Euclidean partition function of the field theory on this manifold can be expanded in a power series expansion in derivatives. This expansion takes the form 
\begin{equation}\label{eq:partition}
\begin{aligned}
    &\ln Z = \int d^3x\, \sqrt{g_3} \, \frac{P(T(x))}{T(x)}  
    - \frac{1}{2} \int d^3x\, \sqrt{g_3} \left( 
        P_1(\sigma)\, R + 
        T^2 P_2(\sigma)\, f_{ij} f^{ij} + 
        P_3(\sigma)\, (\partial \sigma)^2 
    \right)+ \cdots\\
    & T(x) = T e^{-\sigma(x)}, ~~~~T=1/\beta\,,
   \end{aligned} 
\end{equation}
where $R$ is the curvature scalar of the three-dimensional manifold with metric $g_{ij}$, and $f_{ij}$ is the field strength of the KK gauge field, and $T(x)$ can be thought of as the local redshifted temperature. The $\cdots$ in \eqref{eq:partition} denotes terms with more than two derivatives. 

By specializing to the case of a flat spatial manifold (i.e.\  $a_i=0$, $\sigma=0$, and $\beta=1/T$),  we see that the function  $P(T)$ has a simple physical interpretation; it is simply the thermodynamical pressure (negative of the thermal free energy density) 
of the field theory.\footnote{With these choice of parameters, \eqref{eq:partition} receives its full contribution from the first term (all higher derivative terms in this expansion simply vanish).} 
The physical interpretations of the other functions $P_1(\sigma)$, $P_2(\sigma)$ and $P_3(\sigma)$ are more involved. As explained in \cite{Banerjee:2012iz, Jensen:2012jh}, these terms evaluate (and constrain) certain non-dissipative terms in the constitutive relations of the hydrodynamical description of the theory. These interpretations will, however, not play a role in this paper.

\subsection{Constraints from Weyl invariance}

Eqn.~\!\eqref{eq:partition} gives the most general expansion, in derivatives, of the partition function of an arbitrary (in general non-conformal) field theory. In this subsection -- and most of the rest of this paper -- we specialize to the study of CFTs. In this situation, the conformal invariance of the underlying field theory further constrains the expansion 
\eqref{eq:partition}. It follows from the conformal invariance of our theory that the partition function should be Weyl invariant (up to terms involving the Weyl anomaly, which we ignore in this paper, since they are a constant shift in $\beta$ without any $\omega$ dependence). 
Under the Weyl transformation $ \bar{g}_{\mu\nu} = g_{\mu\nu}e^{2\phi(x)}$,  the fields $\sigma$, $g_{ij}$ and $a_i$ transform as \cite{Banerjee:2012iz}
\begin{align}\label{fieldtransf}
\bar{\sigma} &= \sigma + \phi, \quad \bar{a}_i = a_i, \quad \bar{g}_{ij} = e^{2\phi} g_{ij}\,, 
\end{align}
so that 
\begin{equation}
\begin{aligned}
(\nabla \bar{\sigma})^2 &= e^{-2\phi} \left[ (\nabla \sigma)^2 + 2 (\nabla \sigma) \cdot (\nabla \phi) + (\nabla \phi)^2 \right]\,, \\
\bar{R} &= e^{-2\phi} \left[ R - 4 \nabla^2 \phi - 2 (\nabla \phi)^2 \right]\,, \\
\bar{f}_{ij} \bar{f}^{ij} &= e^{-4\phi} f_{ij} f^{ij}\,, \\
\sqrt{\bar{g}_3} &= e^{3\phi} \sqrt{g_3}\,.
\end{aligned}
\end{equation}
The invariance of the partition function under the above transformation imposes the following constraints \cite{Banerjee:2012iz,Jensen:2012jh} on $P_i(\sigma)$.
\begin{equation}\label{eq:conformal_solution}
    P_1(\sigma)=e_1 T(x),~~P_2(\sigma)=\frac{e_2}{T(x)}\,,\quad \mathrm{and}\quad P_3(\sigma) = 2P_1(\sigma)\,,
\end{equation}
where, $e_1,e_2$ are dynamically determined numerical constants, whose value depends on the theory under study. 

Notice that the invariance under Weyl transformations has two effects on the 
coefficients of terms in the derivative expansion. 
\begin{itemize}
\item First it determines the scaling with $e^{\sigma}$ (equivalently $T(x)$) of all unknown functions. These scalings simply follow from scale invariance, i.e.\ the requirement of Weyl invariance under constant shifts of $\sigma$.
\item Second, it tells us that two a priori independent functions, $P_3(\sigma)$ and $P_1(\sigma)$ 
are actually proportional to each other. This relationship follows from the requirement of Weyl invariance under the (more involved) space varying Weyl transformations. 
\end{itemize}

Both these type of relationships continue to hold at every order in the derivative expansion. The 
$e^{\sigma}$ dependence of arbitrary terms in the expansion is easy to determine (see below), and will play an important role in our analysis. In contrast the analogue of the second kind of relationship becomes increasingly complicated at higher orders; fortunately these relations will play no role in our analysis, so their detailed form is unimportant for the purposes of the current paper.

\subsection{Specializing to $S^3 \times S^1$}\label{S3S1}

Let us now consider a quantum field theory on the manifold $S^3 \times S^1$ at a global equilibrium temperature $T$. As above, we use $\tau$ to denote Euclidean time, and $t$ to denote Lorentzian time, with
\begin{equation}\label{taut}
\tau= i t, ~~~~t=-i \tau\,.
\end{equation}
In the Euclidean context, the background metric of interest to us is 
\begin{equation}\label{euclideanmet}
    ds^2 = d\tau^2 + d\theta^2 + \sin^2\theta\, d{\phi'}_1^2 + \cos^2\theta\, d{\phi'}_2^2\,,
\end{equation}
where $\theta \in [0, \pi/2]$, and $(\phi'_1, \phi'_2)$ are angular coordinates on the three-sphere $S^3$, with the identification (this identification evaluates the trace ${\rm Tr}\ e^{-\beta H +i (-i\omega_1 \beta) J_1 +i (-i\omega_2 \beta) J_2} = {\rm Tr}\ e^{-\beta (H - \omega_1  J_1 -\omega_2 J_2)}$) 
\begin{equation}\label{identin}
   (\tau,  \phi'_1, \phi'_2) = (\tau +\beta, \phi'_1 - i\omega_1 \beta, \phi'_2- i \omega_2 \beta)\,.
\end{equation}

Upon making the coordinate change 
\begin{equation}\label{cchange}
    \phi_1= \phi'_1 +i \omega_1 \tau , ~~\phi_2 = \phi'_2 +i \omega_2 \tau, ~~~\tau=\tau\,.
\end{equation}
We find that while the identifications are now of the untwisted thermal circle form 
\begin{equation}\label{identin2}
   (\tau,  \phi_1, \phi_2) = (\tau +\beta, \phi_1, \phi_2)\,, 
\end{equation} 
the spacetime metric now takes a more complicated form 
\begin{equation}\label{eq:boosted_metric}
\begin{split}
    ds^2&= \frac{1}{\gamma (\theta )^2}\left[d\tau - i\gamma (\theta )^2 \left(\omega_1 d\phi_1 \sin ^2(\theta )+\omega_2 d\phi_2 \cos ^2(\theta )\right)\right]^2\\& +d\theta ^2 + \gamma (\theta )^2 \left(\omega_1 d\phi_1 \sin ^2(\theta )+\omega_2d\phi_2 \cos ^2(\theta )\right)^2+d\phi_1^2 \sin ^2(\theta )+d\phi_2^2 \cos ^2(\theta )\,,\\
\end{split}
\end{equation}
where
\begin{equation}\label{gammadef}
\gamma(\theta) = \left(1 - \omega_1^2 \sin^2\theta - \omega_2^2 \cos^2\theta \right)^{-1/2}\,.
\end{equation}
The four-dimensional metric \eqref{eq:boosted_metric} is of  the Kaluza-Klein form \eqref{eq:stationary_metric}, with the three-dimensional base metric $g_{ij}$,  the gauge field $a_i$, and the scalar field $\sigma(x)$ given respectively by 
\begin{equation}\label{eq:fields}
    \begin{split}
        e^{-\sigma(x)} &= \gamma(\theta)\,,\quad a = -\gamma(\theta)^2\left ( \omega_1\sin^2\theta d\phi_1 + \omega_2 \cos^2\theta d\phi_2 \right)\,,\\ g_{ij}dx^{i}dx^{j} &= d\theta ^2 + \gamma (\theta )^2 \left(\omega_1 d\phi_1 \sin ^2(\theta )+\omega_2d\phi_2 \cos ^2(\theta )\right)^2+d\phi_1^2 \sin ^2(\theta )+d\phi_2^2 \cos ^2(\theta )\,.
    \end{split}
\end{equation}
We note for future use that 
\begin{equation}\label{eq:volumeform}
   \sqrt{g_3} = \gamma(\theta)\sin\theta \cos\theta = \gamma(\theta) \sqrt{\Omega_3}\,,
\end{equation}
where $\Omega_3$ is the round metric on $S^3$.

In our new coordinates, the thermal identification \eqref{identin} is in the $\partial_\tau$ direction. For later use we note that the one-form obtained by lowering the indices of vector field $\partial_\tau$, and then normalizing  so that it everywhere has unit norm,  equals 
\begin{equation}\label{eq:velocity_1form}
  \gamma(\theta)  g_{\mu\tau} dx^\mu =\frac{d \tau}{\gamma (\theta )} - i\omega _1 \gamma (\theta ) \sin ^2(\theta ) d \phi_1 - i\omega _2 \gamma (\theta ) \cos ^2(\theta ) d \phi_2 \,.
\end{equation}

\subsection{Interpretation in Lorentzian Space}\label{S3S1l}

In this brief subsubsection, we pause to provide a Lorentzian interpretation of the construction described above. Consider the Lorentzian field theory on the spacetime
\begin{equation}\label{spheor}
    ds^2 = -dt^2 + d\theta^2 + \sin^2\theta\, d{\phi'}_1^2 + \cos^2\theta\, d{\phi'}_2^2,
\end{equation}
As explained in Secs. \ref{tpgm} and \ref{ttpf}, we are interested in studying the equilibrium solution of a conformal fluid whose velocity is proportional to $\partial_t+\omega_1 \partial_{\phi'_1} + \omega_2 \partial_{\phi'_2}$.
In other words, our conformal fluid swirls around the sphere \eqref{spheor} with these real angular speeds $(\omega_1, \omega_2)$.

The coordinate change \eqref{cchange} has a simple Lorentzian interpretation. Using \eqref{taut}, we see that \eqref{cchange} continues, in Lorentzian space, to  
\begin{equation}\label{cchangelor}
    \phi_1= \phi'_1 - \omega_1 t , ~~\phi_2 = \phi'_2 - \omega_2 t, ~~~t=t\,.
\end{equation}
It is easy to check that the fluid velocity $\partial_t$ in the new unprimed coordinates transforms to  $\partial_t+\omega_1 \partial_{\phi'_1} + \omega_2 \partial_{\phi'_2}$ in the old primed coordinates. In other words, the coordinate change \eqref{cchangelor} takes us to `comoving' or Eulerian coordinates. 
As the new coordinates rotate at angular speeds $\omega_1, \omega_2$ compared to the old coordinates (see 
\eqref{cchangelor}), this point is also intuitively clear.

 The velocity of the fluid is 
\begin{equation}\label{velocityoffluid}
    u^\mu = \gamma(\theta)\left(\partial_t + \omega_1\partial_{\phi'_1} + \omega_2 \partial_{\phi'_2}\right)^{\mu},
\end{equation}
The factor $\gamma(\theta)$ was defined in \eqref{gammadef}, and is the usual special relativistic factor  
$(1-v^2)^{-1/2}$. 

The velocity one-form corresponding to \eqref{velocityoffluid} equals 
\begin{equation}\label{veloneform}
\begin{split}
u_\mu dx^\mu &= \gamma(\theta) \left( -dt + \omega_1\cos^2\theta d\phi_1' + \omega_2\sin^2 \theta 
d \phi_2' \right) \\&= -\frac{1}{\gamma (\theta )}dt +\omega _1 \gamma (\theta ) \sin ^2(\theta )d\phi_1+\omega _2 \gamma (\theta ) \cos ^2(\theta )d\phi_2\,,
\end{split}
\end{equation}
and is the analytic continuation (from Euclidean space) of  \eqref{eq:velocity_1form}.

From the Lorentzian point of view, the factors of $\gamma^2(\theta)$ in \eqref{eq:fields} can be understood as follows. Consider the case $\omega_1=\omega_2$. Here, the displacement $d \phi_1=d \phi_2=dx$, with  $d\theta=dt=0$, describes motion along the spatial component of the fluid velocity. As the parameter $x$ varies from $0$ 
to $2 \pi$, we move around an equator on the $S^3$ \footnote{We have focussed on the case $\omega_1=\omega_2$ precisely because motion in the velocity direction in this case takes us around a closed loop (an equator) rather than around a tilted direction on the $S^3$.}.
Restricting the metric $g_{ij}$ (see \eqref{eq:fields}) to this curve, we see that the metric along this curve is $ds^2= \gamma^2 dx^2$. It follows that the full proper length of this `fibre' equator equals $2 \pi \gamma$.  The additional factor of $\gamma$ is simply a consequence of Lorentz contraction: if the rotating observer (who uses the unprimed coordinates) were to lay meter sticks down along the $\partial_{\phi_1} + \partial_{\phi_2}$ direction, the rotating observer would fit $\gamma(\theta)$ times more meter sticks than might at first have been expected, since the sticks appear Lorentz-contracted (by a factor $1/\gamma(\theta)$) to the non-rotating (lab) observer. A very similar explanation was given for the same phenomenon (in a slightly different context) in \cite{Shaghoulian:2015lcn}, see the beginning of \S 4.2 of that paper.

On the other hand, consider the motion on the $S^3$ perpendicular to the 
fluid velocity, i.e.\ along the direction $d\phi_1= \cos^2 \theta dx, d \phi_2 = - \sin^2 \theta dx,  d \theta=0$. For such a motion the term  $(\omega_1 d \phi_1 \sin^2 \theta + \omega_2 d \phi_2 \cos^2 \theta)$
(in $g_{ij}$ given in \eqref{eq:fields}) simply vanishes (recall we are focusing on the case $\omega_1=\omega_2$). Consequently, the length of this `equator' (in the metric listed in the third equation of \eqref{eq:fields}) simply equals $2 \pi$. This reflects the fact that lengths perpendicular to the velocity are not Lorentz-contracted. 

Finally, the factor of $1/\gamma(\theta)
$ (in the first line of 
\eqref{eq:boosted_metric}) precisely captures the time dilation in the lab frame (the passage of one second in the non-rotating lab frame
corresponds to the passage of $1/\gamma(\theta)$ seconds for the proper observer). 

\subsection{The thermal partition function in the large $\gamma(\theta)$ limit}\label{Tetrad}

In this subsection, we will work in the setup of \S \ref{S3S1}, but focus on the limit in which $\gamma(\theta)$ is everywhere large. In this limit we demonstrate that
\begin{itemize}
\item No term in any of the ${\cal L}_n$ in \eqref{newexpn} grows faster than $\gamma^4(\theta)$.
\item Every term in ${\cal L}_n$ (for any value of $n$) that grows as fast as 
$\gamma^4(\theta)$ is, in fact, precisely proportional to $\gamma^4(\theta)$ (rather than, for instance, derivatives of this quantity).
\item One finds terms proportional to $\gamma^4(\theta)$ in ${\mathcal L}_n$ at every value of $n$.  Consequently, the coefficient of $\gamma^4(\theta)$ in 
${\mathcal L}$ has terms proportional to $\beta^n$ for arbitrarily large values of $n$. 
\end{itemize}

The thermal effective action lives on a three-dimensional manifold, whose metric is listed in the second line of \eqref{eq:fields}. In order to work with this spatial metric, we introduce a set of local orthonormal frames (tetrads) $ e^a_{i}(x)$ which, as usual,  relate the curved spacetime metric to the flat Euclidean metric via
\begin{equation}
    g_{ij}(x) = \delta_{ab} \, e^a_{\ i}(x) \, e^b_{\ j}(x),
\end{equation}
where $\delta_{ab} = \text{diag}( +1, +1, +1) $ is the Euclidean metric in the local inertial frame.

As the physical situation involves a revolving fluid, it is natural to choose one of the tetrads -- let us say $e^2$ -- to be the vector along the spatial part of the fluid velocity one-form. From eq.~\eqref{eq:velocity_1form}, we obtain the spatial component of the velocity one-form as
\begin{equation}\label{etwo}
    e^2 = v = \frac{\gamma^2(\theta)}{\sqrt{\gamma^2(\theta)-1}}
    \left( \omega_1 \sin^2\theta\ d\phi_1 + \omega_2 \cos^2\theta\  d\phi_2\right),
\end{equation}
(the prefactor in \eqref{etwo} ensures that $e^2=v$ obeys the normalization condition $g^{ij} v_i v_j = 1$).

The remaining tetrad one-forms must be orthogonal to $e^2$ and to each other (in the metric $g_{ij}$), ensuring that the triplet $(e^1, e^2, e^3)$ forms a complete orthonormal basis of the dual of tangent space at each point on the spatial slice. We find it convenient to choose
\begin{equation}\label{eq:veirbein}
        e^1= d\theta\,,\quad e^3=w= \frac{\omega_2\ d\phi_1-\omega_1\ d\phi_2}{\sqrt{\omega_2^2 \csc ^2(\theta )+\omega_1^2 \sec ^2(\theta )}}\,.
\end{equation}
Recall that we are interested in the limit that $\omega_1$ and $\omega_2$ are both close to unity; in this limit $e^2 \propto (\sin^2\theta\ d\phi_1 + \cos^2\theta\ d\phi_2)$ and $e^3 \propto ( d\phi_1 - d\phi_2)$. 

By definition, the spatial metric listed in the second line of \eqref{eq:fields} takes the simple form in the tetrad basis:
\begin{equation}
    ds^2 =  (e^1)^2 + (e^2)^2 + (e^3)^2\,.
\end{equation}

\subsubsection{Curvature in the large $\gamma(\theta)$ limit}

In the limit of large $\gamma$, we find, by explicit computation, that the only nonzero components of the curvature tensor of the 3-dimensional base manifold (whose metric is listed on the third line of eq.~\eqref{eq:fields}) are
\begin{equation}\label{riemancom}
\begin{split}
    &R_{1212} = \gamma (\theta )^2 + O(1), ~R_{1313} = -3 \gamma (\theta )^2 + O(1),~R_{2323} = \gamma(\theta)^2 + O(1)\,,\\& R_{1213} = -3 \gamma '(\theta ) + O\left(\frac{1}{\gamma(\theta)}\right)\,,
\end{split}
\end{equation}
where all indices are tetrad indices. Since we have assumed $\nu_1$ and $\nu_2$
are of the same order, $\gamma(\theta)$ is a smooth function of $\theta$ (it 
varies on scale unity in the angular coordinate $\theta$) and so $\gamma'(\theta)$ is of the same order as $\gamma(\theta)$, which, to the leading order in $\nu_i$, is given by 
\begin{equation}\label{gammaleading}
\gamma(\theta)= \frac{\sqrt{\beta/2}}{\sqrt{ \nu_1 \sin^2 \theta +\nu_2 \cos^2 \theta }}.
\end{equation} 

We see from \eqref{riemancom} that no component of the curvature scales faster than ${\cal O}(\gamma^2(\theta))$. Moreover, the 
${\cal O}(\gamma^2(\theta))$ contribution to every component of the curvature tensor, is explicitly proportional to $\gamma^2(\theta)$.  In other words, the  leading order contributions to the curvature tensors do not involve derivatives of $\gamma(\theta)$ (like $\gamma'(\theta))$).

The Ricci tensor at leading order is given by,

\begin{equation}\label{rictens}
   R_{ab}=\left(
\begin{array}{ccc}
 -2 \gamma (\theta )^2 + O(1) & 0 & 0 \\
 0 & 2 \gamma (\theta )^2 + O(1) & -3 \gamma '(\theta ) + O(1/\gamma) \\
 0 & -3 \gamma '(\theta ) + O(1/\gamma) & -2 \gamma (\theta )^2 +O(1)
\end{array}
\right)\,.
\end{equation}
By contracting the Ricci tensor with $\delta_{ab}$, we obtain the Ricci scalar as\footnote{\label{fn:exactR}The exact expression for the Ricci scalar is given by $R = 6 (1 - \omega_1^2)(1 - \omega_2^2) \gamma(\theta)^4 - 2 \gamma(\theta)^2 + 2$.}
\begin{equation}\label{eq:ricci_scalar}
    R = -2 \gamma^2(\theta) + \mathcal{O}(1),
\end{equation}

From the vierbein postulate $D_\mu e^a_\nu = 0$, we obtain the spin connection:
\[
\omega_\mu^{ab} = e^a_\nu \left( \partial_\mu e^{\nu b} + \Gamma^\nu_{\mu\lambda} e^{\lambda b} \right),
\]
By explicit computation we find that the only nonzero components of the spin connection are 
\begin{equation}\label{sc}
\begin{split}
     &\omega_1^{\ 23} = - \gamma (\theta )+O\left(\frac{1}{\gamma (\theta )}\right)\,,\\ &\omega_2^{\ 12} =-\frac{\gamma '(\theta )}{\gamma (\theta )} +O\left(\frac{1}{\gamma (\theta )}\right)\,, \quad \omega_{2}^{\ 13}=-\gamma (\theta )+O\left(\frac{1}{\gamma (\theta )}\right)\,,\\ & \omega_3^{\ 12} =-\gamma (\theta )+O\left(\frac{1}{\gamma (\theta )}\right)\,, \quad \omega_3^{\ 13} =-2 \cot (2 \theta )+O\left(\frac{1}{\gamma (\theta )}\right)\,.
\end{split}
\end{equation}
We see that no component of the spin connection scales faster (in the large $\gamma$ limit) than $\gamma(\theta)$, and that all contributions to the 
spin connection at this leading order are explicitly proportional to $\gamma(\theta)$ (terms proportional to $\gamma'(\theta)$ do not show up at 
leading order).

When acting on a tensor, a covariant derivative includes an ordinary derivative and additional terms linear in the spin connection. The only nonzero ordinary derivative is $\partial_3$; by explicit evaluation, one can check that $\partial_3$ neither increases nor decreases the order in $\gamma(\theta)$ in any expression (see \eqref{gammaleading}). As all explicit derivatives are ${\cal O}(\gamma^0)$, while spin connections are generically of order $\gamma(\theta)$, we see that covariant derivatives are generically dominated by their spin-connection contributions, and are 
at most of order $\gamma$; moreover, every covariant derivative that contributes at order $\gamma$ does so with an explicit factor of $\gamma$ at this order (as opposed to derivatives of this quantity). 

It follows from the discussion of the previous paragraphs that a term involving $m$ covariant derivatives of the Riemann tensor scales no faster with $\gamma(\theta)$ than $\gamma^{m+2}$, 
and all such leading-order terms have an explicit factor of $\gamma^{m+2}$ (rather than, for instance, $\gamma^{m+1}\gamma'$ or other expressions involving derivatives of $\gamma$).

\subsubsection{Gauge fields}

The Kaluza-Klein gauge field can be read off from the metric \eqref{eq:boosted_metric}; it is given by 
\begin{equation}\label{gfield}
    \begin{split}
         a &= -\gamma^2(\theta)\left ( \omega_1\sin^2\theta d\phi_1 + \omega_2 \cos^2\theta d\phi_2 \right)=-\sqrt{\gamma(\theta)^2-1}\ e^2\,.
    \end{split}
\end{equation}
The field strength $f_{ij}$, obtained from $a$ is given by,
\begin{equation}\label{eq:field_strength_indiv}
    \begin{split}
         f&=  -\frac{2\gamma'(\theta)\gamma(\theta)}{\sqrt{\gamma^2- 1}}e_1\wedge e_2 -\frac{2 \gamma^3\omega_1\omega_2}{\sqrt{\gamma^2 -1}} e_1\wedge e_3\approx -2\gamma'(\theta)e_1\wedge e_2 - 2\gamma^2(\theta) e_1\wedge e_3\,.
    \end{split}
\end{equation}
As $\gamma'(\theta)$ is of the same order as $\gamma(\theta)$,  we see from \eqref{eq:field_strength_indiv} that no component of the KK field strength scales faster with $\gamma$ than $\gamma^2$, and that the field strength (at this leading order) is simply equal to   $2\gamma^2(\theta) e_1\wedge e_3$, and so is explicitly proportional to $\gamma^2$ (derivatives of $\gamma$ drop out at leading order). As in the previous subsubsection, we conclude that a term involving $n$ covariant derivatives of the KK field strength scales no faster with $\gamma$ than $\gamma^{n+2}$, and that all terms that contribute at this leading order have an 
explicit factor of $\gamma^{n+2}$ (derivatives of $\gamma$ do not show up at leading order).

In the next section, we will find it useful to note that $f\wedge\star
f= \frac{1}{2} f_{ij} f^{ij}\mathrm{vol_3}$, and that
\begin{equation}\label{eq:field_strength_contracted}
    \begin{split}
        f_{ij}f^{ij} &= 8 \left(\frac{\gamma'(\theta)^2}{\gamma(\theta)^2} + \gamma'(\theta)^2 + \omega_1^2 \omega_2^2 \gamma(\theta)^4 + \gamma(\theta)^2 (\omega_2^2 \sin^2(\theta ) + \omega_1^2 \cos^2(\theta ))\right) \\
        &= 8\gamma^4(\theta) + \mathcal{O}\left(\gamma^2(\theta)\right).
    \end{split}
\end{equation}

\subsubsection{The scalar}

The scalar field, $\sigma$, of the metric \eqref{eq:stationary_metric} parameterizes the redshift factor of the spacetime on which our field theory is formulated. The local redshifted temperature is given by, 
\begin{equation}
    T(x) = T e^{-\sigma} =T\gamma(\theta)\,,
\end{equation}
where $T$ is the constant thermodynamical temperature of the 
system.

The second order partition function, \eqref{eq:partition}, involves a term proportional to $(\partial \sigma)^2$. From the first equation of \eqref{eq:fields} it follows that 
\begin{equation}\label{pss}
    (\partial \sigma(x))^2 = \frac{\gamma'(\theta)^2}{\gamma(\theta)^2}= {\cal O}(1).
\end{equation}

A little thought will convince the reader that a term involving $n$ covariant derivatives of $\sigma$ scales no faster with $\gamma$ than ${\cal O}(\gamma^{n-1})$. 
This power is $n-1$ rather than $n$ because the first of the $n$ derivatives (that act on $\sigma$) receives no contribution from the spin connection, as $\sigma$ is a scalar. The fact that terms involving $n$ derivatives of $\sigma$ are of ${\cal O}(\gamma^{n-1})$ rather than ${\cal O}(\gamma^{n})$ will ensure that all terms in ${\cal L}$ that involve scalar derivatives will contribute to the partition function only at subleading order in $\gamma$.

\subsubsection{Weyl scaling}

Every local contribution to ${\cal L}$ (see \eqref{ddimpf}) is a  scalar expression, built out of tensors of the form 
\begin{equation}\label{coordscal}
\nabla_{a_1} \ldots \nabla_{a_{n_1}} R_{a_{n_1+1}\ a_{n_1+2}\ a_{n_1+3}\ a_{n_1+4} }, ~~~
\nabla_{b_1} \ldots \nabla_{b_{n_2}} F_{b_{n_2+1}\ b_{n_2+2} }, ~~~
\nabla_{c_1} \ldots \nabla_{c_{n_3}} \phi\,.
\end{equation}
In previous subsubsections  we have argued that:
\begin{enumerate}
    \item No contribution to any of the ${\cal L}_n$ scales faster with $\gamma$ than $\gamma^{T + 2n_R+2n_F - n_S}$ where $T$ is the total number of covariant derivatives involved in the expression, $n_R$ is the number of Riemann tensors involved, $n_F$ is the number of KK field strengths involved, and $n_S$ is the number of  scalar fields that appear differentiated in the expression. 
\end{enumerate}

Although we have found it convenient to use tensors referred to 
vielbein or tangent space indices at intermediate steps in the computation, our final result 1 above bounds the growth with
$\gamma$ of 
every contribution to the diffeomorphism scalar ${\cal L}$, and is, of course, also true when re-expressed in a coordinate basis (with $g_{ij}$ and its inverse used to contract indices). 

Let us now study the Weyl transformation properties of any contribution to ${\cal L}$. For this purpose we find it convenient to work with coordinate rather than tangent space tensors. 
The Weyl transformation of the four-dimensional metric 
\begin{equation}\label{weylscal}
g_{\mu\nu} \rightarrow g_{\mu\nu} e^{ 2 \chi}
\end{equation}
induces the following action on the three-dimensional metric $g_{ij}$, gauge field $a_i$ and scalar field $\sigma$, 
\begin{equation}\label{3dfieldswt}
g_{ij}\rightarrow e^{2\chi} g_{ij},~ a \rightarrow a, ~ \sigma(x) \rightarrow \sigma(x) + \chi\,.
\end{equation}
For our purposes, it will prove sufficient to study constant Weyl transformations
(transformations in which $\chi$ is not a function of $x$). It follows from 
\eqref{3dfieldswt} that every component of $F_{ij}$ is invariant under such a Weyl transformation, every component of the Riemann tensor scales like $R_{ijkl} \rightarrow e^{2 \chi} R_{ijkl}$, 
and every covariant derivative is left unchanged. In particular, 
covariant derivatives of $\sigma$ are invariant under constant Weyl transformations. (Here and through the rest of this subsection, $i, j$ are spatial coordinate indices rather than local Lorentz indices.) A scalar expression that is built out of a total of $T$ covariant derivatives of $n_R$ copies of the Riemann tensor, 
$n_F$ field strengths and any number of differentiated $\sigma$ fields, also needs $T/2+ 2 n_R+n_F$ copies of $g^{ij}$ (these are needed to contract away all indices). It thus follows that any such expression, 
$E$, transforms, under Weyl transformations, like
\begin{equation}\label{etransf}
E \rightarrow e^{-(T+4 n_R+2 n_F) \chi} e^{ 2\chi n_R} E 
=e^{-(T+2n_R+2 n_F) \chi}E\,.
\end{equation}
The first term in (\ref{etransf}), $e^{-(T+4 n_R+2 n_F) \chi}$, accounts for the Weyl transformations of the copies of $g^{ij}$, while the second term, 
$e^{ 2\chi n_R}$, accounts for the Weyl transformations of the Riemann tensors. Recall that all covariant derivatives of the field strength and $\sigma$ are Weyl invariant.

Now, consider a contribution of such a term to the effective action of the form 
\begin{equation}\label{eecont}
\int \sqrt{g_3} e ^{ \alpha \sigma} E\,,
\end{equation}
where $\alpha$ is a number. Under  a constant Weyl transformation \eqref{weylscal}, it follows from \eqref{etransf} and \eqref{3dfieldswt} that the expression in \eqref{eecont} picks up the factor $e^{(3 + \alpha -T -2n_R -2n_F)\chi}$. From the  requirement that $\int \sqrt{g_3} {\cal L}$ is Weyl invariant, it thus follows that 
\begin{equation}\label{alphaform}
\alpha = T + 2 n_R +2 n_F - 3\,.
\end{equation} 
Thus \eqref{eecont} actually takes the form 
\begin{equation}\label{eecontn}
\int \sqrt{g_3} e ^{ (T + 2 n_R +2 n_F - 3 )\sigma} E\,.
\end{equation}

We have seen above that $E \sim \gamma^{T + 2n_R + 2n_F -n_S}$. 
From the first equation of \eqref{eq:fields} we see that $e^{ (T + 2 n_R +2 n_F - 3 )\sigma} \sim \frac{\gamma^3}{\gamma^{(T+2 n_R+2n_F)}}$ and (from \eqref{eq:volumeform}) that $\sqrt{g_3} \sim \gamma$. Putting it all together, we see that the term in \eqref{eecontn} scales no faster with $\gamma$ than ${\cal O}(\gamma^{4-n_S})$. The leading order contributions to ${\cal L}_n$ all have $n_S=0$,  and so scale like ${\cal O}(\gamma^4)$ (such terms do not involve derivatives of $\sigma$). As covariant derivatives of the Riemann tensor and the field strength do not involve derivatives of $\gamma(\theta)$ at leading order,  we also conclude that all leading order contributions to ${\cal L}_n$ (those that are ${\cal O}(\gamma^4)$) are actually simply proportional to $\gamma^4(\theta)$. This result holds for every value of $n$.  We have thus demonstrated that ${\cal L}_n \propto \gamma^4(\theta)$ at leading order in the small $\nu_i$ limit, and that this result holds at every value of $n$.

As a sanity check of the somewhat abstract arguments of this subsection, in Appendix  \ref{loworder} we explicitly compute the twisted $S^3 \times S^1$ partition function for the most general Weyl invariant  form of ${\cal L}$ up to two-derivative order (see \eqref{eq:partition} with \eqref{eq:conformal_solution} and $P(T) \propto T^4$), and verify the arguments presented in this subsection in that case. 

\subsection{Semi-universal form of the partition function in the limit $\nu_i \to 0$ at fixed $\beta$}\label{subsec:semiuniversal}

In the previous subsection we have shown that each of the ${\mathcal L}_n$  -- and so, we argue, ${\cal L}$ itself -- is proportional to $\gamma^4(\theta)$. The proportionality constant is a function of temperature; let us denote this constant by $2 h(\beta)/( \pi^2)$ so that $${\cal L}= \frac{2 h(\beta)}{ \pi^2}\ \gamma^4(\theta)\,.$$ The thermal partition function, $\ln Z$, is now easily computed; 
\begin{equation}\label{thermpartfn}
\ln Z = \frac{2h(\beta)}{ \pi^2} \int_{S^3} \gamma^4(\theta)= \frac{4 h(\beta)}{(1-\omega_1^2)(1-\omega_2^2)} \approx \frac{\beta^2 h(\beta)}{\nu_1 \nu_2}\,.
\end{equation}
The integral above is easy to evaluate directly, but was also evaluated in \cite{Bhattacharyya_2008}, see footnote 11 of that paper.
Here $h(\beta)$ is a theory-dependent (so non-universal) function that receives contributions from all orders in the derivative expansion of the thermal effective action (see \eqref{lnu}). The reduced chemical potentials $\nu_i$ are defined in \eqref{redchempot}. We have thus `derived' the semi-universal form of the partition function \eqref{sn}.

Eqn \eqref{thermpartfn} holds at small $\nu_i$ and at every value of the temperature $T$. On the other hand, the simpler `universal' formula 
\eqref{uniform} holds at large $T$ (small $\beta$) and all values of $\omega_i$. When $\beta$ and $\nu_i$ are both small, both \eqref{thermpartfn} and \eqref{uniform} are valid, so by matching we conclude that  
\begin{equation}\label{ftlarget}
\lim_{\beta \to 0} \beta^3 h(\beta)=\frac{c'}{4}\,.
\end{equation}
where $c'$ is the dimensionless number that appears in \eqref{uniform} (this point was already briefly mentioned under 
\eqref{sn}).

\subsection{Massive theories} \label{mt}

As we have explained in this section, the  `thermal effective action' of a CFT on a twisted $S^3 \times S^1$ simplifies in the limit \eqref{limnew} because the local value of the temperature (inverse of the local proper size of the thermal $S^1$, i.e.\ $T_0 e^{-\sigma}= \gamma(\theta)$, see \eqref{eq:fields}) blows up in this limit. Despite the fact that the local value of $\beta$ vanishes, the thermal effective action is not well-approximated by the leading term in the expansion \eqref{newexpn}, because curvatures (of the base manifold and the KK gauge field) also blow up in this limit.
These two effects (vanishing $\beta(x)$ and blowing up curvatures) are finely balanced. Upon increasing $n$,  the extra divergences from curvatures sometimes compensate for -- but never overcompensate for -- the increased suppression following from the increased power of $\beta(x)$. The net effect is that while no term in any ${\mathcal L}_n$ scales faster than $\gamma^4$, altogether there are an infinite number of terms (taking the union of all the terms, that appears in ${\mathcal L}_n$, over $n$) that have this scaling.

With these points in mind, let us now turn our attention to a partition function of a general QFT (that is not, in general, a CFT). It is intuitively clear that the partition function of a QFT at parametrically high local temperatures should reduce to the partition function of its parent CFT under the same conditions. It is thus natural to expect that QFT partition functions, in the limit  \eqref{limnew}, to reduce to the partition functions of their parent CFTs in the same limit. In the rest of this subsection, we will present an argument for this conclusion in a more quantitative manner.

Any such theory can be defined as an RG flow (seeded by a relevant operator $O$ of dimension $d-r$, with $r>0$, here $d=4$) away from a UV CFT. Let the parameter that appears behind this relevant operator be $m^r$ (here $m$ is a quantity with the dimension of mass). As a scalar operator of scaling dimension $4-r$ transforms under Weyl transformations like, $O \rightarrow e^{-(4-r)\chi }O$, it follows that the deformation term (in the four-dimensional action) transforms under Weyl transformations as
\begin{equation}\label{massdef}
m^r \int \sqrt{g_4} O  \mapsto m^r \int e^{r \chi} \sqrt{g_4} O \,.
\end{equation}
Consequently, such a massive deformation breaks Weyl invariance. This invariance can formally be restored to our system if we allow the mass to transform as 
$m \rightarrow m e^{-\chi}$ (this maneuver is particularly convenient when studying Weyl transformations that are constant in spacetime and indeed such transformations are of interest to this section,  as such a maneuver  maps one constant mass to another). 

Now consider the thermal effective action \eqref{ddimpf} of this mass-deformed theory in the limit \eqref{limnew}. Let us study the thermal  effective action in a Taylor expansion in the massive parameter $m$
\begin{equation}\label{texp2}
{\mathcal L} = \sum_{k=0}^\infty m^{r k} {\cal L}^{(k)} \,,
\end{equation} 
where we note that the expansion \eqref{texp2} is in powers of mass not in powers of derivatives; each of the ${\mathcal L}^{(k)}$ above admits an expansion in derivatives.

It follows from the discussion above, that, under a constant Weyl transformation,  
\begin{equation} \label{constweyltransf}
 {\cal L}^{(k)}  \rightarrow e^{k r \chi} {\cal L}^{(k)}\,. 
\end{equation}
Reasoning as around \eqref{etransf}, we conclude that a contribution 
${\cal L}^{(k)} $ (i.e.\ a product of $n_R$ Riemann curvatures, 
$n_F$  field strengths and $n_S$ (derivatives of) the scalar $\sigma$, and involving a total of $T$ covariant derivatives) must come with a prefactor 
proportional to  
\begin{equation}\label{termppro}
e^{(T+2n_R+2 n_F)  \sigma} \times e^{rk \sigma}\,.
\end{equation}
The first factor is the same as the prefactor in \eqref{eecontn}, and 
the second factor accounts for the additional Weyl scaling \eqref{constweyltransf}. Imitating the reasoning under \eqref{eecontn}, we conclude that no such term scales faster with $\gamma$ than 
\begin{equation}\label{gammascal1}
\gamma^{4-rk - n_S}\,.
\end{equation}
We conclude, in particular, that the leading order scaling with $\gamma$ is
at $k=0$; at this order the thermal effective action for the mass-deformed theory \eqref{massdef} is identical to the thermal effective action \eqref{ddimpf} for its UV CFT.

In \S  \ref{fms}, we will verify that the general predictions of this subsection are, indeed, borne out in the partition function of a free massive scalar.

\subsection{A semi-universal form for entropy as a function of charges}\label{semient}
Given the partition function of a theory as a function of $\beta$ and $\omega_i$, we can perform inverse Laplace transforms (as we will see, these are effectively the same as performing Legendre transformations) to obtain the entropy of the theory as a function of
the twist $\tau$ and the spin $J$. In this subsection we
perform this exercise on a partition function of the form eq.~\!\eqref{thermpartfn}.

The microcanonical entropy $S(E, J_1, J_2)$ is given by 
\begin{equation}
\begin{aligned}\label{invlap}
 e^{S(E, J_1, J_2)} 
 & =\frac{1}{(2\pi i)^3}\int d\beta \, d\omega_1 \, d\omega_2 \, 
 \exp\left[\beta(E - \omega_1 J_1 - \omega_2 J_2) + \ln Z(\beta, \omega_1, \omega_2) \right] \\
&=\frac{1}{(2\pi i)^3} \int\frac{d\beta}{\beta^2} \, d\nu_1 \, d\nu_2 \, 
 \exp\left[\beta \tau+\nu_1 J_1+\nu_2 J_2 + \frac{\beta^2h(\beta)}{\nu_1 \nu_2} \right].
 \end{aligned}
\end{equation}
where the twist $\tau = E-J_1-J_2$ was defined in eq.~\!\eqref{redchempot} and 
each of $\beta$, $\nu_1$ and $\nu_2$ is integrated over the imaginary axis in the complex plane. We will find it useful to perform the inverse Laplace transform over $\nu_1$, $\nu_2$ at first, while keeping $\beta$ fixed. To organize the calculation, let us first define the entropy in a mixed ensemble:
\begin{equation}
\begin{aligned}\label{invlap2}
 e^{S(\beta, J_1, J_2)} 
&:=\frac{1}{(2\pi i)^2} \int d\nu_1 \, d\nu_2 \, 
 \exp\left[\nu_1 J_1+\nu_2 J_2 + \frac{\beta^2h(\beta)}{\nu_1 \nu_2} \right].
 \end{aligned}
\end{equation}
When $\nu_1$ and $\nu_2$ are both small, the $\nu$ integral in eq.~\!\eqref{invlap} (or equivalently in eq.~\!\eqref{invlap2}) can be performed in the saddle point approximation: in this limit, we are, effectively, instructed to extremize the exponent  over the variables $\nu_1, \nu_2$ to find 
\begin{equation}\label{expon}
S(\beta, J_1, J_2)= {\rm Ext}_{ \nu_1, \nu_2} \left( \nu_1 J_1+\nu_2 J_2 + \frac{\beta^2h(\beta)}{\nu_1\nu_2}\right)\,.
\end{equation} 
The extremum over $\nu_i$ (still at fixed $\beta$) occurs when 
\begin{equation}\label{omegain}
    \nu_1 = \left(\frac{J_2 \beta^2h(\beta)}{J_1^2}\right)^{\frac{1}{3}}, 
 \qquad 
\nu_2 =\left(\frac{J_1 \beta^2h(\beta)}{J_2^2}\right)^{\frac{1}{3}}\,.
\end{equation}
The saddle point approximation that led to \eqref{omegain} is valid when 
$\nu_i$ are small enough; this is the case when both angular momenta are large enough.

Upon substituting eq.~\!\eqref{omegain} into eq.~\!\eqref{expon}, followed by using eqs.~\!\eqref{invlap2} and \eqref{invlap}, we find 
\begin{equation}\label{entfun}
\begin{aligned}
    &     S(\tau, J_1,J_2)\sim \ln\left[\frac{1}{2\pi i}   \int d\beta\ e^{\beta\tau+3 (J_1J_2)^{1/3}\beta^{2/3}h(\beta)^{1/3}}\right]\\
    &=\ln\left[\frac{1}{2\pi i} \int d\beta\ \exp\left((J_1J_2)^{1/3}\left(\beta \zeta+3\beta^{2/3}h(\beta)^{1/3} \right)\right)\right]\,,
\end{aligned}
\end{equation}
where $\zeta= \frac{\tau}{(J_1J_2)^{1/3}}$. Since $J_1,J_2\to\infty$, we can perform the above integral in the saddle point approximation (keeping $\zeta$ fixed), leading to 
\begin{equation}
  \ln\left[\frac{1}{2\pi i} \int d\beta\ \exp\left((J_1J_2)^{1/3}\left(\beta \zeta+3\beta^{2/3}h(\beta)^{1/3} \right)\right)\right] \sim (J_1J_2)^{1/3} S^{\rm int}(\zeta)\,,
\end{equation}
where $S^{\rm int}$ is given by 
\begin{equation}\label{intent}
S^{\rm int}(\zeta):={\rm Ext}_{\beta}\left(
\beta \zeta + 3 \beta^{\frac{2}{3}}h^{\frac{1}{3}} (\beta)\right)=\left(\beta_* \zeta +3\beta^{2/3}h(\beta_*)^{1/3} \right)\,.
\end{equation}
Here, $\beta_*$ satisfies\footnote{Note the numerator in this formula is the negative of the derivative of $f(\beta)$ where $f(\beta)=\beta^2h(\beta)$. Here $\zeta$ is positive as $f(\beta)=\beta^2h(\beta)$ is a monotonically increasing function of the temperature and hence a monotonically decreasing function of $\beta$. That this is the case can be seen as follows. From eq.~\!\eqref{thermpfnew} and the fact that there are a finite number of states with negative twist with the contribution to $\partial_\beta \ln Z$ bounded, it follows that for sufficiently small $\nu_i$ and under the technical assumption that the $\nu_i\to 0$ limit of $\ln Z$ commutes with taking derivative w.r.t $\beta$, we have
\begin{equation}\label{lz}
\partial_\beta \ln Z\Big|_{\nu_i}= -\langle \tau \rangle<0\,, 
\end{equation}
where the expectation value is taken in the thermal ensemble \eqref{thermpfnew}. 
Again, using the technical assumption that the $\nu_i\to 0$ limit of $\ln Z$ commutes with taking derivative w.r.t $\beta$, it then follows that the LHS of eq.~\!\eqref{lz} is proportional to $f'(\beta)$, which, therefore is negative. The technical assumption seems very reasonable. This assumption could be violated, for example, in a situation in which a term that is subleading in amplitude (in an expansion in $\nu_i$) has a frequency that grows as $\nu_i\to 0$. In such a situation, however, the function would not have a good Taylor expansion in 
$\nu_i$ -- but our analysis (using the method of \cite{Banerjee:2012iz, Jensen:2012jh}) strongly suggests that it does.}
\begin{equation}\label{betazeta}
    \zeta= -\frac{\left(\beta^2 h(\beta) \right)'\Big|_{\beta=\beta^*}}{\left(\beta_*^{2} h(\beta_*)\right)^{2/3}}\,,
\end{equation}
Thus we conclude that the microcanonical entropy $S(\tau, J_1, J_2)$ assumes a semi-universal form given by
\begin{equation}\label{finansforent}
S(\tau, J_1, J_2)\sim  (J_1 J_2)^{\frac{1}{3}}S^{\rm int}\left(\frac{\tau}{(J_1 J_2)^{\frac{1}{3}}}\right)\,,\quad \tau,J_1,J_2\to\infty\,, \quad \frac{\tau}{(J_1J_2)^{1/3}}\,\ \mathrm{fixed}\,.
\end{equation}

Eqn.\!~ \eqref{finansforent} can be regarded as the main result of this section from a microcanonical viewpoint. This formula has a familiar structure; the equation 
captures the formula for an `extensive' entropy in terms of an intrinsic 
entropy function. In a more familiar context,  extensivity tells us that the entropy of a quantum field theory in the large volume limit takes the form 
\begin{equation}\label{entlv}
\tilde{S}(E)= V \tilde{S}_{int}(E/V)
\end{equation}
As another familiar context, the entropy of ${\cal N}=4$ Yang-Mills on $S^3$, in the deconfined phase (dual to a black hole in the bulk), in the large $N$ limit takes the form 
\begin{equation}\label{entbh}
\tilde{S}(E)=N^2 S_{BH}(E/N^2)
\end{equation}
Eq.~\!\eqref{finansforent} has the structure of the eqs.~\!\eqref{entlv} and 
\eqref{entbh}, with the quantity $(J_1 J_2)^{1/3}$ playing the role of the volume in eq.~\!\eqref{entlv}, or of $N^2$ in eq.~\!\eqref{entbh}. 

While the twist and the entropy are both of order the effective `volume' $(J_1J_2)^{1/3}$ (see \eqref{finansforent}), angular momenta are themselves `super-extensive' in this limit: recall that if $J_1$ and $J_2$ are both taken to be of order, say $\sim  J$, the effective `volume' (and so the twist and the entropy) scales like $J^{2/3}$, slower than $J$. Reworded in terms of the canonical variables, while $J \sim \nu^{-3} \sim \gamma^6(\theta)$, the twist and entropy scale like $\nu^{-2} \sim \gamma^4(\theta)$.

At general values of $\beta$,  the Legendre transform that appears in eq.~\!\eqref{intent} cannot be performed without knowledge of the detailed form of 
$h(\beta)$. We know, however, that when $\beta \ll 1$, $h(\beta)=c'/(4\beta^3).$ Inserting this form into eq.~\!\eqref{intent} we can explicitly perform 
the Legendre transform, and find $\tau^3= c'/(4\beta^4)$ and the formula
\begin{equation}\label{sint}
S^{\rm int}= (64c')^{1/4}\ \tau^{1/4} \,,
\end{equation}
so that using \eqref{finansforent} we have
\begin{equation}\label{entofff}
S=(64c')^{1/4}\ \tau^{1/4} (J_1J_2)^{1/4}\,.
\end{equation}
The result presented in eq.~\!\eqref{entofff} can also be obtained directly by Legendre transforming the universal fluid-dynamical partition function given by  eq.~\!\eqref{uniform}, and expanding that result to leading order in the small $\tau/J$ expansion. 

\subsection{Conjectured semi-universality of the stress tensor} \label{cust}

In \S \ref{Tetrad} we demonstrated that each of the ${\cal L}_n$ (see \eqref{lnlim}), on the appropriately twisted $S^3 \times S^1$, is proportional to $\gamma^4(\theta)$; this result is universal i.e.\ it holds for any CFT$_4$. 
This striking local result also holds in the high-temperature limit reviewed in \S\ref{ttpf} (see  \cite{Bhattacharyya:2007vs}). This is the case at all values of $\omega_i$, i.e.\ even away from the small $\nu_i$ limit. In the high-temperature limit, moreover, more is known. The authors of \cite{Bhattacharyya:2007vs} also demonstrated that the stress tensor in the high-temperature limit takes the simple, universal and local form 
\begin{equation}\label{st1}
T_{\mu\nu}= \frac{c'}{2\pi^2} T^4 \gamma^4 \left(4 u_\mu u_\nu +g_{\mu \nu} \right) 
\end{equation}
at leading order in the high-temperature limit. In this subsection we conjecture that a similar result holds in the 
limit \eqref{limnew} studied in this paper. Concretely, we conjecture that the stress tensor, in this limit, takes the form
\begin{equation}\label{st2}
T_{\mu\nu}= \frac{2T h(T)}{\pi^2} \gamma^4 \left(4 u_\mu u_\nu +g_{\mu \nu} \right).
\end{equation}

We offer the following evidence for the conjecture
\begin{itemize}
    \item In Appendix \ref{strten}, we have verified that this conjecture is indeed true for the stress tensor arising from all zero-derivative and two-derivative contributions to ${\cal L}$ (i.e.\ those listed in \eqref{eq:partition}).
    \item The conjecture is also true for the free massless scalar theory (see eq. H.29 of \cite{Bajaj:2024utv}).
    \item It also holds in ${\cal N}=4$ Yang-Mills theory at large 
    $N$ and strong coupling (see \S \ref{lgym}). One can further see \S \ref{bhst} for the black hole phase (or the black hole component of the stress tensor of the grey galaxy phase), and see eq. (2.44) of \cite{Bajaj:2024utv}.
\end{itemize}

It may well be possible to construct an order-by-order `proof' of this conjecture along the lines of the analysis of \S \ref{Tetrad}. As mentioned in the first itemized point above, we have so far carried out this exercise only at the first two nontrivial orders in the derivative expansion, but have not attempted a general all orders proof along the lines of the analysis of \S \ref{Tetrad}. We leave an attempt of this nature to future work. 

\section{Generalization to arbitrary dimensions and scalings}\label{gad}

In the previous section we have studied the partition function $Z(\beta, \nu_i)$ for a CFT$_4$, in the limit that $\nu_1$ and $\nu_2$ both go to zero in a manner such that their ratio is held fixed. This discussion 
can be generalized in several directions.

\subsection{Even $d>2$ with all $J_i$ comparable}\label{geneven}

Let us first generalize to the study of CFTs in diverse spacetime dimensions. 
To start with let us continue to study theories in even spacetime dimensions, 
i.e.\ CFT$_{2n}$ where $n$ is an integer. As we have seen in \S \ref{d2int} of the introduction, the case $d=2$ is special in some aspects so we will analyse that case separately; in this subsection we assume $n>1$ so $d>2$    \footnote{It is sometimes conjectured that CFTs without higher spin currents do not exist in spacetime dimensions 
greater than 6. In this paper, we remain agnostic about this claim; our considerations apply to the study of non-trivial CFTs in whichever dimension they turn out to exist. Our considerations also apply to free CFTs in every dimension (e.g. the free massless scalar theory which certainly exists in arbitrary dimensions)}. In this case the analogue of the partition function 
\eqref{thermpfnew} is 
\begin{align}\label{thermpfd}
Z&={\rm Tr} e^{- \beta \tau- \sum_{i=1}^{n}\nu_i J_i}\,,\\
&{\rm where} ~~~\nu_i\equiv \beta(1-\omega_i)\,, ~~~\tau= E-\sum_{i=1}^{n}J_i \,.\label{redchempotd}
\end{align} 
The partition function \eqref{thermpfd} is characterized by the $n$ chemical potentials $\nu_i,~~i=1 \ldots n$, 
in addition to the temperature. Let us now study the limit in which all $\nu_i$ are taken to zero, with the inverse temperature $\beta$, and  all ratios of $\nu_i$ held constant. In other words we study the limit  
\begin{equation}\label{limnewgend}
\nu_i \to 0, ~~~(i, j=1\ldots n), ~~~\left(\beta,~\left(\frac{\nu_i}{\nu_j}\right)\right)~~~{\rm fixed}\,.~~~~
\end{equation}
In Appendix \ref{scaleformgend} we generalize the discussion of \S \ref{Tetrad} to this case, and, again,  study the partition function \eqref{thermpfd} using the formalism of \cite{Banerjee:2012iz,Jensen:2012jh}. The discussion of \S \ref{S3S1}, \S \ref{S3S1l} and \S \ref{Tetrad} 
generalizes to the study of \eqref{thermpfd} with only minor modifications. Once again, $\ln Z$, for any CFT,  can be written as the integral of  a local density. Once again, the leading order behaviour of this density function is rather simple in the limit \eqref{limnewgend}; the density scales like $\gamma^{2n}$. For future convenience, we call the temperature dependent proportionality constant behind this scaling  $\frac{2^{n}  h(T) }{\Omega_{2n-1}}$ (i.e.\ ${\mathcal L} =  \frac{2^{n}  h(T) }{\Omega_{2n-1}} \gamma^{2n}$) so that -- at leading order -- 
\begin{equation}\label{intoflocal}
\ln Z(\beta, \nu_i)= \frac{2^{n}  h(T) }{\Omega_{2n-1}}\int_{S^{2n-1}} \gamma^{2n}\,,
\end{equation} 
where $\Omega_{2n-1}$ is the volume of the unit $2n-1$ sphere (the factors of $2^n$ and $\Omega_{2n-1}$ in \eqref{intoflocal} have been chosen to simplify later formulae like \eqref{pfanycft}).
The integral \eqref{intoflocal} is easy to perform (see footnote 11 of \cite{Bhattacharyya_2008}); in the limit \eqref{limnewgend} we find 
\begin{equation}\label{pfanycft}
\ln Z(\beta, \nu_i)= \frac{\beta^n h(\beta)}{\nu_1\ldots\nu_n}\,.
\end{equation}
By Legendre transforming \eqref{pfanycft} and repeating the steps presented from eq.~\!\eqref{expon} to eq.~\!\eqref{finansforent},
we find that the entropy as a function of twist and angular momenta is given by
 \begin{equation}\label{entscfd}
     S(\tau,J_1,\ldots J_n)=(J_1\ldots J_n)^{\frac{1}{n+1}}S^{\mathrm{int}}\left( \frac{\tau}{(J_1\ldots J_n)^{\frac{1}{n+1}}}\right)  \,,\end{equation}
where
\begin{equation}
    S^{\rm int}(x)=\mathrm{Ext}_{\beta}\left( \beta x+ (n+1)\beta^{\frac{n}{n+1}}h^{\frac{1}{n+1}}(\beta)\right)\,.
\end{equation}
Eqn.~\!\eqref{entscfd} holds in the limit  
\begin{equation}\label{largelimit}
    \tau \rightarrow \infty, ~~J_i \rightarrow \infty, ~~ \left(\left(\frac{J_i}{J_j}\right)~,~\frac{\tau}{(J_1\ldots J_n)^{\frac{1}{n+1}}}~~(i,j=1,\ldots n)\right)~~\mathrm{fixed}\,.
\end{equation}

\subsection{$d=3$}\label{ssdthree}

Let us now turn to the study of odd values of $d$; to start with 
let us focus on $d=3$. The rotational group $SO(3)$ has rank one, so we have only a single rotational chemical potential
$\nu$. The analogue of $\gamma$ is 
\begin{equation}\label{gammaform}
\gamma(\theta)= \frac{1}{\sqrt{1-\omega^2 \sin^2 \theta}}
\end{equation}
where $\theta$ is the usual polar angle (`angle with the $z$ axis') on $S^2$. 
We set $\beta(1-\omega)=\nu$ and study the partition function in the 
limit
\begin{equation}\label{limnewdthree}
\nu \to 0, ~~~\beta~~~{\rm fixed}.
\end{equation}

We see that the function $\gamma(\theta)$ peaks at $\theta= \frac{\pi}{2}$ and that $\gamma(\frac{\pi}{2}) \approx \frac{\sqrt{\beta}}{\sqrt{2  \nu}}$ 
(this formula is valid at small $\nu$). At generic values of 
$\theta$, $\gamma(\theta) \sim {\cal O}(1) \ll \gamma(\frac{\pi}{2})$. $\gamma(\theta)$ is of the same order as $\gamma(\frac{\pi}{2})$ only when
$\theta = \frac{\pi}{2} - \delta \theta$ with $\delta \theta$ of order $\sqrt{\nu}$. For such values of $\delta \theta,$
\begin{equation}\label{gammatheta}
\gamma(\theta)=
\frac{1}{\sqrt{\frac{2\nu}{\beta} + \delta\theta^2}}
\end{equation} 
We see that (at leading order) the function $\gamma(\theta)$ is sharply localized around $\frac{\pi}{2}$ with localization width $\sqrt{\frac{2\nu}{\beta}}$.
\textit{(A detailed analysis of localisation in this case can be found in Appendix \ref{3dlocalisation} and \ref{df}.)}

In Appendix \ref{3dtwisted}, we once again generalize the discussion of \S \ref{S3S1}, \S \ref{S3S1l} and \S \ref{Tetrad} to this case. While much 
of the discussion goes through with minor modifications, there is one key 
difference here. In the discussions of \S \ref{S3S1}, \S \ref{S3S1l} and \S \ref{Tetrad}, $\gamma'(\theta)$ was of the same order as $\gamma(\theta)$. For this reason all occurrences of $\gamma'(\theta)$
(e.g. in \eqref{riemancom}, \eqref{sc} and \eqref{eq:field_strength_indiv}) were 
subleading at small $\nu$. (Similarly, for this reason, all contributions from derivatives of $\sigma$, see e.g. \eqref{pss}, were subleading at small $\nu$).

The $d=3$ analogues of eqs. \eqref{riemancom}, \eqref{sc}, \eqref{eq:field_strength_indiv} and \eqref{pss} are all very similar in structure to their four-dimensional counterparts, and are listed explicitly (in Appendix \ref{3dtwisted}) in eqs. \eqref{sc3d}, \eqref{rieman3d}, \eqref{field3d} and \eqref{pss3d}, respectively. 
There is, however, one qualitative difference. We have already noted that, in the current context,  $\gamma(\theta)$ is substantial only in the neighborhood of $\theta=\frac{\pi}{2}$ with width $\sqrt{2\nu/\beta}$. In this neighbourhood 
the function $\gamma(\theta)$ varies substantially (i.e.\ changes by a fraction of order unity) over the angular scale 
$\sqrt{2\nu/\beta}$. As a consequence, every $\partial_\theta$ derivative of $\gamma(\theta)$ is larger than $\gamma(\theta)$ itself by a factor $\propto \nu^{-1/2}$. While this point does not change the final conclusion that all contributions to the density $\cal L$ are of order ${\cal O}(\gamma^3)$ 
(or equivalently $\nu^{-3/2}$; recall this power was ${\cal O}(\gamma^4)$ or $\nu^{-2}$ in 4 dimensions),
it is no longer the case that all contributions at this order are simply proportional to $\gamma^3(\theta)$. We also find contributions of (for instance) the form $\frac{\left( \gamma'(\theta)\right)^2}{\gamma(\theta)}$ 
that are of the same order as terms that are simply proportional to $\gamma^3(\theta)$.  

In Appendix \ref{3dtwisted}, we have performed explicit computations of the contribution to the partition function from two-derivative terms
in the expansion \eqref{eq:partition}, and do in fact find a term of the sort flagged above (namely a term proportional to  $\frac{\left( \gamma'(\theta)\right)^2}{\gamma(\theta)}$).  This term is, in particular, generated by the part of the effective action \eqref{eq:partition} $ \propto (\partial \sigma)^2$. At higher orders in the derivative expansion,  we also expect to see expressions like $\frac{(\gamma'')^2}{\gamma^3}$ or 
terms like $\frac{(\gamma')^4}{\gamma^5}$. In other words we expect the leading order formula for $\ln Z$ (in the limit \eqref{limnewdthree}) to take the form 
\begin{equation}\label{hyu}
\ln Z = \int_{S^2} \left( a_0 \gamma^3 + a_1 \frac{(\gamma')^2}{\gamma}+ a_2\frac{(\gamma')^4}{\gamma^5}+a_3 \frac{(\gamma'')^2}{\gamma^3}+\cdots   \right) 
\end{equation} 
where each of the coefficients $a_0, $ $a_1$ etc., is a theory-dependent function of the temperature. In this dimension, therefore, the local distribution $\cal L$ on the sphere is not universal (as was the case in $d=2n$ in the limit \eqref{largelimit}).

While the detailed form of the thermal effective action density is thus no longer universal, it is still true that (at leading order in the small $\nu$ expansion), the integral \eqref{hyu} is semi-universal. This is because each of the integrals on the RHS of \eqref{hyu} integrates  to a number times $\nu^{-1}$. The intuitive argument for this point is simply that  every term in \eqref{hyu} is a function of `height' $\nu^{-3/2}$ but has a width proportional to  $\sqrt{\nu}$. As a consequence, each term  integrates to an answer of order $\nu^{-1}$. (See a more formal argument for this result in the appendix \ref{formchange}.)

Let us rewrite \eqref{hyu} in the form 
\begin{equation}\label{hyure}
\ln Z= \int d\theta\ I(\theta)\,.
\end{equation}
If we agree to work with a description that is coarse-grained over angular scales large compared to $\sqrt{\nu}$ (the scale over which the integrand in \eqref{hyu} is localized to the equator), we can simply replace \eqref{hyu} with the simpler `universal' formula  
\begin{equation}\label{hyusimp}
I(\theta) = \frac{ h(T)}{2 \pi \nu} \ 
\delta \left( \theta - \frac{\pi}{2} \right) \,, 
\end{equation} 
for some $h(T)$.  The factor of 
$\nu^{-1}$, on the RHS of \eqref{hyusimp}, captures the fact that every term in \eqref{hyu} integrates to an expression of order $\nu^{-1}$, as explained in the previous paragraph. The sharp localization of $I(\theta)$ on the equator is reminiscent of a similar phenomenon in the study of grey galaxies \cite{Kim:2023sig} and giant vortices 
\cite{Cuomo:2022kio}.

We see from \eqref{hyusimp} that we do continue to have a sort of semi-universality of the thermal effective density ${\cal L}$ in an appropriate coarse-grained sense. (We emphasize that the universality in \eqref{hyusimp} breaks down at 
angular length scales $\sqrt{\nu}$ in the $\theta$ direction. For instance, 
the detailed distribution of $\cal L$ on this angular 
scale is different in a free scalar theory and a grey galaxy \cite{Kim:2023sig}. This necessitates the coarse-graining.) Integrating \eqref{hyusimp} over the sphere gives
\begin{equation}\label{lnzthree}
\ln Z= \frac{h(T)}{\nu}\,.
\end{equation}

The microcanonical version of \eqref{hyure} proceeds in a manner completely analogous to \S \ref{semient}, and we find the entropy formula
\begin{equation}\label{entrfo}
S(\tau, J)= \sqrt{J} S^{\rm int}\left( \frac{\tau}{\sqrt{J}}\right) \left(1+o(1)\right)\,,
\end{equation}
in the limit,
\begin{equation}\label{lj4}
J \to \infty, ~~~\tau \rightarrow \infty, ~~~{\rm such~that} ~~~\tau/\sqrt{J}~~~{\rm fixed.}
\end{equation}

\subsection{Arbitrary $d>2$ with only some $\nu_i$ scaled to zero}\label{adgen}

Let us now consider the general case.  If $d=2n$, we parameterize the $S^{2n-1}$ by the $n$ angles $\phi_i$ and the $n$ direction cosines $\mu_i$,  $(i=1 \ldots n)$. The metric on the unit sphere then takes the form 
\begin{equation}
   ds^2= \sum_{i=1}^{n} (d\mu_i^2 + \mu_i^2 d\phi_i^2)\,,
\end{equation}
with
\begin{equation}
    \sum_{i=1}^{n} \mu_i^2 =1\,.
\end{equation}

 If $d=2n+1$, on the other hand we parameterize the $S^{2n}$ by the $n$ angles $\phi_i$ and the $n+1$ direction cosines $\mu_j$, with $i$ running over the range  $(i=1 \ldots n)$
 but $j$ running over the range  $(j=1 \ldots n+1)$. In this case the metric on the unit sphere 
 takes the form 
\begin{equation}
   ds^2= \sum_{i=1}^{n} (d\mu_i^2 + \mu_i^2 d\phi_i^2) + d\mu_{n+1}^{2}\,,
\end{equation}
with
\begin{equation}
    \sum_{i=1}^{n+1} \mu_i^2= 1\,.
\end{equation}

Let us now study the partition function in the limit
\begin{equation}\label{limnewdfour2}
\begin{aligned}
&\nu_i \to 0\,, ~~~(i=1\ldots p)\,,\\
&{\rm while}~~~\left(\beta,~\nu_j ~{\rm for}~(j=p+1 \ldots n)~,~~\left(\frac{\nu_i}{\nu_k}\right)~{\rm for} ~(i,k=1\ldots p)\right)~~~{\rm fixed.}
\end{aligned}
\end{equation}
Note that even though our system has $n$ distinct rotational chemical potentials $\nu_m$ $(m=1,2, \ldots n)$, we are studying the limit in which 
only $p \leq n$ of them are taken small (in such a manner that the ratios of 
any of these $p$ chemical potentials are held fixed in the limit).

In this case $\gamma(\theta)$ is sharply localized around $\mu_{p+1}=\mu_{p+2}= \ldots \mu_n=0$ (when $d=2n$) and around $\mu_{p+1}=\mu_{p+2}= \ldots \mu_{n+1}=0$ (when $d=2n+1$). As in the analysis of the previous subsection, the 
angular scale of localization around each of these surfaces is $\delta \theta 
\sim \sqrt{\nu}$. 

\subsubsection{The local form of the free energy density}

In a manner similar to the $d=3$ discussion of the  previous subsection,  the detailed distribution of $\ln Z$, as a function of $\mu_{p+1}, \mu_{p+2} \ldots $, on the scale $\sqrt{\nu}$, is theory-dependent, and so is non-universal. As in the previous subsection, however, we will now demonstrate that if we smear on angular scales large compared to $\sqrt{\nu}$,  the $\ln Z$ density function reduces to the universal form 
\begin{equation}\label{locallnz}
\begin{split}
I (\mu_i) =  \left( \frac{2^p}{\Omega_{2p-1}} \right) h(T, \omega_j)\,  \delta^{d-2p}\left(\vec y \right)\, \gamma_L^{2p}, 
\quad 
\end{split}
\end{equation}
 In \eqref{locallnz}, ${\vec y}$ is the $(d-2 p)$-dimensional `vector' in tangent space orthogonal to the nontrivial $S^{2p-1}$ defined below \footnote{This can be understood as follows. The $S^{d-1}$ can be imagined as embedded inside an $R^{d}$. We can divide the $R^d$ into $n$ orthogonal two planes (plus one extra dimension when $d$ is odd). The direction cosines are the radial coordinates on these two planes, while $\phi_i$ are the angular coordinates on the two planes. 
 If $\mu_{p+1}, \mu_{p+2} \ldots $ all vanish, this means that we are looking at the part of the sphere that lies entirely in the first $p$ two planes. At any point on this $S^{2p-1}$, the tangent space of the full sphere splits into the tangent space of the $S^{2p-1}$ and the orthogonal space to the $S^{2p-1}$, which is (globally) simply the original embedding 
$R^{d-2p}$. ${\vec y}$ is a position vector in this Cartesian space.}
and $\gamma_L$ is the usual special relativistic factor for a fluid 
revolving on that $S^{2p-1}$, with angular velocities such that $\beta(1-\omega_1)=\nu_1$, $\beta(1-\omega_2)=\nu_2$  $\ldots$  $\beta(1-\omega_p)=\nu_p$.

Eqn.~\!\eqref{locallnz}  tells us that 
the $\ln Z$ density is nonzero only when all $\mu_i$ for $i=p+1, \ldots \lfloor d/2+1/2\rfloor$ 
are zero. On this submanifold we have 
\begin{equation}\label{mulow}
\sum_{i=1}^p {\tilde \mu}_i^2= 1\,,
\end{equation}
where ${\tilde \mu}_i$ are the direction cosines on this $S^{2p-1}$. 
Consequently, ${\cal L}$ is only nonzero in the neighbourhood of a  unit $S^{2p-1}$ embedded inside 
the $S^{d-1}$. 

Eqn \eqref{locallnz} is the higher-dimensional analogue of \eqref{hyusimp}. Recall that we had two arguments for 
\eqref{hyusimp}: one intuitive (see below \eqref{hyusimp}) and the second algebraic (see 
Appendix \ref{formchange}). Analogues of both these arguments can be made for \eqref{locallnz} as well. 

Roughly, the intuitive explanation for \eqref{locallnz} follows. As we have explained, the function $\gamma$ is highly peaked about the $S^{2p-1}$. 
While its peak value is of order $\gamma_L$, its width (in the $(d-2p)$ directions parametrized by the $\mu_j$) is of order $\gamma_L^{-1}$. In leading order in our scaling limit, 
${\cal L}$ is a function that has homogeneity $d$
in powers of $\gamma$ plus derivatives. Each power of $\gamma$, and each derivative, adds an extra factor of $\gamma_L$, so ${\cal L}$ has a maximum value of the order 
$\gamma_L^d$. Integrating over the $(d-2p)$ different $\mu_j$'s
gives an answer proportional to $\gamma_L^d/\gamma_L^{d-2p}=\gamma_L^{2p}$.

(See Appendix \ref{gendimalg} to see a clear algebraic derivation (along the lines of those presented for $d=3$ in Appendix \ref{formchange}) for \eqref{locallnz}.)

\subsubsection{Thermodynamics}

The remaining integral of $I$ over the $2p-1$ coordinates of the 
$S^{2p-1}$ was already performed in \eqref{geneven}. Using \eqref{intoflocal}
we conclude that the thermodynamical partition function, as $\nu_i\rightarrow 0$ for $i=1,\ldots , p\leq n$ is given by 
\begin{equation}\label{nuppart}
    \ln Z=\frac{\beta^p h(\beta,\omega_{p+1},\ldots,\omega_{[\frac{d}{2}]})}{\nu_1\ldots\nu_p}\,.
\end{equation}
Legendre transforming this result, we find that the entropy now takes the 
scaling form
\begin{equation}\label{genforment}
    \frac{S(\tau, J_1\ldots J_{[\frac{d}{2}]})}{(J_1\ldots J_p)^{\frac{1}{p+1}}}=S^{\mathrm{int}}\left( \frac{\tau}{(J_1\ldots J_p)^{\frac{1}{p+1}}}, \frac{J_{p+1}}{(J_1\ldots J_p)^{\frac{1}{p+1}}},\ldots , \frac{J_{[\frac{d}{2}]}}{(J_1\ldots J_p)^{\frac{1}{p+1}}}\right)\,, 
\end{equation}
where 
\begin{equation}\label{lgen}
     S^{\rm int}(x,y_i)=\mathrm{Ext}_{\beta, \nu_{p+1},\ldots, \nu_{[\frac{d}{2}]}}\left( \beta x+\sum_{i=p+1}^{[\frac{d}{2}]}\nu_iy_i + (p+1)\beta^{\frac{n}{n+1}}h^{\frac{1}{p+1}}(\beta,\nu_{p+1},\ldots,\nu_{[\frac{d}{2}]})\right)\,.
\end{equation}
Eqn.~\!\eqref{genforment} holds when, for $i,j=1,\ldots p$ and $k=p+1,\ldots n$, we take the limit:
\begin{equation}
\begin{cases}
    \tau \rightarrow \infty\\ J_i \rightarrow \infty \!\!
\end{cases}\ni\ \left(\frac{J_i}{J_j}\right)~,~\left(\frac{\tau}{(J_1\ldots J_n)^{\frac{1}{n+1}}}~,~\frac{J_{k}}{(J_1\ldots J_n)^{\frac{1}{n+1}}}\right)~~\mathrm{fixed.}
\end{equation}
The details of the derivation of the Legendre transformations are provided in Appendix \ref{scaleformgend}.

\subsection{$d=2$}\label{d=2section}

As we have already explained in \S \ref{d2int}, the 
case $d=2$ is special in many ways. First, the local part of ${\cal L}$ is uniquely determined on general grounds. Second, this answer receives a series of corrections that are suppressed (compared to the leading order answer) only by powers of $\nu$, rather than exponentially in $\frac{1}{\nu}$. These corrections can be viewed as a consequence of the fact that one encounters gapless modes when evaluating the twisted $S^1\times S^1$ partition function. Finally, $d=2$ is different, because we can use modular invariance to make several precise statements without using the thermal effective action 
\eqref{ddimpf}, giving us a check on all our predictions.  In this subsection we provide more details on each of these points. 

In this subsection we largely review earlier work done in $d=2$.  The discussion of this subsection has a somewhat different flavor from  the rest of this paper, and will not feed into any of the subsequent analysis of this paper. 
(Readers uninterested in $d=2$ intricacies can safely skip this subsection.)

\subsubsection{Precise results from modular invariance}\label{sec:MI}

Consider the usual two-dimensional partition function 
\begin{equation}\label{pftd}
Z_{\text{usual}}={\rm Tr} \left( q^{L_0-c/24} {\bar q}^{{\bar L}_0-c/24} \right), ~~~q=e^{2 \pi i \tau}, ~~~~{\bar q}= e^{-2 \pi i {\bar \tau}}\,.
\end{equation}
Using $\Delta= L_0+{\bar L}_0$, ~$J={L_0-{\bar L}_0}$, and $\tau=2 {\bar L}_0$, we see that \eqref{pftd} reduces to the 2d version of \eqref{thermpf} up to a factor of $e^{\beta c/12}$, i.e.\ 
\begin{equation}
Z(\beta,\omega):=\sum_{\Delta, J}e^{-\beta\Delta+\beta\omega J}\, =  \sum_{\tau, J}e^{-\beta\tau-\nu J}=e^{\frac{\beta c}{12}} Z_{\text{usual}}\,\quad \mathrm{Re}(\beta)>0\,,\ |\mathrm{Re}(\omega)|<1\,~~~~|\nu|<\beta
\end{equation}
if we identify
\begin{equation}\label{identtd}
q=e^{-\nu}, ~~~{\bar q}= e^{\nu - 2 \beta}, ~~~{\rm 
i.e} ~~~\tau=\frac{i \nu}{2 \pi},~~~{\bar \tau}=
\frac{-i(2 \beta -\nu)}{2\pi}\,.
\end{equation}
Note that 
\begin{equation}\label{obt}
-\frac{1}{\tau}= \frac{2 \pi i}{\nu}, ~~~~-\frac{1}{{\bar \tau}}=-\frac{2 \pi i}{2\beta -\nu}\,.
\end{equation}
It follows from modular invariance that 
$Z_{\text{usual}}(\tau, {\bar \tau})= Z_{\text{usual}}( -\frac{1}{\tau}, -\frac{1}{{\bar \tau}})$
so that 
\begin{equation}\label{nonreldt}
Z_{\text{usual}}(\beta, \nu)
={\rm Tr} e^{-\frac{4 \pi^2}{\nu}(L_0-c/24)} e^{- \frac{4 \pi^2}{2 \beta -\nu}({\bar L}_0-c/24)}\,.
\end{equation}
In the small $\nu$ limit of interest to this paper, 
$\frac{4 \pi^2}{\nu}$ in \eqref{nonreldt} is large, so the trace in \eqref{nonreldt} is dominated by the state with the lowest value of $L_0$; if we assume unitarity and a twist gap in the spectrum of primaries, this value is  $h=0$. Subject to this restriction, however, states with all values of ${\bar L}_0 ={\bar h}$ contribute to the trace. In other words, the small $\nu$ limit dominantly receives contributions from states with $(h, {\bar h})= (0, {\bar h})$, where 
${\bar h}$ is arbitrary.

If we parameterize states by their dual operators (via the state operator map), we see that the trace \eqref{nonreldt} is dominated by anti-analytic or anti -chiral operators. In theories without an extended (anti) chiral algebra, the only anti-analytic operators are the Virasoro descendants of the vacuum (the rightmoving part of the vacuum module). As this module is generated by ${\bar L}_2$, 
${\bar L}_3 \ldots$, it follows that in the small $\nu$ limit 
\begin{equation}\label{nonreldt1}
Z_{\text{usual}}(\beta, \nu)
=e^{\frac{4 \pi^2 c}{24 \nu}} \times \left[
e^{\frac{\pi^2 c}{12 \beta}} \prod_{n=2}^\infty \left(\frac{1}{1-e^{- \frac{2 \pi^2 n}{\beta}}} \right) \right]\,.
\end{equation}
If the CFT has a larger (anti) chiral algebra than Virasoro, but continues to have a gap for primaries, then the second factor in \eqref{nonreldt} is simply replaced by the vacuum module of the relevant (anti) chiral algebra.

We have so far worked with a CFT on a spatial circle of length $2\pi$. The partition function for a CFT
on a circle of length $L$ is easily obtained from \eqref{nonreldt} by the replacement $\beta \rightarrow \frac{2 \pi \beta}{L}$. We conclude that the 
partition function of a $c>1$ non-rational unitary CFT with non-zero twist gap $2T>0$ in the spectrum of Virasoro primaries takes the form 
\begin{equation}
\begin{aligned}
Z_{\text{usual}}(\beta,\omega)&\underset{\omega\to1}{\sim}  
\frac{e^{\tfrac{\pi L}{\beta} \tfrac{(c-1)}{24} }}{\eta\!\left(\frac{\pi L}{\beta}\right)} \left(1-e^{-\tfrac{\pi L}{\beta}}\right)\;  \times\; e^{\tfrac{L}{\beta (1-\omega)} \tfrac{2\pi c}{24}}\Bigg(1+O\!\left(e^{-\tfrac{L}{\beta (1-\omega)}T}\right)\Bigg)
\end{aligned}
\end{equation}
(recall $\beta(1-\omega)=\nu$).

Consequently, taking the logarithm,  we obtain
\begin{equation}\label{eq:lightcone1}
\ln Z_{\text{usual}} \underset{\omega\to1}{\sim}  
\frac{L}{\beta (1-\omega)}\, \frac{2\pi c}{24}
+ \underbrace{\frac{\pi L}{\beta}\, \frac{(c-1)}{24} -\ln \eta\!\left(\frac{\pi L}{\beta}\right)+\ln\!  \big(1-e^{-\frac{\pi L}{\beta}}\big)}_{\;\;\ O_\beta(\nu^0)\ \mathrm{terms}}+O\!\left(e^{-\tfrac{L}{\beta (1-\omega)}T}\right)\,.
\end{equation}
Finally note that at leading order in $\nu$, we have 
$\ln Z_{\text{usual}} =\ln Z$.
The results reviewed in this subsubsection were worked out and explored in \cite{Collier:2016cls,Kusuki:2018wpa,Benjamin:2019stq, Ghosh:2019rcj}. The implications of the above are rigorously studied in \cite{Pal:2022vqc, Pal:2023cgk, Pal:2025yvz}  with detailed and precise statements about the large spin density of states. See \S\ref{subsec:MC2d} for a quick survey of the results in the microcanonical ensemble.

\subsubsection{Physical interpretation of the leading piece}

In the rest of this subsection, we will understand various aspects of the formula \eqref{eq:lightcone1} from the viewpoint of the thermal effective partition function \eqref{ddimpf}. In this subsubsection, we focus on the leading term. In \S \ref{d2int}, we have explained how it is obtained from a 2d `thermal effective theory' 
\eqref{ddimpf}. In this section we emphasize a Lorentzian physical interpretation of the $\omega$ dependence of \eqref{hgy}.

We consider a $(1+1)$-dimensional CFT on a spatial circle of length $L$ at inverse temperature $\beta$. In \S \ref{S3S1l} we have already explained that the length of this circle, as measured in the comoving fluid frame, becomes $\gamma L$. Moreover, the time dilation also (see \S \ref{S3S1l}) tells us that the rest frame fluid temperature becomes $T \gamma$,  where $\gamma=(1-\omega^2)^{-1/2}$. Putting these two effects together, we see that 
\begin{equation}
\frac{L}{\beta} \;\to\; \frac{L}{\beta}\,\gamma^2 = \frac{L}{\beta(1-\omega^2)},
\end{equation}
physically explaining the factor of $\frac{1}{1-\omega^2}$ in \eqref{hgy}.

\subsubsection{Gapless modes in the twisted partition function for a free massless scalar field}

In this subsection (and in \S \ref{d2int}) we have explained two facts that may at first seem to contradict each other. First, the local part of the thermal effective action \cite{Banerjee:2012iz,Jensen:2012jh,Bhattacharyya:2007vs,Benjamin:2023qsc} is tightly constrained to take the unique form \eqref{hgy}. Second, though the actual answer for $\ln Z$ does take this unique form at leading order in a small $\nu$ expansion, the first correction to this answer is only suppressed by a single power of $\nu$  compared to this leading piece (rather than exponentially suppressed, as the uniqueness of the local form might initially have suggested). In this subsection we will explain that the resolution to this apparent paradox is the following: the thermal effective action is exceptionally nonlocal in $d=2$ because the dimensional reduction on a twisted thermal circle leads to a plethora of zero modes on an effectively very large spatial $S^1$ (see the previous subsection).  In other words, the $d-1=1$ dimensional effective theory not only effectively lives on a circle of length $\gamma^2$, but also has  massless modes on this huge circle. 

Let us first analyse how these massless modes arise in the context of the simple example of a free massless scalar field.  A torus with modular parameters $(\tau, {\bar \tau})$ is parameterized by the flat coordinates 
$(\sigma_1, \sigma_2)$, subject to the identifications
\begin{equation}\label{idento}
\begin{split}
(\sigma_1, \sigma_2)&\sim (\sigma_1+2 \pi, \sigma_2) \sim \left(\sigma_1+ 2 \pi \left( \frac{\tau + {\bar \tau}}{2} \right) , 
\sigma_2 + 2 \pi \left( \frac{\tau-{\bar \tau}}{2i} \right) \right)\,.
\end{split}
\end{equation}

The Euclidean Laplacian has eigenmodes of the form 
$e^{i k_1 \sigma_1+i k_2 \sigma_2}$. These eigenmodes respect the periodicity \eqref{idento} if and only if 
\begin{equation}\label{kxky}
(k_1, k_2)=\left(n_1,~  \frac{i}{\tau -{\bar \tau}} \left( 2 n_2 - n_1 (\tau +{\bar \tau}) \right) \right) \,,
\end{equation}
where $n_1$ and $n_2$ are integers. In our physical situation, $\tau$ and ${\bar \tau}$ are given by \eqref{identtd}, so that $\frac{\tau+{\bar \tau}}{2}= \frac{-i (\beta-\nu)}{2\pi}$, and $\frac{\tau-{\bar \tau}}{2}= i \frac{\beta}{2 \pi}$. Plugging these into \eqref{kxky} yields 
\begin{equation}\label{kxkypi1}
(k_1, k_2)=\left(n_1,~ -i n_1 (1-\frac{\nu}{\beta}) +\frac{2 \pi n_2 }{\beta} \right) \,.
\end{equation}
Now consider the special case $n_2=0$ (i.e.\ the `dimensional reduction' on the thermal circle). 
In this case, 
\begin{equation}\label{kxkypi2}
(k_1, k_2)=n_1\left(1,~ -i(1-\frac{\nu}{\beta}) \right) = n_1(1, i\omega)\,,
\end{equation}
and 
\begin{equation}
k^2= n_1^2 (1-\omega^2)=  \frac{2 n_1^2\nu}{\beta } +{\cal O}(\nu^2)\,.
\end{equation}
It follows that this `dimensional reduction' gives rise to a massless field that effectively lives on a circle of length 
\begin{equation}\label{lenn}
L^2 = \frac{1}{1-\omega^2}= \gamma^2\,,
\end{equation}
as anticipated in the previous subsubsection.
The gapless modes of this massless field effectively live on a very long circle, and are almost anti-chiral. These complicate the thermal effective field theory, presumably giving rise to the  ${\cal O}(\nu^0)$ corrections described above. (Of course, the free scalar field has a much larger (anti-chiral) symmetry algebra than Virasoro.)

Modular invariance of the partition function, along with the presence of zero twist states in the vacuum module, predict a subleading ${\cal O}(\nu^0)$ correction in a generic interacting CFT. We expect the presence of zero modes, as elucidated above in a generic interacting CFT.

\subsubsection{More about the gapless modes}

In this subsubsection we make some further remarks about the gapless modes, especially in the context of free scalar field theory in $2$ dimensions.

The calculation of the earlier subsubsection asserts that if we have $O(1)$ correction to the leading term in the thermal EFT, we should expect the presence of zero modes under the KK reduction along the twisted thermal circle. A natural question is whether the converse is true as well. In fact, we can ask this question in a simpler set-up i.e.\ by keeping $\omega=0$  and taking the $\beta\to 0$ limit. In this subsubsection, we point out that at least for the compactified free Boson, the converse is not true i.e.\ we cannot claim the following is true:
\begin{equation}
\Delta_{\text{gap}}\underset{Subtle}{=}m_{\text{gap}}\,,
\end{equation}
where $\Delta_{\text{gap}}$ corresponds to the gap in the spectrum of scaling dimensions above the vacuum and $m_{\text{gap}}$ corresponds to the mass gap under the KK reduction.

For the compactified boson, modular invariance of the partition function implies that \eqref{hgy} is true up to exponentially suppressed corrections since $\Delta_{\text{gap}}\neq 0$. This might lead us to believe that the zero modes are entirely absent. But this is clearly not true. First of all, there is a zero mode of the scalar field coming from the classical solution $\phi=0$. In the path integral, this gives rise to a contribution proportional to $R$, which is proportional to the volume of the target space. In addition, there are classical solutions of the equations of motion involving zero winding along the time-circle. Finally, for each of these classical trajectories, we have harmonic oscillations, among which there are some which are flat along the time direction and only have oscillations along the spatial direction. Each of these three kinds of modes individually provides non-exponentially suppressed corrections to \eqref{hgy}. However, altogether, they conspire to cancel each other and ensure that the prediction of modular invariance is true i.e.\ \eqref{hgy} is true up to exponentially suppressed corrections. See Appendix \S\ref{app:freeBoson} for a detailed account of this conspiracy.

\subsubsection{More about the microcanonical ensemble} \label{mme}

In \S \ref{subsec:MC2d} we have pointed out that one of the many special features of $d=2$ is that the leading order twist, i.e.\ the twist obtained by differentiating the leading order partition function \eqref{lnztd} w.r.t. $\beta$, vanishes. 

The twist $\tau$, of course,  vanishes only at leading order in the large $J$ limit (i.e.\ $\tau$ fails to scale like $\sqrt{J}$ as one might have naively expected). However $\tau$ does not completely vanish as  \eqref{lnztd} receives subleading $\beta$-dependent corrections. (For example, these corrections can be given by \eqref{eq:lightcone1} if there is a twist gap in the spectrum of Virasoro primaries. Note that the first term in \eqref{eq:lightcone1} is in fact \eqref{lnztd}).  As there is no freedom in the local structure of ${\cal L}$, these corrections arise from nonlocal contributions to ${\cal L}$ (these are larger than what might have been expected because the dimensional reduction on the $d=2$  twisted thermal circle turns out to have massless modes). The correction to \eqref{lnztd} also leads to
a $\tau$ dependent correction to the leading order entropy formula 
\eqref{enttd}. 

This aforementioned correction is universal in the following sense: it is determined if we know the chiral algebra and impose a twist gap in the spectrum of chiral primaries. For CFTs that have a twist gap for Virasoro primaries (no holomorphic primaries) and $c>1$, this correction takes the universal form given in eq.~\!\eqref{eq:lightcone1} \cite{Kusuki:2018wpa,Benjamin:2019stq, Ghosh:2019rcj,Pal:2022vqc, Pal:2023cgk, Pal:2025yvz}. The first correction to \eqref{enttd} occurs at ${\cal O}(J^0)$ (i.e.\ these corrections are independent of $J$). In order to obtain this correction, one needs to perform an honest inverse Laplace transformation, as opposed to a Legendre transformation. In particular a Legendre transformation is only justified if we can perform the inverse Laplace transformation in the saddle point approximation; however, we cannot do the saddle point approximation here unless $\beta$ is very small.

More concretely, consider the inverse Laplace transformation on $Z_{\text{usual}}$. In this case, the conjugate variable to $\beta$ is $\tau-c/12$ i.e we consider
$$\mathcal{L}^{-1}\left[\frac{e^{\tfrac{2\pi^2}{\beta} \tfrac{(c-1)}{24} }}{\eta\!\left(\frac{2\pi^2}{\beta}\right)} \left(1-e^{-\tfrac{2\pi^2}{\beta}}\right)\right][\tau-c/12]\,,$$
to obtain the correction. From the perspective of $Z=\sum_{J,\tau} e^{-\beta\tau-\nu J}$, the shift in $\tau$ can be understood arising from an extra order one correction in going from \eqref{eq:lightcone1} to the expression for $\ln Z$ i.e  $$\mathcal{L}^{-1}\left[\frac{e^{\tfrac{2\pi^2}{\beta} \tfrac{(c-1)}{24} }}{\eta\!\left(\frac{2\pi^2}{\beta}\right)} \left(1-e^{-\tfrac{2\pi^2}{\beta}}\right)e^{\beta c/12}\right][\tau]\,,$$
which is the same as the above. Now we note that 
\begin{equation}
    \eta\!\left(\frac{2\pi^2}{\beta}\right)^{-1}= \sqrt{\frac{\pi}{\beta}}\eta\!\left(\frac{\beta}{2}\right)^{-1}\,.
\end{equation}
Thus we need to inverse Laplace transform
\begin{equation}
\begin{aligned}
&\mathcal{L}^{-1}\left[\sqrt{\frac{\pi}{\beta}}e^{\tfrac{2\pi^2}{\beta} \tfrac{(c-1)}{24} } \left(1-e^{-\tfrac{2\pi^2}{\beta}}\right)\left(\sum_{k=1}^{\infty}P(k)e^{-k\beta/2+\beta/12}\right)\right][\tau-c/12]\\
&=\sum_{k=1}^{\infty}P(k) \left(\mathcal{L}^{-1}\left[\sqrt{\frac{\pi}{\beta}}e^{\tfrac{2\pi^2}{\beta} \tfrac{(c-1)}{24} } \left(1-e^{-\tfrac{2\pi^2}{\beta}}\right)\right] \left[\tau-\tfrac{c-1}{12}-\tfrac{k}{2}\right]\right) \,,
\end{aligned}
\end{equation}
where $P(k)$ is the number of partitions of $k$. One can verify that the inverse Laplace transformation has support 
\begin{equation}\label{tauineq}
	\tau \geq \frac{c-1}{12}\,.
\end{equation}

To summarize, we find that accounting for the correction term the expectation value of $\tau$ is now an (order unity) function of $\beta$. As $\beta$ is 
varied, $\tau$ takes arbitrary values (of order unity) subject to the inequality \eqref{tauineq}. Subject to this inequality, the order one correction is completely universal. In fact, ref.~\cite{Pal:2025yvz} establishes an even stronger universality; 
with appropriate assumptions and definitions \footnote{This extended universality holds on assuming a twist gap in the spectrum of Virasoro primaries, and  if one smears the fixed-spin–$J$ density of Virasoro primary states against a smooth, compactly supported function of the twist.}, this universality extends to the order $O(J^{-N})$ (for any $N>0$) and can easily be generalized to density of all states.

\subsubsection{Extended domain of validity in theories with a twist gap}

As derived, eq.~\!\eqref{eq:univ2d} holds in the limit 
$\beta^2(1-\omega^2)\to 0$. In the large central charge limit, and under the assumption of sparse low lying spectrum of scaling dimensions and twists, however, the regime of validity of  ~\!\eqref{eq:univ2d} extends to the following larger domain:
\begin{equation}
	\beta^2(1-\omega^2)\leqslant 4\pi^2\,.
\end{equation}
(i.e.\ the regime of temperatures that lies above the Hawking-Page transition). 
This extension of the domain was conjectured (with numerical evidence) in the ref.~\!\cite{Hartman:2014oaa}, and proven recently in the ref.~\!\cite{Dey:2024nje}. In the microcanonical ensemble, this tells us that the  universal entropy formula, described above, applies not just at asymptotically large values of $J$, but, in fact, for all $J$
and $\tau$ that are large enough so that 
\begin{equation}~\!
	\left(\tau-\tfrac{c}{12}\right)
	\left(\tau+2|J|-\tfrac{c}{12}\right)
	\geq \left(\tfrac{c}{12}\right)^2\,,
\end{equation}
which extends the usual universal entropy formula at fixed twist, and large spin.

\section{Free scalar fields} \label{free}

\subsection{Partition function of a free massless scalar field in general dimension} \label{fgd}

In this short subsection, we explicitly compute the partition function of a free scalar field in arbitrary dimension. We then take the scaling limit $\nu_i \to 0$ at fixed $\beta$, and demonstrate that the partition function takes the general form predicted in the previous section. 
Focussing on the symmetric limit $\nu_i \to 0$ $\forall i$, we then read off the functions $h(\beta)$ in this particular case. In subsequent subsections we then go on to discuss the surprisingly intricate properties of these functions. The reader can also use the exact formulae below to take limits in which a few  $\nu_i$ are taken to zero at a fixed ratio, while others are held fixed, and once again verify that the partition function takes the form predicted in the previous section.

Let us start by presenting the well-known exact expression for the partition function of a free scalar field on $S^1\times S^{2n-1}$ (when $d=2n$) and on $S^1\times S^{2n}$ (when $d=2n+1$). In the first case, the single-particle partition function is given by 
\begin{equation}\label{speven}
\begin{split}
    Z_{\rm{sp}}^{S^1\times S^{2n-1}}(\beta)&=\frac{e^{-\beta \Delta}(1-e^{-2\beta})}{\prod_{i=1}^n\left(1-e^{-\beta(1-\omega_i)}\right)\left(1-e^{-\beta(1+\omega_i)}\right)}\\&
    =\frac{e^{-\beta \left(\frac{2n-2}{2}\right)}(1-e^{-2\beta})}{\prod_{i=1}^n\left(1-e^{-\beta(1-\omega_i)}\right)\left(1-e^{-\beta(1+\omega_i)}\right)}\\
    &=\frac{e^{-\beta \left(\frac{2n-2}{2}\right)}(1-e^{-2\beta})}{\prod_{i=1}^n\left(1-e^{-\nu_i}\right)\left(1-e^{-2 \beta +\nu_i}\right)}\,.
    \end{split}
\end{equation}
In the first line of \eqref{speven} we evaluated the partition function over $\partial_{\mu_1} \partial_{\mu_2} \ldots \partial_{\mu_n} \phi$
after subtracting out those terms proportional to the equation of motion $\partial^2 \phi$. In the second line, we have used the conformal dimension for a free massless scalar field $\Delta=\frac{2n-2}{2}$ (as the conformal dimension of a massless scalar on $S^{d-1}\times S^1$ is $d/2-1$).

In the second case we have (using $\Delta=n-1/2$)
\begin{align}\label{spodd}
    Z_{\rm{sp}}^{S^1\times S^{2n}}(\beta)&=\frac{e^{-\beta \Delta}(1-e^{-2\beta})}{(1-e^{-\beta})\prod_{i=1}^n\left(1-e^{-\beta(1-\omega_i)}\right)\left(1-e^{-\beta(1+\omega_i)}\right)}\nonumber\\
    &=\frac{e^{-\beta \left(\frac{2n-1}{2}\right)}(1-e^{-2\beta})}{(1-e^{-\beta})\prod_{i=1}^n\left(1-e^{-\beta(1-\omega_i)}\right)\left(1-e^{-\beta(1+\omega_i)}\right)}\\
    &=\frac{e^{-\beta \left(\frac{2n-1}{2}\right)}(1-e^{-2\beta})}{(1-e^{-\beta})\prod_{i=1}^n\left(1-e^{-\nu_i}\right)\left(1-e^{-2\beta+\nu_i}\right)}\,.
\end{align}
By multiparticling the single-particle partition function \eqref{speven} or \eqref{spodd}, we find
\begin{align}\label{mpodd}
    \ln Z&=\sum_{q=1}^{\infty}\frac{Z_{\rm{sp}}(q\beta, \omega_i )}{q}=\sum_{q=1}^{\infty}\frac{Z_{\rm{sp}}(q\beta, q\nu_i )}{q}\,.
\end{align}
From the partition function \eqref{mpodd}, we can read off the function\footnote{In the limit that all $\nu_i$ are small, but of the same order, we can estimate the summation in \eqref{mpodd} as follows. We first split the sum over $q$ into a sum from $q=1$ to $Q$, and $Q+1$ to $\infty$. We choose $Q$ so that $Q \gg 1$ but 
$\nu Q\ll 1$. The sum from $Q+1$ to $\infty$ is of order $e^{-Q \Delta_n}$, and so should roughly be thought of as capturing contributions of order  $e^{-a/\nu}$ where $a$ is a number of order unity. 
The sum from $1$ to $Q$ can be simplified by Taylor expanding each term in $q \nu_i$. This Taylor expansion is always valid, as $q\nu_i < Q \nu_i \ll 1$. One can then interchange the two summations and perform the summation over $q$ separately for each term in the power series. The factors of $e^{-\beta \Delta_n q}$ ensure that each of these summations over $q$ is `convergent' (i.e.\ the summation over $q$ can be extended to $1,..., \infty$, at the price of introducing errors that are only of order $e^{-\beta Q \Delta_n}$). This procedure generates a Taylor expansion of $\ln Z$ in the variables $\nu_i$. The argument above strongly suggests that the Taylor expansion only has errors of order $e^{-a/\nu}$. This suggests that the Borel transform of the sum over the Taylor series dummy variable has a singularity somewhere in the Borel plane. This singularity may represent an instanton, which would then be very interesting to find.  At any rate, the zeroth order term in the Taylor series is an excellent approximation when all $\nu_i$ are small. One can also use the methods which were used to prove the Meinardus theorem, as explained in chapter $6$ of \cite{Andrews_1984}, and used in \cite{Melia:2020pzd} (see Appendix \ref{app:Rigor} for the rigorous approach).}
 $h(\beta)$ (appearing in \eqref{zint} with $p=n$) for a massless scalar field on $S^1\times S^{2n-1}$ 
\begin{equation}\label{evenh}
  \lim_{\omega_i\to 1}\left[\left(\prod_{i=1}^{n}(1-\omega_i)\right)\ln Z^{S^1\times S^{2n-1}}(\beta)\right]=:  h^{d=2n}(\beta) = \sum_{q=1}^\infty \frac{1}{q}\frac{e^{-q\beta(\frac{2n-2}{2})}}{(q\beta)^n(1-e^{-2q\beta})^{n-1}}\,.
\end{equation}
For example, in $d=4$, we have
\begin{align}\label{hd4}
    h^{d=4}(\beta) &= \sum_{q=1}^{\infty} \frac{\text{csch}(q\beta)}{2q^3\beta^2} = \frac{\pi^4}{180\beta^3} - \frac{\pi^2}{72\beta} + \frac{3\zeta(3)}{8\pi^2} - \frac{7\beta}{1440}+ \sum_{m=1}^\infty \sum_{\ell|m} \frac{(-1)^{\ell+1} }{\pi^2 \ell^3} e^{-\frac{2\pi^2 m}{\beta}}\,,
\end{align}
and in $d=6$, we have
\begin{align} \label{hd6}
    h^{d=6}(\beta) &= \sum_{q=1}^\infty \frac{\text{csch}(q\beta)^2}{4q^4\beta^3} \nonumber \\
    &= \frac{\pi^6}{3780\beta^5} - \frac{\pi^4}{1080\beta^3} + \frac{\pi^2}{360\beta} - \frac{\zeta(5)}{\pi^4} + \frac{\beta}{756} - \sum_{m=1}^\infty \sum_{\ell|m} \left(\frac{m}{\pi^2\ell^5\beta} + \frac{2}{\pi^4\ell^5}\right) e^{-\frac{2\pi^2 m}{\beta}}\,.
\end{align}
See Appendix \ref{app:freeScalarComp} for the derivation of \eqref{hd4} and \eqref{hd6} and a general set of formulae \eqref{eq:alleven}, \eqref{eq:allevenPert} and \eqref{eq:allevenNonPert} valid for all even dimensions. See also Appendix C of \cite{Benjamin:2023qsc} for related expressions at high temperature instead of high angular velocity.

In odd dimensions, $d = 2n+1$, we have 
\begin{equation}
\label{oddh}
    h^{d=2n+1}(\beta) = \sum_{q=1}^\infty \frac{1}{q}\frac{e^{-q\beta(\frac{2n-1}{2})}}{(q\beta)^n(1-e^{-q\beta})(1-e^{-2q\beta})^{n-1}}\,.
\end{equation}
The expression \eqref{oddh} has an asymptotic expansion in $\beta$ (in contrast to the convergent expansions in even dimensions; see the appendix \ref{app:freeScalarComp} for a derivation). For example, in $d=3$,
\begin{equation}
\begin{split}
h^{d=3}(\beta)&=\sum_{q=1}^\infty \frac{\text{csch}(\frac{q\beta}{2})}{2q^2\beta} \\&\sim \frac{\zeta(3)}{\beta^{2}} + \frac{\ln \beta}{24} - \frac{\gamma}{24} 
+ \frac{\zeta'(2)}{4\pi^{2}} - \frac{\ln\pi}{24} 
\\&\quad + \sum_{k=1}^{\infty} (-1)^{k+1} \frac{1}{2\pi^{2}} 
\left( \frac{\beta}{2\pi} \right)^{2k} 
\zeta(1-2k)\,\zeta(2k+2)\,\left(1 - 2^{-1-2k}\right)\,,
\end{split}
\end{equation}
where we use $\sim$ to indicate an asymptotic series (due to the factorial growth in $\zeta(1-2k)$ at large $k$).
Note that the perturbative part of each of these expansions is a power series expansion in $\beta^2$ (rather than $\beta$), reflecting the fact that each subsequent term in the derivative expansion in \eqref{psexpl} has two additional derivatives (this follows from the fact that each curvature tensor itself has two derivatives, and 
KK field strengths -- which are one derivative objects -- appear in the action in pairs in parity invariant theories (i.e.\ in Lorentz invariants that do not involve the $\epsilon$ tensor). The $\ln \beta$ terms come from the gapless mode in the free boson.

In $d\geq 4$, we can also consider taking only some of fugacities close to $1$. For instance, let us consider $d=4$, and only $\omega_1$ close to $1$, leaving $\omega_2$ a free parameter. (This is an example of the general limit studied in Sec. \ref{adgen}.) We see explicitly that in this more complicated limit, we reproduce the semiuniversal form as expected. More precisely, for the 4d free scalar, we get:
\begin{align}
    \lim_{\omega_1\rightarrow 1} \left[(1-\omega_1) Z^{S^1 \times S^3}(\beta)\right] &= \sum_{q=1}^{\infty} \frac{\text{csch}\left(\frac{q\beta(1-\omega_2)}{2}\right) \text{csch}\left(\frac{q\beta(1+\omega_2)}{2}\right)}{4q^2\beta} \nonumber \\
    &= \frac{\pi^4}{90\beta^3(1-\omega_2^2)} - \frac{\pi^2(1+\omega_2^2)}{72\beta(1-\omega_2^2)} + O(\beta^0).
\label{eq:onlyonefugacity}
\end{align}
As expected, (\ref{eq:onlyonefugacity}) indeed takes the general semi-universal form expected when one of the chemical potentials is taken close to $1$.

\subsection{Instanton interpretation of the non-perturbative terms}\label{instnonpert}

As we have already emphasized in \S \ref{fmsint}, it is striking that the small $\beta$ expansion of the even $d$ $h(\beta)$ has (an infinite series of) `non-perturbative' corrections, even though the perturbative (power series)  part of this function truncates at finite order. In this subsection we sketch a physical explanation for these non-perturbative corrections, and outline an argument that suggests that the feature displayed above -- namely that the theory has  non-perturbative corrections unconnected to the high order behaviour of its perturbative expansion  -- is likely special to the free theory. We work in $d=4$, leaving a careful generalization to arbitrary $d$ to future work. 

\subsubsection{Kaluza-Klein reduction}

The partition function $Z$ may be computed by evaluating the Euclidean field theory path integral  on the metric $g_{\mu\nu}$ presented in \eqref{eq:boosted_metric}. As the CFT under study is Weyl invariant, we get the same path integral by studying it on the Weyl rescaled metric $G_{\mu\nu}= \gamma^2(\theta) {\tilde g}_{\mu\nu}$, 
provided we also scale all dynamical fields in the appropriate manner. In the case of the free scalar field, we obtain the new $\phi$ by dividing the old $\phi$ by $\gamma(\theta)$.

Our new rescaled metric is given by  
\begin{equation}\label{gmunudef}
ds^2= G_{\mu\nu}dx^\mu dx^\nu= \left(d\tau+ia_idx^i\right)^2+g_{ij}dx^idx^j\,,
\end{equation}
with
\begin{equation}\label{eq:fieldsnew}
    \begin{split}
        a &=a_idx^i= -\gamma^2(\theta)\left ( \omega_1\sin^2\theta d\phi_1 + \omega_2 \cos^2\theta d\phi_2 \right)\,,\\ g_{ij}dx^{i}dx^{j} &= \gamma(\theta)^2\left(d\theta ^2 + \gamma (\theta )^2 \left(\omega_1 d\phi_1 \sin ^2(\theta )+\omega_2d\phi_2 \cos ^2(\theta )\right)^2\right)\\  &\qquad\qquad+\gamma(\theta)^2\left(d\phi_1^2 \sin ^2(\theta )+d\phi_2^2 \cos ^2(\theta)\right)\,.
    \end{split}
\end{equation}
Note that while the base metric has been Weyl rescaled, the KK gauge field is, of course, unaffected by this scaling.

The Euclidean action for the free scalar field on the metric $G_{\mu\nu}$ takes the form 
\begin{equation}\label{euclidact}
    S_E=\frac{1}{2}\int d^4x\sqrt{G}\left(G^{\mu\nu}\partial_\mu\phi\partial_\nu\phi+\frac{1}{6}R\phi^2\right)\,,
\end{equation}
with $\tau \sim \tau+\beta$. We expand our field as 
\begin{equation}\label{expmodes}
    \phi=\sum_{n=-\infty}^{\infty}\phi_n(x) e^{\frac{2i n\pi\tau}{\beta}}\,.
\end{equation}
Under the coordinate transformation 
\begin{equation}\label{taugauge}
\begin{split}
\tau ={\tilde \tau} + f(x^i)\,,
\end{split}
\end{equation} 
we have 
\begin{equation}
    \phi=\sum_{n=-\infty}^{\infty}{\tilde \phi_n}(x) e^{\frac{2i n\pi\tilde\tau}{\beta}}\,,
\end{equation}
with ${\tilde \phi}_n = e^{ \frac{2\pi i n f(x_i)}{\beta}}\phi_n$. The metric \eqref{gmunudef} can also be recast 
$ds^2 =(d {\tilde \tau} +{\tilde a})^2 + \ldots$ with \begin{equation}\label{tildea}
{\tilde a}_i = a_i -i \partial_i f\,.
\end{equation}
It follows from the requirement of coordinate invariance that the action for $\phi_n$ must be written entirely in terms of the KK $U(1)$ covariant derivatives (with this choice, ${\tilde D}_i {\tilde \phi}_n= e^{\frac{2\pi i nf}{\beta}}D_i\phi_n$) 
\begin{equation}\label{covder}
D_i \phi_n= \left( \partial_i + \frac{2 \pi n}{\beta} a_i\right) \phi_n\,.
\end{equation}
Upon inserting \eqref{expmodes} into \eqref{euclidact} and performing a little algebra we find that the action $S_E$ takes the form 
\begin{equation} \label{actf}
    S_E=\sum_{n=-\infty}^{\infty}\int d^3x\sqrt{g_3}\Bigg(\Big(\frac{4\pi^2n^2}{\beta^2}+\frac{1}{6}R_{4d}\Big)\phi_n\phi_{-n}+g^{ij}D_i\phi_nD_j\phi_{-n}\Bigg)
\end{equation}
The 4-dimensional Ricci scalar, $R_{4d}$, is given in terms of the 3-dimensional Ricci scalar ($R$), and the KK field strengths $f_{ij}$ as
\begin{equation}
    R_{4d}=R+\frac{1}{4}f^2\,,
\end{equation}
where $f_{ij}=\partial_ia_j-\partial_ja_i$. Using 
\eqref{eq:fieldsnew}, and performing some algebra, we find 
\begin{align}\label{3drf}
   R&=\frac{40 \left(\omega _1^2-1\right) \left(\omega _2^2-1\right)}{\left(\omega _1-\omega _2\right) \left(\omega _1+\omega _2\right) \cos (2 \theta )-\omega _1^2-\omega _2^2+2}+6 \omega _1^2+6 \omega _2^2-14\,, \\
   f^2&=\frac{4 \left(\left(\omega _1-\omega _2\right) \left(\omega _1+\omega _2\right) \cos (2 \theta )+\left(1-2 \omega _2^2\right) \omega _1^2+\omega _2^2\right)}{1-\omega _1^2 \sin ^2(\theta )-\omega _2^2 \cos ^2(\theta )}\,.
\end{align}
The final expression for $R_{4d}=R+f^2/4$ is rather ugly in general. Note, however, that this quantity is independent of $\beta$, and so is subleading (compared to the mass term, which is  proportional to $\frac{4 \pi^2 n^2}{\beta^2}$)  in the small $\beta$ expansion. 
For this reason this term can be ignored at leading order in small $\beta$. 

The reader can verify the following. If we take $1-\omega_1$ and $1-\omega_2$ to both be of order $\epsilon$, then, while $R$ and $f^2$ 
are individually of order unity, the combination $R+ f^2/4$ is of order $\epsilon$. Consequently this term also vanishes at leading order in the limit $\omega_i \to 1$. For both these reasons, this term will not enter our saddle point equations (see below for more details) though it could affect subleading corrections. 

\subsubsection{Specializing to $\omega_1=\omega_2$}

In the rest of this subsection, we focus, for simplicity, on the special case $\omega_1=\omega_2$. In this special case 
\eqref{3drf} simplifies dramatically, and we find\footnote{The vanishing of the RHS on the last line of \eqref{smallR} (when $\omega \to 1$) is an illustration of one of the comments made at the end of the last subsubsection.} 
\begin{equation} \label{smallR}
\begin{split}
    R=6-8\, \omega ^2\,,\quad f^2=8\omega^2\,\quad \Rightarrow R+\frac{1}{4}f^2=6\left(1-\omega^2\right)\,,
    \end{split}
\end{equation}
and \eqref{actf} simplifies to 
\begin{equation}\label{fineuclidactn}
    S_E=\sum_{n=-\infty}^{\infty}\int d^3x\sqrt{g_3}\Bigg(\Bigg(\frac{4\pi^2n^2}{\beta^2} + (1-\omega^2) \Bigg)\phi_n\phi_{-n}+g^{ij}D_i\phi_nD_j\phi_{-n}\Bigg).
\end{equation}
Our action describes the motion of a particle of mass 
\begin{equation}\label{masspart}
m_n= \frac{2 \pi |n|}{\beta}\left( 1 + \frac{ \beta^2(1-\omega^2)}{8 \pi^2 n^2} + \ldots \right) \,,
\end{equation}
and charge $\frac{2\pi n}{\beta}$, propagating on the 3d background 
\eqref{eq:fieldsnew}. In \eqref{masspart} the terms denoted by $\ldots$ are suppressed both in $\beta$ and in $1-\omega$. 

We will now proceed to evaluate the  path integral over the fields $\phi_n$ in 
\eqref{fineuclidactn} using the worldline formalism. 
This formalism is well suited to the study of the small  $\beta$ limit, because the fact that $m_n$ and the charge are both large (when $\beta$
is small) will allow us to evaluate the worldline path integral in the saddle point approximation.

\subsubsection{The worldline action}

The `one loop' (determinant) path integral over the field $\phi_n$ may be obtained by performing the worldline path integral over all closed particle  trajectories 
\begin{equation}\label{wlpt}
\ln Z = \int Dx^i e^{-S_E}
\end{equation}
where\footnote{We have fixed the sign (and factor of $i$) in the second term in \eqref{actforpart} as follows. The particles  in question
are created by the field $\phi_n$. Inserting the second term in \eqref{actforpart} into \eqref{wlpt} yields the Wilson line 
\begin{equation}\label{wll}
W= e^{\frac{2\pi n}{\beta}\int_
{x_i}^{x_f}a}
\end{equation}
Using \eqref{tildea}, we see that under the gauge transformation \eqref{taugauge}, the open Wilson line $W$ in \eqref{wll} transforms as 
\begin{equation}\label{wlt}
W \rightarrow e^{\frac{2 \pi i n}{\beta} f(x_i)} W e^{-\frac{2 \pi i n}{\beta}f(x_f)}.
\end{equation} 
which is the correct gauge transformation property for a worldline of a particle of charge $n$ (see around \eqref{taugauge}).
} 
\begin{equation}\label{actforpart}
    S_E=m_n\int\sqrt{g_{ij}dx^idx^j}-\frac{2\pi n}{\beta}\int a.
\end{equation}

Because every term in  \eqref{actforpart} has an overall factor of order  $\frac{1}{\beta}$,  the worldline path integral is well-approximated by  saddle points in the small $\beta$ limit. The leading saddle point is a worldline that maps to a single point. This saddle has vanishing action, and the  expansion around it is generated by closed particle trajectories  of length of order $\beta$.  As these paths are much smaller than the curvature scale of the sphere, their contributions reproduce the local terms in the effective action (the power series expansion \eqref{ddimpf}).

Exponentially suppressed corrections to this leading perturbative answer are obtained from `instantons'; i.e.\ solutions to the equations of motion following from \eqref{actforpart} that extend over a large distance (of order of the length scale of the base manifold). We now turn to a complete characterization of these instantonic solutions.

\subsubsection{Symmetries of instantonic solutions}

The isometry group of a round $S^{3}$ is $SO(4)=SU(2)_L \times SU(2)_R$. Turning on a generic twist ($\omega_1\neq \omega_2$) breaks this isometry down to $U(1)_L  \times U(1)_R$ (where the $U(1)$'s are generated by rotations in the $Q^z_L$ and $Q^z_R$ directions respectively). However, the special case $\omega_1=\omega_2$ preserves the larger subgroup $U(1)_L \times SU(2)_R$. This follows both on general grounds (it is the largest isometry group that commutes with this twist), and from direct computation. Adopting the notation 
$(\sigma_1, \sigma_2, \sigma_3)$ to denote the right-invariant one-forms, in Appendix \ref{formskilling} we verify that the base metric can be written as a linear combination (with numerical coefficients) of  $\sigma_1^2+\sigma_2^2$ and $\sigma_3^2$, the gauge field as proportional to $\sigma_3$, and its field strength as proportional to 
$\sigma_1 \wedge \sigma_2$.

This large amount of symmetry allows us to efficiently find all solutions to the equations of motion following from \eqref{actforpart}.
We can introduce a parameter (let us call it $\tau$) along the worldline of the particle \eqref{actforpart}. Of course, monotonic redefinitions of this parameter are an invariance of the action \eqref{actforpart}, so we can make any choice we like. The particle motion now has conserved 
Noether charges (see Appendix \ref{chargesgen}) corresponding to its $U(1)_L \times SU(2)_R$ symmetry, and solutions to the equations of motion can be parameterized (in part) by these charges. As the $SU(2)_R$ charges, themselves, transform in the three-dimensional vectorial representation of $SU(2)$, it is sufficient to first find all solutions with $Q^x_R=Q^y_R=0$, and then 
obtain the most general solution by performing an $SU(2)_R$ rotation
on these special solutions.

In what follows it is sometimes helpful to use the coordinates 
\begin{equation}\label{phipsi1}
\phi= \phi_1+ \phi_2\,, ~~~\psi=\phi_1-\phi_2\,.
\end{equation}
In Appendix \ref{gengeod} we demonstrate that $Q^x_R=Q^y_R=0$ if and only if 
\begin{equation}\label{thetadot}
{\dot \theta}=0\,,
\end{equation}
and 
\begin{equation}\label{othercond}
{\rm Either} ~~~\sin 2 \theta=0\,, ~~~{\rm or} ~~~\frac{\dot \psi}{\dot \phi}= 
\frac{1}{\omega}\,.
\end{equation}
It follows from \eqref{thetadot} that
all solutions with $Q^x_R=Q^y_R=0$ live at a fixed value of $\theta$. 

\subsubsection{Generic solutions}

The second category of solutions listed in \eqref{othercond} generates generic solutions of
the action \eqref{actforpart}. These solutions appear in a two parameter family, labelled by the fixed value of $\theta$, and the `initial' value of $\psi$ (i.e.\ the value of $\psi$ at any $\phi$). Acting on these solutions with $SU(2)_R$ rotations adds two additional parameters, yielding a four parameter set of solutions to the equations of motion. These are the generic solutions. If we split the three coordinates of the $S^3$ into two spatial and one `time' coordinate, we expect generic solutions to be labelled by their two initial positions and two initial momenta, and hence to appear in a four parameter set.

These generic solutions appear not to be of interest to this paper for the following reason. Let us first suppose that $\omega$ is an irrational number. In that case the fact that $\frac{d \psi}{d \phi}=\frac{1}{\omega}$
tells us that the worldline described by these solutions never closes back on itself, and so is infinite in extent. Such solutions, therefore, have infinite action, and do not contribute to the path integral. 

Since we expect the partition function to be an analytic function of $\omega$,  the corresponding solutions presumably do not contribute even when $\omega$ equals a rational number like $\frac{p}{q}$. \footnote{In this case a finite action solution (which winds $q$ times) does exist, but the expectation of analyticity suggests that, perhaps,  these solutions do not lie on the steepest descent contour. It would be interesting to understand this better.} For this reason we do not consider these solutions any further.

\subsubsection{Special solutions}

The first condition in \eqref{othercond} -- namely 
$\sin 2 \theta=0$ -- is met either when $\theta=0$ or when $\theta=\frac{\pi}{2}$. 

Let us first study the solution that lives at $\theta=0$.
When $\theta=0$, we  see from \eqref{eq:fieldsnew} 
that all values of $\phi_1$ represent the same point (this is like the origin in polar coordinates) and so the value of this coordinate is meaningless. In this case our worldline extends from $\phi_2=0$ to 
$\phi_2=2 \pi$ at $\theta=0$ (this is an `equator' of the warped $S^3$). The charges of this solution turn out to be 
$Q_R^z=-Q_L^z=-\frac{1}{(1-\omega^2)}\left(m_n+\frac{2\pi n \omega}{\beta}\right)$.

On performing an $SU(2)_R$ rotation on this solution, we generate a 
two parameter set of solutions given by 
\begin{equation}\label{thetannotpsinot}
\theta=\theta_0, ~~~\psi=\psi_0\,,
\end{equation}
where $\theta_0$ and $\psi_0$ are both constants. Using the formulae for charges given in Appendix \ref{chargesgen}, the charges of this solution are given by 
\begin{equation}\label{chargessoln}
\begin{split}
&Q_R^x=\frac{1}{(1-\omega^2)}\left(m_n+\frac{2\pi n \omega}{\beta}\right)\cos{\psi_0}\sin{2\theta_0}\,,\\
&Q_R^y=-\frac{1}{(1-\omega^2)}\left(m_n+\frac{2\pi n \omega}{\beta}\right)\sin{\psi_0}\sin{2\theta_0}\,, \\
& Q_R^z= -\frac{\cos{2\theta_0}}{(1-\omega^2)}\left(m_n+\frac{2\pi n \omega}{\beta}\right)\,,\\
&Q_L^z= \frac{1}{(1-\omega^2)}\left(m_n+\frac{2\pi n \omega}{\beta}\right)\,.
\end{split}
\end{equation}
The SU(2)$_R$ Casimir is given by the following
\begin{equation}\label{casimir}
    (Q_R^{x})^{2}+(Q_R^{y})^{2}+(Q_R^{z})^{2}=\frac{1}{(1-\omega^2)^2}\left(m_n+\frac{2\pi n \omega}{\beta}\right)^2=(Q_L^{z})^{2}\,.
\end{equation}
Note that the Casimir is independent of $\theta_0$ and $\psi_0$, as might have been expected from the fact that solutions with different values of $\theta_0$ and $\psi_0$ all lie in the same $SU(2)_R$ orbit of solutions.

These configurations solve the equations of motion for every choice of $\theta_0$ and $\psi_0$. The particular choice $\theta_0=\frac{\pi}{2}$
reproduces the second special solution flagged at the beginning of this subsubsection (second `first case' solution of \eqref{othercond}).

As these solutions all lie in the same $SU(2)_R$ symmetry orbit, they all have the same easily evaluated action, which turns out to equal
\begin{equation}\label{actconfig}
S_E= m_n \frac{2 \pi}{1-\omega^2}  +\frac{2 \pi n}{ \beta} \frac{2\pi \omega}{1-\omega^2}
\end{equation}
where $m_n$ is given in \eqref{masspart}. Let us choose $n<0$. In this case we find 
\begin{equation}\label{hact}
S_E= \frac{4 \pi^2 |n|}{\beta(1+\omega)} + \frac{\beta}{2 |n|} + \ldots
\end{equation}
where the terms $\ldots$ are even further subleading both in $\beta$ and in $1-\omega^2$. 

In the small $\beta$ limit (where our instanton analysis is relevant) the second term in \eqref{hact} can be ignored. Upon setting 
$\omega=1$ we obtain $S_E=  \frac{2\pi^2 |n|}{\beta}$, in perfect agreement with the exponents in \eqref{hd4}. The instantons described in this subsection can, therefore, be viewed as an explanation of the non-perturbative corrections in \eqref{hd4}. 

\subsubsection{Comments on the prefactor}\label{prefac}

Eqn.\!~\eqref{hd4} predicts that the contribution of the leading non-perturbative exponential to $\ln Z$ is 
\begin{equation}\label{yuy}
\ln Z_{\rm non-pert}= \frac{4e^{\frac{-2\pi^2 n}{\beta}}}{\pi^2(1-\omega^2)^2}\,.
\end{equation}

The two most interesting aspects of the prefactor in 
\eqref{yuy} are 
\begin{itemize}
\item Its dependence on $\omega$ (which diverges as $\omega \to 1$).
\item Its (positive) sign.
\end{itemize}
In this brief subsubsection we present some comments on these two aspects. 

To compute the prefactor, we need to expand particle trajectories to quadratic order around the saddle point and perform the Gaussian integral over fluctuations. In the current context, this expansion is easy to perform. Note that our saddle point is given by the worldline particle `moving' from $\phi=0$ to $\phi=4\pi$ (recall, $\phi=\phi_1+\phi_2$) at a fixed value of $\theta$ and $\psi$. Since $\phi$
is periodic with periodicity $4\pi$, we expand the fluctuations, $\delta \theta(\phi)$ and $\delta \psi(\phi)$, around this saddle point as 
\begin{equation}\label{deltathetaphi}
\begin{split}
&\delta \theta(\phi)= \sum_p \delta \theta_p e^{i \frac{p}{2} \phi}\\
& \delta \psi(\phi)= \sum_p \delta \psi_p e^{i \frac{p}{2} \phi}\\
\end{split}
\end{equation}
where $p$ runs over all integers. While $\delta \theta$
and $\delta \psi$ mix with each other at quadratic order, translational invariance in $\phi$ ensures that modes with momentum $p$ couple only to modes with momentum $-p$. Consequently, the fluctuation matrix can be diagonalised separately at each $|p|$. The relevant computations are presented in Appendix \ref{quadfluc}. The generic `eigenvalue' turns out to be finite in the limit $|\omega| \to 1$. However, there are two special cases. 

The eigenvalues at $p=0$ both vanish. This fact has a natural interpretation. Modes at $p=0$ continue to be 
$\phi$ independent, and so represent motion on the two parameter family of solutions labelled by $\theta_0$ and $\psi_0$. The integral over these zero modes is equal to the `volume' of this space of solutions. The space of solutions is our base manifold \eqref{gmunudef} with \eqref{eq:fieldsnew}, modded out by the length of the geodesics. The volume of the base is easily computed and is $\propto \gamma^4$. The length of the geodesic is also easily computed and is $\propto \gamma^2$. Consequently, the `volume' of our space of solutions is proportional to $\gamma^4/\gamma^2=\gamma^2= 
\frac{1}{1-\omega^2}$. 

The second special case happens at $p=1$. At this value of $p$, one of the two `eigenvalues' of the quadratic form turns out to be proportional to 
\begin{equation}\label{proptominus}
-(1-\omega^2)\,.
\end{equation}
The Gaussian integral over the corresponding modes yields a second factor of 
$(1-\omega^2)^{-1}$. We obtain a factor of $(1-\omega^2)^{-1}$
rather than $1/\sqrt{1-\omega^2}$ because the mode in question is complex.

Combining the special factors from $p=0$ and $p=1$ gives the total factor $(1-\omega^2)^{-2}$, explaining the $(1-\omega^2)$ dependence of the singularity of the coefficient of \eqref{yuy} as $|\omega|\to 1$ (recall that all other `eigenvalues' are finite as $|\omega| \to 1$, so the integral over the relevant modes does not generate further singularities).

The modes that gave rise to the first factor of 
$(1-\omega^2)^{-1}$ had a clear interpretation; they were the zero modes on the manifold of solutions. We will now explain that the second special mode -- the mode at $p=1$ -- also has a simple physical interpretation; it describes the `slipping mode' on 
the warped $S^3$. 

In order to see this, consider the instanton located at $\theta_0=0$. As explained above, the instanton at this value consists of a worldline wrapping the $\phi_2$ direction. Now consider the configurations that continue to wrap the $\phi_2$ direction, but this time at a fixed nonzero value of $\theta$, and also at a fixed value of $\phi_1$. As $\theta$ is increased from $0$ to $\frac{\pi}{2}$, 
this worldline evolves from an equator to a zero size loop. It follows, 
in other words, that motion along $\theta$ for this one parameter set of configurations describes the `slipping mode', a mode that follows the rubber band as it slips off the warped $S^3$. 

Let us compute the action of this one parameter set of `slipping configurations' as a function of $\theta$. As $\theta$ and $\phi_1$ are constant on any one of these configurations, the action is easily computed, and is  given by
\begin{equation}
    S_E=\frac{2\pi m_n}{(1-\omega^2)}\sqrt{(1-\omega^2\sin^2{\theta})}\cos{\theta}\,.
\end{equation}
Expanding this near $\theta=0$, we have
\begin{equation}
    S_E=\frac{
    2\pi m_n
    }{1-\omega^2}\left(1-\frac{\delta\theta^2}{2}(1-\omega^2)\right)\,.
\end{equation}
We see that this mode has a {\it negative} Euclidean action, proportional to $1-\omega^2$, exactly as in the case \eqref{proptominus}. We can thus identify the light $p=1$ mode with the 
slipping mode (actually the two slipping modes;  recall the slipping modes are parameterized also by the value of the angle $\phi_1$) described in this paragraph. We emphasize that these modes are light and tachyonic in the limit $\omega \to 1$. As we decrease $\omega$, these modes become increasingly tachyonic, and are the two robustly tachyonic modes on the round $S^3$ (see \cite{Maldacena:2024spf}). 

The light modes of our path integral supply a complete explanation for the factor of $\frac{1}{(1-\omega^2)^2}$ in \eqref{yuy}. However, they also leave us with one puzzle that we have not yet resolved. As the two $p=1$ modes are tachyonic, the determinant of each of these modes naively appears with one factor of $i$, suggesting that the sign of the RHS of \eqref{yuy} should have been negative rather than positive. This prediction, which appears to be how things work at $\omega=0$ \cite{Maldacena:2024spf}, does not match the sign of \eqref{yuy}. The saddle switches signs at $\omega \approx 0.476858$, which we derive in Appendix \ref{app:freeScalarComp}. We are unsure of the physical reason for this sign change, and leave a resolution of this puzzle to future work. Curiously, the quantity $\partial_\beta \ln Z(\beta,\omega)$ (instead of $\ln Z(\beta,\omega)$) undergoes a sign switch in front of $\beta^{-2} e^{-\frac{4\pi^2}{\beta(1+\omega)}}$ at precisely $\omega=\frac13$. It is conceivable that the saddle point contour undergoes a `Stokes shift' at an intermediate value of $\omega$, and that (for values of $\omega$ that are larger than this critical value), the new contour approaches the saddle via imaginary (rather than real) values of the amplitude of the tachyonic mode.

\subsection{Instanton effects in  interacting CFTs}\label{intinst}

In this section we have demonstrated that the function $h(\beta)$ receives a nontrivial contribution from worldline instantons. 
Our instantons contribute to the partition function, even in the strict limit $|\omega|\rightarrow 1$, because their action $\left(\frac{4\pi^2 n}{\beta(1+\omega)}\right)$ is finite in this limit. As we have explained around \eqref{actconfig}, the finiteness of the action is  a consequence of a delicate cancellation between the two terms in that equation, each of which was itself divergent in the limit $|\omega| \rightarrow 1$. 
The cancellation was possible because the mass $m_n$ (coefficient of the first term in \eqref{actconfig}) was numerically almost exactly equal to the charge $\frac{2 \pi n}{\beta}$ of the field $\phi_n$.   

It is now natural to wonder about the following question. Suppose we have an interacting (rather than free) CFT. Let us suppose that our CFT is relatively weakly interacting  (e.g. it could be some version of a Bank-Zaks fixed point), and that one of the fields that enters its Lagrangian (in a duality frame in which the theory is weakly interacting) is a scalar. We should, once again, expect the path integral over this scalar field to have worldline instantonic contributions. Will these contributions differ significantly from their free scalar cousins?

We will now argue that it is reasonable to expect that this is indeed the case. In the situation outlined above, we should, once again, expect the instanton action to be given by 
\begin{equation}\label{intact}
    S_E=\tilde{m}_n\frac{2\pi}{1-\omega^2}+\frac{2\pi n}{\beta}\frac{2\pi\omega}{1-\omega^2}.
\end{equation}
While the charge of the $n^{th}$ mode under the KK gauge field is exactly $\frac{2\pi n}{\beta}$, independent of interactions, its mass 
${\tilde m}_n$ is typically renormalized by interaction effects. In particular, the renormalization generically ensures that 
$m_0^2 = \frac{\alpha^2}{\beta^2}$ (where $\alpha$ is a positive real number). Indeed the fact that interaction effects generate a thermal mass for $m_0$ is what ensures that the effective 3-dimensional field theory is gapped (the starting point in the discussion of `thermal effective actions', \cite{Banerjee:2012iz,Jensen:2012jh,Shaghoulian:2015lcn,Benjamin:2023qsc}). 

Like $m_0$, the masses of the $n^{th}$ KK modes are also renormalized. 
In some computable examples (large $N$ vector-like theories interacting with Chern-Simons theory at large $N$ in 3-dimensions \cite{Giombi:2011kc,Jain:2013gza,Jain:2013py,Jain:2014nza,Minwalla:2015sca,Choudhury:2018iwf,Aharony:2018pjn,Dey:2018ykx,Dey:2019ihe,Halder:2019foo}), the renormalized value of ${\tilde m}_n$ turns out to take the very simple form 
\begin{align}\label{intmass}
    \tilde{m}_n=\sqrt{\frac{\alpha^2}{\beta^2}+\left(\frac{2\pi n}{\beta}\right)^2}. 
\end{align}
More generally, it follows from dimensional analysis that $\beta^2 {\tilde m}_n^2$ is a pure number, and we expect, on physical grounds, 
that this number is larger than the free value $\left(\frac{2\pi n}{\beta}\right)^2$. Inserting a renormalized mass of this form into 
\eqref{intact} (and once again choosing $n<0$), we find that the delicate cancellation is obstructed, and that the instanton action is now proportional to 
\begin{equation}\label{instact}
S_E = \frac{A_n}{\beta(1-\omega^2)}\,,
\end{equation}
where $A_n$ is a positive number set by the effective coupling of the CFT under study. 

The key point is that the action \eqref{instact} now diverges in the 
limit $|\omega| \to 1$, and so no longer contributes to the path integral in the limit \eqref{limnew}. 

The discussion of this subsection suggests that the fact that $h(\beta)$
receives contributions from real instantonic corrections is an artifact of the free approximation; the instanton studied in this subsection should not contribute to $h(\beta)$ in the interacting limit. This fact presumably goes hand in hand with the expectation that the perturbative expansion of the sphere partition function includes terms of all orders in the $\beta$ expansion when the CFT in question is interacting. 

It would be very interesting to make the qualitative discussion of this subsection more quantitative, perhaps in the context of simple examples. We leave this exercise to future work.

\subsection{Free massive scalar field in $3+1$ dimensions}\label{fms}

In this section we check that the general predictions of \S \ref{mt} are, indeed, borne out in the example of the theory of a  free massive scalar field in $3+1$ dimensions.

The action for the minimally coupled massive scalar field on $S^1\times S^3$ is given by
\begin{equation}
S = \frac{1}{2} \int_{S^{1} \times S^{3}} d^{4}x \sqrt{g} \left( \partial^{\mu} \phi \partial_{\mu} \phi + m^2 \phi^{2} \right).
\end{equation}
The partition function takes the form
\begin{align}
    \ln Z&=-\sum_{l=0}^{\infty}\sum_{m_L,m_R=-\frac{l}{2}}^{\frac{l}{2}}\ln \left(1-e^{-\beta E_l + \beta\left(\omega_1 +\omega_2\right)m_L + \beta \left(\omega_1-\omega_2\right) m_R}\right)\nonumber\\
  &  =\sum_{q=1}^{\infty}\sum_{l=0}^{\infty}\sum_{m_L,m_R=-\frac{l}{2}}^{\frac{l}{2}}\frac{1}{q}\,e^{-q\beta\left(E_l-(\omega_1+\omega_2)m_L-(\omega_1-\omega_2)m_R\right)}
\end{align}
where $E_l=\left(l(l+2)+m^2\right)^{\frac{1}{2}}$. In going from the first to the second line, we have Taylor expanded the logarithm. We  can now perform the sum over $m_L$ and $m_R$ easily. We find \begin{align}
    \ln Z&=\sum_{l=0}^{\infty}\sum_{q=1}^{\infty}\frac{e^{-\beta q E_l}}{q}\left(\frac{\cosh \left(\beta  (l+1) q \omega _1\right)-\cosh \left(\beta  (l+1) q \omega _2\right)}{\cosh \left(\beta  q \omega _1\right)-\cosh \left(\beta  q \omega _2\right)}\right),
\end{align}
We can now break up the sum over $l$ into two sums: one from $l=0$ to $l= L-1$, and one from $l=L$ to $l= \infty$, where $L$ is chosen to be a suitably large number (which, however, will not be taken to scale like an inverse power of $\nu_i$). The first summation is a sum over $L$ terms, each of order unity. In the second summation we Taylor expand the expression for $E_l$ in the exponent (this is valid because $m^2/L^2\ll 1$) and then perform the sum. Note that, in the large angular momentum regime,
\begin{align}
    e^{-\beta q E_l}\approx e^{-(\beta  (l+1) q)}-\frac{(m^2-1) \left(\beta  q e^{-(\beta  (l+1) q)}\right)}{2 (l+1)}+O\left(m^4/L^2\right)\label{eq:massexpansion}
\end{align}
Let us take the leading term in the above expansion, and perform the sum over $l$ which leads to
\begin{align}
    \ln Z^0&=\sum_{l=0}^{\infty}\sum_{q=1}^{\infty}\frac{e^{-\beta q (l+1)}}{q}\left(\frac{\cosh \left(\beta  (l+1) q \omega _1\right)-\cosh \left(\beta  (l+1) q \omega _2\right)}{\cosh \left(\beta  q \omega _1\right)-\cosh \left(\beta  q \omega _2\right)}\right)\nonumber\\
    &=\sum_{q=1}^{\infty}\frac{1}{q}\frac{2 e^{\beta  q \left(\omega _1+\omega _2+2\right)} \sinh (\beta  q)}{\left(e^{\beta  q}-e^{\beta  q \omega _1}\right) \left(e^{\beta  q \left(\omega _1+1\right)}-1\right) \left(e^{\beta  q}-e^{\beta  q \omega _2}\right) \left(e^{\beta  q \left(\omega _2+1\right)}-1\right)}.
\end{align}
Expanding the leading order partition function around $\omega_1\rightarrow 1$ and $\omega_2\rightarrow 1$, and collecting the leading order divergent piece gives:
\begin{align}
    \ln Z^0&\approx\sum_{q=1}^{\infty}\frac{1}{(1-\omega^2_1)(1-\omega^2_2)}\frac{2}{ q^3\beta^2\sinh(q\beta)}+O\left(\frac{1}{1-\omega}\right).\label{leadingpart}
\end{align}
In the expression above, we have performed the summation over $l$ from $0$ to 
$\infty$ (rather than $L$ to $\infty$); the error this introduces is of order 
$L$.

Let us now take the subleading correction in \eqref{eq:massexpansion} and obtain its contribution to the partition function.
\begin{align}
    &\ln Z^1=-(\beta\hat{m}^2)\sum_{l=0}^{\infty}\sum_{q=1}^{\infty}\frac{1}{q}\left(\frac{\beta qe^{-\beta q(l+1)}}{2(l+1)}\right)\left(\frac{\cosh \left(\beta  (l+1) q \omega _1\right)-\cosh \left(\beta  (l+1) q \omega _2\right)}{\cosh \left(\beta  q \omega _1\right)-\cosh \left(\beta  q \omega _2\right)}\right)\nonumber\\
    &=-(\beta \hat{m}^2)\sum_{q=1}^{\infty}\frac{1}{4 \left(\cosh \left(\beta  q \omega _1\right)-\cosh \left(\beta  q \omega _2\right)\right)}\Big[\ln \left(1-e^{-\beta  q \left(1-\omega _1\right)}\right)+\ln \left(1-e^{-q\beta   \left(\omega _1+1\right)}\right)\nonumber\\
    &~~~~~~~~~~~~~~~~~~-\ln \left(1-e^{-\beta  q \left(1-\omega _2\right)}\right)-\ln \left(1-e^{-q\beta   \left(\omega _2+1\right)}\right)\Big],
\end{align}
where $\hat{m}^2=m^2-1$.
In the second line, we have performed the sum over $l$; once again this has introduced an error of order $m^2 L$.

Let us expand $\ln Z^1$ around the $\omega_1\rightarrow 1$ and the $\omega_2 \rightarrow 1$ limit. We find
\begin{align}
    \ln Z^1\approx -\sum_q \frac{\hat{m}^2 \left(\ln \left(\beta q \left(1-\omega _1\right)\right)-\ln \left(\beta q\left(1-\omega _2\right)\right)\right)}{2q \beta\sinh(\beta q)\, \left(\omega^2_1-\omega^2_2\right)}\label{subleadingpart}.
\end{align}

We choose the quantity $L$ to be large compared to unity, but small compared to all inverse powers of $1-\omega_i$. With this choice, the errors in our estimates above, which are all of order $L$, are subleading (in $(1-\omega_i)$) compared to both $\ln Z^0$ and $\ln Z^1$, and so can be ignored. 
The $\omega_i$ dependencies of $\ln Z^0$ in \eqref{leadingpart} and $\ln Z^1$ in \eqref{subleadingpart} have a simple interpretation. It is not difficult to verify that 
\begin{equation}\label{ver}
\begin{split}
&\int_{S^3}  \gamma^4 (\theta) = \frac{2\pi^2}{(1-\omega_1^2)(1-\omega_2^2)}\,,\\
 &\int_{S^3} \gamma^2 (\theta) = -2\pi^2\frac{\ln\left(\frac{1-\omega_1^2}{1-\omega_2^2}\right)}{\omega_1^2 - \omega_2^2}\,.
\end{split}
\end{equation}
We see that  $\ln Z^0 \propto \int_{S^3} \gamma^4(\theta)$ and 
$\ln Z^1 \propto \int_{S^3} \gamma^2(\theta)$, in perfect agreement with our general expectation that the leading term in the $m$ independent local part of the effective action should scale like $\gamma(\theta)^4$. The leading part of the action (in the term proportional to $m^2$) should scale like $\gamma(\theta)^2$ (see \eqref{gammascal1} with $r=2$, as appropriate for a mass term).

\section{Large $N$ ${\cal N}=4$ Yang-Mills theory at high $J$}\label{lgym}

In this section we perform an independent study of the entropy (in the limit \eqref{lj}) of ${\cal N}=4$ Yang-Mills at large $N$ and large $\lambda$. We analyze this theory using the dual bulk description provided by the AdS/CFT correspondence. 
Our motivation  for this study is threefold.
\begin{itemize}
\item First to  subject the general predictions of this paper to a nontrivial test.
\item Second, to examine the structural properties of the  function $S^{\rm int}$ in an example of a family of strongly coupled field theories, labelled by the discrete parameter $N$,  in the large $N$ (or large central charge) limit. We will see that $S^{\rm int}$ undergoes sharp phase transitions (as a
function of the scaled twist $\tau/(J_1J_2)^{1/3}$) in this limit.
\item Third, to identify and investigate an interesting new simplifying limit of the quantum numbers of this much studied theory.
\end{itemize}

\subsection{The three relevant phases}\label{Thrphase}

In the large $N$ limit, ${\cal N}=4$ Yang-Mills has three qualitatively distinct thermal phases, namely 
\begin{enumerate}
	\item Black hole phases.
	\item Grey galaxy phases.
	\item Thermal gas phases.
\end{enumerate}

\subsubsection{Preliminary remarks on the black hole phase}

A black hole phase is a bulk configuration whose  energy, angular momentum and entropy are all dominantly carried by a bulk black hole. In such a phase, the bulk gas -- which is always present in thermal equilibrium with any black hole -- carries only ${\cal O}\left(\frac{1}{N^2} \right)$ of all of the thermodynamical charges (e.g.\ angular momentum, twist, entropy) of the system. At large $N$, consequently, this gas (and, presumably, all other quantum corrections)
can be ignored for thermodynamical purposes. The thermodynamics of such phases can be derived from the formulae of classical black hole thermodynamics. 

AdS$_5 \times S^5$ hosts at least two qualitatively distinct black hole phases: those  involving black holes that are smeared on the $S^5$ (i.e.\ black holes whose event horizon has the topology $S^3 \times S^5$) and those involving black holes that are localized on the $S^5$ (i.e.\ black holes whose event horizon has the topology $S^8$). While localized black holes typically dominate over their smeared cousins at 
low energies, the reverse is true at high energies\footnote{Localized black hole solutions presumably do not even exist 
at high enough energies, as they fail to `fit' on the $S^5$ in this limit. Note that the situation is different for localized black hole that carry high angular momentum on the $S^5$ (all throughout this paper we only study states that carry momentum in the  $AdS_5$ directions); such localized black holes can carry high $SO(6)$ charge while being `small enough' to fit on the $S^5$; the large charge simply comes from the fact that they have large momentum on the $S^5$. See the introduction of \cite{Bhattacharyya:2010yg} for 
relevant related remarks.}. As our focus in this paper is on extremely  high angular momenta and twists (see \eqref{lj}), we restrict our attention to smeared black holes, i.e.\ to  Kerr-AdS$_5$ black holes in an appropriate scaling limit. 

As an aside, we note that the entropy of (very) small black holes in $AdS_5 \times S^5$ also takes a scaling form- with some similarities to, but other important differences from  \eqref{finansforent}, see appendix \ref{sbhads} for details. As small black holes in $AdS_5\times S^5$ never dominate the ensemble in the limit \eqref{lj}, the differences between the scalings of Appendix \ref{sbhads} and \eqref{finansforent} are unsurprising and inconsequential.

\subsubsection{Preliminary remarks on the grey galaxy phase}

Grey galaxies were constructed and studied in \cite{Kim:2023sig, Bajaj:2024utv} at total twist and total angular momenta of order $N^2$. Over this range of  charges, grey galaxies consist of a central Kerr-AdS$_5 \times S^5$ black hole in equilibrium with a bulk gas. What distinguishes grey galaxies from black hole configurations is  that at least one of the black hole angular velocities ($\omega_1$ or $\omega_2$) is tuned to be parametrically (in $1/N$) close to unity. In \cite{Kim:2023sig, Bajaj:2024utv}, this scaling  was chosen to ensure that the bulk gas (whose partition function diverges as $\omega_i \to 1$), also carries an angular momentum of order $N^2$; with this scaling, however, the twist and entropy of this gas turn out to be  parametrically smaller than $N^2$ and so thermodynamically negligible\footnote{Grey galaxies have also recently appeared in studies of the supersymmetric spectrum of ${\cal N}=4$ Yang-Mills in \cite{Choi:2025lck}.}. As we explain in more detail later in this subsubsection, in this paper we will also have the occasion to study grey galaxies in a slightly generalized scaling limit - one in which the entropy and twist of the gas is sometimes comparable to that of the black hole.

At leading order in $1/N$,  the thermodynamics of the grey galaxies of \cite{Kim:2023sig, Bajaj:2024utv} is that of a non-interacting mix of a black hole and the bulk thermal gas, with the division of angular momentum  between the black hole and the gas being determined by the principle of entropy maximization.

In this paper we reinvestigate the thermodynamics of grey galaxies at total angular momenta that are parametrically larger than $N^2$. More precisely, we study grey galaxies whose total angular momentum lies in the range 
\begin{equation}\label{jon}
N^2 \ll \sqrt{J_1J_2} \ll N^6.
\end{equation}
We pause now to explain why we restrict attention to the angular momentum range \eqref{jon}.

We focus on angular momenta that obey $ \sqrt{J_1J_2} \gg N^2$, because the limit \eqref{lj}, of interest to this paper, instructs us to make all angular momenta large, and $G_5 \sqrt{J_1 J_2} \sim \frac{\sqrt{J_1J_2}}{N^2}$ is the natural  measure of angular momentum in supergravity units. Consequently, $ \sqrt{J_1J_2} \gg N^2$ is simply the condition that the grey galaxies we study carry large angular momenta in (apparently) natural units. 

The reason we focus on the charge regime  $\sqrt{J_1J_2} \ll N^6$ is less principled. Technical limitations (as we now explain) force us to work with angular momenta that do not violate this inequality. In more detail, it turns out that neither the self interactions of the gas, nor the interaction between the gas and the central black hole, can be ignored when $\sqrt{J_1J_2}$(of the gas) is of order $N^6$ or larger. At these charges the non-interacting approximations of \cite{Kim:2023sig, Bajaj:2024utv} no longer apply, and the solutions \cite{Kim:2023sig, Bajaj:2024utv} need to be reanalyzed, with interactions taken into account. Such a reanalysis promises to be fascinating - but also rather involved. In this paper we do not attempt to grapple with this reanalysis, but simply restrict our attention to angular momenta that are upper bounded as in \eqref{jon}. We emphasize that our technical limitations have a price; the inequality 
$\sqrt{J_1J_2}\ll N^6$ stops us from exploring the true infinite $J$ limit 
\eqref{lj}. In this paper we only explore the grey galaxy phase of ${\cal N}=4$ Yang-Mills in an intermediate charge regime; at angular momenta that are large in supergravity units, but small in units of $N^6$. We hope that future work (on grey galaxy solutions that deal with interactions) improves this unsatisfactory state of affairs.

As we have mentioned above,  gas-gas and gas-black hole interactions are both negligible in the  angular momentum range \eqref{jon}, and grey galaxies continue to be well-approximated as a non-interacting mix of a black hole and a free bulk gas. We demonstrate below that $\ln Z$ and the entropy of grey galaxies do, indeed, take the scaling forms presented in \eqref{sn} 
and \eqref{finansforentint} in this range of angular momentum. 

Even though the non-interacting approximation of \cite{Kim:2023sig, Bajaj:2024utv} applies to the range of charges given by eq.\!~\eqref{jon}, grey galaxies  display qualitatively new features at these parametrically larger values of angular momenta (i.e.\ at angular momenta that are parametrically larger than $N^2$). In particular, it turns out that (as already mentioned above) in this charge regime, the twist and entropy of the gas component are sometimes comparable to or larger than those of the black hole component. \footnote{\label{footln}Whether this happens depends on the scaled twist $\tau/(J_1J_2)^{1/3}$. When the scaled twist is of order $N^{\frac{2}{3}}$ (i.e.\ near the upper end of the range \eqref{ggint}), grey galaxies are qualitatively similar to their $J \sim N^2$ counterparts: while the gas angular momentum is comparable 
to that of the black hole, its twist and entropy are parametrically smaller than those of the black hole. 
When the scaled twist is of order unity (i.e.\ near the lower end of the range \eqref{ggint}), on the other hand, it turns out that while the gas angular momentum is parametrically {\it larger} than that of the black hole, its twist and entropy are comparable to or larger than that of the black hole. This last behaviour is never seen at angular momenta of order $N^2$ (the total angular momentum needs to be at least of order $N^3$ for this second range of behaviours to be displayed).} This is never the case when angular momenta are of order $N^2$. In fact, the smallest value of angular momenta at which this happens is when  \begin{equation}\label{minbh}\sqrt{J_1J_2} \sim N^3.\end{equation} With this scaling (as we will see below) the twist of the grey galaxy phase varies between an upper limit of order $N^{\frac{8}{3}}$ to a lower limit of order $N^2$. Near the lower limit of this range, the twist (and entropy) of the solution are distributed between the black hole and the gas in a comparable manner.

\subsubsection{Preliminary remarks on the gas phase}

Finally, let us turn to the gas phase; i.e.\ the bulk configuration that hosts a bulk gas but no black holes. As in the previous subsection, we are only able to study this phase in the intermediate angular momentum range \eqref{jon}, and the gas self interactions cannot be ignored at $J_i \sim N^6$ or larger. Therefore, going to higher values of $\sqrt{J_1 J_2}$ than \eqref{jon} requires a new supergravity solution, whose construction we leave to future work. 

 In more traditional AdS/CFT studies of ${\cal N}=4$ thermodynamics, the bulk gas is famously the dominant phase at charges that are of order unity. This comes about because black holes simply do not exist at such charges. In this section we are interested not in charges of order unity but, instead, in values of angular momenta that are parametrically large in supergravity units (see  \eqref{jon}). Nonetheless, we will explain below that the gas is the dominant phase at values of the scaled twist that lie below a critical number of order unity (see \eqref{casenowint}). 

It may, at first, seem surprising that a gas dominates over black holes at large charges. The reason this happens, however, is actually rather similar to the one cited above (for charges of order unity); 
Kerr-AdS$_5$ black holes have a twist gap, and simply do not exist at values of the scaled twist below a critical number of order $N^{\frac{2}{3}}$ (see Appendix \ref{extbh}). In contrast, the bulk gas has no significant twist gap; it has states all the way down to arbitrarily low values of the scaled twist.
 
It follows that at values of the scaled twist smaller than a critical number times 
$N^{\frac{2}{3}}$, the bulk system simply cannot be in the black hole phase (because black holes simply do not exist at such charges) but can be in the gas phase (because the gas extends down to arbitrarily small scaled twist).

The discussion of the previous paragraph might make it seem that the system will undergo a phase transition from a black hole to a gas phase at a critical value of scaled twist
of order $N^{\frac{2}{3}}$. The actual situation is, of course, more involved because grey galaxies form in the middle.  
We can understand the formation of the intermediate grey galaxy phase as follows. At  scaled twists below the critical value, the system can partition its twist and angular momentum budget between a black hole 
and the gas, i.e.\ into a grey galaxy.
The partitioning is done in such a manner that while the black hole takes some of the system's angular momentum, it takes much more than its fair share of the total twist, so as to ensure that {\it its} scaled twist lies well above its twist gap.

 Pulling an excess of the twist into the black hole lowers the entropy of the gas; however the gain in black hole entropy more than makes up for this loss, provided the total scaled twist is larger than a critical number of order unity.  Once the total scaled twist descends below this critical value (see \eqref{casenowint}), it simply turns out to be entropically most favourable for the system to completely abandon the black hole, and put all its charges into the gas, yielding the gas phase.  

We will show below that the gas only dominates the ensemble when its temperature is  $\leq \frac{1}{2 \pi}$. In the large $\lambda$ limit in which we work throughout this section, $\frac{1}{2\pi}  \ll \lambda^{\frac{1}{4}}$. Consequently, string oscillator states are never significantly excited at these temperatures, and the gas in question is the 10-dimensional supergravity gas, whose partition function was already presented (in the free regime of relevance to this paper) in \cite{Bajaj:2024utv}.

We now turn to a more detailed study of each of the three phases described above.

\subsection{The black hole phase}\label{bhphase}

As mentioned above, the thermodynamics of the black hole phase is simply the classical thermodynamics of Kerr-AdS$_5$ black holes. Adopting the notation of \cite{Hawking:1998kw,Caldarelli:1999xj,Gibbons:2004ai}, this three parameter family of black holes is parameterized by three real numbers,  $a$, $b$ and $m$. The radial coordinate, $r_+$,  of the event horizon of these black holes is easily computed; it is given as the largest real root of the cubic equation
\begin{equation}\label{rpeqsec}
\left(r_+^2 + a^2\right)\left(r_+^2 + b^2\right)\left(r_+^2 + 1\right) - 2 m r_+^2 = 0.
\end{equation}
In the rest of this section, we use \eqref{rpeqsec} to solve for $m$
in terms of $r_+$, and so trade the parameters $a, b, m$ for the parameters $a, b, r_+$. All thermodynamical quantities associated with these black holes are easily evaluated. In \eqref{thermformmt} we present a listing of all relevant thermodynamical quantities as functions of $a, b, r_+$. We present our formulae in terms of $N$, the rank of the dual field theory (see \S \ref{uncc} for expressions of the formulae for the 5- and 10-dimensional Newton constant in terms of $N$).

For reasons that will soon become clear, in this section we study the family of Kerr-AdS$_5$ black holes in the limit in which 
$a$ and $b$ are scaled to unity. More precisely, we set 
$$a=1-\delta a, ~~~b=1-\delta b\,,$$ 
and study the limit 
\begin{equation}\label{limihb}
\delta a \to 0, ~~~\delta b \to 0, ~~~~\delta a/\delta b~~~{\rm fixed}.
\end{equation} 
In this limit \eqref{rpeqsec} simplifies to 
\begin{equation}\label{leadmassms}
m= \frac{(1+r^2_+)^3}{2 r_+^2}.
\end{equation}

\subsubsection{Microcanonical ensemble}\label{mcens}

At leading order in the limit \eqref{limihb}, the formulae for the  microcanonical thermodynamical quantities, namely the twist $\tau$, the angular momenta $J_i$ and the entropy $S$, simplify to 
\begin{equation}\label{microbh}
\begin{split}
 \frac{J_a}{N^2}=\frac{ \left(r_+^2+1\right){}^3}{16 r^2_+\, \delta a^2 \delta b}\,,
 \quad 
\frac{J_b}{N^2}
&=\frac{\left(r_+^2+1\right){}^3}{16r^2_+\, \delta b^2 \delta a}\,,\quad
    \frac{\tau}{N^2}
    =\frac{1}{16} \frac{  \left(r_+^2+1\right){}^3}{r^2_+\,\delta a\delta b}\,,\\
\frac{S}{N^2}
    &=\frac{\pi  \left(r_+^2+1\right){}^2}{4r_+ \delta a\, \delta b}\,.
\end{split}
\end{equation}
It follows from \eqref{microbh} that, to leading order in the limit \eqref{limihb}, 
\begin{equation}\label{sacalenthb}
\begin{split}
    \frac{S}{(N^2 J_a J_b)^{\frac{1}{3}}}&=
    2^{\frac{2}{3}}\pi r^{1/3}_+\,,\quad 
    \frac{\tau}{(N^2J_a J_b)^{\frac{1}{3}}}=\frac{\left(r_+^2+1\right)}{ 2^{\frac{4}{3}}r^{\frac{2}{3}}_+}\,,\quad 
    \frac{J_a}{J_b}= \frac{\delta b}{\delta a}
    \end{split}
\end{equation}

It follows immediately from \eqref{microbh} that the black hole angular momenta and twist both scale to infinity in the limit \eqref{limihb}, but in such a manner that the quantities on the LHS of the second and third equations of \eqref{sacalenthb} stay fixed. In other words, the limit \eqref{limihb} implements the limit \eqref{lj} for Kerr-AdS$_5$ black holes (this, in fact, is why we chose to study the 
limit \eqref{limihb}).
The first equation of \eqref{sacalenthb} then tells us that the black hole entropy does, indeed, take the scaling form \eqref{finansforentint} in this limit, with 
the function $S^{\rm int}$ given by solving the second equation of \eqref{sacalenthb} for $r_+$ in terms of $\tau_s:=\tau/(N^2J_a J_b)^{1/3}$ (we use the largest positive root of this equation):
\begin{align}
r_+^2 &= 2\tau_s\left[\left(-1+i\sqrt{\frac{64}{27}\tau_s^3-1}\right)^{1/3}+\left(-1-i\sqrt{\frac{64}{27}\tau_s^3-1}\right)^{1/3}\right]-1 \nonumber \\
&= 4 \tau_s^{3/2}-\frac{3}{2}+O(\tau_s^{-3/2})
\,.
\label{eq:rplussolution}
\end{align}
and plugging the result into the first equation of \eqref{sacalenthb}. Note that (\ref{eq:rplussolution}) involves choosing branch cuts so the final answer is real and positive; this is allowed if $\tau_s \geq \frac34$.

Implementing the procedure described above yields an explicit formula for $S^{\rm int}$ as a function of the scaled twist in the Kerr-AdS$_5$ black hole phase
\begin{equation}\label{eq:sintBH}
S^{\text{int}}=N^{\frac{2}{3}} \times 2\pi \times\tau_s^{\frac{1}{6}}\left(\frac{\left(-1+i\sqrt{\frac{64}{27}\tau_s^3-1}\right)^{1/3}+\left(-1-i\sqrt{\frac{64}{27}\tau_s^3-1}\right)^{1/3}}{2}-\frac{1}{2}\right)^{1/6}\,,
\end{equation}
where 
\begin{equation}
    \tau_s:=\tau/(N^2 J_aJ_b)^{1/3}\,.
\end{equation}
Therefore, the entropy can be written as
\begin{align}\label{entlargetist}
S&=( J_a J_b)^{\frac{1}{3}} S^{\rm{int}}\nonumber\\
&=(J_a J_b)^{\frac{1}{3}} N^{\frac{2}{3}} \times 2\pi \times\tau_s^{\frac{1}{6}}\left(\frac{\left(-1+i\sqrt{\frac{64}{27}\tau_s^3-1}\right)^{1/3}+\left(-1-i\sqrt{\frac{64}{27}\tau_s^3-1}\right)^{1/3}}{2}-\frac{1}{2}\right)^{1/6}\,
\end{align} 
This explicit formula simplifies in the limit of large scaled twist, $\tau_s\gg 1$. To leading order in this limit, the entropy as a function of $\tau_s$ is given by 
\begin{align}\label{entlargetist0}
S&=(N^2 J_a J_b)^{\frac{1}{3}} (2\pi)\tau_s^{\frac{1}{4}}=N^{\frac{1}{2}}\left(J_a J_b\right)^{\frac{1}{4}}(2\pi)\tau^{\frac{1}{4}}\,.
\end{align} 

\subsubsection{Canonical ensemble}\label{cens}

It is also interesting to study the thermodynamics of Kerr-AdS$_5$ black holes in the canonical ensemble.
The exact formula for the black hole partition function is presented 
in \eqref{thermformmt}; at leading order in the 
limit \eqref{limihb}, the canonical thermodynamical quantities of Kerr-AdS$_5$ black holes simplify in a manner we now describe. Setting $\omega_a=1-\delta \omega_a$ and $\omega_b = 1-\delta \omega_b$, we obtain 
\begin{equation}\label{cansimp}
\begin{split} 
\delta \omega_a&= \frac{r^2_+-1}{r^2_++1}\delta a\,,\quad 
\delta \omega_b= \frac{r^2_+-1}{r^2_++1}\delta b\,,\quad 
T=\frac{2r^2_+-1}{2\pi r_+}\,,\quad 
\ln Z=\frac{h_{BH}(r_+)}{\delta \omega_a \,\delta \omega_b}\,,
\end{split}
\end{equation}
where \begin{align}\label{hbetaBHmt}
\frac{ h_{BH}(r_+)}{N^2}=\frac{\pi  \left(r_+^2-1\right){}^3}{8 r_+ \left(2 r_+^2-1\right)}.
\end{align}

We can find $r_+$ as a function of $\beta=T^{-1}$ by inverting the third equation \eqref{cansimp} (and choosing the largest real positive root if multiple such roots exist). Plugging this solution into 
\eqref{hbetaBHmt} allows us to verify the prediction \eqref{sn}, and to find an explicit  general formula for the function $h_{BH}(\beta)=h_{BH}(r_+(\beta))$ that applies to the black hole phase. This procedure can be carried out, yielding the surprisingly simple final expression
\begin{equation}
    h_{BH}(\beta)  = \frac{N^2}{64}\left(\frac{2\pi}{\beta^3} \left(2\beta^2+\pi^2\right)^{\frac32} + \frac{2\pi^4}{\beta^3}-\frac{10\pi^2}{\beta}-\beta\right). 
    \label{eq:explicithBH}
\end{equation}
Note that $h(\beta)$ is easily Taylor series expanded at small $\beta$; the relevant expansion is convergent and has no nonperturbative corrections (in contrast with 
the free result \eqref{hd4}). At the first two nontrivial orders in $\beta$ we find 
\begin{equation}\label{leadordhit}
   \ln Z =  \frac{N^2}{\delta \omega_a\delta\omega_b}\Big[\frac{\pi ^4}{16 \beta ^3} - \frac{\pi^2}{16\beta} + O(\beta)\Big] \,.
\end{equation}

As we have explained in \S \ref{semiuniint} (see around \eqref{texp_orig}), \eqref{leadordhit} must agree with the black hole partition function in the fluid limit of \S \ref{ttpf}. That this is indeed the case follows from the fact that \eqref{leadordhit} is in precise agreement with equations (57) and (64) of \cite{Bhattacharyya:2007vs} \footnote{In making the comparison recall that the denominator in Eq. (57) in \cite{Bhattacharyya:2007vs}
reduces to $4 \delta \omega_1 \delta \omega_2$ in the limit under study, and that $G_5 =\pi/(2 N^2)$.}.


As a consistency check of our final results, we have directly demonstrated in \S \ref{eomac} that the formulae for $S^{\text{int}}$ and $h(\beta)$, presented earlier in this subsection, are indeed related by \eqref{intent}, as expected on general grounds. 

\subsubsection{Superradiant instability}

In \eqref{limihb} we have studied the limit in which  $\delta a$ and $\delta b$ are taken to zero from above (i.e.\ $\delta a$ and $\delta b$ should always be viewed as small positive quantities that tend to zero). We see from 
\eqref{cansimp} that $\delta \omega_a$ and 
$\delta \omega_b$ are both positive when  $r_+>1$, while both $\delta \omega_i$ are negative when $r_+<1$. It follows, in other words, that Kerr-AdS$_5$ black holes are stable to the superradiant instability only when $r_+>1$ (see \cite{Kim:2023sig, Bajaj:2024utv} for detailed discussions about this instability).
\footnote{Away from the small $\delta a$ and $\delta b$ limit, the codimension one sheet $\omega_1=1$ differs from the codimension one sheet $\omega_2=1$. As a consequence, the condition for black hole stability takes a complicated and only piecewise-analytic form (see \cite{Bajaj:2024utv} for a discussion). In the limit of interest to this paper, on the other hand, these two sheets (quite remarkably) become identical, and take the extremely simple form $r_+=1$. }

The RHSs of both the second equation of \eqref{sacalenthb} and the third equation of \eqref{cansimp} are increasing functions of $r_+$ for $r_+\geq 1$. It thus follows that Kerr-AdS$_5$ black holes in the limit under study are stable at all values of the scaled twist above a critical value, and at all values of the temperature above a critical temperature. These critical values are obtained by plugging $r_+=1$ into the second equation of \eqref{sacalenthb} and the third equation of \eqref{cansimp}. We conclude that all stable Kerr-AdS$_5$ black holes obey\footnote{In the microcanonical ensemble (and the scaling limit of interest to this paper) one can also ask about the lower bound of the `existence curve' for Kerr-AdS$_5$ black holes. Of course black holes always carry $\tau >0$. In fact, extremal black holes in 
$AdS_5$ oversaturate this bound. In the large angular momentum limit, the extremality bound for black holes can be obtained by setting $T=0$ in the third equation of \eqref{cansimp} (this occurs at $r_+^2= 1/2$). Substituting this value of $r_+$ into the second equation of \eqref{sacalenthb} tells us that the twist of extremal black holes (as a function of their angular momenta) is given by 
\begin{equation}\label{extrememt}
\frac{\tau}{(J_1 J_2)^{\frac{1}{3}}} = \frac{3 }{4}
N^{\frac{2}{3}}=0.75 \times N^{\frac{2}{3}}\,.
\end{equation} 
It is satisfying that this bound is, once again, given in terms of the scaled twist. In fact the inequality \eqref{extrememt} is very similar to \eqref{stabadsf}, but with a smaller coefficient on the RHS. Black holes exist -- but are unstable -- in the twist range between \eqref{extrememt} and \eqref{stabadsf}.}
\begin{equation}\label{stabadsf}
\begin{split}
  \frac{\tau}{(N^2J_1 J_2)^{\frac{1}{3}}} \geq \left(\frac{1}{2}\right)^{\frac{1}{3}}=0.793701\ldots\,,\quad T \geq \frac{1}{2 \pi}
\end{split}
\end{equation}

\subsubsection{Validity of the classical approximation}\label{clapprox}

Is the thermodynamics of Kerr-AdS$_5$ black holes well-approximated by the classical approximation? When all charges 
(i.e.\ $J_i$ and $\tau$) are of order $N^2$, quantum corrections to these classical formulae are famously subleading in $\frac{1}{N^2}$. Indeed, the most important of these quantum corrections captures the contribution of the bulk gas to the black hole partition function. This contribution is of order unity (compared to the order $N^2$ classical contribution of the black hole), confirming its interpretation as a one loop quantum effect. 

In this paper we are interested in black holes with charges governed by \eqref{lj}; these charges are allowed to be parametrically larger than $N^2$. The reader may wonder whether quantum effects at these large charges could be enhanced (so that they compare with or dominate the classical answer). While we have not considered this question very carefully, we see no indication that this is the case. The key point here is that the black hole temperature stays fixed (i.e.\ does not scale to an arbitrarily high 
value) in the scaling limit under consideration (see the third equation of \eqref{cansimp}). For this reason, we expect the bulk gas -- and more generally the determinant around the black hole saddle -- to continue 
to contribute to the partition function, and all thermodynamical charges, at order unity, and so to be subdominant compared to the classical contribution. We leave further verification of this 
physically motivated expectation to future work. Of course the classical approximation fails when the $\omega_i$ approach close enough to unity; in this case quantum corrections are no longer negligible, but instead yield the grey galaxy solutions we study below.

\subsection{The gas phase}\label{tgp}

\subsubsection{Thermodynamics of the non-interacting bulk gas}

An exact formula for the partition function of the free 10d supergravity gas was presented in Eqns. $(4.11)$, $(4.12)$, $(4.13)$ of  \cite{Bajaj:2024utv}, and takes the form 
\begin{equation}\label{pfgas1}
    \ln Z=\frac{4h_{YM}(\beta)}{(1-\omega_1^2)(1-\omega_2^2)}\,,
\end{equation}
where $h_{YM}(\beta)$ takes the explicit (but somewhat complicated) form listed in Eqn. $(4.13)$ of \cite{Bajaj:2024utv}. In the scaling limit of interest to this paper, \eqref{pfgas1} simplifies to 
\begin{equation}\label{pfgas2}
    \ln Z=\frac{\beta^2 h_{YM}(\beta)}{\nu_1 \nu_2},
\end{equation}
in agreement with the general expectation \eqref{sn}, with $h(\beta)$ identified with $h_{YM}(\beta)$. The angular momenta, entropy, and twist of 
our gas are given as functions of $\nu_1, \nu_2$ and $\beta$ by 
\begin{equation}\label{estimates}
\begin{split}
J_1 = \frac{\beta^2h_{YM}(\beta)}{\nu_1^2\nu_2}\,,&\quad 
J_2 =\frac{\beta^2h_{YM}(\beta)}{\nu_2^2\nu_1}\,,\quad \tau =-\frac{1}{\nu_1\nu_2}\partial_{\beta}\Big(\beta^2 h_{YM}(\beta)\Big)
\\
&S=-\frac{\beta^2\Big( \beta \partial_\beta -1\Big) h_{\rm YM}(\beta)}{\nu_1\nu_2}
\end{split}
\end{equation}

See Appendix \ref{supergravitygas} for explicit formulae in the limit $\beta \ll 1$.

\subsubsection{Domain of validity of the non-interacting approximation}\label{nonintmix}

We have, so far, focused our attention on the limit in which the gas is free, and interactions can be ignored. It is of interest to know when this approximation is valid. This question may be answered as follows. Working with grey galaxies in $AdS_5 \times S^5$, the paper  
\cite{Bajaj:2024utv} has already computed the first backreaction of the bulk gas on the metric. It was demonstrated in that paper that the fractional change of the bulk geometry -- brought about by the presence of the bulk gas -- is of order $\frac{\gamma^2}{N^2}$  (see e.g.\ Eqns. $(6.16)-(6.19)$ of that paper), where $\gamma(\theta)$ is  defined in \eqref{gammadef}, and is of order 
$(\nu_1 \nu_2)^{-1/4}$. It follows that the interaction effects in this gas are negligible
\footnote{More quantitatively, we expect
\begin{equation}\label{fractwist}
\frac{\delta \tau}{\tau} \sim \frac{\gamma^2}{N^2}.
\end{equation} 
We expect the RHS of \eqref{fractwist} to capture the fractional change of twist rather than the actual change of twist. When the twist itself vanishes, the backreaction will presumably preserve the Killing nature of  $\partial_t -\partial_{\phi_1}-\partial_{\phi_2}$. The renormalisation of $\tau$ will change the entropy function in the following manner.  The typical state that had twist $\tau$ in the free theory will carry twist 
$\tau^{renorm}= \tau+\delta \tau$ in the interacting theory. As a consequence, to 
first order we expect the entropy as a function of twist, $S_{interacting}(\tau)$, in the interacting theory, to be given by $S_{free}(\tau-\delta \tau)$. }
(in the limit of interest to this paper) provided 
\begin{equation}\label{gaseffects}
(\nu_1 \nu_2)^{-\frac{1}{4}}\ll N, ~~~~{\rm i.e. }~~~\frac{J_1 J_2}{\tau^2} \ll N^{4}\,.
\end{equation}
The estimate \eqref{gaseffects} determines when the `mean field type interactions' of the gas can be ignored (these interactions affect the gas by changing the metric on which it propagates). In Appendix \ref{app:meanfield}, we investigate when more violent `collision' type interactions can be ignored for this gas, and find the condition for this is less exacting than 
\eqref{gaseffects}. In other words collision effects are utterly negligible when \eqref{gaseffects} holds.

At temperatures of order unity (as will be the case in the gas phase whenever 
it is dominant), $\tau \sim (J_1 J_2)^{\frac{1}{3}}$ (with the proportionality constant in this equation of order unity). It follows that gas interactions are 
negligible provided 
\begin{equation}\label{jojt}
\sqrt{J_1 J_2}  \ll N^6
\end{equation} 

\subsection{The grey galaxy phase}

We have seen above (in particular, see \eqref{stabadsf}) that Kerr-AdS$_5$ black holes are unstable when $\tau/\left(J_1J_2\right)^{1/3}<2^{-1/3}N^{2/3}$. 
When  $\tau/\left(J_1J_2\right)^{1/3}$ is lowered to $2^{-1/3}N^{2/3}$, $\omega_1$ and $\omega_2$ both simultaneously hit the instability bound\footnote{We note in \S \ref{extbh} that the black holes hit the extremality bound at the lower value of scaled twist $N^{\frac{2}{3}} \frac{3}{4} \left(\frac{\pi}{2}\right)^{\frac{1}{3}}$.} $\omega_i=1$. As both $\omega_i$ reach unity simultaneously, the 
analysis of  \cite{Kim:2023sig, Bajaj:2024utv} strongly suggests that the black hole phase transits, at lower values of the twist,  to a  grey galaxy of rank 4 (see \cite{Bajaj:2024utv} for terminology), 
i.e.\ a grey galaxy in equilibrium with a bulk gas that carries $J_1$ and $J_2$ that are comparable to each other. 

\subsubsection{Black Holes at the edge of instability: microcanonical formulae}

We see from \eqref{cansimp} that black holes have $\delta\omega_i>0$ when $r_+>1$, but $\delta\omega_i<0$ when $r_+<1$. In this section we focus on black holes at the edge of instability; i.e.\ we set  $r_+=1 + \epsilon$ with $0 < \epsilon \ll 1$. We work with the formulae of \S \ref{mcens} and \S \ref{cens}, and so take the limit
$$\delta a  , \delta b~ \ll \epsilon \ll 1 .$$
We present all formulae to first nontrivial
order in all parameters.

In this scaling regime, the microcanonical thermodynamical formulae for the black holes are given by  \eqref{microbh} and \eqref{sacalenthb} with the replacement $r_+=1+\epsilon$. (In fact, we effectively set  $r_+=1$ rather than $1+\epsilon$, because the order $\epsilon$ corrections turn out to be subdominant for all microcanonical quantities.) The thermodynamic formulae simplify to
\begin{equation}\label{microbhggmc}
\begin{split}
    \frac{J_a}{N^2} &\approx \frac{1}{2(\delta a^2)(\delta b)}\,,\quad 
   \frac{ J_b}{N^2} \approx \frac{1}{2(\delta b^2)(\delta a)}\,,\quad 
   \frac{\tau}{N^2}\approx \frac{1}{2(\delta a)(\delta b)}\,,\quad
   \frac{S}{N^2}\approx \frac{\pi}{(\delta a)(\delta b)}
   \\
    \end{split}
\end{equation}
so that we have
\begin{equation}\label{sacalenthbgg}
\begin{split}
    \frac{S}{(N^2 J_a J_b)^{\frac{1}{3}}}&=
    2^{\frac{2}{3}}\pi \,,\quad 
    \frac{\tau}{(N^2J_a J_b)^{\frac{1}{3}}}= 2^{-\frac{1}{3}}\,,\quad 
    \frac{J_a}{J_b}= \frac{\delta b}{\delta a}\,,\quad 
    T=\frac{1}{2\pi}
    \end{split}
\end{equation}
We pause here to emphasize two points. First, even though the black hole described above has both $\omega_1$ and $\omega_2$ near to unity, $J_a\neq J_b$.
\footnote{In contrast the black holes at the centres of rank 4 grey galaxies with angular momenta and twist both of order $N^2$ have $J_1=J_2$, see \cite{Bajaj:2024utv}.}

Second, it follows immediately from \eqref{sacalenthbgg} that the entropy $S$ of these black holes is given in terms of their twist by 
\begin{equation}\label{entbhtwi}
S= 2 \pi \tau\,.
\end{equation}
The strikingly simple equation \eqref{entbhtwi} can be understood in thermodynamical terms as follows. Any motion on our special (i.e.\ $r_+=1$) surface of black holes changes the value of $\tau$, 
$J_a$ and $J_b$ for our black hole solutions. The usual thermodynamical formulae 
tell us that 
\begin{equation}\label{changeent}
\delta S= \beta \delta \tau + \nu_a \delta J_a + \nu_b \delta J_b\,.
\end{equation} 
Our special surface consists of those black holes with $a=1-\delta a$. 
$b=1-\delta b$ and $r_+=1+\epsilon$. (As before, the small but fixed number $\epsilon$ is a regulator that will be taken to zero at the end of the computation.) $\delta a$ and $\delta b$ parameterize the relevant two-dimensional submanifold of the black hole solution space.  
In the limit $\epsilon \to 0$, $J_a$, $J_b$ and $\tau$ are the smooth nonzero functions of $\delta a$ and $\delta b$ listed in \eqref{microbhggmc}. While $\beta = 2\pi + {\cal O}(\epsilon)$ (see \eqref{texp} below),  $\nu_a= 2 \pi \epsilon \delta a$ and 
$\nu_b = 2 \pi \epsilon \delta b$ (see \eqref{nudeltasmall}).  As $\nu_a \ll \beta$ it follows that the last two terms in \eqref{changeent} are negligible compared to the first term on the RHS of \eqref{changeent} at small $\epsilon$. In the limit $\epsilon \to 0$ it follows that  \eqref{changeent} reduces to the equation $\delta S= 2 \pi \delta \tau$. Integrating this equation gives \eqref{entbhtwi}.

Another way of presenting the argument of the previous paragraph is to 
use \eqref{nudeltasmall} to trade the parameters $\delta a$ and 
$\delta b$ with the parameters $\nu_a$ and $\nu_b$ (which are now taken as the coordinates on our special manifold of black holes at fixed small $\epsilon$). Rewriting in terms of $\nu_a$ and $\nu_b$, we find $\frac{J_a}{N^2} = \frac{4\pi^3\epsilon^3}{\nu_a^2 \nu_b}$, and a similar formula for $J_b$. On the other hand, 
$\tau= \frac{(2\pi^2\epsilon^2)}{\nu_a \nu_b}$. Recalling that each of $\nu_a$ and $\nu_b$ is of order $\epsilon$, we see that while the first term on the RHS of \eqref{changeent} is of order unity, the second and third terms on the RHS of that equation are of order $\epsilon$, and so negligible w.r.t. the first.

\subsubsection{Black Holes at the edge of instability: canonical partition function}

The canonical partition function for the black holes \eqref{microbhggmc} may be obtained as follows. 
Working to linear order in $\epsilon$,  we find 
\begin{equation}\label{texp}
\beta = 2 \pi \left( 1-3 \epsilon \right)
=2 \pi +\delta \beta, ~~~~~\delta \beta =- 6 \pi \epsilon
\end{equation}

The eqs.\!~\eqref{cansimp} and \eqref{hbetaBHmt} 
simplify, respectively, to 
\begin{equation}
\begin{aligned}\label{nudeltasmall}
    \delta\omega_a &\approx \epsilon\, \delta a\,,\quad 
    \nu_a \approx (2\pi)\epsilon\, \delta a\,,\quad 
    \delta\omega_b \approx \epsilon\, \delta b\,,\quad \nu_b \approx (2\pi)\epsilon\, \delta b\,,
\end{aligned}
\end{equation}
and we have
\begin{align}\label{partedgebh}
    \frac{\ln Z}{N^2}&=\frac{\pi \epsilon^3}{\,\delta \omega_a\, \delta \omega_b}=\frac{ \pi (2 \pi)^2 \epsilon^3}{\nu_a\nu_b} = - \frac{(\delta \beta)^3}{54 \nu_a \nu_b}.
\end{align}
The thermodynamical charges of this family of black holes, in terms of $\nu_a, \nu_b$ and $\delta \beta$, can be found by differentiating $\ln Z$ w.r.t. the chemical potentials. For future reference, we note
the result of this exercise at leading nontrivial  order in $\delta \beta$: 
\begin{equation}\label{microbhggcc}
\begin{split}
    \frac{J_a}{N^2} \approx \frac{1}{\nu_a^2\nu_b} \left[ - \frac{(\delta \beta)^3}{54 }\right]\,,&\quad 
   \frac{ J_b}{N^2} \approx \frac{1}{\nu_b^2 \nu_a} \left[ - \frac{(\delta \beta)^3}{54 }\right]\,,\quad 
   \frac{\tau}{N^2}\approx \frac{1}{\nu_a\nu_b}\left[\frac{(\delta\beta)^2}{18}\right]\,,\\
   \frac{S}{N^2}&\approx 
   \frac{1}{\nu_a\nu_b}\left[(2\pi)\frac{(\delta\beta)^2}{18}\right].\\
    \end{split}
\end{equation}

\subsubsection{The free bulk gas near $\beta=2 \pi$: partition function}

In a grey galaxy, the central black hole (studied in the last two subsubsections) is in equilibrium with a gas with the same values of $T$, $\nu_a$ and $\nu_b$ as the black hole itself. Plugging the temperature \eqref{texp} into \eqref{pfgas2}, we see that the gas partition function is 
\begin{align} \label{pfbh}
    \ln Z&=\frac{\beta^2 h_{YM}(\beta)}{\nu_a\nu_b} \equiv \frac{B-C \delta \beta}{\nu_a \nu_b}\,,
\end{align}
where, in the last equality, we have Taylor expanded $\beta^2 h_{YM}(\beta)$ around 
$\beta=2\pi$, and the order unity positive constants $B$ and $C$ are defined via 
\begin{equation}\label{bcdef}
\begin{split}
    B&:=(2\pi)^2 h_{YM}(\beta=2\pi) \approx 1.399\,,\quad C:=-\left(\partial_{\beta}\Big(\beta^2 h_{YM} (\beta)\Big)\right)\Big|_{\beta=2\pi} \approx 0.112.\\
\end{split}
\end{equation}

\subsubsection{Free bulk gas at $\beta=2 \pi$: microcanonical formulae }

Differentiating $\ln Z$ in \eqref{pfbh} w.r.t. $\delta \beta$, $\nu_1$ and $\nu_2$
yields formulae for the twist and angular momenta as functions of $\nu_a$, $\nu_b$ and $\delta \beta$. We find $S$ as a function of the same variables via $S=\ln Z+ \beta \tau +\nu_a J_a +\nu_b J_b$. Using \eqref{pfbh} we find 
\begin{equation}\label{gasbhestimates}
\begin{split}
J_1 &=-\partial_{\nu_a} \ln Z= \frac{B}{\nu_a^2 \nu_b}\,,\quad 
J_2 =-\partial_{\nu_b}\ln Z= \frac{B}{\nu_a \nu_b^2}
\,,\quad \tau =-\partial_{\delta \beta} \ln Z  = \frac{C}{\nu_a \nu_b}\,,\\
&S= \ln Z + \beta \tau +\nu_a J_a + \nu_b J_b=  \frac{2\pi C+3B}{\nu_a \nu_b}\,.
\end{split}
\end{equation}

In this subsection, we have studied the bulk gas at the fixed temperature $\beta = 2 \pi$. At this fixed temperature, the 
twist of the gas is determined  (see \eqref{gasbhestimates}) to be 
\begin{equation}\label{scaledtwist2}
\frac{\tau}{(J_1J_2)^{1/3}}=\frac{C}{B^\frac{2}{3}}=\frac{-\left(\partial_{\beta}\Big(\beta^2 h_{YM}(\beta)\Big)\right)\Big|_{\beta=2\pi}}{\Big[(2\pi)^2 h_{YM}(\beta=2\pi)\Big]^{\frac{2}{3}}} \approx 0.0898,
\end{equation}

Eqn.\!~\eqref{scaledtwist2} tells us that the gas at temperature $T=(2\pi)^{-1}$ carries a scaled twist equal to $C/B^{2/3}$, which we call $A$ for later use.
\begin{equation}\label{adef}
    A:=\frac{C}{B^{2/3}} \approx 0.0898.
\end{equation}
Using \eqref{gasbhestimates} one can easily verify that 
the entropy of the gas scales linearly with the twist. In contrast with the situation described in \eqref{entbhtwi}, however, the proportionality constant between the entropy and the twist does not equal the inverse temperature  $\beta=2 \pi$ of the gas. Instead we find 
\begin{equation}\label{gasym}
\begin{split}
    & S = 
    \left( 2 \pi  + \frac{3 B}{C} \right)  \tau\,.
\end{split} 
\end{equation}
The reader can easily track down the source of the extra term -- proportional to $3B/C$ -- on the RHS of \eqref{gasym}, by 
performing an analysis similar to that under \eqref{changeent}.
In this case all three terms on the RHS of \eqref{changeent} contribute at the same order. The main difference from the black hole case is the absence of explicit factors of $\epsilon$ in the charge relations, when expressed in terms of $\nu_a$. The last two terms on the RHS of \eqref{changeent} give rise to the terms proportional to  $3B/C$ above.

\subsubsection{Grey galaxy phase in the canonical ensemble in the non-interacting approximation}\label{ggphasenonmix}

 As the grey galaxy can be well-modelled by a non-interacting mix of a central black hole and the bulk gas, it follows that its free energy is given by the sum of \eqref{pfbh} and \eqref{partedgebh}, namely
\begin{equation}\label{ggpf}
\ln Z=\frac{1}{\nu_a\nu_b}\left[ -N^2 \frac{(\delta \beta)^3}{54 } + \left(B-C \delta \beta\right)\right].
\end{equation}
The charges and entropy (as a function of chemical potentials and temperature) are also the sum of those of the black hole and gas at the same values of chemical potentials, and so are given (to leading order in $\delta \beta$) by
\begin{equation}\label{mixedcharges0}
    \begin{split}
        J_a&=\frac{1}{\nu^2_a\nu_b}\left[-N^2\frac{(\delta\beta)^3}{54}+B\right]\,,\ J_b=\frac{1}{\nu^2_b\nu_a}\left[-N^2\frac{(\delta\beta)^3}{54}+B\right]\,,\ \tau=\frac{1}{\nu_a\nu_b}\left[N^2\frac{(\delta\beta)^2}{18}+C\right]\,,\\
        S&=\frac{1}{\nu_a\nu_b}\left[N^2\left((2\pi)\frac{(\delta\beta)^2}{18}\right)+2\pi C+3B\right]\,.
   \end{split}
   \end{equation}
Hence we have
\begin{equation}\label{mixedcharges}
\begin{split}
        \frac{\tau}{(J_a J_b)^{\frac{1}{3}}} =\frac{\left[N^2\frac{(\delta\beta)^2}{18}+C\right]}{\left[-N^2\frac{(\delta\beta)^3}{54}+B\right]^{\frac{2}{3}}}\,,\ \ \frac{S}{(J_a J_b)^{\frac{1}{3}}} = \frac{\left[N^2\left((2\pi)\frac{(\delta\beta)^2}{18}\right)+2\pi C+3B\right]}{\left[-N^2\frac{(\delta\beta)^3}{54}+B\right]^{2/3}}
    \end{split}
\end{equation}
In each of these formulae, the first term (proportional to $N^2$) is the contribution of the black hole, while the second term (independent of $N^2$) is the contribution of the gas. We now examine these formulae in several ranges of the parameter $\delta \beta$, moving from higher to lower temperatures (i.e.\ lowering $|\delta \beta|$; recall $\delta \beta$ is negative). 
\begin{enumerate}
\item \label{case1}When $|\delta \beta|^3 \gg N^{-2}$ all black hole charges (and the black hole entropy) are much larger than the corresponding gas charges. At these values of $|\delta \beta|$ the system is effectively in the pure black hole phase (at a value of $\beta$ that approaches $2\pi$ from below). 
\item\label{case2} When $|\delta \beta|^3 \sim N^{-2}$, the black hole and gas contributions to angular momentum are comparable. Over these values of $|\delta \beta|$, however, the black hole contribution to the 
twist and the entropy is much larger than the gas contribution to these charges. \footnote{This is precisely the situation in the `supergravity' 
grey galaxy phases studied in \cite{Bajaj:2024utv} (in that context 
$\nu_a \sim \delta \beta \sim N^{-2/3}$ so all black hole charges -- and the black hole entropy -- are of order $N^2$).}
\item \label{case3} When $N^{-1}\ll |\delta \beta| \ll N^{-2/3}$, the angular momenta of the gas are much larger than those of 
the black hole. However, the entropy and twist of the black hole 
are much larger than those of the gas. In this situation, the analysis of \cite{Bajaj:2024utv} -- which only assumed that the twist and entropy of the gas are negligible compared to those of the black hole -- continues to apply.
\item \label{case4} When $|\delta \beta| \sim N^{-1}$ the black hole and gas carry comparable twist and entropy, while the angular momentum is dominantly carried by the gas. 
\item \label{case5} When $|\delta \beta| \ll N^{-1}$, all charges (and the entropy) are dominantly those of the gas, and the system transits into the pure gas phase at $\beta = 2 \pi$.
\end{enumerate}

\subsubsection{The grey galaxy phase in the microcanonical ensemble }

The thermodynamics of these grey galaxies is best appreciated in the microcanonical ensemble. The first equation of \eqref{mixedcharges} expresses the scaled twist $\zeta=\tau/(J_1J_2)^{1/3}$ as a function of $\delta \beta$ (the fact that $\zeta$ depends only on $\delta \beta$, but not on $\nu_1$ and $\nu_2$ follows on general grounds; see \eqref{betazeta}). 
It is easy to check that the RHS of the first equation of \eqref{mixedcharges} is a monotonic function of $\delta \beta$; as $|\delta \beta|$ is decreased,  the scaled twist $\zeta$ also monotonically decreases. In the rest of this subsubsection we will translate the sequence of behaviours observed upon decreasing $|\delta \beta|$ into the microcanonical ensemble. 

\begin{enumerate}
\item  In the limit $|\delta \beta| 
\gg N^{-2/3}$ (\cref{case1} of the previous subsubsection -- recall the grey galaxy reduces to a pure black hole in this limit) we have
$$\frac{\tau}{(N^2J_1J_2)^{\frac{1}{3}}} = 2^{-1/3}.$$ 
This is precisely the scaled twist below which pure black holes first become unstable (see \eqref{stabadsf}). Consequently, this regime captures grey galaxies at the largest value of scaled twist at which they exist (which equals the lowest value of the scaled twist at which stable black holes exist). We see from the last equation of \eqref{mixedcharges} that the entropy of such grey galaxies takes the form 
\begin{equation}\label{entropygg}
 S= 2 \pi \tau \,,
\end{equation}
in agreement with the expression \eqref{entbhtwi} for the microcanonical entropy for the pure black hole, as might have been expected. Of course, this entropy can be rewritten in the scaling form \eqref{finansforent} as 
\begin{equation}\label{scaledentgg}
    S=(2\pi)(J_1J_2)^{\frac{1}{3}}\left(\frac{\tau}{(J_1J_2)^{\frac{1}{3}}}\right)\,,
\end{equation} 
as anticipated in \eqref{finansforentint}.
\item When $|\delta \beta| 
\sim N^{-2/3}$ (\cref{case2} of the previous subsubsection) we see from \eqref{mixedcharges} that $\tau/(N^2J_1J_2)^{1/3}$ is a number of order unity, less than
$2^{-1/3}$. As both the entropy and the twist of the grey galaxy receive their dominant contribution from the black hole component, we should expect \eqref{entropygg} to continue to hold at these lower values of the scaled twist. It is easy to directly verify (from the last equation of \eqref{mixedcharges}) that this is indeed the case. 
\item When $N^{-1}\ll |\delta \beta| \ll N^{-2/3}$ (\cref{case3} of the previous subsubsection) we see from the last equation of \eqref{mixedcharges} that $ 1 \ll \tau/(J_1J_2)^{1/3} \ll N^{\frac{2}{3}}$.
As both the entropy and the twist of the grey galaxy continue to receive their dominant contribution from the black hole component at these charges,  we should expect \eqref{entropygg} to continue to hold even at these lower values of the scaled twist. Once again, it is 
easily verified, from the last equation of \eqref{mixedcharges}, that this is true. 
\item When $|\delta \beta| \sim N^{-1}$ (\cref{case4} of the previous subsubsection) we see from the last equation of \eqref{mixedcharges} that $\tau/(J_1J_2)^{1/3}$ is of order unity.  At these values of the scaled twist, while the gas makes the dominant contribution to the angular momentum of the solution,  the black hole and the gas make comparable contributions to both the twist and the entropy of the grey galaxy. As the gas contribution to the entropy (and twist) is non-negligible, we should expect \eqref{entropygg} to be modified; indeed we find from \eqref{mixedcharges} that 
\begin{equation}\label{scaledentggmod1}
\begin{split}
        \frac{S}{(J_1J_2)^{\frac{1}{3}}} =(2\pi) \left(\frac{\tau}{(J_1J_2)^{\frac{1}{3}}}\right)   + 3 B^{\frac{1}{3}}\,.
\end{split}     
\end{equation}
\item  When $|\delta \beta| \ll N^{-1}$ (\cref{case5} of the previous subsubsection) we see from the last equation of \eqref{mixedcharges} that $\tau/(J_1J_2)^{1/3} \rightarrow A$ from above 
(recall $A$ is the order-one constant defined in \eqref{adef} that characterizes the scaled twist of a bulk gas at $\beta = 2 \pi$). 
In this case \eqref{scaledentggmod1} continues to apply, 
but can be simplified by substituting in the fact that $\tau/(J_1J_2)^{1/3} =A$, so we get 
\begin{equation}\label{scaledentggmod2}
\begin{split}
 \frac{S}{(J_1J_2)^{\frac{1}{3}}}  &=(2\pi) A  + 3 B^{\frac{1}{3}}\,,
\end{split}     
\end{equation}
Eqn. \!~\eqref{scaledentggmod2} is precisely 
the entropy formula for a pure bulk gas at $\beta =2 \pi$ (see \eqref{scaledtwist2} and \eqref{gasym}). $A$ is 
the lowest value that the scaled twist $\tau/(J_1J_2)^{1/3}$ can take in the grey galaxy phase.
\end{enumerate}

We see that the grey galaxy phase exists when the scaled twist  $\tau/(J_1J_2)^{1/3} $ lies in the range $(A,~N^{2/3}/2^{1/3})$. At the upper end of this range, the grey galaxy reduces to the pure black hole at the edge of instability; at higher values of scaled twist, the system transits into the pure black hole phase. At the lower end of this range, on the other hand, the grey galaxy reduces to the pure bulk gas at $\beta = 2 \pi$. As the scaled twist is further lowered, the system transits into the pure bulk gas phase (at lower values of the temperature), and stays in this phase all the way as the scaled twist is lowered to zero. 

\subsubsection{Regime of validity of the non-interacting approximation}\label{nonintmixgas}

In \S \ref{nonintmix} we have already explained that a bulk supergravity gas in $AdS_5 \times S^5$ (in the scaling limit of interest to this paper) is effectively non-interacting  -- because it has a negligible backreaction on the ambient metric -- provided that 
\begin{equation}\label{inter}
\sqrt{J^g_1 J^g_2} \ll N^6
\end{equation}
The superscript $g$ in \eqref{inter} stands for `gas', i.e.\ $J_i^g$ is the appropriate component of the angular momentum carried by the gas. When the condition \eqref{inter} is satisfied, the gas itself is effectively non-interacting; as it also has a negligible backreaction on the metric, it is also effectively non-interacting with the black hole.

It follows from \eqref{gasbhestimates} that 
\begin{equation}\label{gasin}
\sqrt{J_1^g J_2^g} = \frac{B}{(\nu_1 \nu_2)^{\frac{3}{2}}} 
\end{equation}
Recalling that $B$ is a number of order unity, we conclude that the free approximation (employed in our study of grey galaxies in this subsection) is appropriate only when 
\begin{equation}\label{whenni}
\frac{1}{\nu_1 \nu_2}  \ll N^4.
\end{equation}

In the analysis of this subsection, we have also assumed that 
the black hole components of the grey galaxies studied here are large black holes, and in particular that they carry large twist in units of $N^2$. It follows from the third equation of \eqref{microbhggcc} that this approximation is correct only if 
\begin{equation}\label{whenbig}
\frac{\delta \beta^2 }{\nu_1 \nu_2}  \gg  1
\end{equation}
It follows that it is only possible to obey both \eqref{whenni} and \eqref{whenbig} when
\begin{equation} \label{dmb}
(\delta \beta)^2 \gg \frac{1}{N^4}, ~~~{\rm i.e.\ ~~~when ~~~}
|\delta \beta| \gg \frac{1}{N^2}
\end{equation}
The condition \eqref{dmb} is always satisfied in the parameter ranges of \cref{case1}, \cref{case2} and \cref{case3} in \S\ref{ggphasenonmix}, and is also obeyed over a large part of the parameter range of \cref{case4} of the same subsubsection. While we cannot quite reach the lower-bound value of $\tau/(J_1J_2)^{1/3} $ while staying consistent with 
\eqref{dmb}, we can reach within ${\cal O}(1/N^2)$ of this value (this follows from the first equation of \eqref{mixedcharges}).
\footnote{From a microcanonical viewpoint, this can be understood as follows. As we have mentioned above, we have assumed that the black hole component of our grey galaxy carries a twist that is much larger than $N^2$. Now let us suppose that the scaled twist $\tau/(J_1J_2)^{1/3} $ is of order unity, say around the lower bound value $A$. We know from the analysis above that, in this case, while almost all of the angular momentum of the system is carried by the gas, the gas twist is of the same order as the black hole twist. As the scaled twist is of order unity, it follows that $\sqrt{J_1^g J_2^g} \gg N^3$. There is, however, a lot of room to accommodate this inequality, while obeying 
the condition \eqref{inter}, which guarantees the fidelity of the non-interacting approximation.}

We see that all the phases described in this subsection 
{\it can} be accessed while respecting \eqref{whenni} and 
\eqref{whenbig}, and so are accurately described by the 
approximations used in this paper {\it in an appropriate intermediate range of parameters.}

While the condition \eqref{dmb} is a necessary condition for the accuracy of the free non-interacting approximation of this paper, it is certainly not sufficient. If we take 
$\frac{1}{\nu_1 \nu_2}$ to infinity, the non-interacting approximation \eqref{whenni} always fails, no matter what value of $\delta \beta$ we might sit at. As we have already explained in \S \ref{nonintmix}, the regime of parameters 
$\frac{1}{\nu_1 \nu_2} \gg N^2$ -- a regime that is of great interest to this paper -- cannot be analyzed without accounting for the backreaction of the bulk gas. 
We leave the (intensely interesting) analysis of grey galaxies 
in this regime to future work. 

\subsection{Summary of the phase diagram}

The final phase diagram of our system (in the large $J$ limits \eqref{lj},  equivalently \eqref{limnew}) is rather simple. The phase diagram may be viewed either in the microcanonical or canonical ensemble; the translation between the two is simple as the temperature of the system is a (almost) monotonic function of the scaled twist. 
\begin{itemize}
\item At large enough values of the scaled twist (large enough temperatures), the system is in the black hole phase and its thermodynamics is governed by the formulae of \eqref{bhphase}. As the scaled twist is lowered, the temperature also decreases. At 
the critical value of the scaled twist 
\begin{equation}\label{scaledtwistcrit}
\frac{\tau}{(J_1J_2)^{\frac{1}{3}}}=\left( \frac{N^2}{2} \right)^{\frac{1}{3}}
\end{equation}
the temperature reaches $(2\pi)^{-1}$. 
At this point the black hole becomes unstable, and transitions to the grey galaxy phase. 
\item At the scaled twist \eqref{scaledtwistcrit} (or $T=(2\pi)^{-1}$), the system moves into the grey galaxy phase\footnote{We see from \eqref{partedgebh} that the Kerr-AdS$_5$ black hole function $h(\beta)$ -- and so $\ln Z$ 
in the black hole phase -- switches from positive at $\beta<2\pi$, to negative 
when $\beta > 2 \pi$ (recall that $\delta \beta$ is negative for $\beta<2\pi$, but positive when $\beta>2\pi$). It follows that the Hawking-Page transition for black holes also equals 
$(2\pi)^{-1}$ in the scaling limit. In other words, the Hawking-Page transition temperature coincides with the superradiant instability temperature in the scaling limit \eqref{limnew}. This coincidence ensures that -- in the scaling limit --  black holes are microcanonically unstable below the Hawking-Page transition, a difference from the situation at (say) zero angular momentum, when black holes are microcanonically stable well below the Hawking-Page transition.}, and stays in this phase until 
the scaled twist is lowered to 
\begin{equation}\label{scaledtwistcrittwo}
\frac{\tau}{(J_1J_2)^{\frac{1}{3}}}=A
\end{equation} 
(where the order one number $A$ is defined in \eqref{adef}). All throughout this phase, the system temperature stays at (or parametrically close to) $T=(2\pi)^{-1}$.
The grey galaxy can be thought of as the analogue of the `half water half steam' phase of partially boiled water.
\item As the scaled twist is lowered below \eqref{scaledtwistcrittwo}, the system transits to the pure bulk gas phase, whose thermodynamics is captured by the formulae of \S \ref{tgp}. The temperature of this gas decreases as the scaled twist decreases, and goes to zero as the twist goes to zero.
\end{itemize}

As we have emphasized above, we believe that the formulae of  \eqref{bhphase} give a quantitatively accurate description of the black hole even at arbitrarily large $J$. In contrast, the quantitative description of the other two phases (grey galaxies and bulk gas) is quantitatively accurate only in the intermediate angular momentum limit \eqref{jon}. The quantitative study of these two phases at angular momenta of order (or larger than) $N^6$ is an interesting outstanding challenge for future work.

\subsection{Comparison with generalized free fields}

There is a particular (lightcone bootstrap) sense in which all CFTs become effectively free at asymptotically large angular momenta. One source of intuition for this assertion uses the $AdS/CFT$ correspondence, noting that any two large-angular-momentum bulk particles typically stay very far from each other, and so hardly interact with each other. 

Consider a collection of free particles in the bulk, and let $Z$ denote the second quantized (multiparticle) partition function of this free system. In \eqref{zeo} (see Appendix  \ref{lcform}), we have computed $\ln Z$ for any given choice of bulk particles, and demonstrated that this formula always takes the semi-universal form \eqref{sn}
in the limit \eqref{limnew}. The reader may wonder whether
\eqref{zeo}, together with the lore that  `all CFTs are free at large enough $J$',  constitute a general explanation for the semi-universal behaviour 
\eqref{sn} in every CFT. Our analysis of ${\cal N}=4$ Yang-Mills at strong coupling makes it abundantly clear that this suspicion is very far off the mark. 

The partition function of ${\cal N}=4$ Yang-Mills theory takes the GFF form (\eqref{zeo} with 10d gravitons as seed primaries) when the dual phase is a free bulk gas. As we have explained in this section, this is the case in the $T \leq \frac{1}{2 \pi}$ gas phase (in this case the seed primaries for GFF are the modes of 10 SUGRA in $AdS_5\times S^5$), provided also that 
\begin{equation}\label{angmomcond}
\sqrt{J_1J_2} \ll N^6
\end{equation}

The equation \eqref{angmomcond} emphasizes that the GFF formula fails even in the low temperature gas phase at large enough values of the angular momentum.  As the temperatures are raised above $\frac{1}{2\pi}$, the bulk phase includes a black hole, and any attempt to model the entropy of a black hole by using generalized free fields is simply hopeless. Even though there is simply no sense in which they are well-approximated by the GFF formula, we see that the black hole and grey galaxy phases continue to obey the universal prediction \eqref{sn}. 

The reader may wonder if the difficulties with modeling our theory using a GFF description have their root in the fact that we are working in the large $N$ and strong coupling limit. In fact we believe the opposite is true; the large $N$ limit is one of the most favourable situations for the applicability of the GFF formula, as interactions between bulk particles are proportional to $1/N$, and so are weakest at large $N$. This point can also be seen from \eqref{angmomcond}, which asserts that the  GFF never applies (at finite values of the temperature) when $N$ takes a small finite value. 

The reason that the non-interacting GFF type formula fails even at very large charge is that even in the gas phase, the thermal ensemble involves not a finite number of gravitons, but order $(J_1 J_2)^{1/3}$ of these particles, (each of which typically carries angular momentum of order $(J_1 J_2)^{1/6}$). Even though the interaction of any two of these gravitons is small, the collective backreaction of these particles on the metric becomes large when \eqref{angmomcond} is violated. In a generic (e.g. $N=2$) theory, we expect that these interactions can only be suppressed by reducing the density of these particles, i.e.\ by lowering the temperature. We thus expect that the only regime in which the GFF description of arbitrary CFTs becomes quantitatively accurate is if the temperature (or scaled twist) is scaled to zero in a manner coordinated with the $J \to \infty$ limit.

\section{Spin-refined universality at large spin}

In this section, we would like to study the spin-refined partition function: 
\begin{equation}
\mathrm{Tr}\ e^{-\beta \left(H-\Omega_i J_i\right)+2\pi i \tfrac{p_i}{q_i} J_i}
\end{equation}
in the $\Omega_i\to1$ limit. We note this is equivalent to studying the usual partition function (\ref{thermpf}) with
\begin{equation}
    \omega_i = \Omega_i + \frac{2\pi i}{\beta} \frac{p_i}{q_i}.
\end{equation}
This ``spin-refined" partition function is sensitive to more subtle number-theoretic properties of the partition function or density of states, such as the difference between even and odd spins. In ref.~\!\cite{Benjamin:2024kdg}, the spin-refined partition function is shown to be universal in the high-temperature limit. In particular, 
\begin{equation}
	\ln \mathrm{Tr}\ e^{-\beta \left(H-\Omega_i J_i\right)+2\pi i \tfrac{p_i}{q_i} J_i} \underset{
	\beta\to 0}{\sim} 	\frac{1}{q}\ln \mathrm{Tr}\ e^{-q\beta \left(H-\Omega_i J_i\right)} + \cdots \,,
\end{equation}
where $q=\text{l.c.m}\left[q_1,q_2,\cdots, \right]$.

The central claim of this section is that the above result holds at finite temperature in the $\Omega_i\to 1$ limit. Concretely, we have 
\begin{equation}\label{eq:centralClaim}
	\ln \mathrm{Tr}\ e^{-\beta \left(H-\Omega_i J_i\right)+2\pi i \tfrac{p}{q} J} \underset{
		\Omega\to 1}{\sim} 	\frac{1}{q}\ln \mathrm{Tr}\ e^{-q\beta \left(H-\Omega_i J_i\right)} + \cdots \, \sim \frac{\tfrac{1}{q}h(q\beta)}{\prod_i \left(1-\Omega_i^2\right)}+\cdots
\end{equation}

\subsection{2d CFTs}
Let us first derive the above  in $1+1$-D CFTs. In section~\!\ref{d=2section}, we have shown that 
\begin{equation}\label{eq:2d}
\ln\ \mathrm{Tr}\ e^{-\beta \left(H-\omega J\right)} \underset{\omega\to 1}{\sim} \frac{4\pi^2}{\beta(1-\omega)}\frac{c}{24} + \cdots\,,
\end{equation}
where $\cdots$ has contributions coming from the non-identity chiral states in the vacuum module and non-perturbatively suppressed contribution from the non-vacuum states.  The  size of the non-perturbative correction compared to the leading term is controlled by the twist gap $T_{\text{gap}}$ and given by $$O\left(e^{-\frac{4\pi^2}{\beta(1-\omega^2)}T_{\text{gap}}}\right)\,.$$

Naively,  while computing $$\mathrm{Tr}\ \exp\left(-\beta \left(H-\Omega J\right)+2\pi i \tfrac{p}{q} J\right)\,,$$  we might be tempted to set 
$\omega=\Omega+2\pi i p/(q\beta)$
in the R.H.S of eq.~\!\eqref{eq:2d} and take the $\Omega\to 1$ limit.  This would lead us to conclude that the leading behaviour is not divergent.  However, we must be careful since the contributions coming from a non-zero twist gap are not suppressed anymore if we set $\omega=\Omega+2\pi i p/(q\beta)$ and take $\Omega\to1$.   In other words we can go to the $S$-transformed channel and argue that the vacuum term dominates to compute $\omega\to 1$ limit of  $\ln \mathrm{Tr}\ e^{-\beta \left(H-\omega J\right)} $. This is no longer the case when $\omega=\Omega+2\pi i p/(q\beta)$ and we take $\Omega\to 1$ limit.  The appropriate modular transformation (see e.g. \cite{Benjamin:2019stq}), which makes sure that in the transformed channel, the vacuum dominates, is given by
\begin{equation}
\pm \begin{pmatrix}
	-(p^{-1})_q & b\\
	q & -p
\end{pmatrix}\,,
\label{eq:sl2zguy}
\end{equation}
where $(p^{-1})_q $ is the inverse of $p$ modulo $q$: this exists since $(p,q)=1$ and $b$ is chosen so that the determinant of (\ref{eq:sl2zguy}) is $1$. 

To show the above, let us re-express the quantity of interest in terms of left and right moving inverse temperature
\begin{equation}
\beta_L= \beta(1+\Omega)- 2\pi i \frac{p}{q}\,,\quad \beta_R= \beta(1-\Omega)+2\pi i \frac{p}{q}\,.
\end{equation}
Applying (\ref{eq:sl2zguy}) we get new temperatures $\beta_L', \beta_R'$ which have:
\begin{equation}
\begin{aligned}
\mathrm{Re}(\beta_L')=& \frac{4\pi^2}{q^2\beta(1+\Omega)}\to \frac{2\pi^2}{q^2\beta}\,,\\
\mathrm{Re}(\beta_R')=&  \frac{4\pi^2}{q^2\beta(1-\Omega)} \to \infty\,,
\end{aligned} 
\end{equation}
Therefore, the vacuum term dominates in this frame and we conclude 
\begin{equation}\label{eq:2dOmega}
\ln \mathrm{Tr}\ e^{-\beta \left(H-\Omega J\right)+2\pi i\tfrac{p}{q}J} \underset{\Omega\to 1}{\sim} \frac{4\pi^2}{q^2\beta(1-\Omega)}\frac{c}{24} + \cdots\,.
\end{equation}
Eqn.~\!\eqref{eq:2dOmega} proves the central claim given by eq.~\!\eqref{eq:centralClaim}  for $1+1$-D CFTs.

\subsection{$d>2$ CFTs}

To incorporate a large rotation of $2\pi i p/q$ in the trace, when we go around the thermal circle $S^1_\beta$, we effectively rotate the $S^{d-1}$ by an angle of $2\pi p/q$. We can incorporate this in the path integral using a ``folding trick" discussed in \cite{Benjamin:2024kdg}. In particular, this is equivalent to a $S^1_{q \beta}$ nontrivially fibered over a $S^{d-1}/\mathbb Z_q$ base manifold. (See Fig. 2 of \cite{Benjamin:2024kdg}.)

At this point, we can take the $\Omega_i \rightarrow 1$ limit via a boost that makes $\beta$ small. The folding trick effectively increases the temperature by a factor of $q$ and decreases the volume of the spatial manifold by a factor of $q$ which takes $h(\beta) \rightarrow \frac{h(q\beta)}{q}$.

There is one important subtlety we would like to emphasize in this section. For general large rotations inserted into the partition function, there may be fixed points on the $S^{d-1}$. This will then allow for a defect action to be inserted into the thermal effective action. These defect actions are in general at a lower order in temperature, but not necessarily in angular fugacity. In particular, there can be terms in the defect action that go as $1/(1-\Omega_i^2)$ for a single fugacity (see e.g. Eqn. (3.31) of \cite{Benjamin:2024kdg}). Therefore if the fugacity that we turn on has a fixed point on the $S^{d-1}$, it is important that we take all of the $\Omega_i$'s to $1$ to make (\ref{eq:centralClaim}) trustworthy. If there are no fixed points (for example, turning on incommensurate angles for the spins), then we can take only some of the $\Omega_i$'s to $1$ if need be. 

\section{Discussion}\label{disc}

In this paper, we have argued that the partition function of every conformal field theory takes the semi-universal form \eqref{zint} in the scaling limit \eqref{limnewdfour1}. Upon performing an inverse Laplace transform, we deduce that the entropy of every CFT$_d$ takes the scaling form  given in \eqref{genformint} in the  limit, \eqref{limint}. 
We have independently verified our predictions for free scalar theories in arbitrary dimensions and large $N$, strongly-coupled  ${\cal N}=4$ Yang-Mills theory.

Our principal argument that the partition function takes the semi-universal form \eqref{zint} was constructed as follows. We first expanded the partition function in a Taylor series in  $\beta$. The coefficient of every term in this expansion is a function of $\nu_i$. We then demonstrated that each of these coefficients  scales like \eqref{zint} as $\nu_i$ is taken to zero. As this result holds uniformly for all coefficient functions, at each order in the small $\beta$ expansion, we conjectured that the full partition function also has the same singular scaling behaviour as $\nu_i \to 0$. The use of a Taylor expansion in the (not necessarily small) variable $\beta$, to study behaviour at small $\nu_i$, is clearly awkward. It would be very interesting to find an alternate argument for our central result: one that does not proceed via a small $\beta$ expansion.

In this paper we have tested our general predictions against the (independently computed) twisted thermal partition function of free scalars in every dimension. It would be useful to extend these checks to other tractable examples of conformal field theories, e.g. to free fermionic theories in arbitrary dimensions, to the free $U(1)$ Maxwell theory in four dimensions, and also to computations in vector-type large $N$ models (e.g. $O(N)$ Wilson-Fisher theories, or $\psi^4$ theories or Chern-Simons-matter theories in three dimensions). 

In \S \ref{ttpf} we have noted that the fluid dynamical results of \cite{Bhattacharyya_2008}  on the universal Cardy-like high-temperature behaviour of CFT partition functions -- and the results of this paper -- suggest a nontrivial inequality for the energies of all states in a CFT.  This inequality is stronger than the unitarity bound \eqref{unibound} (see under \eqref{limhold}), but is implied by the ANEC bounds of \cite{Cordova:2017dhq}.  It would be interesting to better understand how two seemingly different principles (ANEC and the applicability of fluid dynamics) both end up implying very similar energy bounds. It would also be useful to generalize the $d=4$ discussion of  \S \ref{ttpf} to arbitrary  dimensions.\footnote{In particular, the arguments of fluid dynamics seem to suggest a bound of the form $\Delta - \sum_i J_i > O(1)$ for some $O(1)$ (i.e. $\Delta$- and $J$-independent) number, in all dimensions. It would be interesting to explore this more, for example if there is a 6d generalization of \cite{Cordova:2017dhq} that gives a bound similar to this. We thank Cyuan-Han Chang and David Simmons-Duffin for emphasizing the higher-dimensional question to us.}

One possibly useful way to explore the connection between ANEC and hydrodynamical constraints might be to further investigate the exotic free theories studied in \cite{Loganayagam:2012zg}\footnote{We thank R. Loganayagam for drawing our attention to this paper, and for a very useful discussion on this point.}. The partition function of these `theories' is defined by specifying the single-particle spectrum to transform in a particular short representation of the conformal algebra, and by multiparticling over the states of this representation.
In particular, one can study theories whose single-particle states transform in the conformal representation $(\Delta=j+1, j_L=j, j_R=0)$
for $j\geq 1$. These quantum numbers saturate the unitarity bound \eqref{unibound}, and so define a short representation\footnote{The null states themselves transform in a short representation with quantum numbers $(\Delta =j+2, j_L=j-\frac{1}{2}, j_R=\frac{1}{2})$. The null states of this representation transform in $(\Delta =j+3, j_L=j-1, j_R=0)$. The null states may be thought of as an `equation of motion'; indeed the partition function of all these theories scales like $T^3$ (rather than $T^4$) in the high-temperature limit. }. 
The important point here is that the highest-weight-state in this representation has $\tau =-(j-1)$, and so has negative $\tau$ 
for $j \geq \frac{3}{2}$. As two of the derivatives carry twist zero, the single-particle spectrum includes an infinite number of states with $\tau =-\frac{1}{2}$, and so it follows that the  multiparticle spectrum has states with arbitrarily negative values of $\tau$, in conflict with the ANEC bounds of \cite{Cordova:2017dhq}. It is thus interesting to ask how the hydrodynamical description of these theories sees this ANEC violation. We study this question in Appendix \ref{exotic} (in the context of simple examples), and explain that while the high-temperature expansion of these theories looks, apparently, perfectly normal, the small $\nu_i$ expansion of these theories is always divergent (e.g. the function $h(\beta)$ diverges), 
and the source of this divergence can be traced to the existence of states with arbitrarily negative $\tau$. Consequently, the small 
$\nu_i$ expansion of this paper is better suited to detecting ANEC violations than the traditional high-temperature expansion of hydrodynamics.\footnote{At the fundamental level, the qualitative structure of $\ln Z$ in these theories is very different depending on whether $j$ is integral (so bosonic) or half-integral (so fermionic). When $j$ is integral, the highest-weight-state of the theory can be occupied an indefinite number of times, and the partition function diverges for $\omega_1=\omega_2=\omega > \frac{j+1}{2j}$. When $j$ is half-integral, on the other hand, states with large 
twists form a Fermi sea, so that states with large negative twist have 
an even larger angular momentum (roughly $|\tau|$ scales like $(j_1j_2)^{\frac{1}{3}}$) so the partition function does not diverge for any value of $\omega<1$. However, this important structural difference is hard to see at finite orders in the low $\beta$ or low
$\nu_i$ expansion.} 

We have also checked the general predictions of this paper against computations performed using the bulk side of the AdS/CFT correspondence, in the context of large $N$ strongly coupled ${\cal N}=4$ Yang-Mills theory in the limit \eqref{limnew}. It would be interesting to generalize this check to the asymmetric high angular momentum limit  $\nu_1 \to 0$ at fixed $\omega_2$, $\beta$. It seems likely that such a study will again find three bulk phases, with the intermediate phase this time being a grey galaxy of rank 2 (see \cite{Bajaj:2024utv} for terminology). 
It would also be useful to extend this holographic test to other dimensions (e.g.\ to  ABJM theory and the theory of $M5$-branes). 

In our study of ${\cal N}=4$ Yang-Mills theory in the limit \eqref{limnew}, we encountered three phases  (see \S \ref{Thrphase}). At temperatures above $1/(2\pi)$, our system lies in the `black hole phase'; i.e.\ is dominated, in the bulk, by a black hole with angular momentum and twist much larger than $N^2$. We have analysed this phase using the classical formulae of black hole thermodynamics. While the classical approximation certainly holds when $J/N^2$ is smaller than any power of $N$ (even when $J/ N^2 \gg 1$), in \S \ref{clapprox}, we have used the fact that the black hole temperature stays fixed in the limit \eqref{lj} to propose that quantum effects will continue to be fractionally negligible even at much larger values of $J $ (e.g.\ $J \sim N^6$). It would be interesting to further develop and tighten this argument, perhaps by computing a bulk determinant. 

Continuing along these lines, $AdS_5 \times S^5$ black holes in the limit \eqref{lj} describe a fascinating family of solutions. While we have studied the thermodynamics of these solutions in this paper, it would also be interesting to study their detailed spacetime metrics. As $ J \gg N^2$, there may be a sense in which these solutions can be usefully viewed in a `tube by tube' manner,  along the lines of \cite{Bhattacharyya:2007vjd}, even though their temperature stays finite. Thinking adventurously, one might even wonder if it might be possible to generalize such tube wise solutions to dynamical situations, obtaining a large generalization of the equations of hydrodynamics. It is possible that dynamical questions will be simplest to answer at large $D$, using the formulae presented in \cite{Bhattacharyya:2017hpj, Dandekar:2017aiv, Bhattacharyya:2018szu, Bhattacharyya:2019mbz} (based on the formalism developed in  \cite{Bhattacharyya:2015dva, Emparan:2015hwa, Bhattacharyya:2015fdk, Dandekar:2016fvw, Bhattacharyya:2016nhn}). We hope to revisit this discussion in the future. 

As we have explained in this paper, ${\cal N}=4$ Yang-Mills at temperatures at or lower than $(2\pi)^{-1}$ lies in a grey galaxy or gas phase. Let us first restrict our attention to values of $J \ll N^2$. We have explained in \S \ref{nonintmixgas} that the bulk gas is effectively free at such angular momenta, and the corresponding gas phase is accurately captured by the analysis of \cite{Bajaj:2024utv}. One of the key findings of \cite{Bajaj:2024utv}
was that the boundary partition function density (and boundary stress tensor, see below) dual to the bulk gas takes the semi-universal form described in this paper. The analysis of the current paper completely explains this observation. However, it was also noted in \cite{Bajaj:2024utv} that the bulk free energy density and bulk gas stress tensor take the semi-universal form (see Eq. (5.41) of \cite{Bajaj:2024utv}). It would be interesting to find an explanation for this point along the lines of the current paper\footnote{Note that the bulk gas lives in an $AdS_5$ 
spacetime (rather than on a sphere as mainly studied in this paper), and also that the bulk fields that make up this gas have a range of masses, including masses that are arbitrarily large. Also note that the proper size of the Euclidean time circle, in this context,  is of order unity (see equation 5.40 of \cite{Bajaj:2024utv}) and so is parametrically comparable to the curvature scale of the $AdS_5$ spacetime.}.

In this paper we have only studied the gas and grey galaxy phases dual to ${\cal N}=4$ Yang-Mills theory at angular momenta $J \ll N^6$. 
At larger angular momenta, the backreaction of the gas on the metric can no longer be neglected. It may be possible -- and would be very interesting -- to find the new correspondingly backreacted solutions (perhaps in a self-consistent manner). It would likely be easiest to first work in the special case $\omega_1=\omega_2$ in order to preserve $SU(2)_R$ symmetry. 

On a related note, while the last three decades have seen an intensive study of  ${\cal N}=4$ Yang-Mills in the large $N$ limit, comparatively little effort has been dedicated to the study of solutions whose charges scale parametrically faster than $N^2$ as $N \to \infty$. The current paper yields a physical context in which such a study is very natural. We have explained in the main text that  $J \sim N^3$ is the smallest value of $J$ at which central black holes are large (compared to the size of $AdS$) at the phase transition point between grey galaxies and the gas (see around \eqref{minbh} and footnote \ref{footln}).  Also, $J \sim N^6$  is the scale at which the backreaction of the gas on the metric first becomes important
(see \S \ref{nonintmixgas}). It would be interesting to better understand the large $N$ physics of such `super-classical' charge regimes, and  search for and study other such physically motivated examples of such scalings. 

Our study of ${\cal N}=4$ Yang-Mills theory has been performed at strong coupling. It would be fascinating to also study this theory at weak coupling in the limit \eqref{limnew}, and, in particular, understand the possible emergence of the new superclassical scalings of the previous paragraph from Yang-Mills perturbation theory. When the Yang-Mills theory is completely free, we expect the gauge singlet condition
\cite{Aharony:2003sx} to trivialize in the large $J$ limit, and for the partition function (in the limit of interest to this paper) to reduce to $N^2$ copies of 6 free scalar fields, one free gauge field and four free fermions. In this free limit we expect the function $h_{YM}(\beta)$ to stay in the high-temperature `black hole' phase (and undergo no phase transitions) all the way down to $T=0$ (because this gluonic phase does not have a twist gap in free ${\cal N}=4$ Yang-Mills theory). Moving away from this strictly free limit, however, it would be interesting to understand how the formulae \eqref{extreme} (for the black hole twist gap at large $\lambda$) and \eqref{stabadsf} (for the lowest twist of a stable black hole at large $\lambda$) generalize to arbitrary values of $\lambda$. The general considerations of this paper suggest that the finite $\lambda$ generalizations of bounds \eqref{extreme} and \eqref{stabadsf} respectively take the form 
\begin{equation}\label{stabin}
\begin{split}
\frac{\tau}{\left(N^2J_1J_2\right)^{\frac{1}{3}}}&  \geq  f_{\rm gap}(\lambda)~~~{\rm for~existence~of~``black~hole"~~ phase}\\ 
& \geq f_{\rm stab}(\lambda)~~~{\rm for~stability~of~``black~hole"~~ phase} \\
\end{split}
\end{equation}
Here $f_{\rm stab}(\lambda)$ and $f_{\rm gap}(\lambda)$ are functions of $\lambda$ with $f_{\rm stab}(\lambda) \geq f_{\rm gap}(\lambda)$, such that $f_{\rm gap}(0)= f_{\rm stab}(0)=0$, and $f_{\rm stab}(\infty)=\left(\frac{1}{2}\right)^{\frac{1}{3}}$ and $f_{\rm gap}(\infty)=\frac{3}{4}$ (see eq. \eqref{stabadsf} and eq. \eqref{extreme}). It is tempting to guess that $f_{\rm stab}(\lambda)$ and $f_{\rm gap}(\lambda)$ are both smooth, monotonically increasing functions of $\lambda$ that admit a (not necessarily convergent) Taylor series expansion around $\lambda=0$. If this is the case, the function $h(\beta)$ will, presumably, develop a grey galaxy and gas phase at any nonzero $\lambda$, no matter how small. This would render these phases perturbatively accessible. It would be very interesting to understand all this better. 

In \S \ref{lgym} of this paper, we have focussed on the study of the thermodynamics of ${\cal N}=4$ Yang-Mills. It would be very interesting to generalize the study of this paper to dynamical considerations. Several such questions suggest themselves. For instance, let us suppose we start out with a thermalized gas in the gas phase at a temperature larger than $(2 \pi)^{-1}$. As the gas is not perturbatively entropically dominant at these charges, it should eventually settle down to a grey galaxy or black hole. It would be very interesting to understand the dynamical mechanism for this process. Does such a gas exhibit the Jeans instability, or do we have to rely on tunneling processes for the corresponding transition? What are the time scales for this transition?

In our study of the twisted partition function of free scalar theories, we discovered that the function $h(\beta)$ admits a surprisingly simple high-temperature expansion in even dimensions. This expansion consists of a truncated (finite) power series in $\beta$, plus an infinite series of non-perturbative corrections. It would be interesting to better understand several aspects of this formula. To start with, why does the perturbative part of the partition function truncate? 
Turning to the non-perturbative corrections, in \S \ref{prefac} we have suggested a physical interpretation for these terms in terms of instantons of the worldline effective action; the actions for our instantons reproduce the exponential factors in \eqref{hd4}, and our rough analysis of one loop corrections around these instantons also reproduces the parametric dependence of the prefactor. It would be interesting to complete this analysis: to reproduce the puzzling positive sign of these coefficients and to compute their numerical values. It would also be interesting to understand why 
these coefficients are one loop exact (i.e.\ why the coefficients of the non-perturbative terms in \eqref{hd4} are numbers independent of $\beta$). Indeed, the formula \eqref{hd4} is so simple that it may be possible -- and would be very interesting -- to completely reproduce this equation, with all factors, from a worldline point of view. Note, in this context, that several earlier studies of partition functions on spheres \cite{Anninos:2020hfj, Fliss:2023muk, Maldacena:2024spf, David:2024pir} exhibit similar simple behaviour.

In \S \ref{intinst} we have conjectured that the worldline instantons, described in the previous paragraph, cease to contribute to the function $h(\beta)$ in interacting theories. It would be very interesting to test this prediction, perhaps in the context of  tractable (e.g. vector large $N$?) models. 

In \S \ref{cust} we have conjectured that, in addition to the partition function, the one point function of the stress tensor takes the semi-universal form \eqref{st2} in the scaling limit studied in this paper. In the language of \cite{Buric:2024kxo}, our conjecture effectively specifies the form of the (one point) stress tensor block.
As already mentioned in \S \ref{cust}, it would be interesting -- and may not be too difficult --
to construct an all orders `proof' for this conjecture along the lines of the discussion of \S \ref{Tetrad}. As 
the stress tensor of an equilibrated rotating perfect conformal fluid on $S^3$ is known to take the analogous form \eqref{st1}, the form \eqref{st2} -- if true -- might suggest that dynamics (and not just equilibrium physics) at large $J$ is governed by equations that are nearer to traditional hydrodynamics than might initially have been suspected. It would be interesting to explore this exciting prospect.

It would be interesting to generalize the work presented in this paper in several directions. A relatively straightforward generalization would be to study theories with a conserved charge, at values of 
the chemical potential that are held fixed (like $\beta$ is) 
as $\nu_i \to 0$. A much more interesting generalization would be to 
set the angular momentum to zero, but to search for a large charge scaling limit analogous to the large angular momentum limit studied in this paper, perhaps in the context of theories with a 
nontrivial Coulomb branch (see e.g. \cite{Yamada:2006rx, Bhattacharyya:2007sa, Cuomo:2024fuy, Choi:2024xnv}). It is possible that such a discussion will make connections with the charge expansion of \cite{Hellerman:2015nra} (see e.g. \cite{Cuomo:2022kio, Choi:2025tql, Lee:2025qim} for possibly relevant recent developments). 

It would also be interesting to investigate if semi-universality continues to hold in an interesting manner for twisted partition functions on manifolds other than $S^{d-1}\times S^1$ (an obvious 
example -- and one that has already come up -- is twisted thermal $AdS$
space). 

In addition to studying the density of states, it would be interesting to study OPE coefficients. In \cite{Benjamin:2023qsc}, ``heavy-heavy-heavy" OPE coefficients in $d>2$ CFTs were studied by building a higher-dimensional analog of a genus-2 partition function. \cite{Benjamin:2023qsc} used a ``hot spot" hypothesis to estimate the path integral on this geometry, by conjecturing it was dominated by a region that locally looked like a circle fibration with a large temperature. It would be very interesting to see if we can take a large $\omega_i$ limit instead, and predict semi-universal OPE coefficients at large spin, in addition to semi-universal entropy at large spin. Similarly, ``heavy-heavy-light" OPE coefficients were studied in \cite{Delacretaz:2020nit,Buric:2025uqt}, and ``heavy-heavy-heavier" in \cite{Simmons-Duffin:2025qox}; it would also be interesting to generalize these OPE coefficients to large spin. Furthermore, it would be interesting to explore whether large spin semi-universality appears in the context of thermal holographic correlators, explored e.g. in recent papers \cite{Buric:2025anb,Buric:2025fye} and in the context of a recent revival \cite{Barrat:2025twb,Barrat:2025nvu,Barrat:2025wbi,Marchetto:2023xap} in the analytic thermal bootstrap, initiated in \cite{Iliesiu:2018fao}.

It has been known that for a $2$D CFT with a global symmetry, the Cardy formula for high energy density of states  holds true individually in each of the charged sectors \cite{Pal:2020wwd}. A similar phenomenon happens in higher dimensional CFTs as well \cite{Harlow:2021trr,Kang:2022orq}. We expect that the large spin semi-universality appears in each of the charged sectors in a similar way.  

To end this discussion, we note that the reduced angular velocities $\nu_i = \beta(1-\omega_i)$ play a key role in recent studies of the superconformal index \cite{Romelsberger:2005eg, Kinney:2005ej}. The recent paper \cite{Choi:2025lck} has pointed out that the manifold of supersymmetric black holes includes black holes with $\nu_i <0$, and has conjectured that these black holes are always thermodynamically unstable to the formation of supersymmetric grey galaxies (or revolving black holes). It is thus natural to wonder whether superconformal indices (like the thermal partition functions studied in this paper) also display some universal scaling in appropriate small $\nu_i$ limits. The authors of \cite{Choi:2018hmj} have already established that this is the case in the limit that $\nu_1$ and $\nu_2$ are both taken to zero at fixed values of other indicial chemical potentials. It would be interesting to study less symmetric limits (e.g. a limit 
in which $\nu_1 \to 0$ at fixed $\nu_2$). It may be possible -- and would be very interesting -- to make universal predictions about such asymmetric limits using the formalism of \cite{DiPietro:2014bca, Choi:2018hmj, Kim:2019yrz,  Cassani:2021fyv}. We hope to return to this point in the future.

\section*{Acknowledgments}

We thank J. Bhattacharya, C.-H. Chang, A. Gadde, D. Jain, S. Kundu, J. P. Martinez, D. Maz\'{a}\v{c}, O. Parrikar, C. Patel, D. Simmons-Duffin, S. Trivedi and especially E. Lee and R. Loganayagam for very useful discussions. We would also like to thank C. Cordova, T. Hartman, D. Jain, K. Jensen, S. Kim, Z. Komargodski,  P. Kovtun, S. Kundu, R. Loganayagam,
G. Mandal, D. Maz\'{a}\v{c}, C. Patel, J. Penedones,  M. Rangamani and  L. Rastelli for their comments on a preliminary version of this manuscript. N.B. is supported by the U.S. Department of Energy, Office of Science, under grant Contract Number DE-SC0026324. The work of H.A., V.K., S.M., J.M. and A.R. was supported by the J C Bose Fellowship JCB/2019/000052 and the Infosys Endowment for the study of the Quantum Structure of Spacetime. H.A, V.K, S.M, J.M., and A.R would also like to acknowledge their debt to the people of India for their steady support of the study of basic science.
\appendix

\section{The equilibrium partition function at low derivative order} \label{loworder}

As an illustration of the general arguments of \S \ref{Tetrad}, in this Appendix we explicitly work out the form of  ${\cal L}$ and $\ln Z$ on a twisted $S^3 \times S^1$, at leading order in the limit 
\eqref{limnew}, in the case that ${\cal L}$ takes the (zero- and two-derivative) form \eqref{eq:partition}.

\subsection{Leading zero derivative order}

At zero derivative order, the partition function is given by the first term on the RHS of \eqref{eq:partition}. By scale invariance 
$P(T) = \kappa T^4$ for some dimensionless constant $\kappa$ (in the case of a single free scalar, $\kappa=\frac{\pi^2}{90}$).
Plugging this relationship into the first part of \eqref{eq:partition}, using  $T(x)=T \gamma$ (where $T$ is the constant thermodynamical temperature)  
and $\sqrt{g_3} = \gamma(\theta)\sin\theta \cos\theta$ (see 
\eqref{eq:volumeform}), we find 
\begin{equation}
    \begin{split}
        \ln Z^{0} &= \int d^3x \sqrt{g_3} \frac{P(T(x))}{T(x)} \\
        &= \int d\phi_1 d\phi_2 d\theta \cos\theta \sin\theta \gamma(\theta) \kappa T^3 \gamma^3(\theta) \\
        &= \frac{ 2\pi^2 \kappa T^3}{(1 - \omega_1^2)(1 - \omega_2^2)}\\
        &\approx \frac{\pi^2 \kappa T}{2 \nu_1\nu_2}.
    \end{split}
\end{equation}
Here we have used
\begin{equation}
  \int_{S_3} \gamma^4(\theta) = \frac{2\pi^2}{(1-\omega_1^2)(1-\omega_2^2)} \approx \frac{\pi^2\beta^2}{2\nu_1\nu_2}
\end{equation}
(the integral is taken over the usual round measure on $S^3$).

\subsection{Subleading second derivative order}

We now compute the contribution of the second derivative terms in 
\eqref{eq:partition} (the second integral on the RHS of that equation).
Substituting the expressions for $R$ and $f_{ij}f^{ij}$ from 
\eqref{eq:ricci_scalar} and \eqref{eq:field_strength_contracted} into the relevant terms in \eqref{eq:partition}, and using the form of 
$P_i(\sigma)$ listed in \eqref{eq:conformal_solution}, we obtain:   
\begin{equation}\label{eq:conc_partition3d}
    \begin{split}
       \ln Z^{1} &= -\frac{1}{2} \int d^3x \sqrt{g_3} \left( P_1(\sigma) R + T^2 P_2(\sigma) f_{ij}f^{ij} + P_3(\sigma) (\partial \sigma)^2 \right) \\
       &= -\frac{1}{2} \int d\phi_1 d\phi_2 d\theta \cos\theta \sin\theta\gamma(\theta)\left[\left( -2 e_1 T \gamma^3(\theta) + 8 e_2 T \gamma^3(\theta) \right) + \mathcal{O}\left(\gamma^2(\theta)\right)\right] \\
       &= 4\pi^2 T (e_1 -4 e_2) \int d\theta \cos\theta \sin\theta \left[\gamma^4(\theta) + \mathcal{O}\left(\gamma^3(\theta)\right)\right] \\
       &= \frac{2\pi^2 T (e_1 -4e_2)}{(1 - \omega_1^2)(1 - \omega_2^2)} + \mathcal{O}\left(\frac{\ln\left[(1-\omega_1)/(1-\omega_2)\right]}{\omega_1^2- \omega_2^2}\right) \\
       &\approx \frac{\pi^2  (e_1 -4e_2)}{2T\nu_1\nu_2} + \mathcal{O}\left(\frac{1}{(\nu_1\nu_2)^{1/2}}\right)
    \end{split}
\end{equation}
Therefore, the full expression for the partition function is
\begin{equation}\label{eq:partition_full}
    \begin{split}
        \ln Z= \ln Z^{0} + \ln Z^{1} +\cdots &=\frac{\pi^2 \kappa T}{2\nu_1\nu_2}\left(1 + \frac{(e_1 -4 e_2)}{\kappa T^2} + \dots \right).
    \end{split}
\end{equation}
where $e_1$ and $e_2$ are theory-dependent terms of order unity (by Taylor expanding $\ln Z$ in \eqref{evenh} with $n=2$, we find $e_1-4e_2=-\frac{1}{36}$ in the case of a single free scalar field). The $\ldots$ in \eqref{eq:partition_full} refers to the contribution of fourth and higher order derivative terms in ${\cal L}$ to the partition function $\ln Z$. 

\section{Details of 3d analysis}\label{3ddetails}
\subsection{Localization around the equator in $d=3$}\label{3dlocalisation}

In three dimensions the Lorentz factor $\gamma(\theta)$ (see \eqref{gammaform}) is sharply peaked around $\theta= \frac{\pi}{2}$. 
In a neighbourhood (of angular size of order $\sqrt{\frac{\nu}{\beta}}$) around the equator, $\gamma(\theta)$ takes the form listed in \eqref{gammatheta}. Using \eqref{gammatheta}, it is easy to check that (to leading order in the small $\nu$ limit)
\begin{equation}\label{gammathetader}
\begin{split}
&\gamma(\delta \theta)= \frac{1}{\sqrt{\frac{2 \nu }{\beta }+\delta \theta ^2}}\\
&\gamma^{'}(\delta \theta)= \frac{\delta \theta }{\left(\frac{2 \nu }{\beta }+\delta \theta ^2\right)^{3/2}}\\
&\gamma^{''}(\delta \theta)=\frac{2 \delta \theta ^2-\frac{2 \nu }{\beta }}{\left(\frac{2 \nu }{\beta }+\delta \theta ^2\right)^{5/2}} \\
&\gamma^{'''}(\delta \theta)=\frac{3 \left(2 \delta \theta ^3-\frac{6 \delta \theta  \nu }{\beta }\right)}{\left(\frac{2 \nu }{\beta }+\delta \theta ^2\right)^{7/2}} \\
\end{split}
\end{equation}
In order to get a sense of the scales involved, we evaluate (\ref{gammathetader})  at  $\delta \theta^2=\frac{2\nu}{\beta}$ (thought of as a typical value within the peaked region of $\gamma$). At this value we find
\begin{equation}
    \gamma=  \sqrt{\frac{\beta}{4\nu}} \qquad
    \gamma'=\frac{\beta }{4 \sqrt{2} \nu }, \qquad
    \gamma''=\frac{1}{16}\left(\frac{\beta}{\nu}\right)^{\frac{3}{2}}, \qquad
    \gamma'''=\frac{3 \beta ^2}{32 \sqrt{2} \nu ^2}.
\end{equation}
These results illustrate the fact that each extra derivative brings in a factor of $\frac{1}{\sqrt{\nu}}$
so that 
\begin{equation}
    \gamma' \sim \gamma^{2}, \qquad
    \gamma'' \sim \gamma^{3}, \qquad
    \gamma^{(n)} \sim \gamma^{\,n+1}.
\end{equation}

It follows, in particular, that $\gamma^3(\theta)$ and $\frac{(\gamma')^2}{\gamma}$ each scale like $\frac{1}{\nu^{\frac{3}{2}}}$. In the main text we predicted that the integrals of both these quantities would be of order $\frac{1}{\nu}$. This is easily verified in a couple of special examples. Using \eqref{gammathetader} we find 
\begin{equation}\label{eq:integrals}
    \begin{split}
        \int_{S^2} \gamma^3(\theta) &= \frac{4\pi}{(1-\omega^2)} \overset{\omega \rightarrow 1}{\sim } \frac{2\pi \beta}{\nu}\\\int_{S^2} \gamma'^2(\theta)/\gamma(\theta) & =2 \pi  \left(\frac{6-4 w^2}{3-3 w^2}-\frac{2 \tanh ^{-1}(w)}{w}\right)\underset{\omega\to1}{
        \sim
        }
        \frac{2 \pi\beta }{3\nu}
    \end{split}
\end{equation}

\subsection{The twisted $S^3 \times S^1$ partition function in $d=3$}\label{3dtwisted}

In this subsection, we rework the analysis of \S \ref{Tetrad} in $d=3.$

\subsubsection{Metric and tetrads}

The metric on $S^2\times S^1$ with twist can be written as
\begin{equation}
    ds^2 = \frac{\big(d\tau-i\,\gamma(\theta)^2\,\omega \sin^2\theta\, d\phi\big)^2}{\gamma(\theta)^2}
    + \gamma(\theta)^2 \big(\omega \sin^2\theta\, d\phi\big)^2
    + \sin^2\theta\, d\phi^2
    + d\theta^2 \, .
\end{equation}
where $\tau$ is identified with $\tau+\beta$.

A convenient choice of tetrads on the two-dimensional base manifold is 
\begin{equation}
\begin{split}
    e_1 &= d\theta, \\
    e_2 &= \omega \sin^2\theta\,\sqrt{\gamma(\theta)^2 + \frac{\csc^2\theta}{\omega^2}}\, d\phi \, .
\end{split}
\end{equation}

With this choice of basis, the spin connection can then be computed by the standard procedure. We find that the only non-vanishing component is
\begin{equation}\label{sc3d}
   \omega^{21}_{\ \ 2} = \frac{\gamma(\theta)\,\gamma'(\theta)}{\gamma(\theta)^2-1}
   \;\approx\; \frac{\gamma'(\theta)}{\gamma(\theta)}
   = \mathcal{O}\!\big(\gamma(\theta)\big),
\end{equation}
together with its antisymmetric counterpart.
\subsubsection{Riemann tensor}
The only nonzero component for the Riemann tensor is
\begin{equation}\label{rieman3d}
R_{1212}=\gamma (\theta )^2-\frac{3 \gamma '(\theta )^2}{\gamma (\theta )^2-1}
\end{equation}
as well as its antisymmetric counterparts.
\subsubsection{Ricci tensor}
The nonzero components are,
\begin{equation}
    R_{11} =R_{22} = \gamma (\theta )^2-\frac{3 \gamma '(\theta )^2}{\gamma (\theta )^2-1}.
\end{equation}
The expression for the Ricci scalar can then be easily obtained as
\begin{equation}\label{R3d}
    R =\delta^{ab}R_{ab}= 2 \left(\gamma (\theta )^2-\frac{3 \gamma '(\theta )^2}{\gamma (\theta )^2-1}\right)
\end{equation}
\subsubsection{Gauge field}
The gauge field one-form and field strength are given by
\begin{equation}
    a= -\gamma^2(\theta)\omega \sin^2\theta d\phi
\end{equation}
\begin{equation}\label{field3d}
    f = da=-\frac{2 \omega  \gamma (\theta )^3}{\sqrt{\frac{\gamma (\theta )^2 \left(\gamma (\theta )^2-1\right)^2}{\gamma '(\theta )^2}+1}} e_1\wedge e_2.
\end{equation}
The expression for $f_{ij}f^{ij}$ is given by
\begin{equation}\label{f23d}
    f_{ij}f^{ij} = \frac{8 \gamma (\theta )^2 \gamma '(\theta )^2}{\gamma (\theta )^2-1}
\end{equation}
It is also easy to check that 
\begin{equation}\label{pss3d}
 (\partial \sigma)^2 = \frac{\gamma'^2(\theta)}{\gamma^2(\theta)}
\end{equation}

\subsubsection{Evaluation of $\ln Z$ keeping terms in ${\cal L}$ up to second order in derivatives}

At zeroth derivative order, using Weyl invariance we obtain
$P(T)= \kappa_1 T^3$. We then substitute $P(T)$ into \eqref{eq:partition} to obtain
\begin{equation}
\begin{split}
   \ln Z^{0}&= \int d^2x \sqrt{g_2}\frac{P(T(x))}{T(x)} \\&= \int d\phi d\theta\sin\theta \gamma(\theta)\kappa_1 T^2 \gamma^2(\theta)\\&\approx \frac{2\pi \kappa_1 T}{\nu}
\end{split}
\end{equation}

At second derivative order, we use \eqref{R3d}, \eqref{f23d}, \eqref{pss3d}, and the expression for $P_i(\sigma)$'s. The expressions for the $P_i(\sigma)$'s are obtained by demanding Weyl invariance. Under the Weyl transformation $g_{\mu\nu}\rightarrow e^{2\chi(x)}g_{\mu \nu}$, in three dimensions, the lower dimension KK fields transform as,
\begin{equation}
    \sigma\rightarrow \sigma + \chi, ~~a\rightarrow a,~~g_{ij}\rightarrow e^{2\chi(x)}g_{ij}
\end{equation}
and consequently, the terms in the partition function (given in Eqn. (\ref{eq:partition})) transform as
\begin{equation}\label{eq:weyl3d}
\begin{split}
    R &\rightarrow e^{-2\chi(x)} \left (R - 2\nabla^2\chi(x)\right)\\ (\nabla \sigma)^2&\rightarrow (\nabla \sigma(x))^2 + 2(\nabla\sigma(x)).(\nabla \chi(x)) + (\nabla \chi(x))^2\\ f_{ij} f^{ij} &\rightarrow e^{-4\chi(x)}f_{ij} f^{ij}\\\sqrt{g_2}&\rightarrow e^{2\chi(x)}\sqrt{g_2}
\end{split}
\end{equation}
The requirement that the partition function in \eqref{eq:partition} be Weyl invariant, using \eqref{eq:weyl3d}, leads to the following expressions for the $P_i$'s
\begin{equation}\label{eq:weyl}
P_1(\sigma)=c_1,~~P_2(\sigma)=c_2\frac{1}{T(x)^2},~~P_3(\sigma)=c_3=0
\end{equation}
After substituting \eqref{R3d}, \eqref{f23d}, \eqref{pss3d}, and \eqref{eq:weyl} in the second derivative terms of the effective thermal partition function given in \eqref{eq:partition}, we obtain:
\begin{equation}\label{thermpf2ap1}
\begin{split}
   &\ln Z^{(1)}= - \frac{1}{2} \int d^2x\, \sqrt{g_2} \left( 
        P_1(\sigma)\, R + 
        T^2 P_2(\sigma)\, f_{ij} f^{ij} + 
        P_3(\sigma)\, (\partial \sigma)^2 
    \right)+ \cdots\\
    &=-\frac{1}{2}\int d\theta d\phi \sin\theta\gamma(\theta)\left(2c_1 \left(\gamma (\theta )^2-\frac{3 \gamma '(\theta )^2}{\gamma (\theta )^2-1}\right)+c_2\frac{8 \gamma '(\theta )^2}{\gamma (\theta )^2-1}+c_3\frac{\gamma'^2(\theta)}{\gamma^2(\theta)}\right)
    \end{split}
\end{equation}
Doing the $\phi$ integral and taking large $\gamma(\theta)$ limit, we get
\begin{equation}
    \ln Z^{(1)}\approx-2\pi\int d\theta\sin\theta\left(c_1\gamma(\theta)^3-\frac{\gamma'(\theta)^2}{\gamma(\theta)}(3c_1-4c_2)\right)
\end{equation}
Using the integrals from \S\ref{eq:integrals}, we obtain:
\begin{equation}
    \ln Z^{(1)} \approx -(4c_2)\frac{2\pi\beta }{3\nu}
\end{equation}
where the constants $c_i$ are defined in \eqref{eq:weyl}.

We observe that the term involving $R$ does not contribute at the leading order to the partition function. This originates from the fact that in $2d$, $\int d^2x ~ \sqrt{g_2} R = \mathcal{O}(1)$, is a topological term.

\subsection{Integral over $\theta$ of the $d=3$ partition function}\label{formchange}

In this Appendix we demonstrate that the integral \eqref{hyu} is proportional to $\frac{1}{\nu}$ at leading order in the small $\nu$ expansion. 

Consider the rewriting of \eqref{hyu} in the form 
\eqref{hyure}. While $I(\delta \theta)$ in \eqref{hyure} is a function of $\delta \theta$, it can also be thought of as a functional of $\gamma(\delta \theta)$. We adopt the second viewpoint below, and use the notation $I[\gamma(\delta \theta)]$. All throughout this subsection, we work in the approximation within which $\gamma(\delta \theta)$ takes the explicit form \eqref{gammatheta}. Note that when $(\delta \theta)^2 \gg \frac{2\nu}{\beta}$, it follows from \eqref{gammatheta} that $\gamma(\delta \theta) \sim \frac{1}{\delta \theta}$. Since $I[\gamma]$ is of homogeneity 3 in powers of $\gamma$ and derivatives together (and as each additional power of $\gamma$ -- and each additional derivative -- adds an additional 
power of $\delta \theta$ in the denominator), it follows that $I(\delta \theta)$ scales like $1/(\delta \theta)^3$ at large $\delta\theta$.\footnote{This is the case when we use the approximation \eqref{gammatheta}. If we worked with the exact expression for $\gamma$, these statements would all hold when $\delta \theta$ was large compared to $\frac{\sqrt{\nu}}{\beta}$, but small compared to unity.}

Even though the formula \eqref{hyu} is not completely universal, every term in this expression (including the terms with the $\ldots$) --  and so the entire integrand $I[\gamma(\theta)]$ on the RHS of \eqref{hyu} -- obeys the scaling property
\begin{equation}\label{termscal}
I\left[\alpha\gamma(\alpha \zeta )\right]\Bigg|_{\zeta= \frac{\delta \theta}{\alpha}}= \alpha^3 I\left[\gamma(\delta \theta)\right].
\end{equation}
Eqn. \eqref{termscal} simply asserts that $I$ is of homogeneity three in derivatives plus powers of $\gamma$. Note the operation of first scaling the coordinate with 
$\alpha$ (using the argument $\gamma(\alpha \zeta)$) taking derivatives (i.e.\ evaluating the function $I$ on this argument), then unscaling the coordinate
(finally replacing $\zeta$ by $\frac{\delta \theta}{\alpha}$), effectively weights every derivative with a factor of $\alpha$.

The scaling property \eqref{termscal}, together with the explicit expression \eqref{gammatheta}, allows 
one to immediately read off the dependence of $\ln Z$ on $\nu$ in a manner we now describe. In what follows, it will be useful to explicitly keep track of the dependence of $\gamma$ on $\nu$, i.e.\ view $\gamma$ as a function of both $\delta \theta$ and $\nu$, and denote it as  $\gamma( \delta \theta, \nu)$. 

From the explicit expression \eqref{gammatheta}, it is easily verified that 
\begin{equation}\label{gammascal2}
\alpha \gamma \left( \alpha \delta\theta, \nu \right)= \gamma\left( \delta \theta, \frac{\nu}{\alpha^2} \right) 
\end{equation}
Consequently, \eqref{termscal} can be rewritten as 
\begin{equation}\label{termscaln1}
I\left[\gamma\left(\zeta, \frac{\nu}{\alpha^2} \right)\right]\Big|_{\zeta=\frac{\delta \theta}{\alpha}}= \alpha^3 I\left[\gamma\left(\delta \theta, \nu)\right) \right]
\end{equation}
Setting $\alpha^2= \nu$ we conclude that 
\begin{equation}\label{termscalnint}
I\left[\gamma\left(\zeta, 1)\right) \right]
\Big|_{\zeta=\frac{\delta \theta}{\sqrt{\nu}}}=
\nu^{\frac{3}{2}} I\left[\gamma\left(\delta \theta, \nu)\right) \right]
\end{equation}
Rearranging, we conclude that
\begin{equation}\label{termscaln2}
    I\left[\gamma\left(\delta \theta, \nu)\right) \right]=\frac{1}{\nu^{\frac{3}{2}}}I\left[\gamma\left(\zeta, 1)\right) \right]
\Big|_{\zeta=\frac{\delta \theta}{\sqrt{\nu}}}
\end{equation}

Let us define the function $g$ by 
\begin{equation}\label{gdef}
g(\delta \theta)= I\left[ \gamma(\delta \theta, 1)\right]
\end{equation}
Note that the $\nu$ independent function $g$ varies over length scale unity (in its argument $\delta \theta$). As the RHS of \eqref{gdef} is independent of $\nu$, it follows that the quantity 
\begin{equation}\label{uui}
\int g(x) dx
\end{equation}
is also a $\nu$ independent (so ${\cal O}(1)$) number.
The discussion earlier in this section tells us that 
$g(x) \sim \frac{1}{x^3}$ at large $x$. Consequently, the integral \eqref{uui} converges at infinity.

Now it follows from \eqref{gdef} that 
\begin{equation}\label{gethe}
g\left(\frac{\delta \theta}{\sqrt{\nu}} \right) = I\left[\gamma\left(\zeta, 1)\right) \right]
\Big|_{\zeta=\frac{\delta \theta}{\sqrt{\nu}}}
\end{equation}
Consequently 
\begin{equation}\begin{split}\label{scalarg}
\ln Z&= \int d\delta \theta  I\left[ \gamma(\delta \theta,\nu) \right]\\
    & =  \int d\delta \theta ~ \frac{1}{\nu^{\frac{3}{2}}}I\left[\gamma\left(\zeta, 1)\right) \right]
\Big|_{\zeta=\frac{\delta \theta}{\sqrt{\nu}}} \\
& = \int d\zeta  ~ \frac{1}{\nu^{\frac{3}{2}}}g\left(\frac{\zeta}{\sqrt{\nu}}\right)\\
    & = \int\sqrt{\nu} d\chi ~ \frac{1}{\nu^{\frac{3}{2}}}g\left(\chi\right)\\ & = \frac{1}{\nu}\int d\chi g\left(\chi\right) \end{split}
\end{equation}
In going from the first to the second line of \eqref{scalarg} we have used \eqref{termscaln2}. In going from the second to the third line we have used \eqref{gethe}. In going from the third to the fourth line, we changed integration variables from $\zeta$
to $\chi=\frac{\zeta}{\sqrt{\nu}}.$
As the integral on the last line of \eqref{scalarg} is 
convergent (see the discussion under \eqref{uui}) and 
independent of $\nu$, it is a number of order unity. 
It thus follows that $\ln Z$ is of order $1/\nu$, as asserted.

\subsection{Illustration of $\delta$ function like behaviour about the equator in $d=3$}\label{df}

As we have explained in the main text, the partition function density ${\cal L}$ is highly localized around the equator, but the functional form of this localization is non-universal in $d=3$ (see around \eqref{hyu}). In this subsection we take one example of this profile (we assume that ${\cal L} \propto \gamma^3$, the first term in \eqref{hyu}), and carefully explain that ${\cal L}$ tends to a delta function localized around the equator in $d=3.$

Consider the integral
\begin{equation}
\int_{\Omega_2} \gamma^3(\theta) g_{\omega}(\theta, \phi)= \int d\theta\, d\phi\; \frac{\sin\theta}{\big(1-\omega^2\sin^2\theta\big)^{3/2}}\,g_\omega(\theta,\phi)\,,
\end{equation}
for some function $g_\omega(\theta,\phi)$ with a regular $\omega\to 1$ limit. The key point is that, in the $\omega\to 1$ limit, the dominant contribution to the integral arises from the region near $\theta=\pi/2$.  

For simplicity, let us first take $g=1$, so that we have 
\begin{equation}\label{eq:int}
2\pi \int^{\pi} d\theta\; \frac{\sin\theta}{\big(1-\omega^2\sin^2\theta\big)^{3/2}}\,.
\end{equation}

\begin{figure}[!ht]
\centering
\includegraphics[scale=0.6]{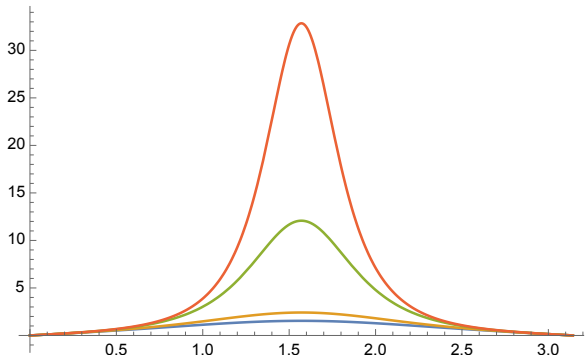}
\caption{The integrand in \eqref{eq:int} for $\omega \in  \{1/2,2/3,0.9,0.95\}$ is plotted.   As $\omega \to 1$, the denominator $(1-\omega^2\sin^2\theta)$ becomes small only in the vicinity of the equator $\theta=\pi/2$.  Thus the integrand is sharply peaked there, and the dominant contribution to the integral comes from a narrow neighborhood around $\theta=\pi/2$.  We make this precise below.}
\label{fig:1}
\end{figure}

This integral can be evaluated exactly:
\begin{equation}\label{eq:fest}
2\pi \int^{\pi} d\theta\; \frac{\sin\theta}{\big(1-\omega^2\sin^2\theta\big)^{3/2}}
= \frac{4\pi}{1-\omega^2}\,.
\end{equation}
We now show that most of the contribution to this integral indeed comes from the neighborhood of $\theta=\pi/2$.  To illustrate, we plot the integrand for various values of $\omega$ (see figure~\ref{fig:1}). The plots make it evident that, in the $\omega\to 1$ limit, the dominant contribution is localized near $\theta=\pi/2$.  

We can also support this conclusion with an analytic argument.  Let us define $I=(\pi/2-\epsilon,\pi/2+\epsilon)$.   We will show that as long as $\epsilon =(1-\omega^2)^{r}$ with $r<1/2$, the integral receives its dominant contribution from $I$ in the $\omega \to 1$ limit.  In order to do that,  we divide the integral into two pieces:
\begin{equation}
\int^{\pi } d\theta\  \frac{\sin\theta}{(1-\omega^2\sin^2\theta)^{3/2}} =\int_{I} d\theta\  \frac{\sin\theta}{(1-\omega^2\sin^2\theta)^{3/2}} + \int_{[0,\pi]\setminus I}  d\theta\  \frac{\sin\theta}{(1-\omega^2\sin^2\theta)^{3/2}}
\end{equation}

We have $1-\omega^2\sin^2\theta \;\geqslant\; \cos^2\theta\,$,  resulting in
\begin{equation}\label{eq:ub}
\bigg| \int_{[0,\pi]\setminus I} d\theta\; 
\frac{\sin\theta}{\big(1-\omega^2\sin^2\theta\big)^{3/2}} \bigg|
\;\leqslant\; \int_{[0,\pi]\setminus I} d\theta\; 
\frac{\sin\theta}{\big(1-\sin^2\theta\big)^{3/2}}
=\cot^2\epsilon
\;\underset{\epsilon\to0}{\sim}\; \,\epsilon^{-2}\,.
\end{equation}

Combining \eqref{eq:fest} and \eqref{eq:ub}, we conclude that
\begin{equation}
\displaystyle \int_{I} d\theta\; 
\frac{\sin\theta}{\big(1-\omega^2\sin^2\theta\big)^{3/2}}
= \frac{2}{1-\omega^2}\Bigg(1 + O\!\left((1-\omega^2)^{1-2r}\right)\Bigg)\,, 
\qquad r<\tfrac{1}{2}\,.
\end{equation}

This exercise demonstrates why EFT methods can still be reliably used to compute partition functions even in $d=3$ dimensions. Although $\gamma(\theta)$ does not diverge everywhere, the partition function integral localizes near the equator, where $\gamma$ becomes large.  In general dimension,  if we turn on one angular fugacity,  we will have $r< 1/(d-1)$. 

\section{General dimensions}\label{twon}

\subsection{Scaling form in general dimension}\label{scaleformgend}

As we have explained in the main text, in the limit \eqref{limnewdfour2}, $\ln Z$ (in $d \geq 2p$ dimensions) takes the form \eqref{nuppart}, i.e.\ is given by
\begin{equation}\label{pgenpart}
    \ln Z=\frac{\beta^p h(\beta,\nu_{p+1},\ldots,\nu_{[\frac{d}{2}]})}{\nu_1\ldots\nu_p}.
\end{equation}
The density of states may be obtained from this partition function by an inverse Laplace transform
\begin{equation}\label{entpgen}
\begin{split}
&\exp{\left(S(\tau,J_i)\right)}\\
&=\frac{1}{(2\pi i)^{[\frac{d}{2}]+1}}\int \frac{1}{\beta^{[\frac{d}{2}]}}d\beta d\nu_1\ldots d\nu_{[\frac{d}{2}]}e^{\left(\beta\tau+\sum_{i=1}^p\nu_iJ_i+\sum_{i=p+1}^{[\frac{d}{2}]}\nu_iJ_i+\ln Z(\beta,\nu_i)\right)}\\
    &=\frac{1}{(2\pi i)^{[\frac{d}{2}]+1}}\int \frac{1}{\beta^{[\frac{d}{2}]}}d\beta d\nu_1\ldots d\nu_{[\frac{d}{2}]}~e^{\left(\beta\tau+\sum_{i=1}^p\nu_iJ_i+\sum_{i=p+1}^{[\frac{d}{2}]}\nu_iJ_i+\frac{\beta^n h\left(\beta,\nu_{p+1},\ldots,\nu_{[\frac{d}{2}]}\right)}{\nu_1\ldots\nu_p}\right)}
    \end{split}
\end{equation}
When $\nu_1 \ldots \nu_p$ are all small, $\frac{1}{\nu_1 \ldots \nu_p}$
is large, and the integral over these chemical potentials may be performed via a saddle point approximation. The relevant saddle point equations are 
\begin{equation}
    J_i=\frac{\beta^p h(\beta,\nu_{p+1},\ldots,\nu_{[\frac{d}{2}]})}{\nu_i(\nu_1\ldots\nu_p)}=\frac{\beta^p h(\beta,\nu_{p+1},\ldots,\nu_{[\frac{d}{2}]})}{\theta \nu_i}~~\mathrm{for}~~ i = 1,\ldots, p
\end{equation}
where $\theta$ is given by 
\begin{equation}\label{deftheta}
    \theta=\nu_1\ldots\nu_p
\end{equation}
Now we can solve for $\theta$ as a function of $J_i$ as follows:
\begin{equation}
\begin{split}
    J_1\ldots J_p&=\frac{\beta^{p^2} h^{p}(\beta,\nu_{p+1},\ldots,\nu_{[\frac{d}{2}]})}{\theta^p \nu_1\ldots\nu_p}\\
    &=\frac{\beta^{p^2} h^{p}(\beta,\nu_{p+1},\ldots,\nu_{[\frac{d}{2}]})}{\theta^{p+1}}.
    \end{split}
\end{equation}
This gives
\begin{equation}
    \theta=\frac{\beta^{\frac{p^2}{p+1}}h^{\frac{p}{p+1}}(\beta,\nu_{p+1},\ldots,\nu_{[\frac{d}{2}]})}{(J_1\ldots J_p)^{\frac{1}{p+1}}}.
\end{equation}
Thus $\nu_i$ is given by the following 
\begin{equation}
    \nu_iJ_i=(J_1\ldots J_p)^{\frac{1}{p+1}}\beta^{\frac{p}{p+1}}h^{\frac{1}{p+1}}(\beta,\nu_{p+1},\ldots,\nu_{[\frac{d}{2}]}).
\end{equation}
Plugging this back in the exponent of \eqref{entpgen}, we see that this exponent simplifies to 
\begin{equation}\label{entpgenfinal}
\begin{aligned}
&S(\tau,J_i)\\
&=
\Bigg[
\beta \tau 
+ \sum_{i=p+1}^{[\frac{d}{2}]}\nu_i J_i 
+ (p+1)(J_1\ldots J_p)^{\frac{1}{p+1}}
  \beta^{\frac{p}{p+1}}
  h^{\frac{1}{p+1}}(\beta,\nu_{p+1},\ldots,\nu_{[\frac{d}{2}]})
\Bigg] \\
&=
(J_1\ldots J_p)^{\frac{1}{p+1}} \\
&\quad \times
\Bigg[
\beta \frac{\tau}{(J_1\ldots J_p)^{\frac{1}{p+1}}}
+ \sum_{i=p+1}^{[\frac{d}{2}]}
  \nu_i \frac{J_i}{(J_1\ldots J_p)^{\frac{1}{p+1}}}+(p+1)\beta^{\frac{p}{p+1}}h^{\frac{1}{p+1}}(\beta,\nu_{p+1},\ldots,\nu_{[\frac{d}{2}]})\Bigg].
\end{aligned}
\end{equation}
In the microcanonical ensemble, the limit \eqref{limnewdfour2} corresponds to having, for $i,j=1,\ldots p$ and $k=p+1,\ldots n$, the following limit:
\begin{equation}
\begin{cases}
    \tau \rightarrow \infty\\ J_i \rightarrow \infty \!\!
\end{cases}\ni\ \left(\frac{J_i}{J_j}\right)~,~\left(\frac{\tau}{(J_1\ldots J_n)^{\frac{1}{n+1}}}~,~\frac{J_{k}}{(J_1\ldots J_n)^{\frac{1}{n+1}}}\right)~~\mathrm{fixed.}
\end{equation}
As the RHS of the second line of \eqref{entpgenfinal} is multiplied by a large number, $(J_1\ldots J_p)^{\frac{1}{p+1}}$, the integral over $\beta$ and $\nu_j$, $j=p+1 \ldots n$, can be evaluated in  
the saddle point limit. It follows that the entropy as a function of charges is given by 
\begin{equation}\label{entgenchar}
\begin{split}
   S(\tau, J_i)&= (J_1\ldots J_p)^{\frac{1}{p+1}}
\, S^{\mathrm{int}}\!\left(
\frac{\tau}{(J_1\ldots J_p)^{\frac{1}{p+1}}},
\frac{J_i}{(J_1\ldots J_p)^{\frac{1}{p+1}}}
\right) 
\end{split}
\end{equation}

where $S^{\mathrm{int}}$ is given by
\begin{equation}
    S^{\rm int}(x,y_i)=\mathrm{Ext}_{\beta, \nu_{p+1},\ldots, \nu_{[\frac{d}{2}]}}\left( \beta x+\sum_{i=p+1}^{[\frac{d}{2}]}\nu_iy_i + (p+1)\beta^{\frac{n}{p+1}}h^{\frac{1}{p+1}}(\beta,\nu_{p+1},\ldots,\nu_{[\frac{d}{2}]})\right)
\end{equation}

In the special case that all $\nu_i$'s tend to zero with fixed ratios (see 
\eqref{limnewgend}), the partition function \eqref{pgenpart} simplifies to 
\begin{equation}\label{pfanycftap}
\ln Z(\beta, \nu_i)= \frac{\beta^n h(\beta)}{\nu_1\ldots\nu_n}\,.
\end{equation}
(see \eqref{pfanycft}) and the formula for the entropy \eqref{entgenchar} simplifies to 
\begin{equation}
\frac{S(\tau,J_i)}{(J_1\ldots J_n)^{\frac{1}{n+1}}} = S^{\mathrm{int}}\left( \frac{\tau}{(J_1\ldots J_n)^{\frac{1}{n+1}}}\right) 
\end{equation}
where 
\begin{equation}
    S^{\rm int}(x)=\mathrm{Ext}_{\beta}\left( \beta x+ (n+1)\beta^{\frac{n}{n+1}}h^{\frac{1}{n+1}}(\beta)\right)
\end{equation}

\subsection{Effective thermal partition function in $d=4$ with $\nu_1$ small but $\omega_2$ finite}\label{4drank2}

In this subsection we study the simplest example of an asymmetric scaling limit; namely the theory in $d=4$, in which one of the two angular velocities (let us say $\omega_1$) is scaled to unity, while the other is held fixed. In the terminology of \cite{Bajaj:2024utv}, this is a `Rank 2' situation (the grey galaxies that will be relevant to ${\cal N}=4$ Yang-Mills in this limit are rank 2 grey galaxies).

We use the same conventions as \S \ref{Tetrad} (i.e.\ same metric and vielbeins from \S \ref{Tetrad}). While the  exact formulae for curvatures, spin connections, and field strengths are unchanged, a few additional terms now survive at leading order, ${\cal L}$ localizes near one of the equators of $S^3$, and  every derivative orthogonal to the localization surface picks up an additional factor of $\frac{1}{\nu}$. (In many ways the situation is very similar to the 3d analysis, see  \S \ref{ssdthree}.)

With this point in mind, we see that the exact formulae for spin connections, namely \eqref{sc} are replaced by
\begin{equation}
    \begin{split}
        &\omega_2^{12} \approx -\frac{\gamma '(\theta )}{\gamma (\theta )}\\ &\omega_1^{23}=\omega_2^{13}=\omega_3^{12} \approx-\omega _2 \gamma (\theta )\\ &\omega_{3}^{\ 13}\approx\frac{\left(1-\omega _2^2\right) \gamma (\theta )^3}{\gamma '(\theta )}\\
    \end{split}
\end{equation}
together with their antisymmetric counterparts.

The exact expression for the Ricci scalar is given in footnote \ref{fn:exactR}
\begin{equation}\label{ricsca}
    R = 6 (1 - \omega_1^2)(1 - \omega_2^2) \gamma(\theta)^4 - 2 \gamma(\theta)^2 + 2.
\end{equation}
The leading order gauge field and field strength consist of all terms listed in \eqref{gfield} and \eqref{eq:field_strength_indiv}. The term $f_{ij}f^{ij}$ is now exactly
\begin{equation}\label{fielstr}
    f_{ij}f^{ij}=8 \gamma(\theta)^4\left(1-(1-\omega_1^2)(1-\omega_2)^2\gamma(\theta )^2\right).
\end{equation}
The third term in \eqref{thermpf2ap1} is given exactly by
\begin{equation}\label{4dr2sig}
(\partial\sigma)^2=\frac{\gamma'(\theta)^2}{\gamma(\theta)^2}=\gamma(\theta)^4(\omega_1^2-\omega_2^2)^2\sin^2\theta\cos^2\theta
\end{equation}

\subsubsection{Twisted partition function on $S^3 \times S^1$ at second order in derivatives}

Let us work with the second order truncation of the general expansion of  ${\cal L}$ (i.e.\ the form for ${\cal L}$ listed in \eqref{eq:partition}), namely
\begin{equation}\label{thermpf2ap2}
\begin{split}
&\ln Z^{(0)}=\int d^3x \sqrt{g_3} \frac{P(T(x))}{T(x)}\\
   &\ln Z^{(1)}= - \frac{1}{2} \int d^3x\, \sqrt{g_3} \left( 
        P_1(\sigma)\, R + 
        T^2 P_2(\sigma)\, f_{ij} f^{ij} + 
        P_3(\sigma)\, (\partial \sigma)^2 
    \right)+ \cdots\\
    & T(x) = T e^{-\sigma(x)}=T\gamma(\theta), ~~~~T= \frac{1}{\beta},~~~\sqrt{g_3}=\gamma(\theta)\sin\theta\cos\theta
    \end{split}
\end{equation}
where, from the requirement of invariance under Weyl scaling
\begin{equation}\label{4dpi}
    \begin{split}
        &P_1(\sigma)=c_1Te^{-\sigma}\\
        &P_2(\sigma)=\frac{c_2e^{\sigma}}{T}\\
        &P_3=2P_1=2c_1Te^{-\sigma}
    \end{split}
\end{equation}

We now specialize \eqref{thermpf2ap2} to the twisted partition function on $S^3 \times S^1$, in the limit of interest to this subsection. At zero derivative order (see the first line of \eqref{thermpf2ap2}), using Weyl invariance, we obtain
$P(T)= \kappa_2 T^4$. We then substitute $P(T)$ in \eqref{eq:partition} to obtain
\begin{equation}
\begin{split}
   \ln Z^{(0)}&= \int d^3x \sqrt{g_3}\frac{P(T(x))}{T(x)} \\&= \int d\phi_1 d\phi_2 d\theta\sin\theta\cos\theta \gamma(\theta)\kappa_2 T^3 \gamma^3(\theta)\\&\approx \frac{\pi ^2 T^2}{\nu _1 \left(1-\omega _2^2\right)}.
\end{split}
\end{equation}
Using \eqref{ricsca}, \eqref{fielstr}, and \eqref{4dr2sig}, we find that the effective action at second derivative order is (after performing the integral in the $\phi_1$ and $\phi_2$ directions)
\begin{equation}
\begin{split}
    \ln Z^{(1)}&=-2T\pi^2\int d\theta\sin\theta\cos\theta\Bigg(c_1\gamma(\theta)^2\bigg(6 (1 - \omega_1^2)(1 - \omega_2^2) \gamma(\theta)^4 - 2 \gamma(\theta)^2 +2\\&+2\gamma(\theta)^4(\omega_1^2-\omega_2^2)^2\sin^2\theta\cos^2\theta\bigg) +8c_2 \gamma(\theta)^4\left(1-(1-\omega_1^2)(1-\omega_2)^2\gamma(\theta )^2\right)\Bigg)\\
    &\approx\frac{h^{(1)}(\omega_2)}{\nu_1}
    \end{split}
\end{equation}
where we have taken the limit $\omega_1\rightarrow 1$ at the end. The coefficients $c_i$ were defined in \eqref{4dpi} and the function $h^{(1)}(\omega_2)$ is given by
\begin{equation}
    h^{(1)}(\omega_2)=
     \frac{\pi ^2 \left(c_1 \left(2 \omega _2^2-1\right)-2 c_2\left(1+\omega _2^2\right)\right)}{\left(1-\omega _2^2\right)}.
\end{equation}
If we further take the limit $\nu_2\rightarrow 0$, we get
\begin{equation}
    \ln Z^{(1)} \approx \frac{\pi ^2 \left(c_1 -4 c_2\right)}{2T\nu_1 \nu_2 }
\end{equation}
which indeed matches with our result in \eqref{eq:conc_partition3d}.

\subsection{Coefficient of the $\delta$ function in ${\cal L}$ in general dimensions }\label{gendimalg}

The argument for Eqn. \eqref{locallnz} parallels arguments used in Appendix \ref{formchange} (where we demonstrated that every leading order term in the $d=3$ version of 
\eqref{eq:partition} yields a contribution to $\ln Z \propto \frac{1}{\nu}$). We now present a sketch of this argument. 

We have explained above that the direction cosines 
$\mu_{p+1} \ldots \mu_n$ are all small when $\nu_i$ are small. Working to leading nontrivial order in these quantities, we see that 
\begin{equation}\label{expgama}
    \gamma=\frac{1}{\sqrt{\left(1-\sum_{i=1}^{p}\mu_i^2(1-\frac{\nu_i}{\beta})^2- \sum_{j=p+1}^{n}\mu_j^2\omega_j^2\right)}}
\end{equation}
The reader can easily verify that $\gamma$ is peaked when each of the $\mu_j$ 
(with $j=p+1 \ldots u$) is of order $\sqrt{\nu_i}$ or smaller. 

The $\mu_i$ that appear in \eqref{expgama} obey 
\begin{equation}\label{sumovem}
    \sum_{i=1}^{p}\mu_i^2=1-\sum_{j=p+1}^{u}\mu_j^2
\end{equation}
For this reason we work with the rescaled variables 
\begin{equation}\label{tm}
\mu_i= {\tilde \mu}_i \sqrt{1-\sum_{j=p+1}^{u} \mu_j^2}~~ ,~~i=1,\ldots,p
\end{equation}
As $\mu_j^2$ are all small, this redefinition is trivial at leading order, and so does not affect the metric of $S^{2p-1}$ at leading order. 

This redefinition turns \eqref{sumovem} into \eqref{mulow}. Inserting 
\eqref{tm} into \eqref{expgama}, and working to leading order in ${\cal O}(\nu_i)={\cal O} (\mu_j^2)$, and doing a little algebra, we find 
\begin{equation}\label{gammasimp}
    \gamma=\frac{1}{\left(\sum_{j=p+1}^{u}\mu^2_j(1-\omega_j^2) + \left(\frac{1}{\gamma_L}\right)^2\right)^{\frac{1}{2}}}
\end{equation}
The quantity $\gamma_L$ in equation \eqref{gammasimp} is defined by 
\begin{equation}\label{gammal}
\frac{1}{\gamma_L^2}= 1- \sum_{i=1}^p 
(1-\frac{\nu_i}{\beta}) {\tilde \mu}_i^2. 
\end{equation}
It has the following interpretation: $\gamma_L$ is simply the 
effective special relativistic $\gamma$ factor on the $S^{2p-1}$, 
as explained under \eqref{locallnz}. 

We observe from \eqref{gammasimp} that 
\begin{equation}\label{scalegamma}
    \alpha \gamma(\alpha\, \mu_j,\gamma_L)=\gamma(\mu_j,\gamma_L\alpha)
\end{equation}

We have already seen that the $\ln Z$ density 
${\cal L}$, in this case, is highly peaked around $\mu_j=0$. In the limit of interest to this subsection, therefore, the free energy can be rewritten as
\begin{equation}\label{lnz}
\ln Z= \int_{S^{2p-1}} \left[ \int d^{d-2p} \mu_j I(\mu_j, \gamma_L) \right] 
\end{equation}
where the $\mu_j$ only need be integrated over values of order $\frac{1}{\gamma_L} \ll 1$ (because the integrand is comparatively negligible at larger values of $\mu_j$) and so effectively are vectors 
in the part of the tangent space of $S^d$ that is orthogonal to $S^p$. (Eqn. \eqref{gammasimp} explains why $I$ depends on the coordinates of the $S^p$ only through $\gamma_L$.)

The analysis of Appendix \ref{formchange} -- when generalized to arbitrary dimensions, tells us that 
\begin{equation}\label{integscale}
\begin{split}
    I\left [\alpha \gamma\left(\alpha \zeta_i,\gamma_L\right)\right]\Big|_{\zeta_j =\frac{\mu_j}{\alpha}} = \alpha^{d}I\left[\gamma\left(\mu_j,\gamma_L\right)\right]\\
    \end{split}
\end{equation}
Using \eqref{scalegamma}, \eqref{integscale} can be rewritten as
\begin{equation}\label{integscalen}
\begin{split}
    I\left [\gamma\left(\zeta_i, \alpha \gamma_L\right)\right]\Big|_{\zeta_j =\frac{\mu_j}{\alpha}} = \alpha^{d}I\left[\gamma\left(\mu_j,\gamma_L\right)\right]\\
    \end{split}
\end{equation}
\eqref{integscalen} holds for any choice of the constant $\alpha$. Making the choice $\alpha = \frac{1}{\gamma_L}$ yields 
\begin{equation}\label{integscalenn}
\begin{split}
    I\left [\gamma\left(\zeta_i, 1\right)\right]\Big|_{\zeta_j =\gamma_L\mu_j} = \gamma_L^{-d}I\left[\gamma\left(\mu_j,\gamma_L\right)\right]\\
    \end{split}
\end{equation}
Upon rearranging, we obtain 
\begin{equation}\label{integscalennn1}
\begin{split}
I\left[\gamma\left(\mu_j,\gamma_L\right)\right]=\gamma_L^{d} I\left [\gamma\left(\zeta_i, 1\right)\right]\Big|_{\zeta_j =\gamma_L\mu_j} \\
    \end{split}
\end{equation}
Following the analysis of Appendix \ref{formchange}, we define the function 
\begin{equation}\label{fdef}
g_d(\mu_j)=I\left [\gamma\left(\zeta_i, 1\right)\right].
\end{equation} 
The function $g_d$ is independent of $\gamma_L$, and so all angular scales of variation in this function are of order unity. \eqref{integscalennn1} can be re-expressed, in terms of the function $g_d$ and using the substitution $\zeta_j = \gamma_L \mu_j$, as 
\begin{equation}\label{integscalennn2}
\begin{split}
I\left[\gamma\left(\mu_j,\gamma_L\right)\right]=\gamma_L^{d}  g_d(\gamma_L\mu_j) \\
    \end{split}
\end{equation}

Let ${\vec y}$ represent a Cartesian coordinate in the $(d-2p)$-dimensional part of tangent space that lies orthogonal to the $S^{2p-1}$. We see that 
\begin{equation}\begin{split} \label{manman}
\int d^{d-2p}y I(\mu_j, \gamma_L) 
&=\int d^{d-2p}y \gamma_L^d g_d(\gamma_L y_j)\\
&=\gamma_L^{2p}\int d^{d-2p} z  g_d(z_j)
\end{split}
\end{equation}
where, in going from first to second line of eq. \eqref{manman}, we have made the variable change $z_j=\gamma_L y_j$. We also have used the facts that there are $d-2p$ of the $\mu_j$'s, and that $\gamma_L$ is a constant as far as the integral over the (tangent space of) $\mu_j$ is concerned. 

Imitating the analysis of the analogous question in 
$d=3$ (see Appendix \ref{formchange}), it is easy to argue that the integral on the RHS of \eqref{manman} converges. As the function $g_d$ is independent of 
$\gamma_L$, it follows that the integral converges to a number of order unity. We thus conclude that the integral on the RHS of \eqref{manman} scales like $\gamma^{2p}$. 
Plugging this result into \eqref{lnz} establishes \eqref{locallnz}.

\section{Asymptotics in $d=4$ using Meinardus technique}\label{app:Rigor}
In this appendix, we will  use the methods which were used to prove the Meinardus theorem, as explained in Chapter $6$ of \cite{Andrews_1984}, and used in \cite{Melia:2020pzd} to study the $\nu_i\to 0$ limit of the partition function. For simplicity, we will set $\omega_1=\omega_2=\omega$,  and define 
\begin{equation}
\beta_{L}:=\beta(1+\omega)\,,\quad\quad \beta_R:=\beta(1-\omega)\,.
\end{equation}
Here we will be using $\beta_R$ as a proxy for $\nu$. The logarithm of the partition function for the massless free scalar field theory in $d=4$ is given by 
\begin{equation}
\ln Z = \sum_{q}\frac{1}{q} \frac{e^{-\beta q}\left(1-e^{-q(\beta_L+\beta_R)}\right)} {(1-e^{-q\beta_L})^2(1-e^{-q\beta_R})^2}\,.
\end{equation}
We are interested in the $\beta_R\to 0$ and $\beta_L\to 2\beta$ limit.  

First of all, note that 
\begin{equation}\label{eq:sum}
\begin{aligned}
\ln Z &= \sum_q\frac{e^{-\beta q}}{q} \left(1-e^{-q(\beta_L+\beta_R)}\right) \left(\sum_{k=0}^{\infty} (k+1)e^{-kq\beta_L}\right) \left(\sum_{k'=0}^{\infty} (k'+1)e^{-k'q\beta_R}\right)\\
&=\sum_q\frac{e^{-q\beta_L/2}}{q}\left(\sum_{k=0}^{\infty} (k+1)e^{-kq\beta_L}\right) \left(\sum_{k'=0}^{\infty} (k'+1)e^{-(k'+1/2)q\beta_R}\right)\\
&\qquad\qquad- \sum_q\frac{e^{-\beta_L q/2}}{q}\left(\sum_{k=1}^{\infty} ke^{-kq\beta_L}\right) \left(\sum_{k'=1}^{\infty} k'e^{-(k'+1/2)q\beta_R}\right)\,.
\end{aligned}
\end{equation}

Now we will use the following identity
\begin{equation}\label{eq:Mellin}
e^{-(k'+1/2)q\beta_R}= \frac{1}{2\pi i}\int_{3-i\infty}^{3+i\infty}\ ds\  \frac{\Gamma(s)}{\beta_R^{s}}\frac{1}{q^{s}}\ (k'+1/2)^{-s} \,,\quad \mathrm{Re}\left(q\beta_R(k'+1/2)\right)>0\,,
\end{equation}
to replace the exponential factors appearing in the sums of \eqref{eq:sum}. To proceed, let us define 
\begin{equation}
\begin{aligned}
w_1(s)&:=\sum_q \frac{1}{q^{s+1}}\ \frac{e^{-q\beta_L/2}}{(1-e^{-q\beta_L})^2}\\
w_2(s)&:=\sum_q \frac{1}{q^{s+1}}\ \frac{e^{-3q\beta_L/2}}{(1-e^{-q\beta_L})^2}\\
D_1(s)&:=\sum_{k'=0}^{\infty} \frac{k'+1}{(k'+1/2)^s}=\frac{1}{2} \left(\left(2^s-2\right) \zeta (s-1)+\left(2^s-1\right) \zeta (s)\right)\\
D_2(s)&:=\sum_{k'=1}^{\infty} \frac{k'}{(k'+1/2)^s}=\frac{1}{2} \left(\left(2^s-2\right) \zeta (s-1)-\left(2^s-1\right) \zeta (s)\right)
\end{aligned}
\end{equation}
We plug the identity \eqref{eq:Mellin}, coming from Mellin transformation, back into the sum \eqref{eq:sum}, exchange the sum over $k',q$ and the integral over $s$ (the exchange is allowed since the sum defining $D_i(s)$ is absolutely convergent), and do the sum over $k',q$ to obtain 
\begin{equation}
\ln Z=\frac{1}{2\pi i}\int_{3-i\infty}^{3+i\infty}\ ds\ w_1(s) \Gamma(s) D_1(s) \beta_R^{-s}-\frac{1}{2\pi i}\int_{3-i\infty}^{3+i\infty}\ ds\ w_2(s)  \Gamma(s) D_2(s) \beta_R^{-s}
\end{equation}

Now $\beta_R^{-s}D_i(s)w_i(s)$ has first order poles at $s=2$, $s=1$ and $s=0$. We use the methods which were used to prove the Meinardus theorem, as explained in the Chapter $6$ of \cite{Andrews_1984} and \cite{Melia:2020pzd}, to deduce\footnote{Essentially, we shift the vertical contour to the left to a vertical line given by $-1<\text{Re}(s)<0$. We note that once we move the contour to $-1<\text{Re}(s)<0$, $\beta_R^{-s}$ is not a divergent piece on this contour. This helps tame the integral and show that it is subdominant, see \eqref{eq:subD} and \eqref{eq:subD2} below. While shifting the contour, the contribution at $\text{Im}(s)=\pm\infty$ can be dropped owing to the presence of the $\Gamma(s)$ factor. However, when we shift the contour,  we do encounter poles and they contribute to the integral.} 
\begin{equation}
\begin{aligned}
 \ln Z &\sim \frac{w_1(2)\Gamma(2)}{\beta_R^2}+\frac{w_1(1)\Gamma(1) }{2\beta_R}+w_1(0)D_1(0)-\frac{w_2(2)\Gamma(2)}{\beta_R^2}+\frac{w_2(1)\Gamma(1) }{2\beta_R}-w_2(0)D_2(0)\\
  &\sim \frac{1}{\beta^2(1-\omega)^2}\sum_{q}\frac{1}{q^3}\frac{e^{-\tfrac{q\beta_L}{2}}(1-e^{-q\beta_L})}{(1-e^{-q\beta_L})^2}+ \frac{1}{2\beta(1-\omega)}\sum_{q}\frac{1}{q^2}\frac{e^{-\tfrac{q\beta_L}{2}}(1+e^{-q\beta_L})}{(1-e^{-q\beta_L})^2}\\
&\qquad\qquad\qquad\qquad\qquad\qquad\qquad\qquad\qquad+\frac{1}{24}\sum_{q}\frac{1}{q}\frac{e^{-\tfrac{q\beta_L}{2}}(1-e^{-q\beta_L})}{(1-e^{-q\beta_L})^2}.
 \end{aligned}
\end{equation}

The error term can be estimated by bounding the following integral 
\begin{equation}\label{eq:subD}
   \left(\int_{-C-i\infty}^{-C+i\infty}\ ds\ \sum_i \left|\frac{1}{2\pi i}w_i(s)\Gamma(s)D_i(s)\beta_R^{-s} \right| \right)\,,\quad  0<C<1\,.
\end{equation}
Now we use the estimates \begin{equation}\label{eq:subD2}
    \Gamma(s)=O\left(e^{-\frac{\pi}{2}|\mathrm{Im}(s)|}|\mathrm{Im}(s)|^{C_1}\right)\,,\quad w_i(s)= O\left(|\mathrm{Im}(s)|^{K_i}\right)\,,\quad D_i(s)= O\left(|\mathrm{Im}(s)|^{R_i}\right)
\end{equation}
for some $C_1, K_i, R_i>0$ to show that the above integral is $O(\beta_R^{C})$ and goes to $0$ as $\beta_R\to0$.

\section{Details of the instanton path integral}\label{symmetries}

In this appendix we present several details relating to the worldline 
instantons discussed in \S \ref{instnonpert}. 

\subsection{The space of solutions from the viewpoint of $SU(2)_R$}\label{formskilling}

When $\omega_1=\omega_2=\omega$, the  Weyl rescaled base metric \eqref{eq:fieldsnew} can be written in terms of the right-invariant one-forms
\begin{equation}
ds^2=\frac{1}{\left(1-\omega^2\right)^2}\left[{(\sigma_1^2+\sigma_2^2)(1-\omega^2)+\sigma_3^2}\right]
\end{equation}
where
\begin{equation}\label{oneforms1}
    \begin{split}
    \sigma_1 &= \frac{1}{2}\left(\sin(\phi) d (2\theta) - \cos(\phi)\sin(2\theta)d\psi\right)\\ \sigma_2 &=\frac{1}{2}\left(\cos(\phi) d(2{\theta}) + \sin(\phi)\sin(2{\theta})d\psi\right)\\\sigma_3 &=\frac{1}{2}(d\phi - \cos(2{\theta})d\psi)\\
    \end{split}
\end{equation}
 and 
\begin{equation} \label{phipsi2}
    \begin{split}
     \phi&= \phi_1 + \phi_2, ~~~~\psi = \phi_1 - \phi_2\\
    \end{split}
\end{equation}
(see \eqref{phipsi1}). 
The gauge field in \eqref{eq:fieldsnew} turns out to be proportional to 
$\sigma_3$, 
\begin{equation}\label{aa}
    a=-\left(\frac{\omega}{1-\omega^2}\right)\sigma_3
\end{equation}
 and its field strength is
 \begin{equation}\label{fsaa}
     f=-\left(\frac{\omega}{1-\omega^2}\right)\sigma_1\wedge \sigma_2
 \end{equation}
It follows, therefore, that \eqref{aa} and \eqref{fsaa} are 
$SU(2)_R \times U(1)_L$ invariant. 

The Killing vectors that generate $SU(2)_L$ are the right-invariant vector fields\footnote{
These vector fields induce the following coordinate changes: 
\begin{equation}\label{changes}
    \begin{split}
        &\delta\phi_1=-\frac{2\cos{\phi}}{\tan{2\theta}},~~~\delta\theta_1=\sin{\phi},~~~\delta\psi_1=-\frac{2\cos{\phi}}{\sin{2\theta}}\\
        &\delta\phi_2=\frac{2\sin{\phi}}{\tan{2\theta}},~~~\delta\theta_2=\cos{\phi},~~~\delta\psi_2=\frac{2\sin{\phi}}{\sin{2\theta}}\\
        &\delta\phi_3=2,~~~\delta\theta_3=0,~~~\delta\psi_3=0
    \end{split}
\end{equation} }  \footnote{ These vector fields are right-invariant simply because $SU(2)_R$ and $SU(2)_L$ commute.} 
\begin{equation}\label{rightinvkill}
\begin{split}
    &R_1=-\frac{2\cos{\phi}}{\tan{2\theta}}\partial_\phi+\sin{\phi\partial_{\theta}}-\frac{2\cos{\phi}}{\sin{2\theta}}\partial_\psi\\
    &R_2=\frac{2\sin{\phi}}{\tan{2\theta}}\partial_\phi+\cos{\phi\partial_{\theta}}+\frac{2\sin{\phi}}{\sin{2\theta}}\partial_\psi\\
    &R_3=2\partial_\phi\\
    \end{split}
\end{equation}
These right-invariant vector fields are dual to right-invariant one-forms \eqref{oneforms1}, in the sense that the contraction $\sigma_R^i (R_j)= \delta^i_j$.

On the other hand, the Killing vectors that generate $SU(2)_R$ are 
the left-invariant vector fields and are given by
\begin{equation}\label{linv}
    \begin{split}
        L_1 & = \sin \psi \partial_\theta + \frac{2 \cos\psi}{\sin 2\theta}\partial_\phi + \frac{2 \cos \psi}{\tan 2\theta}\partial_\psi\\
        L_2 & = \cos\psi\partial_\theta - \frac{2\sin\psi}{\sin 2\theta}\partial_\phi
 - \frac{2\sin \psi}{\tan 2\theta}\partial_\psi\\
L_3 & = 2 \partial_\psi\\
 \end{split}
\end{equation}

These Killing vectors are dual to the left-invariant one-forms
\begin{equation}
    \begin{split}
        &\sigma_1^L=\frac{1}{2}\left(\sin{\psi} d(2\theta)+\cos{\psi}\sin{2\theta}d\phi\right)\\
        &\sigma_2^L=\frac{1}{2}\left(-\cos{\psi} d(2\theta)+\sin{\psi}\sin{2\theta}d\phi\right)\\
        &\sigma_3^L=\frac{1}{2}\left(d\psi-\cos{2\theta}d\phi\right)
    \end{split}
\end{equation}
(once again this duality means that $\sigma_L^i(L_j)= \delta^i_j$).

As the base metric and gauge field in \eqref{eq:fieldsnew} are both written entirely in terms of right-invariant one-forms, it follows that coordinate transformations generated by \eqref{linv} are symmetries (and thus conserved charges) of our particle motion. The $z$ component of $SU(2)_L$ is also a symmetry. The Killing vector for this symmetry simply generates a shift of $\phi$. 

\subsection{Noether charges}\label{chargesgen} 

In this subsection we present the expressions for the Noether charges corresponding to the preserved symmetries described in the previous subsection. Consider the worldline action \eqref{actforpart}. We use the coordinate $\tau$ to parameterize positions on this worldline (the precise choice of $\tau$ will not matter as all our formulae below will be invariant under monotonic coordinate redefinitions of $\tau$). Let $p_{\mu}$ be the canonical momenta for the action \eqref{actforpart}.  Specifically:
\begin{equation}\label{momenta}
\begin{split}
    &p_{\theta}=\frac{\gamma \,m_n\, \dot{\theta}}{\mathcal{L}_0},\\
    &p_{\phi}=\frac{\gamma^3\, m_n\left(\dot{\phi}-\dot{\psi}\cos(2\theta)\right)}{4\mathcal{L}_0}+\frac{\gamma^2n\pi \omega}{\beta}\\
    &p_{\psi}= \gamma m_n\left(-\,\frac{\omega^{2}}{4(1-\omega^{2})}\frac{(\dot{\phi}-\dot{\psi}\cos2\theta)\cos2\theta}{\mathcal{L}_0}
\;+\;\frac{\dot{\psi}-\dot{\phi}\cos2\theta}{4\,\mathcal{L}_0}
\right)-\frac{\pi n \omega\gamma^2}{\beta}\cos2\theta.
    \end{split}
\end{equation}
where
\begin{equation}
  \begin{split}&\mathcal{L}_0=\left[\dot{\theta}^{2}
+\frac{w^{2}}{4(1-w^{2})}\big(\dot{\phi}-\dot{\psi}\cos2\theta\big)^{2}
+\frac{1}{4}\big(\dot{\phi}^{2}+\dot{\psi}^{2}-2\dot{\phi}\dot{\psi}\cos2\theta\big)\right]^{\frac{1}{2}}\\
&\gamma=\frac{1}{\sqrt{1-\omega^2}}.
\end{split}
\end{equation}
The Noether charge corresponding to any target space Killing vector $K^\mu \partial_\mu $ is given by  $Q= K^\mu p_\mu$. The  
conserved charges corresponding to the $SU(2)_R$ and $U(1)_L$ Killing vectors are given by plugging \eqref{linv} and the third equation of 
\eqref{rightinvkill} into this formula and we find 
 \begin{align}\label{q1}
Q^x_{R} &= p_{\theta}\sin\psi
  + \frac{2\cos\psi}{\sin2\theta}\,p_{\phi}
  + \frac{2\cos\psi\cos2\theta}{\sin2\theta}\,p_{\psi}, \\[4pt]\label{q2}
Q^y_{R} &= p_{\theta}\cos\psi
  - \frac{2\sin\psi}{\sin2\theta}\,p_{\phi}
  - \frac{2\sin\psi\cos2\theta}{\sin2\theta}\,p_{\psi}, \\[4pt]
Q^z_{R} &= 2\,p_{\psi}, \\[4pt]
Q^z_{L}&=2p_{\phi}.
\end{align}

\subsection{Solutions with $ Q^x_R=Q^y_R=0$}\label{gengeod}
In this subsection,  we find all solutions with 
\begin{equation}\label{q0}
Q^x_R=Q^y_R=0
\end{equation}
From \eqref{q1} and \eqref{q2}, we can immediately see that both the charges $Q^x_R$ and $Q^y_R$ can vanish if and only if
\begin{equation}\label{constgeo}
  \qquad p_{\theta}=0 ~~~~~ \rm{and}~~~~ p_{\phi}+\cos{2\theta} p_{\psi}=0 
\end{equation}
It follows immediately from the first condition in \eqref{constgeo} (see \eqref{momenta})  that
\begin{align}
    \dot{\theta}=0.
\end{align}

In order to understand the implications of the second equation in \eqref{constgeo}, we plug the expressions of $p_{\phi}$ and $p_{\psi}$ from the second and the third equations of \eqref{momenta} and square the second condition in \eqref{constgeo}. We find 
\begin{equation}\label{congeo2}
    \frac{\sin ^4(2\theta )  \left(1-\frac{\dot{\psi}^2}{\dot{\phi}^2}  \omega^2\right)}{16\left(1-\omega^2\right)}=0.
\end{equation}
It follows, therefore that the second condition in \eqref{constgeo} is satisfied if
\begin{equation}\label{othercondapp}
{\rm Either} ~~~\sin 2 \theta=0, ~~~{\rm or} ~~~\frac{\dot \psi}{\dot \phi}= 
\frac{1}{\omega}
\end{equation}
The charges $Q^z_R$ and $Q^z_L$ on the second of the above-mentioned geodesic configurations are given by
\begin{equation}
    \begin{split}
        &Q^z_R=\frac{\, m_n\left((1-\omega^2\sin^2{2\theta})-\omega\cos(2\theta)\right)}{\left(1-\omega^2\right)\left(1-\omega \cos (2 \theta )\right)}-2\frac{\gamma^2n\pi \omega\cos{2\theta}}{\beta}\\
        &Q^z_L=\frac{\, m_n\left(\omega-\cos(2\theta)\right)}{\left(1-\omega^2\right)(1-\omega  \cos (2 \theta ))}+2\frac{\gamma^2n\pi \omega}{\beta}
    \end{split}
\end{equation}
We can see easily that when $\theta=0$, the Casimir of SU(2)$_R$ reduces to the one given in \eqref{casimir}.

\subsection{Quadratic fluctuations about the special solutions}\label{quadfluc}

As we have mentioned in the main text, the $SU(2)_R$ orbits of the special solutions at $\theta=0$ (and $\theta=\frac{\pi}{2}$) are solutions with $\theta=\theta_0$ and $\psi$ both constant. That these configurations are solutions may also be verified by expanding the worldline action around these configurations. Our action takes the form  
\begin{equation}
    \begin{split}
        S_E &= \frac{2 \pi |n|}{\beta\sqrt{(1-\omega^2)}} \int \left(d\theta^2 + \frac{w^2}{1-w^2}(d\phi_1 \sin^2\theta + d\phi_2 \cos^2 \theta)^2 + d\phi_1^2 \sin ^2 \theta + d\phi_2^2 \cos^2 \theta \right )^{\frac{1}{2}}\\
    &\quad +\frac{2\pi\omega n}{\beta(1-\omega^2)} \int \left (\sin^2\theta d\phi_1 + \cos^2\theta d\phi_2 \right).
    \end{split}
\end{equation}
Rewriting in terms of $\phi$ and $\psi$, and expanding to quadratic order in the small parameters $\delta \theta$ and $\delta \psi$, 
we find 
\begin{equation} \label{see}
    \begin{split}
        S_E=&\frac{\pi |n|}{\beta(1+\omega)}\left(\int d\phi-\int d\phi(\cos{2\theta_0}-2\delta\theta\sin{2\theta_0})\frac{d\delta\psi}{d\phi}\right)\\
        &+\frac{2\pi n}{\beta}\left[\int d\phi\left(\frac{d\delta\theta}{d\phi}\right)^2+\frac{1}{4}\int d\phi\left(\sin^22\theta_0+2\delta\theta\sin{4\theta_0}\right)\left(\frac{d\delta\psi}{d\phi}\right)^2\right]
    \end{split}
\end{equation}
(we have assumed that $\omega>0$ and  $n<0$).

The integral over $\phi$ gives $4 \pi$. The integral 
$\int d\psi$ is zero (while $\psi$ can wind, it cannot do so while always remaining small). The last term on the last line of \eqref{see} is cubic in fluctuations, so we ignore it. It follows that \eqref{see} can be simplified to 
\begin{equation} \label{seesimp}
    \begin{split}
        S_E-\frac{4\pi^2 |n|}{\beta(1+\omega)}= \frac{2\pi |n|}{\beta} \int d\phi  \left(
        \left(\delta \theta'\right)^2 +\frac{\sin^22\theta_0}{4} \left(\delta \psi'\right)^2  -\frac{\sin 2 \theta_0 }{1+\omega}
        \delta \theta  \delta \psi ' \right) 
    \end{split}
\end{equation}
where $'$ means derivative w.r.t. $\phi$. Since $\phi$ is periodic with a period of $4\pi$, in momentum space, we have

\begin{equation}\label{delphi}
    \delta \theta(\phi)= 
\displaystyle\sum_{p}  e^{\frac{ip\phi}{2}} \delta \theta_p
\end{equation}

\begin{equation}\label{delpsi}
    \delta \psi(\phi)=\displaystyle\sum_{p}e^{\frac{ip \phi}{2}} \delta \psi_p
\end{equation}
where $p \in \mathbb{Z}$.
 
We find that the fluctuation part of the action takes the form 
\begin{equation}\label{fluclag}
\frac{2 \pi n}{\beta}
\begin{pmatrix} \delta \theta_{-p}  & \sin 2 \theta_0 \delta \psi_{-p}  \end{pmatrix}
\begin{pmatrix} \frac{p^2}{4} & -  \frac{ip}{4(1+\omega)} \\  \frac{ip}{4(1+\omega)} &  \frac{p^2}{16} \end{pmatrix}
\begin{pmatrix} \delta \theta_p \\ \sin 2 \theta_0 \delta \psi_p  \end{pmatrix}.
\end{equation}
The eigenvalues of this matrix are 
\begin{equation}\label{em}
    \frac{1}{32} p \left(5 p-\frac{\sqrt{9 p^2 (\omega +1)^2+64}}{\omega +1}\right)\,,\quad \frac{1}{32} p \left(5 p+\frac{\sqrt{9 p^2 (\omega +1)^2+64}}{\omega +1}\right).
\end{equation}
In the limit  $\omega \to 1$ they simplify to 
\begin{equation}\label{psimp}
    \frac{1}{32} p \left(5 p-\sqrt{9p^2+16}\right)\,,\quad \frac{1}{32} p \left(\sqrt{9p^2+16}+5 p\right).
\end{equation}
These eigenvalues both clearly vanish at $p=0$. As explained in the 
main text, these zero eigenvalues move us on the two-dimensional manifold of solutions on which $\theta$ and $\psi$ are both constant. 

Both the eigenvalues in \eqref{psimp} are positive  for $p\geq 2$. 
When $p=1$, the second eigenvalue in \eqref{psimp} continues to be positive, but the first eigenvalue vanishes. This vanishing is an artifact of sitting at $\omega =1$. Expanding around $\omega=1$ and working to first order in the expansion, we find that the first term of \eqref{em} at $p=1$ reduces to
\begin{equation}
    -\frac{1}{20}(1-\omega)
\end{equation}
Note that this eigenvalue is small and tachyonic. The eigenvalue becomes further negative as $\omega$ is further reduced, taking its minimum value at $\omega=0$. 
For $\omega=0$ the eigenvalues are
\begin{equation}
    \frac{1}{32} p \left(5 p-\sqrt{9 p^2+64}\right),\frac{1}{32} p \left(\sqrt{9 p^2+64}+5 p\right)
\end{equation}

\section{Details of the black hole phase} \label{dbhp}

\subsection{Units, Newton constant and conventions} \label{uncc}

Recall that the low energy effective action of IIB theory in flat 10-dimensional space is 
$$\frac{1}{(2 \pi)^7 (\alpha^{'})^4 g_s^2} 
\int \sqrt{-g} \left( e^{-2\phi} R+ \ldots \right). $$
From this, we see that 
\begin{equation}\label{spg}
16 \pi G_{10}=(2 \pi)^7 (\alpha')^4 g_s^2
, ~~~~{\text{i.e.}}~ G_{10}= \frac{(2 \pi)^6 (\alpha')^4 g_s^2}{8}.
\end{equation}
The radius of $AdS_5 \times S^5$ equals 
\begin{equation}\label{rads}
R_{AdS}^2=\alpha'\sqrt{4\pi gN}
\end{equation}
Throughout this paper, we work in units in which $R_{Ads}$ is set to unity.
In these units, effectively, 
\begin{equation}\label{effalphap}
\frac{1}{\sqrt{\alpha'}}=(4\pi g N)^{\frac{1}{4}}
\end{equation}
and the IIB action takes the 
form
\begin{equation}\label{tendtb}
\frac{N^2}{8 \pi^5}\int d^{10}x \sqrt{-g} \left( e^{-2 \phi} R \ldots \right) 
\end{equation} 
and 
\begin{equation}\label{constants}
G_{10}=\frac{\pi^4}{ 2 N^2}.
\end{equation}

Using the relation $S_5=\pi^3$, and dimensionally reducing, we find that the 5d action takes the form 
\begin{equation}\label{fdbh}
\frac{N^2}{8 \pi^2}\int d^{5}x \sqrt{-g} \left( e^{-2 \phi} R \ldots \right) 
\end{equation} 
while the 5 dimensional Newton constant is given by 
\begin{equation}\label{gfive}
G_5= \frac{G_{10}}{\Omega_5}= \frac{\pi}{2 N^2}
\end{equation}
Working in the same units, the  effective `Hagedorn temperature' 
equals
\begin{equation}\label{hagtemp} 
T_H= \frac{1}{2 \pi \sqrt{ 2 \alpha'}} = \frac{(4\pi)^{\frac{1}{4}}}{2\pi\sqrt{2}}\lambda^{\frac{1}{4}}=\frac{\lambda^{\frac{1}{4}}}{2\pi^{\frac{3}{4}}}
\end{equation}
(see e.g. \cite{Aharony:2003sx}); in going from the first to the second expression in \eqref{hagtemp}, we have used \eqref{effalphap}.

\subsection{Exact thermodynamical formulae for the Kerr-AdS$_5$ black hole}
\label{etfka}

The energy, angular momenta and angular speeds of 5-dimensional Kerr-AdS$_5$ black hole are given (in field theory units) by \cite{Gibbons:2004ai}
\begin{equation} \label{thermformmt}
\begin{split}&\frac{E}{G_5}=\frac{m\pi(2\Xi_a+2\Xi_b-\Xi_a\Xi_b)}{4\Xi_{a}^2\Xi_{b}^2}\\
    &\frac{J_a}{G_5}=\frac{2\pi a m}{4\Xi_{a}^2\Xi_{b}}\\
    &\frac{J_b}{G_5}=\frac{2\pi b m}{4\Xi_{a}\Xi_{b}^2}\\
    &\frac{\tau}{G_5} \;=\; 
\frac{\pi m \,\big(2\Xi_a(1-b) + 2\Xi_b(1-a) - \Xi_a \Xi_b}{4 \,\Xi_a^2 \,\Xi_b^2}\\
    &\omega_a=\frac{a(r_{+}^2+1)}{(r_{+}^2+a^2)}\\
    &\omega_b=\frac{b(r_{+}^2+1)}{(r_{+}^2+b^2)}\\
   &m=\frac{(r_+^2 + a^2)(r_+^2 + b^2)(r_+^2 + 1)}{2r_+^2}\\
   &\frac{S}{G_5}=\frac{\pi^2(r_{+}^2+a^2)(r_{+}^2+b^2)}{2\Xi_{a}\Xi_{b}r_{+}}\\
   &T=\frac{r_+}{2\pi}(1+r^2_{+})\left(\frac{1}{r^2_{+}+a^2}+\frac{1}{r^2_{+}+b^2}\right)-\frac{1}{2 \pi r_{+}}\\ 
\frac{\ln Z}{G_5} 
&= \frac{S - \beta E + \beta \omega_i J^i}{G_5} \\
&= \frac{\pi^2 \left(a^2 + r_+^2\right)\left(b^2 + r_+^2\right)}
{4 \, \Xi_a^2 \, \Xi_b^2 \left(r_+^5 \left(a^2 + b^2 + 1\right) - a^2 b^2 r_+ + 2 r_+^7\right)} \\
&\quad \times \Bigg[ 
2 \left(b^2 - 1\right)\left(r_+^2 + 1\right) r_+^2 \Xi_a \left(a^2 + r_+^2\right) + 2 \left(a^2 - 1\right)\left(r_+^2 + 1\right) r_+^2 \Xi_b \left(b^2 + r_+^2\right) \\
&\qquad + \Xi_a \Xi_b \Big(3 r_+^4 \left(a^2 + b^2 + 1\right) 
+ r_+^2 \left(a^2 \left(b^2+1\right) + b^2\right) - a^2 b^2 + 5 r_+^6 \Big)
\Bigg]\\
   & \Xi_{a} = 1-a^2\\
   &\Xi_{b} = 1-b^2
   \end{split}
\end{equation}
where $G_5= \frac{\pi}{2 N^2}$ (see \eqref{gfive}) and  
$m$, $a$ and $b$ are 3 parameters that parameterize the black hole family, and 
$r_+$, in  \eqref{thermformmt},  is determined in terms of $a$, $b$ and $m$ by the equation 
\begin{equation}\label{rpeq}
\left(r_+^2 + a^2\right)\left(r_+^2 + b^2\right)\left(r_+^2 + 1\right) - 2 m r_+^2 = 0
\end{equation}

\subsection{Equivalence of the microcanonical and canonical formulae in the black hole phase} \label{eomac}

In \S \ref{mcens}, we demonstrated that the entropy as a function of the scaled twist $\tau_s$ (recall $\tau_s =\frac{\tau}{(N^2J_a J_b)^{\frac{1}{3}}}$), in 
the black hole phase, is given in the first equation of \eqref{sacalenthb} (with $r_+$ given as a function of $\tau_s$ by the second equation of \eqref{sacalenthb}). In other words, we demonstrated that in this phase
\begin{equation}\label{entropygn}
\frac{S_{BH}}{(N^2 J_a J_b)^{\frac{1}{3}}}= {\rm Ext}_{r_+, \lambda_1} \left[
    2^{\frac{2}{3}}\pi r_+^{1/3} -
   \lambda_1 \left(  \tau_s - \frac{\left(r_+^2+1\right)}{ 2^{\frac{4}{3}}r^{\frac{2}{3}}_+} \right)  \right]
\end{equation}
where the extremization over the Lagrange multiplier $\lambda_1$ enforces the relationship between $\tau_s$ and $r_+$.

In \S \ref{cens}, on the other hand, we also demonstrated that the function $h(\beta)$ that appears in the thermal partition function for the black hole phase, is given by \eqref{hbetaBHmt}, where $r_+$ is given as a function of $\beta$ by the third equation of \eqref{cansimp}. In other words we demonstrated that 
\begin{equation}\label{partfnygn}
  \frac{\beta^2  h_{BH}(\beta)}{N^2}=\mathrm{Ext}_{r_+,\lambda_2}\Big[\frac{\pi ^3 r_+ \left(r_+^2-1\right){}^3}{2 \left(2 r_+^2-1\right){}^3}-\lambda_2\left(\beta-\frac{2\pi r_+}{2r^2_+-1}\right)\Big]
\end{equation}
where the extremization over $\lambda_2$ enforces the relationship between 
$\beta$ and $r_+$. 

In this brief subsection we will verify that \eqref{entropygn} and \eqref{partfnygn} are indeed related via \eqref{intent}. In other words, we demonstrate that 
\begin{equation}\label{extpf} 
\begin{split}
&{\rm Ext}_{r_+, \lambda_1} \left[
    2^{\frac{2}{3}}\pi r_+^{1/3} -
   \lambda_1 \left(  \tau_s - \frac{\left(r_+^2+1\right)}{ 2^{\frac{4}{3}}r^{\frac{2}{3}}_+}  \right)  \right] \\
   &={\rm Ext}_{\beta, r_+, \lambda_2}\left(
\beta \tau_s + 3 \Big[\frac{\pi ^3 r_+ \left(r_+^2-1\right)^3}{2 \left(2 r_+^2-1\right)^3}-\lambda_2\left(\beta-\frac{2\pi r_+}{2r^2_+-1}\right)\Big]^{\frac{1}{3}} \right)
\end{split}
\end{equation}
We start with the RHS of \eqref{extpf}, and perform the extremization w.r.t. $\beta$. This sets $\lambda_2= \tau_s$ (so
we have also effectively performed the extremization over $\lambda_2$) and the RHS of \eqref{extpf} reduces to 
\begin{equation}\label{extint}
{\rm Ext}_{r_+}\left(
\tau_s\left( \frac{2\pi r_+}{2r^2_+-1} \right) + 3 \Big[\frac{\pi ^3 r_+ \left(r_+^2-1\right)^3}{2\left(2 r_+^2-1\right)^3} \Big]^{\frac{1}{3}} \right)
\end{equation}
The extremization over $r_+$ now gives 
\begin{equation} \label{taus}
    \tau_s=\frac{\left(r_+^2+1\right)}{ 2^{\frac{4}{3}}r^{\frac{2}{3}}_+}
\end{equation}
which we recognize as the condition for the vanishing of the term proportional to $\lambda_1$ in \eqref{extpf}. Plugging \eqref{taus} into \eqref{extint} simplifies the quantity to: \eqref{extint}
\begin{equation}\label{srp}
  2^{\frac{1}{3}}\pi^{\frac{4}{3}}r_+^{1/3} 
  \end{equation}
where $r_+$ should be viewed as the function of $\tau_s$ (obtained by solving \eqref{taus}). However, the LHS of \eqref{extpf} is precisely the same quantity, namely \eqref{srp}, with $r_+$ given as a function of $\tau_s$ by \eqref{taus} 
(on the LHS of \eqref{extpf}, \eqref{taus} follows by varying w.r.t. $\lambda_1$). We conclude that the LHS and RHS are indeed equal as expected. 

\subsection{Extremal Kerr-AdS$_5$ black holes in the scaling limit}\label{extbh}

It follows from \eqref{cansimp} that our scaled black holes become extremal at $r_+^2=\frac{1}{2}$. Plugging this value into
\eqref{microbh}, \eqref{sacalenthb}, yields
\begin{equation}\label{extremalcharges}
    \begin{split}
    \frac{\tau}{N^2} &\approx \frac{27}{64(\delta a)(\delta b)}\\
    \frac{J_a}{N^2} &\approx \frac{27  }{64 (\delta b)(\delta a^2)}\nonumber\\
    \frac{J_b}{N^2}&\approx \frac{27  }{64(\delta a)(\delta b^2)}\nonumber\\
    S&\approx \frac{9 \pi }{8 \sqrt{2} (\delta a)(\delta b)}\\ 
    \end{split}
\end{equation}

Note, in particular, that at extremality,
\begin{equation}\label{extreme}
\begin{split}
    \frac{S}{(N^2 J_a J_b)^{\frac{1}{3}}}&=
    \sqrt{2}\pi\\
    \frac{\tau}{(N^2J_a J_b)^{\frac{1}{3}}}&=\frac{3}{4}
    \end{split}
\end{equation}

\subsection{Thermodynamics of small rotating black holes in $AdS_5 \times S^5$}\label{sbhads}

In this subsection (which is only tangentially connected to the main flow of this paper), we note that the entropy of very small black holes in $AdS_5\times S^5$ (i.e.\ black holes that are much smaller than the radius of $AdS$ and so are well-approximated by black holes in flat space) also takes a scaling form.

Consider a black hole in a flat $D$-dimensional spacetime. If we increase all lengths (and times) by a factor of $\zeta$, the energy, entropy, and angular momentum of a black hole solution (in $D$ dimensional flat space) scale, respectively, like $\zeta^{D-3},$ $\zeta^{D-2}$ and $\zeta^{D-2}$.  (Other $U(1)$ charges scale in the same way as the energy. This reflects the fact that both the Newton and Coulomb laws scale like  $\frac{1}{r^{D-3}}$.)

Small black holes in $AdS_5 \times S^5$ (with radius of the event horizon $\ll R_{AdS}$) behave like black holes in flat space. Setting $D=10$ in the estimates above, we conclude that the entropy of a rotating black hole of this nature must take the form 
\begin{equation}\label{entofenang}
S= \frac{ \pi^{\frac{12}{7}}E^{\frac{8}{7}}}{2 \left( N^2 \Omega_8 \right)^{\frac{1}{7}} }
\times f\left( \frac{J_i (N^2 \Omega_8)^\frac{1}{7}}{E^{8/7}} \right) 
\end{equation} 
for some function $f$ (whose explicit form is easily determined from the exact analytic solutions of these black holes, but this exact form will not concern us here). The scaling form \eqref{entofenang} has some similarities with \eqref{lj}. The main difference is in the value of the powers, and also the fact that the scaling function \eqref{entofenang} involves the energy $E$ rather than the twist $\tau$. 

Of course, the fact that the scaling form (and powers) in \eqref{entofenang} differ from the semi-universal prediction is no contradiction. Small black holes only dominate the ensemble for $\frac{J_i}{N^2}\ll 1$, and so are never relevant for the large $J$ study performed in most of this paper.

\subsection{Black hole stress tensor in the $\omega_i\rightarrow 1$ limit}\label{bhst}
The boundary stress-tensor of a Kerr-$AdS_5$ black hole is given by \cite{Bhattacharyya:2007vs} : 
\begin{equation} \label{bcs1} 
    \begin{split}
        T^{tt} &= \frac{m}{8\pi G_{5}}(4 \gamma_0^6-\gamma_0^4)\\T^{\phi_1\phi_1} &= \frac{m}{8\pi G_5}\gamma_0^4\left ( 4\gamma_0^2 a^2 + \frac{1}{\sin^2({\theta})}\right)\\ T^{\phi_2\phi_2} &= \frac{m}{8\pi G_5}\gamma_0^4\left ( 4\gamma_0^2 b^2 + \frac{1}{\cos^2({\theta})}\right)\\T^{t\phi_1} &= \frac{4m}{8\pi G_5} a\gamma_0^6, ~~T^{t\phi_2} = \frac{4m}{8\pi G_5} b\gamma_0^6\\T^{\phi_1\phi_2}&=\frac{4m}{8\pi G_5} ab\gamma_0^6,~~T^{{\theta}{\theta}} = \frac{m}{8\pi G_5}\gamma_0^4\,,
    \end{split}
\end{equation}
where $\gamma_0^{-2} = 1-a^2\sin^2{\theta}-b^2\cos^2{\theta}$ and 
$G_5=\pi/(2 N^2$).\\
To the leading order in the limit $a \rightarrow 1$ and $b \rightarrow 1$, 
\begin{align}
    m&= \frac{(r^2_++1)^3}{2r^2_+}\\
    \gamma^{-2}_0&=2\left(\delta a\sin^2\theta+\delta b\cos^2\theta\right)\nonumber\\
    &=2\frac{r^2_++1}{r^2_+-1}\left(\delta\omega_a\sin^2\theta+\delta\omega_b \cos^2\theta\right)\\
    &=\frac{r^2_++1}{r^2_+-1}\, \gamma^{-2}
    \label{newmgammabhdef}
\end{align}
Here we have defined 
\begin{equation}
    \gamma^{-2}=\left(1-\omega^2_a\sin^2\theta-\omega^2_b\cos^2\theta\right)\approx 2\left(\delta\omega_a\sin^2\theta+\delta\omega_b \cos^2\theta\right)
\end{equation}
In the last line of \eqref{newmgammabhdef}, we have used the relation of $\delta a_i$ with $\delta \omega_i$ given in the first two equations of \eqref{cansimp}.

From \eqref{newmgammabhdef}, it is clear that for $r_+\gg 1$, $\gamma_0\approx\gamma$.
The components of the stress tensors in the leading order in $\delta \omega_i\rightarrow 0$ are given by
\begin{equation} \label{bcs1_final}
\begin{aligned}
T^{tt}&=T^{\phi_1\phi_1}=T^{\phi_2\phi_2}=T^{t\phi_1}=T^{t\phi_2}=T^{\phi_1\phi_2}
= \frac{N^2 (r_+^2-1)^3}{2\pi^2 r_+^2}\,\gamma(\theta)^6,\\[4pt]
T^{\theta\theta}
&= \frac{N^2 (r_+^2+1)(r_+^2-1)^2}{32\pi^2 r_+^2}\,\gamma(\theta)^4.
\end{aligned}
\end{equation}
From the components of the stress tensor, we can write in the leading order
\begin{equation}\label{fst1}
\begin{split}
T^{\mu\nu}
&= \frac{N^2 (r_+^2-1)^3}{2\pi^2 r_+^2}\,\gamma(\theta)^6\\
&=\frac{2 T h_{BH}(r_+)}{\pi^2}\left(4 u^{\mu}u^{\nu}\right)
\end{split}
\end{equation}
where $h_{BH}(r_+)$ and $T$ in the limit $\omega_i\rightarrow 1$, are given in \eqref{cansimp}.
We can impose the tracelessness of the stress tensor by adding a subleading term proportional to $g^{\mu\nu}$.
\begin{equation}\label{fst2}
\begin{split}
T^{\mu\nu}
&=\frac{2 T h_{BH}(r_+)}{\pi^2}\left(4 u^{\mu}u^{\nu}+g^{\mu\nu}\right)
\end{split}
\end{equation}
The above form of the stress tensor in \eqref{fst2} provides a test for the conjectured stress tensor in \eqref{st2}.
\section{Estimating collision-type interactions in the bulk gas}\label{app:meanfield}
 
It was explained in \cite{Bajaj:2024utv}, that the effective bulk stress tensor of the gravitational gas at high angular momenta but at temperatures of order unity  extends up to $r \sim \frac{1}{(\nu_1 \nu_2)^{\frac{1}{4}}}$ (where $r$ is usual radial coordinate of $AdS_5$). The bulk stress tensor is a nontrivial function of the angular coordinate $\theta$ on $S^3$, and varies over angular scales of order unity, and therefore over proper scales of order $\frac{1}{(\nu_1 \nu_2)^{\frac{1}{4}}}$.

Interactions occurring in this gas are of two sorts 
\begin{itemize}
    \item Coulomb-type interaction effects (involving the exchange of virtual gravitons) sourced by the average value of the gravitational stress tensor and other charges of the bulk gas (which is, itself, assumed to be free on the much shorter length scales of order the $AdS$ radius). These interactions are characterized by the length scale $\left(\frac{1}{(\nu_1 \nu_2)^{\frac{1}{4}}}\right)$. The effects of such interactions were estimated in \S \ref{nonintmix} (we expect them to be of order $\frac{\gamma^2}{N^2}$, see e.g. \eqref{fractwist}.)
    \item Collision-type interactions between individual gravitons. These  
    interactions occur at much smaller length scales than their Coulombic counterparts. Very roughly, these interactions renormalize the intrinsic (short distance) properties (like the equation of state and other intrinsic properties) of the bulk gas (whose stress tensor mediates the `Coulombic interactions' of the previous item). 
\end{itemize}

In this Appendix we estimate the conditions under which the second kind of (collision-type) interactions can be ignored, so that the bulk gas can (locally) be thought of as free.

Consider a gas revolving in $AdS_5$ space at angular velocities $\omega_1$ and $\omega_2$, 
so that $u^\mu \partial_\mu= \Gamma\left( \partial_t +\omega_1 \partial_{\phi_1} + \omega_2\partial_{\phi_2}\right)$
where $\Gamma$ is `the bulk special relativity factor of the fluid velocity'  whose value we now determine. 
At large $r$, the metric of $AdS_5$ space reduces to 
\begin{equation}\label{metadsf}
ds^2=-(1+r^2) dt^2 + \frac{dr^2}{r^2}+ r^2 d \Omega_3^2
\end{equation}
$\Gamma$ is chosen to ensure that $u^2=-1$; a short computation yields 
\begin{equation}\label{Gammaval}
\begin{split}
&\Gamma = \frac{1}{\sqrt{1 +2x^2}} \\
& x= \frac{r}{\sqrt{2}\gamma}
\end{split} 
\end{equation}
As was explained in \cite{Bajaj:2024utv}, the typical value of the radial coordinate for the bulk gas is $r \sim \gamma$, at which values,  $x \sim {\cal O}(1)$. Consequently, the value of the special relativistic $\Gamma$ for our bulk gas is typically of order unity. 

The effect of collisions on the bulk gas is most conveniently analysed in the 
gas rest frame. The local rest frame temperature of this gas is the inverse of the proper length of the time circle identification in Euclidean circle, 
a quantity that can easily be verified to be $\frac{1}{T \Gamma}$ where 
$T$ is the global temperature of our solution (i.e.\ the temperature that appears in \eqref{thermpf}). As the $S$ matrix between two gravitons is of order 
$1/N^2$, it follows that 5 dimensional bulk collision cross section between two gravitons is of order 
\begin{equation}\label{sigmaorder}
\sigma \sim \frac{1}{N^4 (T\Gamma)^3}
\end{equation}
(where we have used the fact that the collision cross section is proportional to the square of an S matrix, and have  fixed the powers of $T \Gamma$ by dimensional analysis, assuming that the local value of the temperature is the only relevant scale in the problem). 

The mean free path $l_{mfp}$ of two bulk gravitons can now be estimated as
\footnote{This estimate can be performed using the formula $\rho l_{mfp} = \sigma$, and the dimensional estimate $\rho\sim (T \Gamma)^4$, where $\rho$ is the number density of gravitons.} 
\begin{equation}\label{lmfp}
l_{mfp}\sim \frac{N^4}{T \Gamma}
\end{equation} 

The bulk gas is an effectively weakly-coupled fluid when its mean free path is larger than the largest length scale of variation that appears in the `fluid stress tensor', which happens to be $\gamma$ in the case of interest. 
Consequently, collision effects in our gas are negligible when  
\begin{equation}\label{lsfm}
l_{mfp} \gg \gamma, ~~~~{\rm i.e.}  ~~~ T \gamma \ll N^4 
\end{equation}
(where we have dropped the $\Gamma$ dependence in this estimate, using that 
$\Gamma$ is of order unity).

When temperatures are of order unity (as is always the case for the bulk gas when it dominates the ensemble), $\gamma\sim \frac{1}{(\nu_1\nu_2)^{\frac{1}{4}}} 
\sim (J_1J_2)^{\frac{1}{12}}$ (see \eqref{estimates}), and \eqref{lsfm} 
turns into the condition
\begin{equation}\label{jgio}
\sqrt{J_1 J_2} \ll N^{24},
\end{equation}
a condition that is comfortably obeyed when \eqref{jojt} holds. 
\footnote{The interested reader can easily generalize \eqref{jgio} to the case in which the bulk gas has a temperature much larger than unity (in this case the RHS of \eqref{jgio} is dressed by an appropriate power of the scaled twist (recall that the scaled twist is of order unity when the temperature is of order unity, as we have assumed throughout this Appendix).}

\section{The 10d gas at high temperatures}\label{supergravitygas}

\subsection{The bulk gas at intermediate temperatures}\label{expform}

In \S \ref{tgp} we have studied the thermodynamics of the bulk gas at angular momenta that are large, but not so large that they violate \eqref{jojt} (so that the gas is effectively non-interacting), and at temperatures such that $T \ll \lambda^{\frac{1}{4}}$
(so that stringy oscillator states are  effectively unexcited). In this subsection we specialize the study of the bulk gas (in the same range of angular momenta) to temperatures in the range\footnote{As we have emphasized in \eqref{temprange}, the temperature $T$ is far from the largest parameter in the problem in the regime in which we perform our analysis. Our analysis throughout this subsection is performed in a regime in which $T$ is much less than $\lambda^{\frac{1}{4}}$ which, in turn is much less than both any positive power of $N$ as well as any positive power of $\sqrt{J_1J_2}$. }
\begin{equation}\label{temprange}
1 \ll T \ll \lambda^{\frac{1}{4}}
\end{equation}

As the bulk gas phase is dominant only at temperatures less than $\frac{1}{2\pi}$, the high-temperature gas studied in this subsection never appears as a dominant phase in the phase diagram described in \S \ref{lgym}.
In this Appendix we nonetheless present formulae that characterize this high-temperature gas, for the following two reasons
\begin{itemize}
\item While the gas never dominates the thermal ensemble at these temperatures, it may turn out to be metastable, and so appear as a long-lived transitory state in some dynamical processes of physical interest. 
\item While the function $h_{YM}(\beta)$, which computes gas thermodynamics in the large $J$ limit, is complicated at finite temperatures, it greatly simplifies (to a power law) at large temperatures. 
\end{itemize}

In the limit $T\gg 1$, the function $h_{YM}(\beta)$ takes a power law form, proportional to  $\frac{1}{\beta^9}$,  as is appropriate for a massless 10-dimensional gas. The precise coefficient of this scaling is given in Eqns. $(4.11)$ and $(4.13)$ of \cite{Bajaj:2024utv}. \footnote{Note that this scaling differs from the $\frac{1}{\beta^3}$ that \eqref{ftlarget} predicts for the high-temperature limit of $h(\beta)$
in a 3+1 dimensional field theory. This is far from surprising; the prediction 
\eqref{ftlarget} applies to the high-temperature limit of the phase that is dominant at these temperatures, i.e.\ to the black hole phase, and does hold in that phase at temperatures $T \gg 1$ (see \eqref{leadordhit}). \eqref{ftlarget} makes no predictions for the high energy scaling of $h(\beta)$ in subdominant phases (also the $1/\beta^9$ scaling for the supergravity gas only applies at intermediate -- and not arbitrarily high -- temperatures, see \eqref{temprange}.} Using the value of this coefficient, we find that \eqref{estimates} simplifies to 
\begin{equation}\label{estimatesap}
\begin{split}
&J_1 \approx \left(\frac{73 \pi ^{10}}{106920}\right)\frac{T^{10}}{(1-\omega_1)^2(1-\omega_2)}\\
&
J_2 \approx \left(\frac{73 \pi ^{10}}{106920}\right)\frac{T^{10}}{(1-\omega_2)^2(1-\omega_1)}\\&
\tau \approx \frac{73 \pi ^{10}}{9720}\frac{T^{10}}{(1-\omega_1)(1-\omega_2)}
\end{split}
\end{equation}
It follows from \eqref{estimatesap} that in this limit 
\begin{equation}\label{temp}
T \sim  \left( \frac{\tau}{(J_1J_2)^{\frac{1}{3}}} \right)^{\frac{3}{10}}
\end{equation}
Plugging \eqref{temp} into \eqref{temprange}, (and taking the $(10/3)^{th}$ power of that equation), we conclude that the formulae of this subsection apply only when the scaled twist obeys 
\begin{equation}\label{scaltwistoo}
1 \ll \frac{\tau}{(J_1J_2)^{\frac{1}{3}}} \ll \lambda^{\frac{5}{6}}
\end{equation} 
We see that the supergravity gas starts behaving like an extensive 10d gas in a box once the scaled twist is much larger than unity, and that stringy oscillator states are negligibly excited provided the scaled twist much smaller than 
$\lambda^{\frac{5}{6}}$.

It also follows from \eqref{estimatesap} that in the same limit, 
\begin{equation}\label{oneome}
1-\omega_i \sim  \frac{\tau}{J_i}.
\end{equation}
Consequently, $1-\omega_i$ is small (so the semi-universal form holds) provided
\begin{equation}\label{whegasemi}
J_i \gg \tau
\end{equation}
We conclude that 
\begin{equation}\label{semiunivesal}
{\rm When}~ \tau\lesssim J^{\frac{2}{3}}  ~~{\rm Gas ~semi~universal} ~~{\rm if} ~~~
J_i \gg 1.
\end{equation}
It is 
not difficult to show that the entropy as a function of twist and angular momenta takes the simple explicit form 
\begin{equation}\label{entexp}
S(\tau, J_1,J_2)=\left(\frac{\left(\frac{2}{7}\right)^{7/10} 5^{9/10} (\frac{73}{11})^{\frac{1}{10}} \pi }{\sqrt{3}}\right)(J_1J_2)^\frac{1}{3}\left(\frac{\tau}{(J_1J_2)^\frac{1}{3}}\right)^{\frac{7}{10}}
\end{equation}
in the same limit, matching the prediction in \eqref{finansforent}. 

The twist of a typical graviton is always of order $T$, while $J_i$ for a typical graviton is of order $\frac{T}{(1-\omega_i)}$. In the high-temperature limit, $T \gg 1$, a typical graviton thus has twist of order
$\left( \frac{\tau^3}{J_1 J_2} \right)^{\frac{1}{10}}$, and has  $i^{th}$ angular momentum of order 
$$j_i\sim \frac{\left( \frac{\tau^3}{J_1 J_2} \right)^{\frac{1}{10}}}{(1-\omega_i)}\sim J_i\left(  \frac{1}{\tau^7 J_1 J_2} \right)^{\frac{1}{10}}.$$ 
By dividing the total twist by the number of gravitons (or dividing the total $J_i$ by the $j_i$ carried by each graviton), we conclude that the average number of gravitons is of order $\tau^{\frac{7}{10}} (J_1J_2)^{\frac{1}{10}}$. 

We have explained above that the partition function and entropy of the gas above takes the semi-universal form \eqref{thermpartfn} and \eqref{finansforent} provided $1-\omega_i \ll 1$, i.e.\ provided $\tau \ll J_i$. We have also explained that string oscillator states can be only be ignored provided $\tau^3 \ll J^2 \lambda^{\frac{5}{2}}$, i.e.\ $\tau \ll \lambda^{\frac{5}{6}} J^{\frac{2}{3}}$. At this upper bound of $\tau$, we have $\frac{\tau}{J}= \frac{\lambda^{\frac{5}{6}}}{J^{\frac{1}{3}}}$. It follows that the semi-universality condition 
holds all the way up to this $\tau$ (i.e.\ holds whenever the twist is small enough so that string oscillators can be ignored) provided that 
\begin{equation}\label{jrange}
J \gg  \lambda^{\frac{5}{2}}
\end{equation}

\section{The fluid stress tensor in the large $\gamma$ limit}\label{strten}

In \S \ref{su} we have demonstrated that, in the symmetric large $\gamma$ (i.e.\ $\omega_i \rightarrow 1$) limit, the partition function 
$\ln Z$ of every four-dimensional CFT takes the form 
\begin{equation}\label{pff}
\ln Z = \frac{4 h(\beta)}{(2\pi^2)} \int_{S^3} \gamma^4(\theta) = \frac{4 h(\beta)}{(1-\omega_1^2)(1-\omega_2^2)}
\end{equation} 
As we have explained in the discussion section, we are also tempted to conjecture that in this limit the field theory stress tensor always takes the universal form 
\begin{equation}\label{stresstens}
T_{\mu\nu}= \frac{2h(T)T}{\pi^2} \gamma^4(\theta)\left(4 u_\mu u_\nu + g_{\mu\nu} \right) \,,
\end{equation}
where $g_{\mu\nu}$ is the metric on $S^3 \times $ time, and $u_\mu$ is the velocity vector \eqref{velocityoffluid}.

If the conjecture \eqref{stresstens} is indeed valid, we should, in particular, be able to demonstrate that \eqref{stresstens} holds at every order in the $\beta$ expansion \eqref{psexpl}. While we think it may well be possible to find a relatively simple proof of this order-by-order claim (along the lines of the similar but simpler argument we presented for the partition function density in \S \ref{su}), in this Appendix 
we do not attempt to construct such an all orders proof, but content ourselves with explicitly verifying that our conjecture does indeed hold 
at leading and first subleading order in $\beta$. 

\subsection{Structure of the stress tensor from the effective action}\label{stressten2}

The stress-energy tensor in a quantum field theory is defined as the response of the partition function to variations in the background metric. Formally, this relation is given by:
\begin{equation}\label{eq:def_stress1}
    \delta \ln Z = \int d^{p+1}x\, \sqrt{-g_{p+1}} \left( -\frac{1}{2} T_{\mu\nu} \delta g^{\mu\nu} \right),
\end{equation}
In \eqref{eq:def_stress1}, we see that the left-hand-side is a scalar quantity; now the right-hand-side, although written in Lorentzian coordinates, can be written in any coordinates as it is an invariant scalar. In our case, we can use our analytically continued metric in the coordinates $\tau, \theta, \phi_1, \phi_2$. In these coordinates, we have to change all the relevant velocity vectors and related quantities, $u^{\mu}\partial_{\mu} = i \gamma(\theta)\partial_\tau$.

In the context of finite-temperature field theory, the Euclidean time coordinate is compactified on a circle of circumference $\beta = 1/T$, where $T$ is the temperature. For stationary backgrounds, all metric components are independent of the Euclidean time coordinate. This allows us to integrate out the time direction:
\begin{equation}\label{eq:def_stress_tensor}
    \delta \ln Z = \frac{1}{T} \int d^p x\, \sqrt{g_{p+1}} \left(-\frac{1}{2} T_{\mu\nu} \delta g^{\mu\nu} \right).
\end{equation}
From this expression, the stress-energy tensor can be obtained as the functional derivative of the partition function with respect to the metric:
\begin{equation}\label{eq:def_stress}
    T_{\mu\nu} =-\frac{2 T}{\sqrt{g_{p+1}}} \frac{\delta \ln Z}{\delta g^{\mu\nu}}.
\end{equation}

As our thermal effective partition function is written as a function of $\sigma, a^i, g^{ij}$, and $T$ (see eq. 2.15 of \cite{Banerjee:2012iz}), we have:
\begin{equation}
    \ln Z = W(e^{\sigma}, a_i, g^{ij},T).
\end{equation}
The metric is given by  
\begin{equation}
    g_{\mu\nu} =
\begin{pmatrix}
e^{2\sigma} & i  e^{2\sigma}a_j \\
i e^{2\sigma}a_i & g_{ij} - e^{2\sigma}a_ia_j
\end{pmatrix}
\end{equation}
and the inverse metric by \begin{equation}\label{eq:inverse_metric}
    g^{\mu \nu} = \begin{pmatrix}
        e^{-2\sigma} - a^2 & - i a^{j}\\ - i a^{j}& g^{ij}
    \end{pmatrix}.
\end{equation}

From  \eqref{eq:def_stress_tensor}, by explicitly expanding $\delta g^{\mu\nu}$ in terms of the differential of $a,\sigma$ and $g^{ij}$, and then taking the functional derivative of \eqref{eq:def_stress_tensor} w.r.t $a,\sigma$ and $g^{ij}$, we obtain the following expression for the components of the stress tensor (these formulae are the analytic continuation to Euclidean space of Eqns. $(2.15)$ and $(2.16)$ of \cite{Banerjee:2012iz}):
\begin{equation}\label{eq:stress_fields}
    \begin{split}
         &T^{ij} = \frac{2T }{\sqrt{g_{p+1}}}\frac{\delta W}{\delta g_{ij}}\\&
        T^{0k} = -i \left(\frac{Te^{-2\sigma }}{\sqrt{g_{p+1}}} \frac{\delta W}{\delta a_k} +T^{kj}a_j\right)\\&T^{00} =\frac{Te^{-2\sigma}}{\sqrt{g_{p+1}}}\frac{\delta W}{\delta \sigma} - 2i T^{0i}a_i +  T^{ij}a_ia_j.
    \end{split}
\end{equation}

\subsection{Explicit computation}
\subsubsection{Zeroth order}

The thermal partition function at the zero derivative order is given by:
\begin{equation}
    \ln Z^0 = \int d^3x \sqrt{g_3}\frac{P(T(x))}{T(x)}
\end{equation}
Using the expressions for the stress tensor from \eqref{eq:stress_fields}\footnote{\begin{align}
    T^{00}&=\frac{T\,e^{-2\sigma}}{\sqrt{g_{4}}}\frac{\delta W}{\delta \sigma}-2i T^{0i}a_i+T^{ij}a_ia_j\nonumber\\
    &=\frac{T\,e^{-2\sigma}}{e^{\sigma}\sqrt{g_{3}}}\left(\frac{P\sqrt{g_3}}{T\,e^{-\sigma}}-\frac{\sqrt{g_3 }\,T\,\,e^{-\sigma}}{T\, e^{-\sigma}}\frac{\partial P(T(x))}{\partial T(x)}\right)-2i\left(-i P (T(x))a.a\right)+P(T(x)) a.a\nonumber\\
    &=e^{-2\sigma}\left(P(T(x))-T(x)\frac{\partial P(T(x))}{\partial T(x)}\right)-P(T(x))\, a.a
\end{align}
For a conformal fluid, $P(T(x))\propto T(x)^4$, therefore, $T^{00}$ can written as
\begin{align}
    T^{00}&=e^{-2\sigma}\left(-3 P(T(x))\right)-P(T(x)) a.a\nonumber\\
    &=-4P (T(x))e^{-2\sigma}+P(T(x))(e^{-2\sigma}-a.a)
\end{align}}, we obtain:
\begin{equation}
   T^{0k} = -i P(T(x)) a^{k}, \quad T^{ij} = P(T(x)) g^{ij}, \quad T^{00} =  -4 P(T(x))e^{-2\sigma} +P(T(x))\left(e^{-2\sigma} - a.a\right)
\end{equation}
where the index of $a_k$ is raised with respect to the spatial metric $g^{ij}$, the inverse of $g_{ij}$.

\begin{equation}
    a_i dx^i =  -\gamma^2(\theta)\left(\omega_1 \sin^2\theta\, d\phi_1 + \omega_2 \cos^2\theta\, d\phi_2\right)
\end{equation}

In fully covariant form, the stress tensor is:
\begin{equation}
    T^{\mu\nu}= P(T(x))\left[4u^{\mu}u^{\nu} + g^{\mu\nu}\right]
\end{equation}
as anticipated in \eqref{stresstens}. Here $u^{\mu}\partial_{\mu} = i \gamma(\theta)\partial_\tau$, where the indices $\mu,\nu$ run from $\tau, \theta, \phi_1$ and $ \phi_2$. This is a covariant expression that we can easily use for the other coordinates as well, namely for the case of the Lorentzian metric. 

\subsubsection{Stress tensor at second derivative order}

To compute the stress tensor at second order, we vary the partition function with respect to $a_i$, $\sigma$, and $g_{ij}$:
\begin{equation}
    \begin{split}
        W^{1} &= \ln Z^{1} = - \frac{1}{2} \int d^3x\, \sqrt{g_3}\, \Big( 
        P_1(\sigma)\, R + 
        T^2 P_2(\sigma)\, f_{ij} f^{ij} + 
        P_3(\sigma)\, (\partial \sigma)^2 
    \Big) + \cdots,\\
    & T(x) = T e^{-\sigma(x)} \, .
    \end{split}
\end{equation}
Using \eqref{eq:stress_fields}, we obtain
\begin{equation}\label{eq:stress_tensor_2nd}
\begin{split}
    -T^{0k} &= 2i T^{0} e^{-3\sigma} T^2_{0} P_2\big(\nabla_{i}f^{ik}\big) 
    - 2i T^{0} e^{-3\sigma} T^2 P'_2 f^{ki} \nabla_i \sigma(x) \\
    &\quad - i T(x)a_{j}\Big[ P_1\!\left(R^{kj}- \tfrac{1}{2}R g^{kj}\right) 
    + T^2 P_2\!\left(2f^{ji}f^{k}_{\ i} - \tfrac{1}{2} f^2 g^{kj}\right)\Big]\\
    &\quad + i T(x) a_j\Big[ 
    P_1(-2\nabla^{j}\sigma \nabla^{k}\sigma + (\nabla\sigma)^2 g^{jk}) 
    + P_1'(\nabla^{j}\nabla^{k}\sigma - \nabla_i\nabla^i g^{jk}) \\
    &\qquad + P_1'' (\nabla^{j}\sigma \nabla^{k}\sigma - (\nabla \sigma)^2 g^{jk}) 
    \Big],\\[1em]
    -T^{ij} &= T(x)\Big[ 
    P_1\!\left(-R^{ij}+\tfrac{1}{2}R g^{ij}\right) 
    + T^2 P_2\!\left(-2f^{ik}f^{j}_{\ k} +\tfrac{1}{2} f^2 g^{ij}\right) \\
    &\qquad + P_1(-2\nabla^{i}\sigma \nabla^{j}\sigma + (\nabla\sigma)^2 g^{ij}) 
    + P_1'(\nabla^{i}\nabla^{j}\sigma - \nabla_k\nabla^k g^{ij}) \\
    &\qquad + P_1'' (\nabla^{i}\sigma \nabla^{j}\sigma - (\nabla \sigma)^2 g^{ij})
    \Big],\\[1em]
    -T^{00} &= \tfrac{1}{2}T e^{-3\sigma}\Big(
    P_1'R+ T^2 P_2' f^2 - 2 P_1' \nabla_i \sigma \nabla^i \sigma - 4 P_1 \nabla_i\nabla^i\sigma
    \Big) + 2i T^{0i} a_i - T^{ij}a_i a_j \, .
\end{split}
\end{equation}

\vspace{1em}
\subsection*{Explicit components form}
At leading order in $\gamma$, the following approximations are useful:
\begin{equation}
\begin{split}
    a_j R^{ij} &= 2 \gamma^2(\theta)\, a^i + O(\gamma(\theta)),\\ 
    \nabla_i f^{ik} &= -4 \gamma^2(\theta)\, a^k + O(\gamma(\theta)),\\ 
    f^{ji}f^{k}_{\ i} &= 4\gamma^4(\theta)\, g^{jk} + O(\gamma^2(\theta)),\\ 
    f^{ji}f^{k}_{\ i} a_i a_j &= 4 \gamma'^2(\theta)\, \gamma^2(\theta).
\end{split}
\end{equation}

With these, the explicit form of the stress tensor at leading order in $\gamma$ becomes:
\begin{equation}\label{eq:stress_tensor_com}
\begin{split}
    -T^{0k} &\approx i\, 3T^2 \gamma^4 (e_1 -4e_2)\, a^{k} 
    - 4i (e_1-4e_2) T^2 \gamma' \gamma^3 w^{k},\\
    -T^{ij} &\approx (e_1 - 4 e_2) T^2 \gamma^4(\theta)\, g^{ij}
    = -3(e_1 - 4 e_2) T^2 \gamma^4(\theta)\, g^{ij} 
    + 4(e_1 - 4 e_2) T^2 \gamma^4(\theta)\, g^{ij},\\
    -T^{00} &\approx 4T^2 \gamma^6 (e_1- 4e_2) + \mathcal{O}(\gamma^4)
\end{split}
\end{equation}
where $w_k$ is defined in \eqref{eq:veirbein}, and its index $k$ is raised with metric $g^{ij}$.

 In computing $T^{0k}$ and $T^{ij}$ at leading order in $\gamma$, it is necessary to retain the $\sigma(x)$-dependent terms in the thermal partition function. Unlike $T^{00}$, the components $T^{0k}$ and $T^{ij}$ are subleading in $\gamma$, so their leading contributions arise precisely from terms that would otherwise be regarded as subleading in the partition function.

\subsection*{Covariantization}

Up to two derivatives in the derivative expansion, the stress tensor components can be organized as linear combinations of $u^{\mu}u^{\nu}$, $g^{\mu\nu}$, and the vorticity tensor $(\omega^2)^{\mu\nu}$, defined as
\begin{equation}
    \omega^{\mu\nu} = P^{\mu\rho} P^{\nu \sigma}\left( \nabla_{\rho} u_{\sigma} - \nabla_{\sigma}u_{\rho} \right)
\end{equation}
where $P^{\mu\nu}$ is the projection tensor
\begin{equation}
    P^{\mu\nu} = g^{\mu\nu} + u^{\mu}u^{\nu}.
\end{equation}
In the large-$\gamma$ limit, the components of the quadratic vorticity tensor
$$
   (\omega^2)^{\mu\nu} := \omega^{\mu}_{\ \lambda}\,\omega^{\lambda\nu}
$$
take the form
\begin{equation}\label{eq:vorticity_square}
\begin{split}
   {(\omega^2)}^{00} &\approx -4\gamma'(\theta)^2 \,, \\
   {(\omega^2)}^{0k} &\approx -4\,i\,\gamma(\theta)\,\gamma'(\theta)\,w^{k} \,, \\
   {(\omega^2)}^{ij} &\approx 4\,\gamma^2(\theta)\, g^{ij} \,.
\end{split}
\end{equation}
Using \eqref{eq:inverse_metric} and \eqref{eq:vorticity_square}, the stress tensor can be written as
\begin{equation}\label{eq:stress_tensor_cov}
    T^{\mu\nu} \overset{\gamma\rightarrow\infty}{=} 
    T^2\,(e_1 - 4e_2)\,\gamma^4(\theta)\,\Big(4u^{\mu}u^{\nu} +3g^{\mu\nu}\Big) 
    - T^2\,(e_1 - 4e_2)\,\gamma^2(\theta)\,\omega^{\mu}_{\ \lambda}\,\omega^{\lambda\nu}\,.
\end{equation}
\subsection*{Weyl invariance: Tracelessness}
Since we started from a Weyl-invariant Lagrangian, the stress tensor is expected to be traceless order-by-order in the $\gamma$-expansion in each derivative order of the thermal partition function. Indeed, at leading order, using
\begin{equation}
    \mathrm{Tr}(\omega^2) = 8\gamma^2(\theta) + O(1), ~~u^{\mu}u_{\mu} = -1,~~ g^{\lambda}_{~\lambda} =4
\end{equation}
we confirm that the stress tensor is traceless.

\subsection{Final expression for stress tensor from the partition function up to two-derivative order in the derivative expansion}

At leading order in $\gamma$, the stress tensor obtained from the partition function in \eqref{eq:partition} reads
\begin{equation}
\begin{split}
    T^{\mu\nu} &= \Big(P(T(X)) + T^2\,(e_1 - 4e_2)\,\gamma^4(\theta) + \ldots\Big)\,(4u^{\mu}u^{\nu}) \\ 
    &\quad \Big(P(T(X)) + 3T^2\,(e_1 - 4e_2)\,\gamma^4(\theta) + \ldots\Big)\,g^{\mu\nu} \\ 
    &\quad - \Big(T^2\,(e_1 - 4e_2) + \ldots\Big)\,\gamma^2(\theta)\,\omega^{\mu}_{\ \lambda}\,\omega^{\lambda\nu} 
    + \ldots
\end{split}
\end{equation}
which can also be written as
\begin{equation}
\begin{split}
    T^{\mu\nu} &= \Big(P(T(X)) + T^2\,(e_1 - 4e_2)\,\gamma^4(\theta) + \ldots\Big)\,\left(4u^{\mu}u^{\nu} + g^{\mu\nu}\right) \\ 
    &\quad -\Big(T^2\,(e_1 - 4e_2) + \ldots\Big)\left(\gamma^2(\theta)\,\omega^{\mu}_{\ \lambda}\,\omega^{\lambda\nu} -2\gamma^4(\theta)g^{\mu\nu}\right) + \ldots.
\end{split}
\end{equation}

This expression can also be written in terms of the function $h(\beta)$ appearing in the thermodynamic partition function in \eqref{sn}. Expressed in terms of $h(T)$, the stress tensor takes the form
\begin{equation}
    T^{\mu\nu} = \frac{2T\,h(T)\,\gamma^4(\theta)}{\pi^2}\,\big(4 u^{\mu}u^{\nu}+g^{\mu\nu}\big) 
    + h_1(T)\left(\gamma^2(\theta)\,\omega^{\mu}_{\ \lambda}\,\omega^{\lambda\nu} -2\gamma^4(\theta)g^{\mu\nu}\right) 
    + \ldots
\end{equation}

\section{Partition Function with the free/GFF approximation}\label{lcform}

\subsection{General Representations}
\label{sec:generalrep4d}

As explained in the introduction \ref{sec:LCboot}, within the GFF approximation (which we seemingly expect from lightcone bootstrap) we simply multi particle over a collection of primaries and their descendants of $\widetilde{SO}(2,d-1)$.

Consider a primary, with scaling dimension $\Delta$ and $SU(2)\times SU(2)$
spins $(j_L, j_R)$. We define the `primary twist partition function', 
$\chi_{\Delta, j_L, j_R}(\beta)$, of this representation as follows. Define the `Verma module twist partition function' associated with a primary of scaling dimension $\Delta$ and angular momentum quantum numbers $(j_L, j_R)$ by 
$\chi^{\rm L}_{\Delta, j_L, j_R}(\beta)$
\begin{equation}\label{tpf}
\chi^{\rm L}_{\Delta, j_L, j_R}(\beta)= {\rm Tr}_{\rm Primaries} e^{-\beta \tau}
=(2 j_R+1) e^{-\beta\left(\Delta -2 j_L \right)} \left( \frac{1-e^{-(2j_L+1)(2\beta)}}{1-e^{-2 \beta}} \right) 
\end{equation}
(The trace in \eqref{tpf} is taken over all primary states)

If the representation question is long (i.e.\ has no null states), then 
we define $\chi_{\Delta, j_L, j_R}(\beta)$ to equal $\chi^{\rm L}_{\Delta, j_L, j_R}(\beta)$. If, on the other hand, the primary is short, then its null states 
themselves lie in one or more irreducible conformal multiplets. These multiplets are, themselves, always long. Let the quantum number of the primaries for these multiplets be denoted by  $(\Delta^i, j_L^i, j_R^i).$
In this case we define 
\begin{equation}\label{tpff}
\chi_{\Delta, j_L, j_R}(\beta)=\chi^L_{\Delta, j_L, j_R}(\beta) - \sum_i \chi^{L}_{\Delta^i, j_L^i, j_R^i}(\beta)
\end{equation}
With all these definitions in place (and using the usual formulae of Bose 
and Fermi statistics), the reader can easily verify that the partition function 
of our theory, in the `free lightcone approximation', is given (at leading order, in the limit that $\omega_i \to 1$  
\begin{equation}\label{iffree}
\begin{split}
\ln Z =&  \sum_q \frac{1}{q} \frac{ \sum_\alpha  \chi_{\Delta_\alpha, j_L^\alpha, j_R^\alpha}(q\beta)}{(1-e^{-q \beta (1-\omega_1)})(1-e^{-q \beta (1+\omega_1)})(1-e^{-q \beta (1-\omega_2)})(1-e^{-q \beta (1+\omega_2)})}\\
+&\sum_q \frac{(-1)^{q+1}}{q} \frac{ \sum_\gamma  \chi_{\Delta_\gamma, j_L^\gamma, j_R^\gamma}(q\beta)}{(1-e^{-q \beta(1-\omega_1})(1-e^{-q \beta(1+\omega_1)})(1-e^{-q \beta(1-\omega_2)})(1-e^{-q \beta(1+\omega_2)})}
\end{split}
\end{equation}
where the summation over $\alpha$ in the first line of \eqref{iffree} 
runs over all bosonic primaries of the theory, and the 
summation over $\gamma$ in the second line of \eqref{iffree} 
runs over all fermionic primaries of the theory, and the 

As our formulae only apply at leading order in the limit of small $1 -\omega_1$ and $1-\omega_2$, \eqref{iffree} can be rewritten in the simpler form 
\begin{equation}\label{flpt}
\ln Z =  \frac{1}{ (1-\omega_1)(1-\omega_2)}\sum_q \frac{1}{q^3} \frac{\left(\sum_\alpha  \chi_{\Delta_\alpha, j_L^\alpha, j_R^\alpha}(q \beta) + (-1)^{q+1}\sum_\gamma  \chi_{\Delta_\gamma, j_L^\gamma, j_R^\gamma}(q\beta)\right)}{\beta^2(1-e^{-2q \beta})^2}
\end{equation}

We see that the toy model described above does, indeed, produce a partition function of the form $\frac{h(\beta)}{(1-\omega_1)(1-\omega_2)}$ as we have predicted in \eqref{sn}, with  
\begin{equation}\label{zeo}
h(\beta) = \sum_q \frac{1}{q^3} \frac{\left(\sum_\alpha  \chi_{\Delta_\alpha, j_L^\alpha, j_R^\alpha}(q \beta) + (-1)^{q+1}\sum_\gamma  \chi_{\Delta_\gamma, j_L^\gamma, j_R^\gamma}(q\beta)\right)}{\beta^2(1-e^{-2q \beta})^2}
\end{equation}
As the functions $\chi_{\Delta, j_L, j_R}(\beta)$ take the form $e^{-\beta \Delta_i}$ times a polynomial in $e^{-2 \beta}$,  \eqref{zeo} predicts that the Taylor expansion of $h(\beta)$ in the variable $e^{-\beta}$ has very specific integrality properties. We note that the $\beta\to 0$ limit of $h(\beta)$ will generally depend on the cardinality of the collection of basic fields, in particular, whether there are finite or infinite of them.

\subsection{Exotic Free theories}\label{exotic}

In the previous subsection we have studied the multiparticling of states in a single representation of the conformal algebra. Familiar free theories, like the $4d$ free scalar studied earlier in section \ref{sec:generalrep4d}, are examples of such partition functions with one provision: the single-particle states in such theories are chosen to be short (i.e. they saturate the unitarity bound \eqref{unibound}) in such a way that the null states represent an equation of motion, cutting down the number of states in the representation. For example, this mechanism makes $\ln Z$ for the 4d free scalar grow like $T^3$ (rather than $T^4$) at high temperatures. 

The sphere partition function of several `exotic' free theories was studied in  \cite{Loganayagam:2012zg}. These theories have states with arbitrarily negative $\tau$, and so violate the ANEC bounds of 
\cite{Cordova:2017dhq}. In this Appendix we investigate the sphere partition functions of two of these theories at leading order in the high-temperature and small $\nu_i$ limits. We demonstrate that while these theories appear to have a completely vanilla small $\beta$ expansion, the function $h(\beta)$, which characterizes the small $\nu_i$ expansion in these theories, diverges. This divergence is a direct consequence of the existence of states with arbitrarily negative twist. We conclude that the small $\nu_i$ limit, studied in this paper, is a more sensitive diagnostic of the ANEC-violating sickness of these theories.

\subsubsection{The chiral gravitino}

Consider the `chiral gravitino' partition function studied in \cite{Loganayagam:2012zg}, i.e. the partition function obtained by multiparticling states in the representation $(\Delta =5/2, J_L=3/2, J_R=0)$. Note that this representation is short. Its null state has quantum numbers $(\Delta =7/2, J_L=1, J_R=1/2)$. These null states themselves constitute a short representation, whose null state has quantum numbers 
$(\Delta =9/2, J_L=1/2, J_R=0)$. Consequently, the effective single-particle partition function is 
\begin{align}\label{sppf1}
Z_{sp}&=\Big[x^{5/2} \left( (z_1 z_2)^{3/2} + (z_1 z_2)^{1/2}+ (z_1 z_2)^{-1/2}+ (z_1 z_2)^{-3/2} \right) \nonumber\\
&- x^{7/2}  \left( (z_1 z_2)^{1} + 1 +  (z_1 z_2)^{-1}\right)  \left( (z_1/z_2)^{1/2} +  (z_1/z_2)^{-1/2} \right) \nonumber\\
&+  (x)^{9/2}\left( (z_1 z_2)^{1/2} +  (z_1 z_2)^{-1/2}\right)\Big]\nonumber\\
&\times\frac{1}{(1-xz_1)(1-\frac{x}{z_1}) (1-xz_2)(1-\frac{x}{z_2})} 
\end{align}
where $z_1=e^{\beta \omega_1}$, $z_2=e^{\beta \omega_2}$ and $x=e^{-\beta}$, and the first, second and third lines of \eqref{sppf1} capture, respectively, the primaries, the null state, and the null states of the null states. 

Notice that the numerator has 6 positive terms and 6 negative terms. This reflects the fact that the null states represent an equation of motion, and so ensure that our partition is of order $T^3$ (rather than $T^4$) at large $T$. This may be seen more explicitly as follows. Taylor expanding \eqref{sppf1} in $\beta$ and keeping the leading nontrivial term, we find  
\begin{align}\label{exhtemp}
    Z_{sp}&\approx\frac{2}{\beta ^3 \left(1-\omega _1^2\right) \left(1-\omega _2^2\right)}
\end{align}

Multiparticling the single-particle partition function, we get
\begin{align}\label{exparth}
    \ln Z\approx\frac{2}{\beta^3(1-\omega^2_1)(1-\omega^2_2)}\left(\sum_{k=1}^{\infty}\frac{(-1)^{k+1}}{k^4}\right)
\end{align}

Let us now turn to the small $\nu_i$ limit. If, as usual, we define $\nu_i=\beta(1-\omega_i)$ and take the limit 
$\nu_i \to 0$. In the numerator, this amounts to setting $z_1=z_2=\frac{1}{x}$. In the denominator, it amounts to setting $z_1=z_2=\frac{1}{x}$ in the second and fourth terms, but Taylor expanding in the first two terms. We obtain
\begin{align}\label{sp2}
Z_{sp}&=\frac{1}{\beta^2\nu_1\nu_2\,x^{\frac{1}{2}}(1-x^2)}, 
\end{align}
where $z_1=e^{\beta \omega_1}$, $z_2=e^{\beta \omega_2}$ and $x=e^{-\beta}$. 
`Multiparticling' the single-particle partition function in \eqref{sp2}, we find\footnote{In the expression \eqref{spfinal}, we formally substitute $x=e^{-\beta}$ and expanding it around $\beta\rightarrow 0$ limit, we find perfect agreement with \eqref{exparth}.}
\begin{equation}\label{spfinal}
    \ln Z=\frac{1}{\beta^2\nu_1\nu_2}\sum_{k=1}^{\infty}\frac{(-1)^{k+1}}{k^3}\frac{1}{x^{\frac{k}{2}}(1-x^{2k})}
\end{equation} 
The important point in \eqref{spfinal} is the factor of $\frac{1}{x^{k/2}} = e^{ \frac{k \beta}{2}}$ in the summand of \eqref{spfinal}. This term is a direct reflection of the fact that our primary state carries $\tau=-\frac{1}{2}$. This term causes the summation over $k$ to diverge, leading to an ill-defined result for the function $h(\beta)$.

\subsubsection{$j_L=2$}

We can now perform a similar analysis for the `free theory' whose single-particle states transform in the $(\Delta=3, j_L=2, j_R=0)$ representation. Once again this representation has null states: the primary for the null states transforms in $(\Delta=4, j_L=\frac{3}{2}, j_R=\frac{1}{2})$. Once again, the null states themselves transform in 
a short representation, whose primary lies at $(\Delta=5, j_L=1=, j_R=0)$. In this representation we have 
\begin{align}
    Z_{sp}&=\Big[x^{3} \left( (z_1 z_2)^{2} + (z_1 z_2) +1 +  (z_1 z_2)^{-1}+ (z_1 z_2)^{-2} \right)  \\
    &- x^{4} \left( (z_1 z_2)^{\frac{3}{2}} + (z_1 z_2)^{\frac{1}{2}} +  (z_1 z_2)^{-\frac{1}{2}}+ (z_1 z_2)^{-\frac{3}{2}} \right) \left((z_1/z_2)^{1/2}+(z_1/z_2)^{-1/2}\right) \nonumber\\
&+ x^{5}  \left( (z_1 z_2)^{1} + 1 +  (z_1 z_2)^{-1}\right)\Big]\times\frac{1}{(1-x z_1)(1-x/z_1)(1-x z_2)(1-x/z_2)}  \nonumber\\
\end{align}
Note that, once again, the numerator in $Z_{sp}$ has 8 positive and 8 negative terms, and so will give rise to a high-temperature partition function that scales like $T^3$ (rather than $T^4$). As above, if we 
set $z_1=e^{\beta \omega_1}$, $z_2=e^{\beta \omega_2}$ and $x=e^{-\beta}$
and work to leading order in $\beta\ll 1$ we find 
\begin{align}
    Z_{sp}&\approx\frac{2}{\beta^3(1-\omega_1^2)(1-\omega^2_2)}
\end{align}
Multiparticling the single-particle partition function
\begin{align}
    \ln Z&=\frac{2}{\beta^3(1-\omega^2_1)(1-\omega^2_2)}\sum_{k=1}^{\infty}\frac{1}{k^4}
\end{align}
Defining $\nu_i=\beta(1-\omega_i)$, and setting $z_1=z_2=1/x$, in the numerator and the two terms in the denominator 
\begin{align}
    Z_{sp}&=\frac{1}{\beta^2\nu_1\nu_2\, x(1-x^2)}
\end{align}
    Multiparticling the single-particle partition function, we get
    \begin{align}
        \ln Z&=\frac{1}{\beta^2\nu_1\nu_2}\sum_{k=1}^{\infty}\frac{1}{k^3}\frac{1} {x^k(1-x^{2k})}
    \end{align}
Once again, the factor of $\frac{1}{x^k}= e^{\beta k}$ in the summand 
causes the sum over $k$ to diverge. The divergence in this case is even clearer than in the previous subsection (all terms have the same sign, so there is no possibility of phase cancellation in the summation).  We see that the negative $\tau$ `sickness' manifests itself in a divergent (so ill-defined) $h(\beta)$.

\section{Lightcone bootstrap in 2d}
\label{sec:lightconeboot2d}

A famous result of the lightcone bootstrap \cite{Komargodski:2012ek, Fitzpatrick:2012yx} asserts that there is a sense in which all CFTs become free in a high spin sector. However, as explained in the introduction,  there are two main differences between the usual lightcone bootstrap and the content of this work:
\begin{enumerate}
    \item The lightcone bootstrap is generally about an $N$-point correlation function (e.g. a four-point-function), and gives universal results about operators that contribute to the correlation function. In contrast, this work is about the entire partition function, and predicts universal features of the entire spectrum.
    \item In order to see the free spectrum in the lightcone bootstrap, one first takes the $\Omega \rightarrow 1$ limit and then the $\beta \rightarrow \infty$ limit. In this work, we do not require the $\beta \rightarrow \infty$ limit, which lets us probe an expanded regime in parameter space (e.g. black hole physics in Sec. \ref{lgym}).
\end{enumerate}

In this appendix, we provide a brief discussion of this comparison for $d=2$ CFTs. The lightcone bootstrap in 2d is qualitatively different than in $d>2$ due to the presence of infinitely many operators of twist $0$ (i.e.\ $h=0$ or $\bar{h}=0$). In particular, the entire Virasoro Verma module built from acting $L_{-n}$'s on the vacuum state has twist $0$. CFTs with an enhanced chiral algebra, or with accumulation points to twist $0$ (e.g. any theory with a $U(1)$ global symmetry \cite{Benjamin:2020swg}) have further states that contribute in the lightcone limit. 

The simplest 2d CFTs would be compact, unitary 2d CFTs with $c>1$ and nonzero twist gap under the Virasoro algebra. It is worth noting that although we expect such theories to be generic, we do not yet explicitly have an example of such a theory \cite{Collier:2016cls}. Nonetheless, if such theories exist (as expected), the lightcone bootstrap tells us a huge amount of universal information at the large spin finite twist limit. The four-point function lightcone bootstrap for 2d CFTs with only Virasoro symmetry was pioneered in \cite{Kusuki:2018wpa, Collier:2018exn}. For completeness, we summarize their result here; for derivation see \cite{Kusuki:2018wpa, Collier:2018exn}. 

If we define the following ``Liouville variables":
\begin{equation}
c = 1 + 6 Q^2, ~~~h = \alpha(Q-\alpha)\,,
\end{equation}
then instead of the twists $h$ adding at large spin, the Liouville momenta $\alpha$ add instead, so long as $\alpha$ remains real. In particular, if there are light operators with Liouville momenta $\alpha_1, \alpha_2$, then there are Regge trajectories that at large spin have twist approaching 
\begin{equation}
    \alpha_1 + \alpha_2 + mb, ~~~m\in \mathbb{N}\,,
\label{eq:2dtwistaddition}
\end{equation}
where $b+b^{-1}=Q$. Eqn (\ref{eq:2dtwistaddition}) is valid until the sum becomes larger than $Q/2$. For all twists with Liouville momenta beyond $Q/2$, i.e.\ $h >  \frac{c-1}{24}$ (where the Liouville momentum obeys $\alpha \in Q/2 + i \mathbb R$), the twist accumulation points become dense. This result is derived from the structure of the Virasoro fusion kernel; note that a key assumption is that the twist gap under the Virasoro algebra is a finite, nonzero number for this to hold. 

As discussed above, in $d>2$ dimensions, the lightcone bootstrap result captures different physics (GFF-like features in the $N$-point correlator) than that of computing the partition function in the $\omega_i\to1$ limit. In contrast, in two dimensions, the situation is slightly different. The presence of dense twist accumulation points can also be shown by studying the limit of $\omega\to1$ of the partition function \cite{Collier:2016cls,Kusuki:2018wpa,Benjamin:2019stq,Pal:2022vqc,Pal:2023cgk}. In fact, any point $x\geq \frac{c-1}{12}$ is an accumulation point, as rigorously shown in \cite{Pal:2025yvz}. However, just like in higher dimension, the twisted partition function in $\omega\to1$ fails to see the existence of the states described in eq.~\!\eqref{eq:2dtwistaddition}: note that these states are analogous to double twist states appearing in the large spin limit in $\phi\phi$ OPE in higher dimensional CFTs.

\section{Zero modes in compactified free Boson in $2$d}\label{app:freeBoson}

The compactified free boson in two dimensions satisfies 
\begin{equation}
\phi(z+2k\pi+2ik'\beta)=\phi(z)+2\pi R\left(k m+ k'm'\right) 
\end{equation}
The zero winding modes along the time circle are labelled by $m'=0$ and $m$ a nonzero integer. 
The partition function of the compactified free boson is given by
	\begin{equation}\label{PI}
		Z(\beta)= \frac{R}{\sqrt{2}} \frac{1}{\left(\sqrt{\frac{\beta}{2\pi}}\right) \eta\left(\frac{\beta}{2\pi}\right)^2}\sum_{m,m'}\exp\left[-\frac{R^2(m\beta-2\pi m')^2}{4\beta}\right]
	\end{equation}
Let us write it in the following suggestive form:
		\begin{equation}
				\begin{aligned}
	Z(\beta)&= \underbrace{\frac{R}{\sqrt{2}}}_{\text{constant mode on Torus}} \times {\color{blue}\underbrace{\sqrt{\frac{\beta}{2\pi}}}_{\text{zero mode along the time direction}}} \times  {\color{magenta}\underbrace{\frac{1}{\left(\frac{\beta}{2\pi}\right)\eta\left(\frac{\beta}{2\pi}\right)^2}}_{\text{integrating out the non-zero modes}}}\\
	&\qquad\qquad \qquad \qquad\qquad \times \left({\color{brown}\underbrace{Z_{m'=0}}_{\text{zero winding}}}+\underbrace{Z_{m'\neq 0}}_{\text{non-zero winding}}\right)
	\end{aligned}\,,
\end{equation} 
where we have 
\begin{equation}
Z_{m'}:=\sum_{m}\exp\left[-\frac{R^2(m\beta-2\pi m')^2}{4\beta}\right]\,.
\end{equation}

Let us compute the contributions coming from $m'=0$ modes (zero winding modes along the time circle); there are multiple such modes, parametrized by $m$, the winding number along the spatial circle. Their cumulative contribution is given by 
	\begin{equation}
	Z_{m'=0}=	\sum_{m}\exp\left[-\frac{R^2  \beta}{4}m^2\right] \approx \int dm\ \exp\left[-\frac{R^2  \beta}{4}m^2\right]= \frac{\sqrt{2}}{R}\sqrt{\frac{2\pi}{\beta}}\,.
\end{equation}

Thus in the $\beta\to 0$ limit, we obtain
\begin{equation}
\begin{aligned}
 Z&= \frac{R}{\sqrt{2}} \times{\color{blue} \sqrt{\frac{\beta}{2\pi}}} \times {\color{magenta}\frac{1}{\left(\frac{\beta}{2\pi}\right)\eta\left(\frac{\beta}{2\pi}\right)^2}} \times   \left[{\color{brown}  \frac{\sqrt{2}}{R}\sqrt{\frac{2\pi}{\beta}}}+ \underbrace{O\left(e^{-a\pi^2R^2/\beta}\right)}_{\text{non-zero temporal winding}}\right]\\
   &=\exp\left[\frac{4\pi^2}{\beta}\frac{1}{12}\right]\left[1+O\left( e^{-\frac{4\pi^2}{\beta}\mathrm{Min}\left(1, \frac{aR^2}{4}\right)}\right)\right]\,,
\end{aligned}
\end{equation}
where $a$ is some order one number. 

Now we make some important remarks about the expression above.
\begin{enumerate}
\item The contribution coming from the non-zero temporal winding modes is a non-local effect owing to the compact nature of the spatial circle.
\item Integrating out the non-zero modes of the harmonic part of $\phi$ (the part which characterizes the fluctuation around classical saddle) gives $ \frac{1}{\left(\frac{\beta}{2\pi}\right)\eta\left(\frac{\beta}{2\pi}\right)^2}$. The leading behaviour of this is captured by the thermal EFT.  However, this has exponentially suppressed corrections:
$${\color{magenta} \frac{1}{\left(\frac{\beta}{2\pi}\right)\eta\left(\frac{\beta}{2\pi}\right)^2} \underset{\beta\to 0}{\sim} \exp\left[\frac{4\pi^2}{\beta}\frac{1}{12}\right] \times  \left[1+ O\left(e^{-\frac{4\pi^2}{\beta}}\right)\right]\,.}$$
This non-perturbative correction is of different nature than what is coming from the non-zero winding modes discussed above.
\item Finally, we see that there is a perfect cancelation in $ \frac{R}{\sqrt{2}} \times{\color{blue} \sqrt{\frac{\beta}{2\pi}}} \times  {\color{brown} \frac{\sqrt{2}}{R}\sqrt{\frac{2\pi}{\beta}}}$ giving us $1$. This is what makes sure that in the $\beta\to 0$ limit, the suppression to the leading piece coming from the thermal EFT is indeed of the type $\exp[-\#/\beta]$, a result that is consistent with the prediction of the modular invariance.
\end{enumerate}

\section{Evaluation of $h(\beta)$ in free scalar field theories in various dimensions}\label{app:freeScalarComp}
\subsection{Evaluation of $h(\beta)$ for free scalar field theory in dimension $4$}
In $4$ dimensions,  the function $h^{d=4}(\beta)$ is given by $\sum_{q=1}^{\infty} \frac{		\mathrm{csch}(q\beta)}{2 q^3}$.  In this appendix,  we explain how to obtain a convergent expansion for $h(\beta)$,  adapted to $\beta\to 0$ limit.  In particular,  we provide a derivation of Eqn (\ref{hd4}). 

We first note that 
\begin{equation}\label{eq:coshId4}
	\mathrm{csch}(\beta)=\frac{1}{\beta}+2\beta\sum_{n=1}^{\infty}\frac{(-1)^n}{\beta^2+\pi^2 n^2}\,.
\end{equation}

By elementary manipulations,  we have
\begin{equation}
	\begin{aligned}
	\sum_{q=1}^{\infty} \frac{		\mathrm{csch}(q\beta)}{2 q^3}&= \frac{\zeta(4)}{2\beta}+\sum_{q=1}^{\infty}\frac{\beta}{q^2}\sum_{n=1}^{\infty}\frac{(-1)^n}{q^2\beta^2+\pi^2 n^2}\\
	&=\frac{\zeta(4)}{2\beta}+\sum_{n=1}^{\infty}(-1)^n\sum_{q=1}^{\infty}\frac{\beta}{q^2}\frac{1}{q^2\beta^2+\pi^2 n^2}\\
	&=\frac{\zeta(4)}{2\beta} - \frac{\zeta(2)\beta}{12}-\frac{7\zeta(4)\beta^3}{16\pi^4}-\sum_{n=1}^{\infty}(-1)^n\frac{\beta^2}{2n^3\pi^2}\left(1+2\sum_{k=1}^{\infty}e^{-\frac{2 \pi^2 n k}{\beta}}\right)\\
	&=\frac{\zeta(4)}{2\beta} - \frac{\zeta(2)\beta}{12}-\frac{7\zeta(4)\beta^3}{16\pi^4}+\frac{3\zeta(3)\beta^2}{8\pi^2}+\sum_{n=1}^{\infty}(-1)^{n+1}\frac{\beta^2}{n^3\pi^2}\sum_{k=1}^{\infty}e^{-\frac{2 \pi^2 n k}{\beta}}\,.
	\end{aligned}
\end{equation}
Going from the first line to the second line,  we have exchanged the summation over $n$ and the summation over $q$.  This is allowed since the sum over $n$ is absolutely convergent.  Going from the second line to the third line,  we have used
\begin{equation}
\sum_{q=1}^{\infty}\frac{\beta}{q^2}\frac{1}{q^2\beta^2+\pi^2 n^2}=\frac{\beta ^3}{2 \pi ^4 n^4}-\frac{\beta ^2 \coth \left(\frac{\pi ^2 n}{\beta }\right)}{2 \pi ^2 n^3}+\frac{\beta }{6 n^2} \,,
\end{equation}
expanded $\coth\left(\frac{\pi ^2 n}{\beta }\right)$ as a sum over $e^{-2\pi^2 n k/\beta}$,  and done the sum over $n$.  The final line follows from perforning the $n$ sum over the $k=0$ modes coming from the expansion of $\coth$. 

In summary,  we derive the following equality
\begin{equation}
		\begin{aligned}
		h(\beta)=\sum_{q=1}^{\infty} \frac{		\mathrm{csch}(\beta)}{2 q^3\beta^2}&=\frac{\zeta(4)}{2\beta^3} - \frac{\zeta(2)}{12\beta}-\frac{7\zeta(4)\beta}{16\pi^4}+\frac{3\zeta(3)}{8\pi^2}+\sum_{n=1}^{\infty}(-1)^{n+1}\frac{1}{n^3\pi^2}\sum_{k=1}^{\infty}e^{-\frac{2 \pi^2 n k}{\beta}}\\
		&=\frac{\pi^4}{180\beta^3}-\frac{\pi^2}{72\beta}-\frac{7\beta}{1440}+\frac{3\zeta(3)}{8\pi^2}+\sum_{m=1}^{\infty}\sum_{d|m}(-1)^{d+1}\frac{1}{d^3\pi^2}e^{-\frac{2 \pi^2 m}{\beta}}
			\end{aligned}
\end{equation}

The above formula is also checked to be true by numerical evaluation of the sum over $q$,  defining $h^{d=4}(\beta)$.

\subsection{Evaluation of $h(\beta)$ for free scalar field theory in dimension $6$}
For dimension $6$,  $h^{d=6}(\beta)$ is given by $	\sum_{q=1}^{\infty} \frac{	\mathrm{csch}(\beta)^2}{4 q^4\beta^3}$.  We follow the same procedure as above.  The identity that we start with in this case is the analogue of \eqref{eq:coshId4}:
\begin{equation}\label{eq:coshId6}
\mathrm{csch}(\beta)^2= \frac{1}{\beta^2}-2\sum_{n=1}^{\infty} \frac{1}{\beta^2+\pi^2 n^2} + 4\beta^2 \sum_{n=1}^{\infty} \frac{1}{(\beta^2+\pi^2n^2)^2}\,.
\end{equation}

By performing similar manipulations as above, we find that 
\begin{equation}
\begin{aligned}
	\sum_{q=1}^{\infty} \frac{	\mathrm{csch}(\beta)^2}{4 q^4\beta^3}&=\frac{\zeta(6)}{4\beta^5}-\frac{1}{2\beta^3} 	\sum_{n=1}^{\infty}\sum_{q=1}^{\infty}  \frac{1}{q^4}\frac{1}{q^2\beta^2+\pi^2 n^2} +  \frac{1}{\beta}\sum_{n=1}^{\infty} \sum_{q=1}^{\infty}\frac{1}{q^2} \frac{1}{(q^2\beta^2+\pi^2n^2)^2}\\
	&=\frac{\zeta(6)}{4\beta^5}-\frac{\pi ^4}{1080 \beta ^3}+\frac{\beta }{3780}+\frac{\pi ^2}{1080 \beta }-\frac{\zeta (5)}{4 \pi ^4}+\frac{\beta }{945}+\frac{\pi ^2}{540 \beta }-\frac{3 \zeta (5)}{4 \pi ^4} + R\\
	&=\frac{\pi ^6}{3780 \beta ^5}-\frac{\pi ^4}{1080 \beta ^3}+\frac{\beta }{756}+\frac{\pi ^2}{360 \beta }-\frac{\zeta (5)}{\pi ^4}+R\,,
\end{aligned}
\end{equation}

where $R$ is given by
\begin{equation}
R=-\sum_{n=1}^{\infty} \frac{2}{ \pi ^4 n^5}\sum_{k=1}^{\infty} e^{-\frac{2n k\pi^2}{\beta}}-\sum_{n=1}^{\infty} \frac{1}{\pi ^2 \beta  n^4}\sum_{k=1}^{\infty}k \exp \left(-\frac{2 \pi ^2 k n}{\beta }\right)\,.
\end{equation}

Here we have used 
\begin{equation}
\begin{aligned}
\sum_{q=1}^{\infty}\frac{1}{q^2} \frac{1}{(q^2\beta^2+\pi^2n^2)^2}&=\frac{\beta ^2}{\pi ^6 n^6}-\frac{3 \beta  \coth \left(\frac{\pi ^2 n}{\beta }\right)}{4 \pi ^4 n^5}-\frac{\text{csch}^2\left(\frac{\pi ^2 n}{\beta }\right)}{4 \pi ^2 n^4}+\frac{1}{6 \pi ^2 n^4}\,,\\
\sum_{q=1}^{\infty}  \frac{1}{q^4}\frac{1}{q^2\beta^2+\pi^2 n^2}&=-\frac{\beta ^4}{2 \pi ^6 n^6}+\frac{\beta ^3 \coth \left(\frac{\pi ^2 n}{\beta }\right)}{2 \pi ^4 n^5}-\frac{\beta ^2}{6 \pi ^2 n^4}+\frac{\pi ^2}{90 n^2}\,.
\end{aligned}
\end{equation}

Once again,  the expression for $h^{d=6}(\beta)$ is verified numerically.  Under a change of variable for the terms in the exprssions for $R$,  the expression becomes the one presented in the main text, i.e.\ \eqref{hd6}.

\subsection{$h^{d=2n}(\beta)$ for any even dimension}
Define $r:=d/2-1=n-1$. We have
\begin{equation}\label{eq:alleven}
	\begin{aligned}
			h^{d=2n}(\beta)
		&=\frac{1}{2^r}\left(h_{\text{pert}}(\beta)+h_{\text{non-pert}}(\beta)\right)\,,
	\end{aligned}
\end{equation}
where $h_{\text{pert}}(\beta)$ is given by
\begin{equation}\label{eq:allevenPert}
    \begin{aligned}
h_{\text{pert}}(\beta)=&\sum_{p=0}^{\lfloor \frac{r-1}{2}\rfloor}\frac{a_{r,p}\zeta(2r-2p+2)}{\beta^{2r-2p+1}}\\
        &\qquad+ \sum_{p=0}^{\lfloor \frac{r-1}{2}\rfloor}2a_{r,p}  \left(\sum _{\ell=1}^{r/2+1} \frac{(-1)^{\ell+p+1}  \binom{-2\ell+2r-2p+1}{r-2p-1}\zeta(2\ell)}{\beta^{2\ell-1}(\pi)^{-2\ell+2r-2p+2}}\text{Li}_{2 (-\ell-p+r+1)}\left((-1)^r\right) \right) \\
			&\qquad+ \sum_{p=0}^{\lfloor \frac{r-1}{2}\rfloor}a_{r,p}\frac{ (-1)^{p-1} }{\pi^{2r-2p}} \binom{2r-2p}{r+1} \text{Li}_{-2 p+2 r+1}\left((-1)^r\right)\\
            &\qquad+\sum_{p=0}^{\lfloor \frac{r-1}{2}\rfloor}(-1)^{p}a_{r,p} \left( \frac{\Gamma (-2 p+2 r+2)}{\Gamma (r+3) \Gamma (r-2 p)}\frac{ \text{Li}_{-2 p+2 r+2}\left((-1)^r\right)}{(\pi)^{2r-2p+2}} \right)\beta\,,
    \end{aligned}
\end{equation}

and $h_{\text{non-pert}}(\beta)$ is given by

\begin{equation}\label{eq:allevenNonPert}
	\begin{aligned}
			&h_{\text{non-pert}}(\beta)\\
&=\sum_{n,k\in\mathbb{Z}_+}e^{-\frac{2\pi^2 nk}{\beta}}\times\\
&\times\left(\sum_{p=0}^{\lfloor \frac{r-1}{2}\rfloor}2(-1)^{p-1}a_{r,p} \left( \sum _{\ell=1}^{r-2p}  \frac{1}{\beta^{\ell-1}}\frac{\binom{-\ell+2r-2p+1}{r+1}}{\Gamma(\ell)} (-1)^{nr}\frac{(n\pi^2 )^{\ell-2}n^{-2}}{(n\pi)^{2r-2p-2}} (2k)^{\ell-1}  \right)\right)\,,
	\end{aligned}
\end{equation}
and finally the coefficients $a_{r,p}$ can be extracted from the expansion of $\mathrm{csch}^r(x)$ i.e 
\begin{equation}
\mathrm{csch}^r(x)=\sum_{p=0}^{\infty} a_{r,p}\ x^{-r+2p}  \,.
\end{equation}

We conclude this subsection with a remark and brief sketch of the derivation. First of all, we have numerically verified the above result for $d=4,6,8,10,12$.
Furthermore, one can derive \eqref{eq:alleven} using methods similar to what has been used to derive the ones for $d=4$ and $d=6$. The key identity (the analog of \eqref{eq:coshId4} and \eqref{eq:coshId6}) that we would have to use is given by 
\begin{equation}
	\mathrm{csch}^{r}(q\beta)= \sum_{n\in\mathbb{Z}} (-1)^{nr} \sum_{p=0}^{\lfloor \frac{r-1}{2}\rfloor} \frac{a_{r,p}}{(q\beta-\imath \pi n)^{r-2p}}\,.
\end{equation}
Hence we have 
\begin{equation}
	\begin{aligned}
		2^{r}	h(\beta)&=\beta^{-r-1} \sum_{p=0}^{\lfloor \frac{r-1}{2}\rfloor} \left(\sum_{n\in\mathbb{Z}} (-1)^{nr}  \sum_q \frac{a_{r,p}}{q^{r+2}(q\beta-\imath \pi n)^{r-2p}}\right)\\
		&=\sum_{p=0}^{\lfloor \frac{r-1}{2}\rfloor}\frac{a_{r,p}\zeta(2r-2p+2)}{\beta^{2r-2p+1}}\\		&+ \sum_{p=0}^{\lfloor \frac{r-1}{2}\rfloor}\frac{a_{r,p}}{\beta^{2r-2p+1}} \left(\sum_{n\in\mathbb{Z}_+} (-1)^{nr}  \sum_q \frac{1}{q^{r+2}}\left(\frac{1}{(q-\imath ny)^{r-2p}}+\frac{1}{(q+\imath ny)^{r-2p}}\right)\right)\,,
	\end{aligned}
\end{equation}
where in the last line we have used $y:=\pi/\beta$.

Now we use the identity
\begin{equation}
	\frac{1}{q^{m}(q+u)^k }=\sum _{j=1}^m \frac{(-1)^{m-j} \binom{-j+k+m-1}{k-1}}{q^j u^{-j+k+m}}+\sum _{j=1}^k \frac{(-1)^m \binom{-j+k+m-1}{m-1}}{(q+u)^j u^{-j+k+m}}\,,
\end{equation}
followed by performing the $q$ sum.  The first kind of terms in the above sum results in $\zeta$ function while the second kind produces Hurwitz $\zeta$ function.  In particular, we will encounter terms such as
$$\mathrm{Re}\left[\frac{e^{\frac{\imath \pi\ell}{2}} \zeta (\ell ,1+\imath n\frac{\pi}{\beta})}{(n\pi)^{-\ell+2r-2p+2}} \right]\,.$$
We deal with this by noting that 
\begin{equation}
\frac{1}{(n\pi)^{-\ell+2r-2p+2}}	e^{\frac{\imath \pi\ell}{2}}  \zeta (\ell,1+\imath n y)= \frac{(-1)^{\ell+1}}{\Gamma(\ell)}\frac{1}{(n\pi)^{2r-2p}} \frac{d^{\ell-2}}{dy^{\ell-2}} \zeta (2,1+\imath n \pi y)\,,
\end{equation}
so that we have 
\begin{equation}
		\begin{aligned}
&\mathrm{Re}\left[\frac{e^{\frac{\imath \pi\ell}{2}} \zeta (\ell ,1+\imath n\frac{\pi}{\beta})}{(n\pi)^{-\ell+2r-2p+2}} \right]\\
&= \frac{(-1)^{\ell+1}}{\Gamma(\ell)}\frac{1}{(n\pi)^{2r-2p}} \frac{d^{\ell-2}}{dy^{\ell-2}} \mathrm{Re}\left[\zeta (2,1+\imath n \pi y)\right]\bigg|_{y=\beta^{-1}}\\
&=-\frac{1}{2(n\pi)^{2r-2p+2}} \beta^{\ell} +\frac{(-1)^{\ell}}{2\Gamma(\ell)}\frac{n^{-2}}{(n\pi)^{2r-2p-2}} \frac{d^{\ell-2}}{dy^{\ell-2}} \left[\text{csch}^2\left(\pi ^2 n y\right)\right]\bigg|_{y=\beta^{-1}}\\
&=-\frac{1}{2(n\pi)^{2r-2p+2}} \beta^{\ell} -\frac{1}{\Gamma(\ell)}\frac{(\pi^2n )^{\ell-2}n^{-2}}{(n\pi)^{2r-2p-2}} \left[ \sum_k (2k)^{\ell-1} e^{-\frac{2\pi^2 n k}{\beta}}  \right]\,.
\end{aligned}
\end{equation}
Here we have used
\begin{equation}
\mathrm{Re}\left[\zeta (2,1+\imath n \pi y)\right]= -\frac{\pi^2}{2}\text{csch}{}^2(\pi^2 ny) +\frac{1}{2\pi^2n^2 y^2}\,.
\end{equation}

\subsection{Evaluation of $h(\beta)$ for free scalar field theory in dimension $3$}
In $d=3$ dimension, we would like to do the sum $\sum_q \frac{1}{q} \frac{\mathrm{csch}(q\beta/2)}{2q\beta}$.  
\begin{equation}\label{oddhcal}
	\begin{aligned}
\sum_{q=1}^{\infty} \frac{1}{q} \frac{\mathrm{csch}(q\beta/2)}{2q\beta}
&=\sum_q \frac{1}{2q^2\beta} \left(\frac{1}{q\beta}+\frac{q\beta}{2}\sum_n \frac{(-1)^n}{\frac{q^2\beta^2}{4}+\pi^2 n^2}\right)\\
&=\frac{\zeta(3)}{\beta^2}+\sum_n (-1)^n \sum_q \frac{1}{2q\left(\frac{q^2\beta^2}{4}+\pi^2 n^2\right)}\\
&=\frac{\zeta(3)}{\beta^2} +\sum_n (-1)^n\left[ \frac{\psi ^{(0)}\left(\frac{i (2 n \pi -i \beta )}{\beta }\right)}{4\pi ^2 n^2}+\frac{\psi ^{(0)}\left(-\frac{i (2 \pi  n+i \beta )}{\beta }\right)}{4 \pi ^2 n^2}+\frac{\gamma }{2\pi ^2 n^2}\right]\\
&=\frac{\zeta(3)}{\beta^2} -\frac{\gamma }{24}+\sum_n (-1)^n\left[ \frac{\psi ^{(0)}\left(\frac{i (2 n \pi -i \beta )}{\beta }\right)}{4 \pi ^2 n^2}+\frac{\psi ^{(0)}\left(-\frac{i (2 \pi  n+i \beta )}{\beta }\right)}{4 \pi ^2 n^2}\right]\\
&=\frac{\zeta(3)}{\beta^2} + \frac{\ln \beta}{24} -\frac{\gamma }{24} +\frac{\zeta '(2)}{4 \pi ^2}-\frac{\ln \pi}{24}+\sum_n (-1)^n r(n,\beta)
\end{aligned}
\end{equation}
where $r(n,\beta)$ is given by 
\begin{equation}
	r(n,\beta)= \frac{\psi ^{(0)}\left(\frac{i (2 n \pi -i \beta )}{\beta }\right)+\psi ^{(0)}\left(-\frac{i (2 \pi  n+i \beta )}{\beta }\right)+\ln (\beta^2/(4\pi^2n^2))}{4 \pi ^2 n^2} 
\end{equation}

Now we note that for $x\in \mathbb{R}$, $\psi ^{(0)}(1+i n x)+\psi ^{(0)}(1-i n x)-\ln(n^2x^2)$ has an asymptotic expansion in $1/(nx)$. Here $x=2\pi/\beta$. This procedure will give us the full asymptotic expansion, given by
\begin{equation}\label{asymp}
	\sum_n (-1)^n r(n,\beta) = \sum_{k=1}^{\infty} (-1)^{k+1}\frac{1}{2\pi^2} \left(\frac{\beta}{2\pi}\right)^{2k}\zeta(1-2k)\zeta(2k+2)\left(1-2^{-1-2k}\right)
\end{equation}

Pluggin in \eqref{asymp} in \eqref{oddhcal}, we derive the asymptotic expansion \eqref{oddh} presented in the main text:
\begin{align}
h^{d=3}(\beta) &\sim \frac{2\zeta(3)}{\beta^2} + \frac{\ln \beta}{12} -\frac{\gamma }{12} +\frac{\zeta '(2)}{2 \pi ^2}-\frac{\ln \pi}{12} \nonumber \\&+ \sum_{k=1}^{\infty} (-1)^{k+1}\frac{1}{\pi^2} \left(\frac{\beta}{2\pi}\right)^{2k}\zeta(1-2k)\zeta(2k+2)\left(1-2^{-1-2k}\right)\,.
\end{align}
We remark that we have verified this asymptotic expansion numerically.  Note that expansion coefficients of the $(\beta/2\pi^2)^{2k}$ for the above asymptotic expansion grow like $(2k)!$,  signaling the presence of non-perturbative terms of the form $e^{-4\pi^2 nk/\beta}$, via resurgence,  albeit with an imaginary coefficient. 
\subsection{$\ln Z$ for free scalar field theory in $d=4$ dimensions on $(\beta,\omega)$ plane}
Here we consider  a free scalar field theory in $4$ dimensions.  We set $\omega_1=\omega_2:=\omega$. 

\begin{equation}\label{fullform}
\begin{aligned}
\ln Z(\beta,\omega)&=\frac{1}{\left(1-\omega ^2\right)^2} \left(\frac{\pi ^4}{45 \beta ^3}-\frac{\pi ^2 \omega ^2}{18 \beta }-\frac{3 \zeta (3)}{16 \pi ^2}\left(1-6 \omega ^2-3 \omega ^4\right)+\frac{\beta }{720}  \left(3-6 \omega ^2-11 \omega ^4\right)\right)\\
&\qquad+ \sum_{n=1}^{\infty}\frac{(-1)^n}{16n\pi^2}\left[R_n(\omega)+R_n(-\omega)\right]\\
&\qquad+ \sum_{n,k\in\mathbb{Z}_+}  (-1)^n \left(f_{n,k}(-\omega)+\frac{g_{n,k}(-\omega)}{\beta}\right)\exp\left[-\frac{4\pi^2 nk}{\beta(1-\omega)}\right]\\
&\qquad+  \sum_{n,k\in\mathbb{Z}_+}(-1)^n \left(f_{n,k}(\omega)+\frac{g_{n,k}(\omega)}{\beta}\right) \exp\left[-\frac{4\pi^2 nk}{\beta(1+\omega)}\right]\,,
\end{aligned}
\end{equation}

where we have 
\begin{equation}
R_n(\omega):=-\psi ^{(1)}\left(\frac{n (1-\omega )}{2 (1+\omega )}+1\right)+\psi ^{(1)}\left(\frac{n (1-\omega )}{2 (1+\omega )}+\frac{1}{2}\right)\,,
\end{equation}

and the non-perturbative pieces are given by 
\begin{equation}
\begin{aligned}
f_{n,k}(\omega)&=\left[\frac{\csc \left(\pi n\frac{1-\omega}{1+\omega}\right)}{4 \pi  n^2}-\frac{\cot \left(\pi n\frac{1-\omega}{1+\omega}\right) \csc \left(\pi n \frac{ 1-\omega}{1+\omega}\right)}{4 n}\left(\frac{2\omega}{1+\omega}\right) \right]\,,\\
g_{n,k}(\omega)&=\frac{k}{(1+\omega)}\frac{\pi  \csc \left(\pi n\frac{1-\omega }{1+\omega }\right)}{n }\,.
\end{aligned}
\end{equation}

In the $\omega\to 1$ limit,  $\exp\left[-\frac{4\pi^2 nk}{\beta(1+\omega)}\right]$ becomes dominant over $\exp\left[-\frac{4\pi^2 nk}{\beta(1-\omega)}\right]$ and $f_{n,k}(\omega)$ becomes dominant over $g_{n,k}(\omega)$.  One can check that $f_{1,1}(\omega)$ changes sign at $\omega\approx 0.476858$,  as mentioned in the main text. Similarly if we take a derivative with respect to $\beta$, one can check that the analogous sign change occurs at $\omega=\frac13$.

We note that we have crosschecked \eqref{fullform} reproduces $h^{d=4}(\beta)$ in the $\omega\to 1$ limit.

It turns out that the $\omega\to 0$ limit is very rich.  Notice that there is an $1/\omega$ divergence in $f_{n,k}(\omega)$ in the $\omega\to 0$ limit: 
\begin{equation}
f_{n,k}(\omega)\underset{\omega\to 0}{\sim}-\frac{(-1)^n}{4 n^3 \pi^2}\left(\omega^{-1} + 1\right)\,,
\end{equation} 
However,  in this limit $\exp\left[-\frac{4\pi^2 nk}{\beta(1\pm\omega)}\right]$ saddles become comparable and combining their contribution,  we have 
\begin{equation}
\begin{aligned}
&(-1)^n\left[f_{n,k}(\omega) \exp\left(-\frac{4\pi^2 nk}{\beta(1+\omega)}\right)-f_{n,k}(-\omega) \exp\left(-\frac{4\pi^2 nk}{\beta(1-\omega)}\right)\right]\\
& \underset{\omega\to 0}{\sim}-\left[\frac{2}{n^2\beta}k +\frac{1}{2n^3\pi^2}\right]\exp\left(-\frac{4\pi^2 nk}{\beta}\right)
\end{aligned}
\end{equation}
A similar effect happens for $g_{n,k}(\omega)$ as well
\begin{equation}
\begin{aligned}
&(-1)^n \left(\frac{g_{n,k}(\omega)}{\beta} \exp\left[-\frac{4\pi^2 nk}{\beta(1+\omega)}\right]+\frac{g_{n,k}(-\omega)}{\beta} \exp\left[-\frac{4\pi^2 nk}{\beta(1-\omega)}\right]\right)\\
&\quad \sim -\frac{4 \pi ^2  }{n\beta^2}\  k^2 \exp\left(-\frac{4\pi^2 nk}{\beta}\right) 
\end{aligned}
\end{equation}

Now using the above limits,  we have verified that \eqref{fullform} reproduces  Eq. \!~C.20 of \cite{Benjamin:2023qsc} in the $\omega\to 0$ limit.  As we take $\omega\to 0$ limit to reach $\omega=0$ point,  an enhancement happens in terms of $\beta\to0$ singularity of the coefficient of the instanton saddles.  This is in fact important to reproduce the $\omega=0$ result.  

A similar collision of the saddle happens whenever $(1+\omega)/(1-\omega)$ is rational number.  These are precisely the points where two of the following four functions $f_{n,k}(\omega)$,  $f_{n,k}(-\omega)$,  $g_{n,k}(\omega)$, $g_{n,k}(-\omega)$ diverge.  Just like in the $\omega=0$ case,  the collision of the saddle cures the divergence,  leading to  an enhancement of $\beta\to 0$ singularity.  

To be explicit, we spell out the non-perturbative pieces for  $\omega=1/3$.
\begin{equation}\label{fullformNon}
\begin{aligned}
&\ln Z_{\text{non-pert}}(\beta,1/3)\\
=&\sum_{n,k\in\mathbb{Z}_+}  (-1)^{n+1}\left(\frac{27 \pi ^2}{8 n\beta ^2 } k^2+\frac{9 }{8n^2  \beta  } k+\frac{3}{16 \pi ^2 n^3}+\frac{1}{32 n}\right)\exp\left[-\frac{3\pi^2 (2n)k}{\beta}\right]\\
&\qquad+ \sum_{n,k\in\mathbb{Z}_+}(-1)^n \left(\frac{ 3 \pi }{4 \beta  (2 n-1)} k+\frac{1}{4 \pi  (2 n-1)^2}\right)\exp\left[-\frac{3\pi^2 (2n-1)k}{\beta}\right]
\end{aligned}
\end{equation}
In the above, the first line captures the collision of saddles and cures the apparent divergences in $f_{n,k}(-\omega)$ and $g_{n,k}(-\omega)$ near $\omega=1/3$.

The above discussion demonstrates that the singularity in the coefficient of the non-perturbative pieces in $\ln Z$ in the $\beta\to 0$ limit is sensitive to whether $\frac{1+\omega}{1-\omega}$ is rational or not. In particular if $\frac{1+\omega}{1-\omega} = \frac pq$, then the instantons $\exp{\left[-\frac{4\pi^2p}{1+\omega}\right]}$ and $\exp{\left[-\frac{4\pi^2q}{1-\omega}\right]}$ both have diverging ``one-loop" pieces multiplying them, but with opposite sign, so that their sum is finite. This renders the partition function, of course, completely continuous in $\omega$.

\subsubsection{A quick sketch of the derivation of \eqref{fullform}}
We define $\beta_L:=\beta(1+\omega)$ and $\beta_R:=\beta(1-\omega)$.    The logarithm of the partition function is given by 
\begin{equation}\label{definit}
\ln Z(\beta,\omega)= \sum_{q} \frac{1}{q} \left[W(q\beta_L,q\beta_R)+W(q\beta_R,q\beta_L)\right]\,,
\end{equation}
where $W$ is given by 
\begin{equation}
W(x,y):= \frac{1}{8}\mathrm{csch}\left(x/2\right)\mathrm{csch}^2\left(y/2\right)\cosh\left(y/2\right)
\end{equation}

We use the following identities
\begin{equation}
	\begin{aligned}
	\mathrm{csch}(x/2)&=\frac{2}{x}+x\sum_{n=1}^{\infty}\frac{(-1)^n}{x^2/4+\pi^2 n^2}\\
	\mathrm{csch}^2\left(y/2\right)\cosh\left(y/2\right)&=\frac{4}{y^2}+2\sum_{m=1}^{\infty}\frac{(-1)^m\left(y^2/4-\pi^2 m^2\right)}{\left(y^2/4+\pi^2 m^2\right)^2}\,,
	\end{aligned}
\end{equation}
and plug these in \eqref{definit}, exchange the sum over $m,n$ with sum over $q$ and perform the sum over $q$.  The computation is similar in nature to what we did to determine $h^{d=4}(\beta)$ earlier in this section.  The distinct feature here is that for few terms,  we have sum over $m,n$ and there can be apparent divergences.  To avoid these potential divergences,  we can choose $\omega$ to be irrational for the purpose of this computation.  Finally,  one can use the continuity of $\ln Z$ in the $\omega$ variable to determine it when $\omega $ is rational.

\bibliographystyle{JHEP}
\bibliography{biblio.bib}

@article{Minwalla:1997ka,
    author = "Minwalla, Shiraz",
    title = "{Restrictions imposed by superconformal invariance on quantum field theories}",
    eprint = "hep-th/9712074",
    archivePrefix = "arXiv",
    reportNumber = "PUPT-1748",
    doi = "10.4310/ATMP.1998.v2.n4.a4",
    journal = "Adv. Theor. Math. Phys.",
    volume = "2",
    pages = "783--851",
    year = "1998"
}

@article{Harlow:2021trr,
    author = "Harlow, Daniel and Ooguri, Hirosi",
    title = "{A universal formula for the density of states in theories with finite-group symmetry}",
    eprint = "2109.03838",
    archivePrefix = "arXiv",
    primaryClass = "hep-th",
    reportNumber = "MIT-CTP 5323, CALT-TH 2021-032, IPMU 21-0055",
    doi = "10.1088/1361-6382/ac5db2",
    journal = "Class. Quant. Grav.",
    volume = "39",
    number = "13",
    pages = "134003",
    year = "2022"
}

@article{Barrat:2025twb,
    author = "Barrat, Julien and Bozkurt, Deniz N. and Marchetto, Enrico and Miscioscia, Alessio and Pomoni, Elli",
    title = "{Analytic thermal bootstrap meets holography}",
    eprint = "2510.20894",
    archivePrefix = "arXiv",
    primaryClass = "hep-th",
    reportNumber = "DESY-25-139 , YITP-SB-2025-16",
    month = "10",
    year = "2025"
}

@article{Cassani:2021fyv,
    author = "Cassani, Davide and Komargodski, Zohar",
    title = "{EFT and the SUSY Index on the 2nd Sheet}",
    eprint = "2104.01464",
    archivePrefix = "arXiv",
    primaryClass = "hep-th",
    doi = "10.21468/SciPostPhys.11.1.004",
    journal = "SciPost Phys.",
    volume = "11",
    pages = "004",
    year = "2021"
}

@article{Romelsberger:2005eg,
    author = "Romelsberger, Christian",
    title = "{Counting chiral primaries in N = 1, d=4 superconformal field theories}",
    eprint = "hep-th/0510060",
    archivePrefix = "arXiv",
    doi = "10.1016/j.nuclphysb.2006.03.037",
    journal = "Nucl. Phys. B",
    volume = "747",
    pages = "329--353",
    year = "2006"
}

@article{Choi:2018hmj,
    author = "Choi, Sunjin and Kim, Joonho and Kim, Seok and Nahmgoong, June",
    title = "{Large AdS black holes from QFT}",
    eprint = "1810.12067",
    archivePrefix = "arXiv",
    primaryClass = "hep-th",
    reportNumber = "SNUTP18-005, KIAS-P18097",
    month = "10",
    year = "2018"
}

@article{Kim:2019yrz,
    author = "Kim, Joonho and Kim, Seok and Song, Jaewon",
    title = "{A 4d $ \mathcal{N} $ = 1 Cardy Formula}",
    eprint = "1904.03455",
    archivePrefix = "arXiv",
    primaryClass = "hep-th",
    reportNumber = "KIAS-P19015, SNUTP19-002",
    doi = "10.1007/JHEP01(2021)025",
    journal = "JHEP",
    volume = "01",
    pages = "025",
    year = "2021"
}

@article{Banihashemi:2024yye,
    author = "Banihashemi, Batoul and Shaghoulian, Edgar and Shashi, Sanjit",
    title = "{Flat space gravity at finite cutoff}",
    eprint = "2409.07643",
    archivePrefix = "arXiv",
    primaryClass = "hep-th",
    doi = "10.1088/1361-6382/ada2d7",
    journal = "Class. Quant. Grav.",
    volume = "42",
    number = "3",
    pages = "035010",
    year = "2025"
}

@article{Banihashemi:2025qqi,
    author = "Banihashemi, Batoul and Shaghoulian, Edgar and Shashi, Sanjit",
    title = "{Thermal effective actions from conformal boundary conditions in gravity}",
    eprint = "2503.17471",
    archivePrefix = "arXiv",
    primaryClass = "hep-th",
    doi = "10.1088/1361-6382/adee72",
    journal = "Class. Quant. Grav.",
    volume = "42",
    number = "15",
    pages = "155004",
    year = "2025"
}

@article{Iliesiu:2018fao,
    author = "Iliesiu, Luca and Kolo{\u{g}}lu, Murat and Mahajan, Raghu and Perlmutter, Eric and Simmons-Duffin, David",
    title = "{The Conformal Bootstrap at Finite Temperature}",
    eprint = "1802.10266",
    archivePrefix = "arXiv",
    primaryClass = "hep-th",
    reportNumber = "CALT-TH-2018-013, PUPT-2550",
    doi = "10.1007/JHEP10(2018)070",
    journal = "JHEP",
    volume = "10",
    pages = "070",
    year = "2018"
}

@article{Marchetto:2023xap,
    author = "Marchetto, Enrico and Miscioscia, Alessio and Pomoni, Elli",
    title = "{Sum rules {\&} Tauberian theorems at finite temperature}",
    eprint = "2312.13030",
    archivePrefix = "arXiv",
    primaryClass = "hep-th",
    reportNumber = "DESY-23-224",
    doi = "10.1007/JHEP09(2024)044",
    journal = "JHEP",
    volume = "09",
    pages = "044",
    year = "2024"
}

@article{Barrat:2025wbi,
    author = "Barrat, Julien and Marchetto, Enrico and Miscioscia, Alessio and Pomoni, Elli",
    title = "{Thermal Bootstrap for the Critical O(N) Model}",
    eprint = "2411.00978",
    archivePrefix = "arXiv",
    primaryClass = "hep-th",
    reportNumber = "DESY-24-167",
    doi = "10.1103/PhysRevLett.134.211604",
    journal = "Phys. Rev. Lett.",
    volume = "134",
    number = "21",
    pages = "211604",
    year = "2025"
}

@article{Barrat:2025nvu,
    author = "Barrat, Julien and Bozkurt, Deniz N. and Marchetto, Enrico and Miscioscia, Alessio and Pomoni, Elli",
    title = "{The analytic bootstrap at finite temperature}",
    eprint = "2506.06422",
    archivePrefix = "arXiv",
    primaryClass = "hep-th",
    reportNumber = "DESY-25-078",
    month = "6",
    year = "2025"
}

@article{Buric:2025anb,
    author = "Buri{\'c}, Ilija and Gusev, Ivan and Parnachev, Andrei",
    title = "{Thermal holographic correlators and KMS condition}",
    eprint = "2505.10277",
    archivePrefix = "arXiv",
    primaryClass = "hep-th",
    doi = "10.1007/JHEP09(2025)053",
    journal = "JHEP",
    volume = "09",
    pages = "053",
    year = "2025"
}

@article{Buric:2025fye,
    author = "Buri{\'c}, Ilija and Gusev, Ivan and Parnachev, Andrei",
    title = "{Holographic Correlators from Thermal Bootstrap}",
    eprint = "2508.08373",
    archivePrefix = "arXiv",
    primaryClass = "hep-th",
    month = "8",
    year = "2025"
}

@article{Buric:2025uqt,
    author = "Buri{\'c}, Ilija and Mangialardi, Francesco and Russo, Francesco and Schomerus, Volker and Vichi, Alessandro",
    title = "{Heavy-Heavy-Light Asymptotics from Thermal Correlators}",
    eprint = "2506.21671",
    archivePrefix = "arXiv",
    primaryClass = "hep-th",
    month = "6",
    year = "2025"
}

@article{Pal:2020wwd,
    author = "Pal, Sridip and Sun, Zhengdi",
    title = "{High Energy Modular Bootstrap, Global Symmetries and Defects}",
    eprint = "2004.12557",
    archivePrefix = "arXiv",
    primaryClass = "hep-th",
    doi = "10.1007/JHEP08(2020)064",
    journal = "JHEP",
    volume = "08",
    pages = "064",
    year = "2020"
}

@article{Kang:2022orq,
    author = "Kang, Monica Jinwoo and Lee, Jaeha and Ooguri, Hirosi",
    title = "{Universal formula for the density of states with continuous symmetry}",
    eprint = "2206.14814",
    archivePrefix = "arXiv",
    primaryClass = "hep-th",
    reportNumber = "CALT-TH-2022-020, IPMU 22-0026, CALT-TH-2022-020; IPMU 22-0026;",
    doi = "10.1103/PhysRevD.107.026021",
    journal = "Phys. Rev. D",
    volume = "107",
    number = "2",
    pages = "026021",
    year = "2023"
}

@article{Choi:2025lck,
    author = "Choi, Sunjin and Jain, Diksha and Kim, Seok and Krishna, Vineeth and Kwon, Goojin and Lee, Eunwoo and Minwalla, Shiraz and Patel, Chintan",
    title = "{Supersymmetric Grey Galaxies, Dual Dressed Black Holes and the Superconformal Index}",
    eprint = "2501.17217",
    archivePrefix = "arXiv",
    primaryClass = "hep-th",
    reportNumber = "TIFR/TH/25-3, LCTP-25-02",
    month = "1",
    year = "2025"
}

@article{Simmons-Duffin:2025qox,
    author = "Simmons-Duffin, David and Xu, Yixin",
    title = "{A genus-2 crossing equation in $d\geq 2$}",
    eprint = "2511.07569",
    archivePrefix = "arXiv",
    primaryClass = "hep-th",
    month = "11",
    year = "2025"
}

@article{Delacretaz:2020nit,
    author = "Delacretaz, Luca V.",
    title = "{Heavy Operators and Hydrodynamic Tails}",
    eprint = "2006.01139",
    archivePrefix = "arXiv",
    primaryClass = "hep-th",
    reportNumber = "EFI 20-10",
    doi = "10.21468/SciPostPhys.9.3.034",
    journal = "SciPost Phys.",
    volume = "9",
    number = "3",
    pages = "034",
    year = "2020"
}

@article{Allameh:2024qqp,
    author = "Allameh, Kuroush and Shaghoulian, Edgar",
    title = "{Modular invariance and thermal effective field theory in CFT}",
    eprint = "2402.13337",
    archivePrefix = "arXiv",
    primaryClass = "hep-th",
    doi = "10.1007/JHEP01(2025)200",
    journal = "JHEP",
    volume = "01",
    pages = "200",
    year = "2025"
}

@article{Jensen:2012jh,
    author = "Jensen, Kristan and Kaminski, Matthias and Kovtun, Pavel and Meyer, Rene and Ritz, Adam and Yarom, Amos",
    title = "{Towards hydrodynamics without an entropy current}",
    eprint = "1203.3556",
    archivePrefix = "arXiv",
    primaryClass = "hep-th",
    reportNumber = "CCTP-2012-03",
    doi = "10.1103/PhysRevLett.109.101601",
    journal = "Phys. Rev. Lett.",
    volume = "109",
    pages = "101601",
    year = "2012"
}

@article{Collier:2016cls,
    author = "Collier, Scott and Lin, Ying-Hsuan and Yin, Xi",
    title = "{Modular Bootstrap Revisited}",
    eprint = "1608.06241",
    archivePrefix = "arXiv",
    primaryClass = "hep-th",
    doi = "10.1007/JHEP09(2018)061",
    journal = "JHEP",
    volume = "09",
    pages = "061",
    year = "2018"
}

@article{Diatlyk:2024qpr,
    author = "Diatlyk, Oleksandr and Khanchandani, Himanshu and Popov, Fedor K. and Wang, Yifan",
    title = "{Effective Field Theory of Conformal Boundaries}",
    eprint = "2406.01550",
    archivePrefix = "arXiv",
    primaryClass = "hep-th",
    doi = "10.1103/PhysRevLett.133.261601",
    journal = "Phys. Rev. Lett.",
    volume = "133",
    number = "26",
    pages = "261601",
    year = "2024"
}

@article{Kusuki:2018wpa,
    author = "Kusuki, Yuya",
    title = "{Light Cone Bootstrap in General 2D CFTs and Entanglement from Light Cone Singularity}",
    eprint = "1810.01335",
    archivePrefix = "arXiv",
    primaryClass = "hep-th",
    reportNumber = "YITP-18-106",
    doi = "10.1007/JHEP01(2019)025",
    journal = "JHEP",
    volume = "01",
    pages = "025",
    year = "2019"
}

@article{Kim:2023sig,
    author = "Kim, Seok and Kundu, Suman and Lee, Eunwoo and Lee, Jaeha and Minwalla, Shiraz and Patel, Chintan",
    title = "{Grey Galaxies\textquoteright{} as an endpoint of the Kerr-AdS superradiant instability}",
    eprint = "2305.08922",
    archivePrefix = "arXiv",
    primaryClass = "hep-th",
    doi = "10.1007/JHEP11(2023)024",
    journal = "JHEP",
    volume = "11",
    pages = "024",
    year = "2023"
}

@article{Hawking:1998kw,
    author = "Hawking, S. W. and Hunter, C. J. and Taylor, Marika",
    title = "{Rotation and the AdS / CFT correspondence}",
    eprint = "hep-th/9811056",
    archivePrefix = "arXiv",
    doi = "10.1103/PhysRevD.59.064005",
    journal = "Phys. Rev. D",
    volume = "59",
    pages = "064005",
    year = "1999"
}

@article{Caldarelli:1999xj,
    author = "Caldarelli, Marco M. and Cognola, Guido and Klemm, Dietmar",
    title = "{Thermodynamics of Kerr-Newman-AdS black holes and conformal field theories}",
    eprint = "hep-th/9908022",
    archivePrefix = "arXiv",
    reportNumber = "UTF-434",
    doi = "10.1088/0264-9381/17/2/310",
    journal = "Class. Quant. Grav.",
    volume = "17",
    pages = "399--420",
    year = "2000"
}

@article{Gibbons:2004ai,
    author = "Gibbons, G. W. and Perry, M. J. and Pope, C. N.",
    title = "{The First law of thermodynamics for Kerr-anti-de Sitter black holes}",
    eprint = "hep-th/0408217",
    archivePrefix = "arXiv",
    reportNumber = "DAMTP-2004-87, MIFP-04-17",
    doi = "10.1088/0264-9381/22/9/002",
    journal = "Class. Quant. Grav.",
    volume = "22",
    pages = "1503--1526",
    year = "2005"
}

@article{Bhattacharyya_2008,
    author = "Bhattacharyya, Sayantani and Lahiri, Subhaneil and Loganayagam, R. and Minwalla, Shiraz",
    title = "{Large rotating AdS black holes from fluid mechanics}",
    eprint = "0708.1770",
    archivePrefix = "arXiv",
    primaryClass = "hep-th",
    doi = "10.1088/1126-6708/2008/09/054",
    journal = "JHEP",
    volume = "09",
    pages = "054",
    year = "2008"
}

@article{Kinney:2005ej,
    author = "Kinney, Justin and Maldacena, Juan Martin and Minwalla, Shiraz and Raju, Suvrat",
    title = "{An Index for 4 dimensional super conformal theories}",
    eprint = "hep-th/0510251",
    archivePrefix = "arXiv",
    doi = "10.1007/s00220-007-0258-7",
    journal = "Commun. Math. Phys.",
    volume = "275",
    pages = "209--254",
    year = "2007"
}

@article{Choi:2024xnv,
    author = "Choi, Sunjin and Jain, Diksha and Kim, Seok and Krishna, Vineeth and Lee, Eunwoo and Minwalla, Shiraz and Patel, Chintan",
    title = "{Dual dressed black holes as the end point of the charged superradiant instability in $\mathcal{N} = 4$ Yang Mills}",
    eprint = "2409.18178",
    archivePrefix = "arXiv",
    primaryClass = "hep-th",
    reportNumber = "TIFR/TH/24-19, LCTP-24-17",
    doi = "10.21468/SciPostPhys.18.4.137",
    journal = "SciPost Phys.",
    volume = "18",
    number = "4",
    pages = "137",
    year = "2025"
}

@article{Mack:1975je,
    author = "Mack, G.",
    title = "{All unitary ray representations of the conformal group SU(2,2) with positive energy}",
    reportNumber = "DESY-75-50",
    doi = "10.1007/BF01613145",
    journal = "Commun. Math. Phys.",
    volume = "55",
    pages = "1",
    year = "1977"
}

@article{Banerjee:2012iz,
    author = "Banerjee, Nabamita and Bhattacharya, Jyotirmoy and Bhattacharyya, Sayantani and Jain, Sachin and Minwalla, Shiraz and Sharma, Tarun",
    title = "{Constraints on Fluid Dynamics from Equilibrium Partition Functions}",
    eprint = "1203.3544",
    archivePrefix = "arXiv",
    primaryClass = "hep-th",
    reportNumber = "TFR-TH-12-05, IPMU12-0037",
    doi = "10.1007/JHEP09(2012)046",
    journal = "JHEP",
    volume = "09",
    pages = "046",
    year = "2012"
}

@article{Shaghoulian:2015lcn,
    author = "Shaghoulian, Edgar",
    title = "{Black hole microstates in AdS}",
    eprint = "1512.06855",
    archivePrefix = "arXiv",
    primaryClass = "hep-th",
    doi = "10.1103/PhysRevD.94.104044",
    journal = "Phys. Rev. D",
    volume = "94",
    number = "10",
    pages = "104044",
    year = "2016"
}

@article{Benjamin:2023qsc,
    author = "Benjamin, Nathan and Lee, Jaeha and Ooguri, Hirosi and Simmons-Duffin, David",
    title = "{Universal asymptotics for high energy CFT data}",
    eprint = "2306.08031",
    archivePrefix = "arXiv",
    primaryClass = "hep-th",
    reportNumber = "CALT-TH 2023-014, IPMU 23-0020",
    doi = "10.1007/JHEP03(2024)115",
    journal = "JHEP",
    volume = "03",
    pages = "115",
    year = "2024"
}

@article{Bajaj:2024utv,
    author = "Bajaj, Kabir and Kumar, Vipul and Minwalla, Shiraz and Mukherjee, Jyotirmoy and Rahaman, Asikur",
    title = "{Grey Galaxies in $AdS_5$}",
    eprint = "2412.06904",
    archivePrefix = "arXiv",
    primaryClass = "hep-th",
    reportNumber = "TIFR/TH/24-25",
    month = "12",
    year = "2024"
}

@article{Aharony:2003sx,
    author = "Aharony, Ofer and Marsano, Joseph and Minwalla, Shiraz and Papadodimas, Kyriakos and Van Raamsdonk, Mark",
    editor = "Doebner, H. D. and Dobrev, V. K.",
    title = "{The Hagedorn - deconfinement phase transition in weakly coupled large N gauge theories}",
    eprint = "hep-th/0310285",
    archivePrefix = "arXiv",
    reportNumber = "WIS-29-03-DPP",
    doi = "10.4310/ATMP.2004.v8.n4.a1",
    journal = "Adv. Theor. Math. Phys.",
    volume = "8",
    pages = "603--696",
    year = "2004"
}

@article{Bhattacharyya:2007vs,
    author = "Bhattacharyya, Sayantani and Lahiri, Subhaneil and Loganayagam, R. and Minwalla, Shiraz",
    title = "{Large rotating AdS black holes from fluid mechanics}",
    eprint = "0708.1770",
    archivePrefix = "arXiv",
    primaryClass = "hep-th",
    doi = "10.1088/1126-6708/2008/09/054",
    journal = "JHEP",
    volume = "09",
    pages = "054",
    year = "2008"
}

@article{Lahiri:2007ae,
    author = "Lahiri, Subhaneil and Minwalla, Shiraz",
    title = "{Plasmarings as dual black rings}",
    eprint = "0705.3404",
    archivePrefix = "arXiv",
    primaryClass = "hep-th",
    doi = "10.1088/1126-6708/2008/05/001",
    journal = "JHEP",
    volume = "05",
    pages = "001",
    year = "2008"
}

@article{Aharony:2005bm,
    author = "Aharony, Ofer and Minwalla, Shiraz and Wiseman, Toby",
    title = "{Plasma-balls in large N gauge theories and localized black holes}",
    eprint = "hep-th/0507219",
    archivePrefix = "arXiv",
    reportNumber = "WIS-18-05-JUL-DPP, HUTP-05-A0035",
    doi = "10.1088/0264-9381/23/7/001",
    journal = "Class. Quant. Grav.",
    volume = "23",
    pages = "2171--2210",
    year = "2006"
}

@article{Bhattacharyya:2007vjd,
    author = "Bhattacharyya, Sayantani and Hubeny, Veronika E and Minwalla, Shiraz and Rangamani, Mukund",
    title = "{Nonlinear Fluid Dynamics from Gravity}",
    eprint = "0712.2456",
    archivePrefix = "arXiv",
    primaryClass = "hep-th",
    reportNumber = "TIFR-TH-07-44, DCPT-07-73, NI07097",
    doi = "10.1088/1126-6708/2008/02/045",
    journal = "JHEP",
    volume = "02",
    pages = "045",
    year = "2008"
}

@article{Benjamin:2024kdg,
    author = "Benjamin, Nathan and Lee, Jaeha and Pal, Sridip and Simmons-Duffin, David and Xu, Yixin",
    title = "{Angular fractals in thermal QFT}",
    eprint = "2405.17562",
    archivePrefix = "arXiv",
    primaryClass = "hep-th",
    reportNumber = "CALT-TH 2024-021",
    doi = "10.1007/JHEP11(2024)134",
    journal = "JHEP",
    volume = "11",
    pages = "134",
    year = "2024"
}

@article{Pal:2022vqc,
    author = "Pal, Sridip and Qiao, Jiaxin and Rychkov, Slava",
    title = "{Twist Accumulation in Conformal Field Theory: A Rigorous Approach to the Lightcone Bootstrap}",
    eprint = "2212.04893",
    archivePrefix = "arXiv",
    primaryClass = "hep-th",
    doi = "10.1007/s00220-023-04767-w",
    journal = "Commun. Math. Phys.",
    volume = "402",
    number = "3",
    pages = "2169--2214",
    year = "2023"
}

@article{Pal:2023cgk,
    author = "Pal, Sridip and Qiao, Jiaxin",
    title = "{Lightcone Modular Bootstrap and Tauberian Theory: A Cardy-Like Formula for Near-Extremal Black Holes}",
    eprint = "2307.02587",
    archivePrefix = "arXiv",
    primaryClass = "hep-th",
    doi = "10.1007/s00023-024-01441-2",
    journal = "Annales Henri Poincare",
    volume = "26",
    number = "3",
    pages = "787--844",
    year = "2025"
}

@article{Kusuki:2025pgx,
    author = "Kusuki, Yuya and Ooguri, Hirosi and Pal, Sridip",
    title = "{Universality of R{\'e}nyi Entropy in Conformal Field Theory}",
    eprint = "2503.24353",
    archivePrefix = "arXiv",
    primaryClass = "hep-th",
    reportNumber = "CALT 2025-008, IPMU 25-0015, KYUSHU-HET-314, RIKEN-iTHEMS-Report-25",
    doi = "10.1103/fsg7-bs7q",
    journal = "Phys. Rev. Lett.",
    volume = "135",
    number = "6",
    pages = "061603",
    year = "2025"
}

@article{Pal:2025yvz,
    author = "Pal, Sridip and Qiao, Jiaxin and van Rees, Balt C.",
    title = "{Universality of the Microcanonical Entropy at Large Spin}",
    eprint = "2505.02897",
    archivePrefix = "arXiv",
    primaryClass = "hep-th",
    doi = "10.1007/s00220-025-05442-y",
    journal = "Commun. Math. Phys.",
    volume = "406",
    number = "12",
    pages = "302",
    year = "2025"
}

@article{Benjamin:2019stq,
    author = "Benjamin, Nathan and Ooguri, Hirosi and Shao, Shu-Heng and Wang, Yifan",
    title = "{Light-cone modular bootstrap and pure gravity}",
    eprint = "1906.04184",
    archivePrefix = "arXiv",
    primaryClass = "hep-th",
    reportNumber = "CALT-TH 2019-020, IPMU19-0086, PUPT-2586",
    doi = "10.1103/PhysRevD.100.066029",
    journal = "Phys. Rev. D",
    volume = "100",
    number = "6",
    pages = "066029",
    year = "2019"
}

@article{Ghosh:2019rcj,
    author = "Ghosh, Animik and Maxfield, Henry and Turiaci, Gustavo J.",
    title = "{A universal Schwarzian sector in two-dimensional conformal field theories}",
    eprint = "1912.07654",
    archivePrefix = "arXiv",
    primaryClass = "hep-th",
    doi = "10.1007/JHEP05(2020)104",
    journal = "JHEP",
    volume = "05",
    pages = "104",
    year = "2020"
}

@article{Hartman:2014oaa,
    author = "Hartman, Thomas and Keller, Christoph A. and Stoica, Bogdan",
    title = "{Universal Spectrum of 2d Conformal Field Theory in the Large c Limit}",
    eprint = "1405.5137",
    archivePrefix = "arXiv",
    primaryClass = "hep-th",
    reportNumber = "CALT-68-2889, RUNHETC-2014-07",
    doi = "10.1007/JHEP09(2014)118",
    journal = "JHEP",
    volume = "09",
    pages = "118",
    year = "2014"
}

@article{Dey:2024nje,
    author = "Dey, Indranil and Pal, Sridip and Qiao, Jiaxin",
    title = "{A universal inequality on the unitary 2D CFT partition function}",
    eprint = "2410.18174",
    archivePrefix = "arXiv",
    primaryClass = "hep-th",
    reportNumber = "CALT-TH 2024-039",
    doi = "10.1007/JHEP07(2025)163",
    journal = "JHEP",
    volume = "07",
    pages = "163",
    year = "2025"
}

@article{Gubser:1998nz,
    author = "Gubser, Steven S. and Klebanov, Igor R. and Tseytlin, Arkady A.",
    title = "{Coupling constant dependence in the thermodynamics of N=4 supersymmetric Yang-Mills theory}",
    eprint = "hep-th/9805156",
    archivePrefix = "arXiv",
    reportNumber = "IASSNS-HEP-98-47, PUPT-1794, IMPERIAL-TP-97-98-47",
    doi = "10.1016/S0550-3213(98)00514-8",
    journal = "Nucl. Phys. B",
    volume = "534",
    pages = "202--222",
    year = "1998"
}

@article{Gubser:1996de,
    author = "Gubser, S. S. and Klebanov, Igor R. and Peet, A. W.",
    title = "{Entropy and temperature of black 3-branes}",
    eprint = "hep-th/9602135",
    archivePrefix = "arXiv",
    reportNumber = "PUPT-1598",
    doi = "10.1103/PhysRevD.54.3915",
    journal = "Phys. Rev. D",
    volume = "54",
    pages = "3915--3919",
    year = "1996"
}

@article{Kravchuk:2024qoh,
    author = "Kravchuk, Petr and Radcliffe, Alex and Sinha, Ritam",
    title = "{Effective theory for fusion of conformal defects}",
    eprint = "2406.04561",
    archivePrefix = "arXiv",
    primaryClass = "hep-th",
    doi = "10.1088/1751-8121/ae14c5",
    journal = "J. Phys. A",
    volume = "58",
    number = "46",
    pages = "465402",
    year = "2025"
}

@article{Komargodski:2012ek,
    author = "Komargodski, Zohar and Zhiboedov, Alexander",
    title = "{Convexity and Liberation at Large Spin}",
    eprint = "1212.4103",
    archivePrefix = "arXiv",
    primaryClass = "hep-th",
    doi = "10.1007/JHEP11(2013)140",
    journal = "JHEP",
    volume = "11",
    pages = "140",
    year = "2013"
}

@article{Luo:2022tqy,
    author = "Luo, Conghuan and Wang, Yifan",
    title = "{Casimir energy and modularity in higher-dimensional conformal field theories}",
    eprint = "2212.14866",
    archivePrefix = "arXiv",
    primaryClass = "hep-th",
    doi = "10.1007/JHEP07(2023)028",
    journal = "JHEP",
    volume = "07",
    pages = "028",
    year = "2023"
}

@article{Bhattacharyya:2010yg,
    author = "Bhattacharyya, Sayantani and Minwalla, Shiraz and Papadodimas, Kyriakos",
    title = "{Small Hairy Black Holes in $AdS_5 x S^5$}",
    eprint = "1005.1287",
    archivePrefix = "arXiv",
    primaryClass = "hep-th",
    reportNumber = "TIFR-TH-ITFA-10-13",
    doi = "10.1007/JHEP11(2011)035",
    journal = "JHEP",
    volume = "11",
    pages = "035",
    year = "2011"
}

@article{Fitzpatrick:2012yx,
    author = "Fitzpatrick, A. Liam and Kaplan, Jared and Poland, David and Simmons-Duffin, David",
    title = "{The Analytic Bootstrap and AdS Superhorizon Locality}",
    eprint = "1212.3616",
    archivePrefix = "arXiv",
    primaryClass = "hep-th",
    doi = "10.1007/JHEP12(2013)004",
    journal = "JHEP",
    volume = "12",
    pages = "004",
    year = "2013"
}

@article{vanRees:2024xkb,
    author = "van Rees, Balt C.",
    title = "{Theorems for the lightcone bootstrap}",
    eprint = "2412.06907",
    archivePrefix = "arXiv",
    primaryClass = "hep-th",
    doi = "10.21468/SciPostPhys.18.6.207",
    journal = "SciPost Phys.",
    volume = "18",
    number = "6",
    pages = "207",
    year = "2025"
}

@article{Benjamin:2020swg,
    author = "Benjamin, Nathan and Ooguri, Hirosi and Shao, Shu-Heng and Wang, Yifan",
    title = "{Twist gap and global symmetry in two dimensions}",
    eprint = "2003.02844",
    archivePrefix = "arXiv",
    primaryClass = "hep-th",
    reportNumber = "CALT-TH 2020-002, IPMU20-0023, PUPT-2609",
    doi = "10.1103/PhysRevD.101.106026",
    journal = "Phys. Rev. D",
    volume = "101",
    number = "10",
    pages = "106026",
    year = "2020"
}

@article{Collier:2018exn,
    author = "Collier, Scott and Gobeil, Yan and Maxfield, Henry and Perlmutter, Eric",
    title = "{Quantum Regge Trajectories and the Virasoro Analytic Bootstrap}",
    eprint = "1811.05710",
    archivePrefix = "arXiv",
    primaryClass = "hep-th",
    doi = "10.1007/JHEP05(2019)212",
    journal = "JHEP",
    volume = "05",
    pages = "212",
    year = "2019"
}

@article{Melia:2020pzd,
    author = "Melia, Tom and Pal, Sridip",
    title = "{EFT Asymptotics: the Growth of Operator Degeneracy}",
    eprint = "2010.08560",
    archivePrefix = "arXiv",
    primaryClass = "hep-th",
    doi = "10.21468/SciPostPhys.10.5.104",
    journal = "SciPost Phys.",
    volume = "10",
    number = "5",
    pages = "104",
    year = "2021"
}

@inbook{Andrews_1984, place={Cambridge}, series={Encyclopedia of Mathematics and its Applications}, title={The Asymptotics of Infinite Product Generating Functions}, booktitle={The Theory of Partitions}, publisher={Cambridge University Press}, author={Andrews, George E.}, year={1984}, pages={88–102}, collection={Encyclopedia of Mathematics and its Applications}}

@article{Maldacena:2024spf,
    author = "Maldacena, Juan",
    title = "{Real observers solving imaginary problems}",
    eprint = "2412.14014",
    archivePrefix = "arXiv",
    primaryClass = "hep-th",
    month = "12",
    year = "2024"
}

@article{Halder:2019foo,
    author = "Halder, Indranil and Minwalla, Shiraz",
    title = "{Matter Chern Simons Theories in a Background Magnetic Field}",
    eprint = "1904.07885",
    archivePrefix = "arXiv",
    primaryClass = "hep-th",
    doi = "10.1007/JHEP11(2019)089",
    journal = "JHEP",
    volume = "11",
    pages = "089",
    year = "2019"
}

@article{Dey:2019ihe,
    author = "Dey, Anshuman and Halder, Indranil and Jain, Sachin and Minwalla, Shiraz and Prabhakar, Naveen",
    title = "{The large N phase diagram of $ \mathcal{N} $ = 2 SU(N) Chern-Simons theory with one fundamental chiral multiplet}",
    eprint = "1904.07286",
    archivePrefix = "arXiv",
    primaryClass = "hep-th",
    reportNumber = "TIFR/TH/19-10",
    doi = "10.1007/JHEP11(2019)113",
    journal = "JHEP",
    volume = "11",
    pages = "113",
    year = "2019"
}

@article{Dey:2018ykx,
    author = "Dey, Anshuman and Halder, Indranil and Jain, Sachin and Janagal, Lavneet and Minwalla, Shiraz and Prabhakar, Naveen",
    title = "{Duality and an exact Landau-Ginzburg potential for quasi-bosonic Chern-Simons-Matter theories}",
    eprint = "1808.04415",
    archivePrefix = "arXiv",
    primaryClass = "hep-th",
    reportNumber = "TIFR/TH/18-26",
    doi = "10.1007/JHEP11(2018)020",
    journal = "JHEP",
    volume = "11",
    pages = "020",
    year = "2018"
}

@article{Aharony:2018pjn,
    author = "Aharony, Ofer and Jain, Sachin and Minwalla, Shiraz",
    title = "{Flows, Fixed Points and Duality in Chern-Simons-matter theories}",
    eprint = "1808.03317",
    archivePrefix = "arXiv",
    primaryClass = "hep-th",
    doi = "10.1007/JHEP12(2018)058",
    journal = "JHEP",
    volume = "12",
    pages = "058",
    year = "2018"
}

@article{Choudhury:2018iwf,
    author = "Choudhury, Sayantan and Dey, Anshuman and Halder, Indranil and Jain, Sachin and Janagal, Lavneet and Minwalla, Shiraz and Prabhakar, Naveen",
    title = "{Bose-Fermi Chern-Simons Dualities in the Higgsed Phase}",
    eprint = "1804.08635",
    archivePrefix = "arXiv",
    primaryClass = "hep-th",
    reportNumber = "TIFR/TH/18-10, TIFR-TH-18-10",
    doi = "10.1007/JHEP11(2018)177",
    journal = "JHEP",
    volume = "11",
    pages = "177",
    year = "2018"
}

@article{Minwalla:2015sca,
    author = "Minwalla, Shiraz and Yokoyama, Shuichi",
    title = "{Chern Simons Bosonization along RG Flows}",
    eprint = "1507.04546",
    archivePrefix = "arXiv",
    primaryClass = "hep-th",
    reportNumber = "TIFR-TH-15-19",
    doi = "10.1007/JHEP02(2016)103",
    journal = "JHEP",
    volume = "02",
    pages = "103",
    year = "2016"
}

@article{Jain:2014nza,
    author = "Jain, Sachin and Mandlik, Mangesh and Minwalla, Shiraz and Takimi, Tomohisa and Wadia, Spenta R. and Yokoyama, Shuichi",
    title = "{Unitarity, Crossing Symmetry and Duality of the S-matrix in large N Chern-Simons theories with fundamental matter}",
    eprint = "1404.6373",
    archivePrefix = "arXiv",
    primaryClass = "hep-th",
    reportNumber = "TIFR-TH-14-12, HRI-ST-1405, ICTS-2014-04",
    doi = "10.1007/JHEP04(2015)129",
    journal = "JHEP",
    volume = "04",
    pages = "129",
    year = "2015"
}

@article{Jain:2013gza,
    author = "Jain, Sachin and Minwalla, Shiraz and Yokoyama, Shuichi",
    title = "{Chern Simons duality with a fundamental boson and fermion}",
    eprint = "1305.7235",
    archivePrefix = "arXiv",
    primaryClass = "hep-th",
    reportNumber = "TIFR-TH-13-17",
    doi = "10.1007/JHEP11(2013)037",
    journal = "JHEP",
    volume = "11",
    pages = "037",
    year = "2013"
}

@article{Jain:2013py,
    author = "Jain, Sachin and Minwalla, Shiraz and Sharma, Tarun and Takimi, Tomohisa and Wadia, Spenta R. and Yokoyama, Shuichi",
    title = "{Phases of large $N$ vector Chern-Simons theories on $S^2 \times S^1$}",
    eprint = "1301.6169",
    archivePrefix = "arXiv",
    primaryClass = "hep-th",
    reportNumber = "TIFR-TH-13-02, ICTS-2012-14",
    doi = "10.1007/JHEP09(2013)009",
    journal = "JHEP",
    volume = "09",
    pages = "009",
    year = "2013"
}

@article{Giombi:2011kc,
    author = "Giombi, Simone and Minwalla, Shiraz and Prakash, Shiroman and Trivedi, Sandip P. and Wadia, Spenta R. and Yin, Xi",
    title = "{Chern-Simons Theory with Vector Fermion Matter}",
    eprint = "1110.4386",
    archivePrefix = "arXiv",
    primaryClass = "hep-th",
    doi = "10.1140/epjc/s10052-012-2112-0",
    journal = "Eur. Phys. J. C",
    volume = "72",
    pages = "2112",
    year = "2012"
}

@article{Bhattacharyya:2007sa,
    author = "Bhattacharyya, Sayantani and Minwalla, Shiraz",
    title = "{Supersymmetric states in M5/M2 CFTs}",
    eprint = "hep-th/0702069",
    archivePrefix = "arXiv",
    doi = "10.1088/1126-6708/2007/12/004",
    journal = "JHEP",
    volume = "12",
    pages = "004",
    year = "2007"
}

@article{Fliss:2023muk,
    author = "Fliss, Jackson",
    title = "{A 3d perspective on de Sitter quantum field theory}",
    doi = "10.22323/1.436.0129",
    journal = "PoS",
    volume = "CORFU2022",
    pages = "129",
    year = "2023"
}

@article{Anninos:2020hfj,
    author = "Anninos, Dionysios and Denef, Frederik and Law, Y. T. Albert and Sun, Zimo",
    title = "{Quantum de Sitter horizon entropy from quasicanonical bulk, edge, sphere and topological string partition functions}",
    eprint = "2009.12464",
    archivePrefix = "arXiv",
    primaryClass = "hep-th",
    doi = "10.1007/JHEP01(2022)088",
    journal = "JHEP",
    volume = "01",
    pages = "088",
    year = "2022"
}

@article{David:2024pir,
    author = "David, Justin R. and Kumar, Srijan",
    title = "{The large N vector model on S$^{1}$ {\texttimes} S$^{2}$}",
    eprint = "2411.18509",
    archivePrefix = "arXiv",
    primaryClass = "hep-th",
    doi = "10.1007/JHEP03(2025)169",
    journal = "JHEP",
    volume = "03",
    pages = "169",
    year = "2025"
}

@article{Hellerman:2015nra,
    author = "Hellerman, Simeon and Orlando, Domenico and Reffert, Susanne and Watanabe, Masataka",
    title = "{On the CFT Operator Spectrum at Large Global Charge}",
    eprint = "1505.01537",
    archivePrefix = "arXiv",
    primaryClass = "hep-th",
    doi = "10.1007/JHEP12(2015)071",
    journal = "JHEP",
    volume = "12",
    pages = "071",
    year = "2015"
}

@article{Lee:2025qim,
    author = "Lee, Eunwoo",
    title = "{Extremal AdS Black Holes as Fluids: A Matrix Large-Charge EFT Approach}",
    eprint = "2507.21240",
    archivePrefix = "arXiv",
    primaryClass = "hep-th",
    month = "7",
    year = "2025"
}

@article{Choi:2025tql,
    author = "Choi, Jaehyeok and Lee, Eunwoo",
    title = "{Large charge operators at large spin from relativistically rotating vortices}",
    eprint = "2501.07198",
    archivePrefix = "arXiv",
    primaryClass = "hep-th",
    doi = "10.1007/JHEP05(2025)087",
    journal = "JHEP",
    volume = "05",
    pages = "087",
    year = "2025"
}

@article{Buric:2024kxo,
    author = "Buric, Ilija and Russo, Francesco and Schomerus, Volker and Vichi, Alessandro",
    title = "{Thermal one-point functions and their partial wave decomposition}",
    eprint = "2408.02747",
    archivePrefix = "arXiv",
    primaryClass = "hep-th",
    doi = "10.1007/JHEP12(2024)021",
    journal = "JHEP",
    volume = "12",
    pages = "021",
    year = "2024"
}

@article{Cuomo:2022kio,
    author = "Cuomo, Gabriel and Komargodski, Zohar",
    title = "{Giant Vortices and the Regge Limit}",
    eprint = "2210.15694",
    archivePrefix = "arXiv",
    primaryClass = "hep-th",
    doi = "10.1007/JHEP01(2023)006",
    journal = "JHEP",
    volume = "01",
    pages = "006",
    year = "2023"
}

@article{DiPietro:2014bca,
    author = "Di Pietro, Lorenzo and Komargodski, Zohar",
    title = "{Cardy formulae for SUSY theories in $d =$ 4 and $d =$ 6}",
    eprint = "1407.6061",
    archivePrefix = "arXiv",
    primaryClass = "hep-th",
    doi = "10.1007/JHEP12(2014)031",
    journal = "JHEP",
    volume = "12",
    pages = "031",
    year = "2014"
}

@article{Haehl:2013hoa,
    author = "Haehl, Felix M. and Loganayagam, R. and Rangamani, Mukund",
    title = "{Effective actions for anomalous hydrodynamics}",
    eprint = "1312.0610",
    archivePrefix = "arXiv",
    primaryClass = "hep-th",
    reportNumber = "DCPT-13-41",
    doi = "10.1007/JHEP03(2014)034",
    journal = "JHEP",
    volume = "03",
    pages = "034",
    year = "2014"
}

@article{Haehl:2015pja,
    author = "Haehl, Felix M. and Loganayagam, R. and Rangamani, Mukund",
    title = "{Adiabatic hydrodynamics: The eightfold way to dissipation}",
    eprint = "1502.00636",
    archivePrefix = "arXiv",
    primaryClass = "hep-th",
    reportNumber = "DCPT-15-01",
    doi = "10.1007/JHEP05(2015)060",
    journal = "JHEP",
    volume = "05",
    pages = "060",
    year = "2015"
}

@article{Haehl:2015foa,
    author = "Haehl, Felix M. and Loganayagam, R. and Rangamani, Mukund",
    title = "{The Fluid Manifesto: Emergent symmetries, hydrodynamics, and black holes}",
    eprint = "1510.02494",
    archivePrefix = "arXiv",
    primaryClass = "hep-th",
    doi = "10.1007/JHEP01(2016)184",
    journal = "JHEP",
    volume = "01",
    pages = "184",
    year = "2016"
}

@article{Haehl:2014zda,
    author = "Haehl, Felix M. and Loganayagam, R. and Rangamani, Mukund",
    title = "{The eightfold way to dissipation}",
    eprint = "1412.1090",
    archivePrefix = "arXiv",
    primaryClass = "hep-th",
    reportNumber = "DCPT-14-65",
    doi = "10.1103/PhysRevLett.114.201601",
    journal = "Phys. Rev. Lett.",
    volume = "114",
    pages = "201601",
    year = "2015"
}

@article{Jensen:2012kj,
    author = "Jensen, Kristan and Loganayagam, R. and Yarom, Amos",
    title = "{Thermodynamics, gravitational anomalies and cones}",
    eprint = "1207.5824",
    archivePrefix = "arXiv",
    primaryClass = "hep-th",
    doi = "10.1007/JHEP02(2013)088",
    journal = "JHEP",
    volume = "02",
    pages = "088",
    year = "2013"
}

@article{Jensen:2013kka,
    author = "Jensen, Kristan and Loganayagam, R. and Yarom, Amos",
    title = "{Anomaly inflow and thermal equilibrium}",
    eprint = "1310.7024",
    archivePrefix = "arXiv",
    primaryClass = "hep-th",
    reportNumber = "YITP-SB-35, YITP-SB-13-35",
    doi = "10.1007/JHEP05(2014)134",
    journal = "JHEP",
    volume = "05",
    pages = "134",
    year = "2014"
}

@article{Bhattacharyya:2012xi,
    author = "Bhattacharyya, Sayantani and Jain, Sachin and Minwalla, Shiraz and Sharma, Tarun",
    title = "{Constraints on Superfluid Hydrodynamics from Equilibrium Partition Functions}",
    eprint = "1206.6106",
    archivePrefix = "arXiv",
    primaryClass = "hep-th",
    reportNumber = "TFR-TH-12-26",
    doi = "10.1007/JHEP01(2013)040",
    journal = "JHEP",
    volume = "01",
    pages = "040",
    year = "2013"
}

@article{Bhattacharya:2011tra,
    author = "Bhattacharya, Jyotirmoy and Bhattacharyya, Sayantani and Minwalla, Shiraz and Yarom, Amos",
    title = "{A Theory of first order dissipative superfluid dynamics}",
    eprint = "1105.3733",
    archivePrefix = "arXiv",
    primaryClass = "hep-th",
    reportNumber = "PUPT-2375",
    doi = "10.1007/JHEP05(2014)147",
    journal = "JHEP",
    volume = "05",
    pages = "147",
    year = "2014"
}

@article{Bhattacharyya:2012nq,
    author = "Bhattacharyya, Sayantani",
    title = "{Constraints on the second order transport coefficients of an uncharged fluid}",
    eprint = "1201.4654",
    archivePrefix = "arXiv",
    primaryClass = "hep-th",
    doi = "10.1007/JHEP07(2012)104",
    journal = "JHEP",
    volume = "07",
    pages = "104",
    year = "2012"
}

@article{Bhattacharya:2012zx,
    author = "Bhattacharya, Jyotirmoy and Bhattacharyya, Sayantani and Rangamani, Mukund",
    title = "{Non-dissipative hydrodynamics: Effective actions versus entropy current}",
    eprint = "1211.1020",
    archivePrefix = "arXiv",
    primaryClass = "hep-th",
    reportNumber = "IPMU12-0196, DCPT-12-43",
    doi = "10.1007/JHEP02(2013)153",
    journal = "JHEP",
    volume = "02",
    pages = "153",
    year = "2013"
}

@article{Bhattacharyya:2013lha,
    author = "Bhattacharyya, Sayantani",
    title = "{Entropy current and equilibrium partition function in fluid dynamics}",
    eprint = "1312.0220",
    archivePrefix = "arXiv",
    primaryClass = "hep-th",
    doi = "10.1007/JHEP08(2014)165",
    journal = "JHEP",
    volume = "08",
    pages = "165",
    year = "2014"
}

@article{Bhattacharyya:2014bha,
    author = "Bhattacharyya, Sayantani",
    title = "{Entropy Current from Partition Function: One Example}",
    eprint = "1403.7639",
    archivePrefix = "arXiv",
    primaryClass = "hep-th",
    doi = "10.1007/JHEP07(2014)139",
    journal = "JHEP",
    volume = "07",
    pages = "139",
    year = "2014"
}

@article{Crossley:2015evo,
    author = "Crossley, Michael and Glorioso, Paolo and Liu, Hong",
    title = "{Effective field theory of dissipative fluids}",
    eprint = "1511.03646",
    archivePrefix = "arXiv",
    primaryClass = "hep-th",
    reportNumber = "MIT-CTP-4734",
    doi = "10.1007/JHEP09(2017)095",
    journal = "JHEP",
    volume = "09",
    pages = "095",
    year = "2017"
}

@article{Glorioso:2017fpd,
    author = "Glorioso, Paolo and Crossley, Michael and Liu, Hong",
    title = "{Effective field theory of dissipative fluids (II): classical limit, dynamical KMS symmetry and entropy current}",
    eprint = "1701.07817",
    archivePrefix = "arXiv",
    primaryClass = "hep-th",
    reportNumber = "MIT-CTP-4860, EFI-17-2",
    doi = "10.1007/JHEP09(2017)096",
    journal = "JHEP",
    volume = "09",
    pages = "096",
    year = "2017"
}

@article{Crossley:2015tka,
    author = "Crossley, Michael and Glorioso, Paolo and Liu, Hong and Wang, Yifan",
    title = "{Off-shell hydrodynamics from holography}",
    eprint = "1504.07611",
    archivePrefix = "arXiv",
    primaryClass = "hep-th",
    reportNumber = "MIT-CTP-4668",
    doi = "10.1007/JHEP02(2016)124",
    journal = "JHEP",
    volume = "02",
    pages = "124",
    year = "2016"
}

@article{Dandekar:2017aiv,
    author = "Dandekar, Yogesh and Kundu, Suman and Mazumdar, Subhajit and Minwalla, Shiraz and Mishra, Amiya and Saha, Arunabha",
    title = "{An Action for and Hydrodynamics from the improved Large D membrane}",
    eprint = "1712.09400",
    archivePrefix = "arXiv",
    primaryClass = "hep-th",
    doi = "10.1007/JHEP09(2018)137",
    journal = "JHEP",
    volume = "09",
    pages = "137",
    year = "2018"
}

@article{Bhattacharyya:2017hpj,
    author = "Bhattacharyya, Sayantani and Biswas, Parthajit and Chakrabarty, Bidisha and Dandekar, Yogesh and Dinda, Anirban",
    title = "{The large D black hole dynamics in AdS/dS backgrounds}",
    eprint = "1704.06076",
    archivePrefix = "arXiv",
    primaryClass = "hep-th",
    doi = "10.1007/JHEP10(2018)033",
    journal = "JHEP",
    volume = "10",
    pages = "033",
    year = "2018"
}

@article{Bhattacharyya:2018szu,
    author = "Bhattacharyya, Sayantani and Biswas, Parthajit and Dandekar, Yogesh",
    title = "{Black holes in presence of cosmological constant: second order in $ \frac{1}{D} $}",
    eprint = "1805.00284",
    archivePrefix = "arXiv",
    primaryClass = "hep-th",
    doi = "10.1007/JHEP10(2018)171",
    journal = "JHEP",
    volume = "10",
    pages = "171",
    year = "2018"
}

@article{Bhattacharyya:2019mbz,
    author = "Bhattacharyya, Sayantani and Biswas, Parthajit and Dinda, Anirban and Patra, Milan",
    title = "{Fluid-gravity and membrane-gravity dualities - Comparison at subleading orders}",
    eprint = "1902.00854",
    archivePrefix = "arXiv",
    primaryClass = "hep-th",
    doi = "10.1007/JHEP05(2019)054",
    journal = "JHEP",
    volume = "05",
    pages = "054",
    year = "2019"
}

@article{Cuomo:2024fuy,
    author = "Cuomo, Gabriel and Rastelli, Leonardo and Sharon, Adar",
    title = "{Moduli spaces in CFT: large charge operators}",
    eprint = "2406.19441",
    archivePrefix = "arXiv",
    primaryClass = "hep-th",
    doi = "10.1007/JHEP09(2024)185",
    journal = "JHEP",
    volume = "09",
    pages = "185",
    year = "2024"
}

@article{Yamada:2006rx,
    author = "Yamada, Daiske and Yaffe, Laurence G.",
    title = "{Phase diagram of N=4 super-Yang-Mills theory with R-symmetry chemical potentials}",
    eprint = "hep-th/0602074",
    archivePrefix = "arXiv",
    doi = "10.1088/1126-6708/2006/09/027",
    journal = "JHEP",
    volume = "09",
    pages = "027",
    year = "2006"
}

@article{Cordova:2017dhq,
    author = "Cordova, Clay and Diab, Kenan",
    title = "{Universal Bounds on Operator Dimensions from the Average Null Energy Condition}",
    eprint = "1712.01089",
    archivePrefix = "arXiv",
    primaryClass = "hep-th",
    doi = "10.1007/JHEP02(2018)131",
    journal = "JHEP",
    volume = "02",
    pages = "131",
    year = "2018"
}

@article{Loganayagam:2012zg,
    author = "Loganayagam, R.",
    title = "{Anomalies and the Helicity of the Thermal State}",
    eprint = "1211.3850",
    archivePrefix = "arXiv",
    primaryClass = "hep-th",
    doi = "10.1007/JHEP11(2013)205",
    journal = "JHEP",
    volume = "11",
    pages = "205",
    year = "2013"
}

@article{Bhattacharyya:2015dva,
    author = "Bhattacharyya, Sayantani and De, Anandita and Minwalla, Shiraz and Mohan, Ravi and Saha, Arunabha",
    title = "{A membrane paradigm at large D}",
    eprint = "1504.06613",
    archivePrefix = "arXiv",
    primaryClass = "hep-th",
    reportNumber = "TIFR-TH-15-11",
    doi = "10.1007/JHEP04(2016)076",
    journal = "JHEP",
    volume = "04",
    pages = "076",
    year = "2016"
}

@article{Bhattacharyya:2015fdk,
    author = "Bhattacharyya, Sayantani and Mandlik, Mangesh and Minwalla, Shiraz and Thakur, Somyadip",
    title = "{A Charged Membrane Paradigm at Large D}",
    eprint = "1511.03432",
    archivePrefix = "arXiv",
    primaryClass = "hep-th",
    doi = "10.1007/JHEP04(2016)128",
    journal = "JHEP",
    volume = "04",
    pages = "128",
    year = "2016"
}

@article{Dandekar:2016fvw,
    author = "Dandekar, Yogesh and De, Anandita and Mazumdar, Subhajit and Minwalla, Shiraz and Saha, Arunabha",
    title = "{The large D black hole Membrane Paradigm at first subleading order}",
    eprint = "1607.06475",
    archivePrefix = "arXiv",
    primaryClass = "hep-th",
    reportNumber = "TIFR-TH-16-25",
    doi = "10.1007/JHEP12(2016)113",
    journal = "JHEP",
    volume = "12",
    pages = "113",
    year = "2016"
}

@article{Hartman:2015lfa,
    author = "Hartman, Thomas and Jain, Sachin and Kundu, Sandipan",
    title = "{Causality Constraints in Conformal Field Theory}",
    eprint = "1509.00014",
    archivePrefix = "arXiv",
    primaryClass = "hep-th",
    doi = "10.1007/JHEP05(2016)099",
    journal = "JHEP",
    volume = "05",
    pages = "099",
    year = "2016"
}

@article{Maldacena:2015waa,
    author = "Maldacena, Juan and Shenker, Stephen H. and Stanford, Douglas",
    title = "{A bound on chaos}",
    eprint = "1503.01409",
    archivePrefix = "arXiv",
    primaryClass = "hep-th",
    doi = "10.1007/JHEP08(2016)106",
    journal = "JHEP",
    volume = "08",
    pages = "106",
    year = "2016"
}

@article{Emparan:2015hwa,
    author = "Emparan, Roberto and Shiromizu, Tetsuya and Suzuki, Ryotaku and Tanabe, Kentaro and Tanaka, Takahiro",
    title = "{Effective theory of Black Holes in the 1/D expansion}",
    eprint = "1504.06489",
    archivePrefix = "arXiv",
    primaryClass = "hep-th",
    reportNumber = "KEK-TH-1810, AP-GR-122, OCU-PHYS-422, KUNS-2557, YITP-15-33, KEK-TH-1810, AP-GR-122, OCU-PHYS-422, KUNS 2557, YITP-15-33",
    doi = "10.1007/JHEP06(2015)159",
    journal = "JHEP",
    volume = "06",
    pages = "159",
    year = "2015"
}

@article{Bhattacharyya:2016nhn,
    author = "Bhattacharyya, Sayantani and Mandal, Anup Kumar and Mandlik, Mangesh and Mehta, Umang and Minwalla, Shiraz and Sharma, Utkarsh and Thakur, Somyadip",
    title = "{Currents and Radiation from the large $D$ Black Hole Membrane}",
    eprint = "1611.09310",
    archivePrefix = "arXiv",
    primaryClass = "hep-th",
    doi = "10.1007/JHEP05(2017)098",
    journal = "JHEP",
    volume = "05",
    pages = "098",
    year = "2017"
}

@article{Chowdhury:2019kaq,
    author = "Chowdhury, Subham Dutta and Gadde, Abhijit and Gopalka, Tushar and Halder, Indranil and Janagal, Lavneet and Minwalla, Shiraz",
    title = "{Classifying and constraining local four photon and four graviton S-matrices}",
    eprint = "1910.14392",
    archivePrefix = "arXiv",
    primaryClass = "hep-th",
    doi = "10.1007/JHEP02(2020)114",
    journal = "JHEP",
    volume = "02",
    pages = "114",
    year = "2020"
}

@article{Chandorkar:2021viw,
    author = "Chandorkar, Deeksha and Chowdhury, Subham Dutta and Kundu, Suman and Minwalla, Shiraz",
    title = "{Bounds on Regge growth of flat space scattering from bounds on chaos}",
    eprint = "2102.03122",
    archivePrefix = "arXiv",
    primaryClass = "hep-th",
    doi = "10.1007/JHEP05(2021)143",
    journal = "JHEP",
    volume = "05",
    pages = "143",
    year = "2021"
}

@article{Hofman:2008ar,
    author = "Hofman, Diego M. and Maldacena, Juan",
    title = "{Conformal collider physics: Energy and charge correlations}",
    eprint = "0803.1467",
    archivePrefix = "arXiv",
    primaryClass = "hep-th",
    doi = "10.1088/1126-6708/2008/05/012",
    journal = "JHEP",
    volume = "05",
    pages = "012",
    year = "2008"
}

\end{document}